\RequirePackage{fix-cm}
\documentclass[smallextended]{svjour3}       % onecolumn (second format)
\smartqed  % flush right qed marks, e.g. at end of proof
\usepackage{graphicx}
\usepackage{natbib}
\usepackage{subfigure}
\usepackage{longtable}
\usepackage{color} 
\usepackage{xcolor}
\usepackage{ulem}
\begin{document}

\title{Pulsating white dwarfs: new insights%\thanks{Grants or other notes
}
%\subtitle{New insights}

\titlerunning{Pulsating white dwarfs}        % if too long for running head

\author{Alejandro H. C\'orsico,
       Leandro G. Althaus,
       Marcelo M. Miller Bertolami,
       S. O. Kepler}

\authorrunning{C\'orsico et al.} % if too long for running head

\institute{Alejandro H. C\'orsico,
           Leandro G. Althaus,
           Marcelo M. Miller Bertolami\at
           Facultad de Ciencias Astron\'omicas y Geof\'{\i}sicas,
           Universidad Nacional de La Plata,
           Paseo del  Bosque s/n,
           (1900) La Plata, Argentina\\
           Instituto de Astrof\'{\i}sica La Plata, IALP, CONICET-UNLP \\
           \email{acorsico, althaus, mmiller@fcaglp.unlp.edu.ar} \\
           \and
           S. O. Kepler \at
           Departamento de Astronomia, Universidade Federal do Rio
           Grande do Sul, Av. Bento Goncalves 9500, Porto Alegre 91501-970,
           RS, Brazil\\
           \email{kepler@if.ufrgs.br}}

\date{Received: \today / Accepted: \today}

\maketitle

\def\lesssim{\mathrel{\hbox{\rlap{\hbox{\lower4pt\hbox{$\sim$}}}\hbox{$<$}}}}
\def\gtrsim{\mathrel{\hbox{\rlap{\hbox{\lower4pt\hbox{$\sim$}}}\hbox{$>$}}}}
\def\teff{$T\rm_{\rm eff}$}
\def\kms{$\mathrm {km\, s}^{-1}$}

\begin{abstract}
Stars are  extremely important astronomical objects that constitute
the pillars  on which  the Universe is  built, and  as such, their
study has  gained increasing  interest over  the years.  White dwarf
stars are not the exception. Indeed, these stars constitute the  final
evolutionary stage  for  more  than  95  per cent  of  all stars. The
Galactic population of white dwarfs conveys a wealth of  information
about several  fundamental  issues  and are  of  vital importance to
study the  structure, evolution and chemical enrichment of  our Galaxy
and its components --- including the  star formation history  of the
Milky  Way. Several  important  studies have emphasized the
advantage of using  white dwarfs as  reliable clocks to  date a
variety of stellar populations in  the solar neighborhood and in the
nearest stellar clusters, including the thin and  thick disks,
the  Galactic spheroid and the system  of globular and open   clusters. In addition, white dwarfs are tracers of the  evolution of planetary systems  along several phases of stellar  evolution.  Not less
relevant than  these applications, the study of matter at  high
densities  has benefited  from our  detailed knowledge about
evolutionary and observational properties of white dwarfs. In this
sense, white dwarfs are used as laboratories for astro-particle
physics,  being their interest focused on  physics beyond the standard
model,  that is, neutrino physics, axion physics  and also radiation
from ``extra dimensions'', and even crystallization.

The last decade has witnessed a great progress in the study of white
dwarfs. In particular, a wealth of information of these stars from
different surveys has allowed us to make meaningful comparison of
evolutionary models with observations. While some information like
surface chemical composition, temperature and gravity of  isolated white dwarfs
can be inferred from spectroscopy, and the total mass and radius can be derived as well when they are in  binaries, the internal structure of these
compact stars  can be unveiled only by means of asteroseismology, an
approach based on the  comparison between the observed pulsation
periods of variable stars and  the periods predicted by appropriate 
theoretical models. The asteroseismological  techniques allow us to infer details of the internal chemical stratification,  the total mass, and even the stellar rotation profile. 

In this review, we first revise the evolutionary channels currently
accepted that lead to the formation of white-dwarf stars, and then, 
we give a detailed account of  the different sub-types of pulsating white
dwarfs known so far,  emphasizing the recent observational and theoretical
advancements in the study of these  fascinating variable stars.


%PACS{PACS code1 \and PACS code2 \and more}
%\subclass{MSC code1 \and MSC code2 \and more}
\end{abstract}

\section{Introduction}
\label{intro}

White dwarf (WD) stars represent the final evolutionary stage of the
majority of stars. Indeed, all stars with stellar masses lower than
$\sim 10 - 11 M_{\odot}$, depending on their initial metallicity
\citep[e.g.][]{2015ApJ...810...34W}, will end their lives as WDs,
earth-sized electron-degenerate stellar configurations. As such, they
play a unique and fundamental role for our understanding of the
formation and evolution of stars, evolution of planetary systems, and
the history of our Galaxy itself. The study of WDs results thus of
central relevance for a vast variety of topics of modern astrophysics,
ranging from the final fate of planetary systems to the
characterization of dark matter
\citep{2016NewAR..71....9F,2018PhyS...93d4002S}. The present
population of WDs keeps a detailed record  of the early star formation
in the Galaxy. Therefore, accurate WD  luminosity functions can be
used to infer the age, structure and evolution of the Galactic disk
and the nearest open and globular clusters
\citep{2001PASP..113..409F,2009ApJ...697..965B,2010Natur.465..194G,2015MNRAS.448.1779B,2013MNRAS.433..243C,2016MNRAS.456.3729C,2016NewAR..72....1G,2017ApJ...837..162K}\footnote{\cite{2018IAUS..334...11C} describes other methods that use WDs to infer ages of stellar populations.}.
In a different context, the host stars of most planetary systems,
including our Sun, will evolve into WDs, and nowadays observational
evidence convincingly demonstrates that numerous WDs foster remnants
of planetary systems --- even planetary matter, shedding light on
the chemical composition of extra-solar  planets
\citep{2012MNRAS.424..333G,2018MNRAS.477...93H}. Also, WDs are
found in binary systems, thus offering a test bed to explore complex
stellar interactions amongst  stars, including WDs exploding as type
Ia supernovae \citep{2014ARA&A..52..107M}. In addition, WDs
can be used as cosmic laboratories  of extreme physics, ranging from atomic and
molecular physics in strong magnetic  fields, and high-density plasmas
and even solid-state physics \citep[through crystallization;][]{2009ApJ...693L...6W,2019Natur.565..202T},  
to exotic physics, like constraining
the axion mass and the possible variation of the gravitational constant
\citep{1992ApJ...392L..23I,2012MNRAS.424.2792C,2013JCAP...06..032C}, and 
also variations of the fine-structure constant \citep{2019MNRAS.485.5050H}. Last
but not least, fundamental properties of WDs, either individually or
collectively, like the mass  distribution, core chemical composition,
and cooling times  are key to place constraints on the stellar
evolution theory, including third dredge up and mass loss on the 
Asymptotic Giant Branch (AGB),
the efficiency of extra-mixing during core helium burning, and nuclear
reaction rates
\citep{2002ApJ...567..643K,2009ApJ...692.1013S,2016ApJ...823...46F}. Excellent
review papers describing the evolutionary properties of  WDs are those
of \cite{1990ARA&A..28..139D}, \cite{1990RPPh...53..837K}, \cite{2001PASP..113..409F}, \cite{2002A&ARv..11...33K},
\cite{2003ARA&A..41..465H}, \cite{2004PhR...399....1H}, \cite{2008PASP..120.1043F}, \cite{2008ARA&A..46..157W}, and \cite{2010A&ARv..18..471A}.

Like many stars, when relevant layers are required to 
transport energy through high
opacity, WDs exhibit periodic brightness variations which are due to
global pulsations associated to their normal modes
\citep{1958HDP....51..353L,1980tsp..book.....C,1989nos..book.....U}. The existence of
these intrinsic luminosity variations implies that, in principle, we
have available a unique window to ``look'' inside these stars,
otherwise  inaccessible  by other means. The analysis of pulsations of
a variety of  stars has led, in the last decades, to the development
of novel techniques which, taken together, are known today as {\it
  asteroseismology}  \citep{2010aste.book.....A,2010csp..book.....B,
  2015pust.book.....C}. In principle, a key factor for a 
successful asteroseismological analysis is the number of  periods visible in the star, 
i.e., the more periods a pulsating star exhibits, the stronger the constraints that 
asteroseismology could place. It must be emphasized, however, that the crucial point for a successful asteroseismological exercise is not the absolute number of modes itself, 
but the diversity in the eigenfunctions of these modes. Putting it in 
other terms: the information provided by a few periods corresponding 
to low-order modes is generally richer than the information that 
provides a larger set of modes with periods in the asymptotic regime 
(high-order modes). 

 Most WD stars go through at least one stage of pulsational instability
during their lives\footnote{ An exception are the high-field magnetic WDs, that represents a significant fraction of the local population of WDs, and for which there is no observational evidence of variability due to pulsations.}, which turns them into multi-periodic pulsating
variable stars and therefore, it is possible to analyze their internal
structure employing the tools of asteroseismology
\citep{1988IAUS..123..305W, 1995BaltA...4..166K,
  2008PASP..120.1043F,2008ARA&A..46..157W,2010A&ARv..18..471A, 
  2013EAS....63..175V, 2017EPJWC.15201011K,2018tjae.conf...13C}. 
  Pulsations in WDs manifest  themselves as
periodic brightness variations  in the optical and also  in the
ultraviolet (UV)  regions of  the electromagnetic spectrum. 
These variations are generated by global nonradial $g$(gravity)-mode
pulsations which are a subclass of spheroidal
modes\footnote{Spheroidal modes  are characterized by $(\vec{\nabla}
  \times \vec{\xi})_r= 0$ and $\sigma \neq 0$, where $\vec{\xi}$ is
  the  Lagrangian displacement and $\sigma$ the pulsation frequency
  \citep{1989nos..book.....U}.} whose main restoring force is gravity
through buoyancy.  The pulsations are characterized by peak-to-peak
amplitudes  between 0.1 mmag and 0.4 mag in typical optical light
curves. 

\begin{figure}
\includegraphics[width=0.95\textwidth]{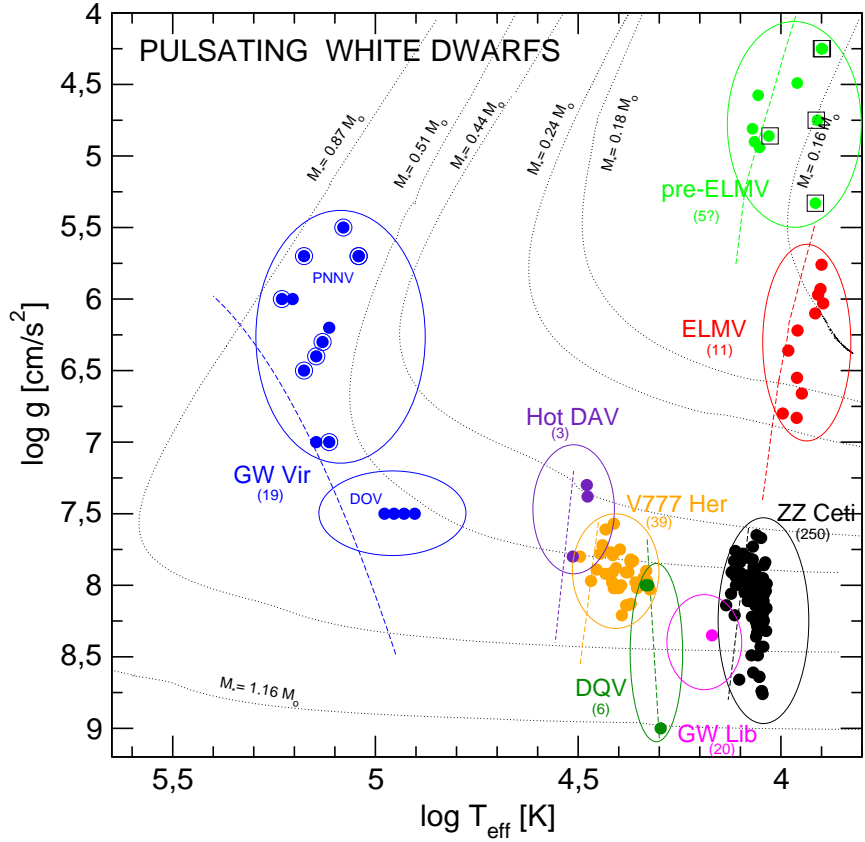}
\caption{Location of the different classes of confirmed and tentative pulsating  WD and
  pre-WD stars (circles of different colors) in the $\log T_{\rm eff}-
  \log g$ diagram. This figure is an update of Fig. 15 of
  \cite{2010A&ARv..18..471A}.  Stars emphasized with squares
  surrounding the light  green circles can be identified as pre-ELMV
  stars as well as SX Phe  and/or $\delta$ Scuti stars. GW Vir stars
  indicated with blue circles surrounded by  blue circumferences are
  PNNVs.  In the case of GW Lib stars, only the location of the prototypical 
  object, GW Librae, has been included (magenta dot).
  Two post-VLTP (Very Late Thermal Pulse) evolutionary tracks
  for H-deficient WDs  \citep[$0.51$ and $0.87
    M_{\odot}$;][]{2006A&A...454..845M}, four evolutionary tracks  of
  low-mass He-core H-rich WDs \citep[$0.16$, $0.18$, $0.24$, and
    $0.44 M_{\odot}$;][]{2013A&A...557A..19A}, and one evolutionary
  track for  ultra-massive H-rich WDs \citep[$1.16
    M_{\odot}$;][]{2019A&A...625A..87C} are  plotted for
  reference. Dashed lines indicate the theoretical blue edge of the
  instability domains for the different classes of pulsating WDs.}
\label{fig01}
\end{figure}

An increasing  number of categories of WD pulsators has been
discovered since 1968.   At present, there are  six  classes
of confirmed pulsating WDs known  (see Fig. \ref{fig01}).  They are:

\begin{itemize}

\item The variables ZZ Ceti
or DAVs ---pulsating WDs  with almost pure H atmospheres--- which are the
most numerous ones \citep{2017EPJWC.15201011K}.  They are located at
low effective temperatures and high gravities  ($10\,400$ K $\lesssim
T_{\rm eff} \lesssim  12\,400$ K and $7.5 \lesssim \log g \lesssim
9.1$). It was the first class of pulsating WDs to be detected
\citep{1968ApJ...153..151L}. Many DA WDs and so, some DAV stars, have an 
atmosphere polluted by the accretion of heavy elements from a debris disk \citep{2014A&A...566A..34K,2017A&A...601A..13W}.

\item The GW Lib stars, which are accreting pulsating WDs in cataclysmic variables
($10\,500$  K $\lesssim T_{\rm  eff} \lesssim  16\,000$ K and $8.3 \lesssim \log g \lesssim
8.7$). They have H-dominated atmospheres, but due to accretion from a 
solar composition or He-enriched low-mass companion, they can have an 
enhanced He abundance \citep{2010ApJ...710...64S}. The first object of this kind, 
GW Librae, was discovered by \citet{1998IAUS..185..321W}.

\item The variables V777 Her  or DBVs (atmospheres almost pure in
He, $22\,400$ K $\lesssim T_{\rm eff} \lesssim 32\,000$ K and $7.5
\lesssim \log  g \lesssim  8.3$),  the existence  of  which was
theoretically  predicted  by  \cite{1982ApJ...252L..65W} before their
discovery \citep{1982ApJ...262L..11W}. 

\item The pulsating PG1159 stars or GW Vir variable
stars, after the prototype of the class, PG 1159$-$035
\citep{1979wdvd.coll..377M}. This is the hottest known
class of pulsating WDs and pre-WDs ($80\,000$ K $\lesssim T_{\rm eff}
\lesssim 180\,000$ K  and $5.5 \lesssim \log g \lesssim  7.5$), 
constituted by variable H-deficient, C-, O- and He-rich atmosphere WD
and pre-WD stars. This  group  includes objects  that  are
still  surrounded  by  a  nebula ---the variable planetary nebula
nuclei, designed as PNNVs--- and stars that lack a nebula ---called
DOVs. 

\item The ELMVs (Extremely Low-Mass WDs variable, $7\,800$ K $\lesssim
T_{\rm eff} \lesssim 10\,000$ K  and $6 \lesssim \log g \lesssim
6.8$, pure H atmospheres), discovered by \cite{2012ApJ...750L..28H}.

\item The pre-ELMVs
($8\,000$ K $\lesssim  T_{\rm eff} \lesssim 13\,000$ K  and $4
\lesssim \log g \lesssim  5$), the probable precursors of ELMVs
\citep{2013Natur.498..463M}.  

\end{itemize}

Also, there are two additional classes of tentative WD pulsators, that
need confirmation:

\begin{itemize}

\item The hot DQ  variable WDs, or  DQVs ($19\,000$  K $\lesssim
T_{\rm  eff} \lesssim  22\,000$ K and $8 \lesssim \log  g \lesssim
9$). They are  WDs  with  C- and He-rich  atmospheres.  The
prototype  of  this class, SDSS  J142625.71+575218.3,   was discovered
by \cite{2008ApJ...678L..51M}. We caution that the variability  of
some of these objects could be explained by other  effects than 
pulsations \citep{2013ApJ...769..123W}. 

\item The so-called ``hot DAVs'' \citep[$T_{\rm eff} \sim 30\,000$ K, $7.3
  \lesssim \log  g \lesssim
  7.8$;][]{2008MNRAS.389.1771K,2013MNRAS.432.1632K}, whose existence
was anticipated by the theoretical calculations of
\cite{2005EAS....17..143S,2007AIPC..948...35S}.  The pulsating nature
of the variability of these stars needs to be confirmed with further observations.

\end{itemize}

In Table \ref{tabla-properties}, we  present  a compact summary  of
the   main pulsation characteristics  of  each class  of pulsating WD
stars. This is an update of Table 2 of \cite{2010A&ARv..18..471A}
\citep[see, also, Table 13.1 of][]{2015pust.book.....C}. The
first  column corresponds to the name of each  class, the second one
gives the  year  of discovery  of the first member  of the group and
the number of known members (in parenthesis),  columns 3 to 8
correspond to  the effective temperature, the logarithm of the surface
gravity, the range of observed periods, the range of the rates of
period change, the interval of amplitudes, and  the surface chemical
composition of each class, respectively.

\begin{table*}
\scriptsize
\caption{Properties of the different families of variable pulsating WDs, sorted  by decreasing effective temperature. The tentative classes of pulsators are labeled with a question mark in parentheses.}
\begin{tabular}{llcccccc}
\hline
\noalign{\smallskip}
Class    &   Year      &  $T_{\rm eff}$      & $\log g$     &  Periods & $\dot{P}\equiv dP/dt$ & Amplitudes &   Main surface  \\
        &    of disc. (\#)     &   [$\times$ 1000 K]  & [C.G.S.]          & [s]      &  [s/s]                & [mag]      & composition\\
\noalign{\smallskip}
\hline
\noalign{\smallskip}
GW Vir (PNNV) & 1984 (10)  & $100-180$  & $5.5-7$  & $420-6000$ &  $10^{-10}-10^{-11}$ & $0.01-0.15$ & He, C, O \\
GW Vir (DOV)  & 1979 (9)   & $80-100$   & $7.3-7.7$  & $300-2600$ &  $10^{-10}-10^{-12}$ & $0.02-0.1$  & He, C, O \\
&&&&&&&\\
Hot DAV  (?)      & 2013 (3)   & $30-32.6$  & $7.3-7.8$  & $160-705$  &  $10^{-13}$$^{\ddag}$  & $0.001-0.015$ & H        \\ 
&&&&&&&\\
V777 Her (DBV) & 1982 (39) & $22.4-32$  & $7.5-8.3$  &  $120-1080$ & $10^{-13}-10^{-14}$  & $0.05-0.3$  & He (H)   \\
&&&&&&&\\
DQV  (?)       &  2008 (6)  & $19-22$    & $8-9$  & $240-1100$ &  $10^{-14}$$^{\ddag}$ & $0.005-0.015$& He, C   \\
&&&&&&&\\
GW Lib & 1998 (20) & $10.5-16$ & $8.35-8.7$ & $100-1900$ & $10^{-8}$$^{\ddag}$ & $0.007-0.07$ &  H, He \\
&&&&&&&\\
ZZ Cet (DAV) & 1968 (250) &  $10.4-12.4$   & $7.5-9.1$  & $100-1400$ & $(1-4) \times 10^{-15}$          & $0.01-0.3$  & H     \\
&&&&&&&\\
pre-ELMV      & 2013 (5)   & $8-13$     & $4-5$  &  $300-5000$ & $10^{-13}$$^{\ddag}$                & $0.001-0.05$ & He, H   \\
&&&&&&&\\
ELMV          & 2012 (11)  & $7.8-10$   & $6-6.8$  & $100-6300$ & $10^{-14}$$^{\ddag}$            & $0.002-0.044$  &  H    \\
\noalign{\smallskip}
\hline
\end{tabular}
{\footnotesize $^{\ddag}$ Estimated from theoretical pulsation models.} 
\label{tabla-properties}
\end{table*}

%
% The rate of period change of GW Lib stars is estimated according to 
% Townsley et al. (2004): d\nu/dt= -10**{-12} Hz/s
%

Regarding the driving mechanisms involved in the excitation of the
pulsations in WDs, there is a strong consensus that they correspond to
thermal processes that give place to {\it self-excited}
pulsations\footnote{This, at variance with the {\it forced}
  pulsations such as stochastic excitation by turbulent convection,
  in which the modes, that are intrinsically stable, are actually
  excited by convective motions.}. The more relevant mechanisms are
the $\kappa-\gamma$ mechanism, that involves an increase in the
opacity of the material due to the partial ionization of the dominant
chemical species
\citep{1981A&A...102..375D,1982ApJ...252L..65W,2005A&A...438.1013G},
and the ``convective driving'' mechanism
\citep{1991MNRAS.251..673B,1999ApJ...511..904G}  that acts efficiently
when the outer convection zone deepens. In the case of GW Vir stars,
which lack a surface convection zone due to their very-high
effective temperatures, only the $\kappa-\gamma$ mechanism appears to
be the one responsible for pulsations. Finally,  the $\varepsilon$
mechanism due to stable nuclear burning could be able to
excite short-period $g$ modes in GW Vir stars
\citep{1986ApJ...306L..41K},  ELMVs \citep{2014ApJ...793L..17C}, and
ZZ Ceti stars evolved from  low-metallicity progenitors
\citep{2016A&A...595A..45C}. Also,  $g$-mode excitation by the
$\varepsilon$  mechanism has been predicted for  very hot H-rich
pre-WDs \citep{2014PASJ...66...76M}. In spite of the fact that the
origin  of pulsations in WDs is known to a large extent, little is
known about the agent  that causes the red edge of the instability
strips\footnote{Although see \cite{2012ApJ...755..128Q} for the case of 
GW Vir stars and \cite{2018ApJ...863...82L}  for the  case of ZZ Ceti stars.}, 
  neither  why many pulsating WDs
---particularly DAVs--- exhibit so few periods. Fortunately,  this
incomplete knowledge of the physics of mode excitation and damping of WD
pulsations does not  prevent us from advancing in asteroseismological
studies based on adiabatic calculations,  in which the physical agent
that gives rise to the pulsations is not relevant\footnote{The sound
  of the bells (their eigenfrequencies) does not depend  on {\it how}
  the bells are rung \citep{1992RvMA....5..125B}.},  but rather the
value of the periods themselves, which depends sensitively on the
internal structure of WDs.

The small number of detected periods in WDs does not allow for the
application of any inversion technique, as in the case of the Sun
(through {\it helioseismology}), for which nearly the whole structure
of the star can be determined from the  eigenfunctions\footnote{Note, however, that
in many cases the studies of solar-type pulsators seem to be limited 
to using the frequency separations and frequency maximum to derive the astrophysical parameters of the stars, using the so-called ``scaling relations'' \citep{2017ApJ...835..172L}.}. Essentially,
WD asteroseismology consists in the comparison of the individual
periods and  period spacings observed in variable WDs with
adiabatic period and period spacings computed for a set of pulsation
models ---the so-called {\it forward method}. 
%The
%  forward method in asteroseismology consists in the comparison
%  between the  theoretical periods computed numerically from
%  stellar models (generally under the assumption of adiabaticity of the %pulsations), and the periods observed in the star under study,
%  with the aim of mimicking as best as possible the observed period
%  pattern. 
%  
Ideally, when the
differences between the theoretical and observed periods are small, an
asteroseismological solution ---the asteroseismological model--- can
be found. As the number of detected normal modes is usually small,
this solution is not unique, and therefore external constraints such
as the surface gravity and effective temperature of the star derived
through spectroscopy ---or distance from parallax--- have to be
considered to break the degeneracy of solutions. In the case of
success in obtaining an asteroseismological model, one automatically
has information about the internal structure, stellar mass, surface
gravity, effective temperature, luminosity, radius, etc. of the star
under study.  With the luminosity it is possible to infer the
seismological distance, which can be compared with the distance
derived through the trigonometric parallax. If it is not possible to
obtain an asteroseismological model, but if it is feasible to derive a
mean separation of the observed periods, it can be compared with the
theoretical one, allowing us to infer the stellar mass. This last
technique, however, is difficult to apply in many cases due to the
simultaneous dependence of the mean period spacing on the stellar mass, 
the effective temperature, and the thickness of the outer envelopes of 
WDs\footnote{However, the dependence of the period spacing on the thickness 
of the outer envelope of DA and DB WDs is generally weaker than 
 its dependence upon the effective temperature and the stellar mass 
\citep{1990ApJS...72..335T}.}. In
connection with this, the departures of the period separation from a
constant period spacing tell us about the steep variations in density 
caused by chemical transition regions of WDs. Another tool of WD asteroseismology 
is the analysis of the
splitting of the frequencies, which can give clues about the nature
and magnitude of rotation (angular velocity) and magnetic fields. In particular, 
\cite{1999ApJ...516..349K} were the first to explore the potential 
of the inversion methods employed in helioseismology to infer the internal 
rotation of GW Vir and V777 Her stars. Finally, there is the secular drift of
the pulsation periods. Although the measurement of this quantity is
extremely difficult,  the rate of period change   can give valuable
information about the core chemical composition of WDs and the cooling
rate. 

In this review, we will not focus on the details about the
asteroseismological tools applied to  pulsating WDs, which can be
found in \cite{2008PASP..120.1043F,2008ARA&A..46..157W,2010A&ARv..18..471A},
\cite{2017EPJWC.15201011K}, and \cite{2017ApJ...834..136G}. 
Instead, we will concentrate on the advancements made in the  study of 
pulsating WDs in the last decade.

\section{Evolutionary channels and uncertainties in progenitor evolution}
\label{sec:2}

More than 60 years after its humble beginnings
\citep{1958ses..book.....S}, stellar-evolution theory is nowadays a
well established and predictive theory \citep{2012sse..book.....K}. 
After more than half a century
of continuous development, its main predictions have been confirmed by
a variety of different observational tests. This is even more true in
the case of the low- and intermediate-mass stars, which are the
progenitors of WD stars, where rotation and magnetic fields do not
play a major role in determining the internal structure. As a
consequence, we can now use the predictions of the stellar-evolution theory
to learn about the
evolution of WD progenitors and make educated guesses about
their internal structure and composition.

\subsection{Single evolution}
\label{sec:2.1}
%acacacacac M3B 

In the simplest picture, WD stars are formed once winds remove most
of the H-rich  envelope below  the  critical  value  required  to
sustain a giant-like structure. For  most single low- and
intermediate-mass stars, this happens once the star finishes burning
He in the core, and evolves to the Thermally Pulsating (TP) Asymptotic
Giant Branch (AGB).  On the TP-AGB phase, stars undergo intense
radiation dust-driven winds as high as  $\dot{M}\sim
10^{-4}M_\odot$/yr, where most of the envelope can be removed in  less
than one million years, leaving a carbon-oxygen (CO) core 
WD \citep{2005ARA&A..43..435H, 2018A&ARv..26....1H}. While
most low- and  intermediate-mass stars will reach the TP-AGB phase,
progenitors with the lower masses  might lose enough mass on the first
ascent of the red giant branch (RGB) to populate the extreme horizontal  branch (sdB
stars) and directly evolve to the WD cooling track without reaching
the  AGB phase (the so-called ``AGB-Manqu\'e'' stars), see \cite{1990ApJ...364...35G} and references therein. Although a
CO-core WD is the most common end  state for single stars, other
possibilities exist. In fact, on the one hand, the heaviest
intermediate mass stars ($M_i\gtrsim 8M_\odot$) might reach
temperatures in the core  high enough to ignite C, enter the Super-AGB
phase and end up as oxygen/neon (ONe)-core WDs \citep{1996ApJ...460..489R, 2010A&A...512A..10S}\footnote{\cite{2010Sci...327..188G}  
discovered  two WDs  exposing
  dredged-up,   O-rich  core  material  that  could  have been
  produced in the interior  of a Super-AGB star. Recently,
  \cite{2016Sci...352...67K} identified a WD having an
  O-dominated atmosphere  with traces of Ne and Mg, that could be the
  bare core of a  Super-AGB star. Finally, another O- and Ne-rich WD
  but with a very low mass was discovered by
  \cite{2017Sci...357..680V}.}.  On the other hand, stars with low
initial masses ($M_i\lesssim 0.65 M_\odot$), low initial
metallicities and high initial He-contents are able to evolve away
from the main sequence  in timescales shorter than a Hubble time, and
might lose their H-rich envelopes  already on the first RGB, leading to the formation of  relatively massive He-core
WDs \citep[$M_f\gtrsim 0.4M_\odot$, see][and references therein]{2001ApJ...550L..65V,2004ApJ...612L..25N,2008ApJ...673L..29C,2009ApJ...699...40S,2013ApJ...769L..32B,2017A&A...597A..67A}.

Although the overall picture of the evolution of single WD progenitors
was well established many decades ago, some uncertainties remain
about the details of the pre-WD evolution. Among them, convective
boundary mixing remains the largest one. While extra-mixing in
H-burning cores during the main sequence is relatively well
constrained and calibrated
\citep{2004ApJ...612..168P,2012A&A...537A.146E}, the extent of
convective boundary mixing during the He-core burning
\citep{2011A&A...530A...3C,2015MNRAS.452..123C} and TP-AGB phases
(Wagstaff et al. 2019, submitted) is somewhat uncertain. The lack
of a complete understanding of convective boundary mixing, together
with the uncertainties in the intensity of winds during the TP-AGB
phase, lead to uncertainties in the Initial-Final Mass Relationship
(IFMR) of stellar evolution models (see \citealt{2009ApJ...692.1013S}
for a detailed study of the uncertainties). Among other things, this
implies that IFMRs cannot be reliably predicted by current stellar evolution
models, but instead semi-empirical IFMRs  must be used (together with other
observables) to calibrate macrophysics processes in stars
\citep{2016A&A...588A..25M}. In fact, the very existence of a tight
IFMR is not supported by semi-empirical determinations of the IFMR,
which give a  significant scatter in the mass of the WD for a given initial
mass
\citep{2009MNRAS.395.1795C,2009ApJ...692.1013S,2018ApJ...866...21C}.
In addition, as most semi-empirical determinations of the IFMR have
been performed for solar-like metallicities, we currently do not  know
how IFMRs depend on metallicity or He content.  Uncertainties in the
IFMRs and convective boundary mixing processes of the models impact
the chemical profiles of WD stars of a given mass (see later).

In the simple scenario discussed above, stellar evolution theory
usually predicts the formation of WDs with pure H atmospheres and with
a total H content of about $M_{\rm H} \sim 10^{-3}
-10^{-5}M_\odot$. About 80\% of the spectroscopically identified WDs
are characterized by H-rich atmospheres. The remaining $\sim 20$\% of
WDs are characterized by He-dominated atmospheres  (spectral types
PG1159, DO, DB, DQ, DZ, DC; see \citealt{2010A&ARv..18..471A}). In
addition, systematic spectroscopic and asteroseismological studies
of DA stars indicate that between 15\% and 20\% of DAs have thin
H-envelopes with $M_{\rm H} \lesssim 10^{-6} M_\odot$
\citep[see][]{2008ApJ...672.1144T,2009MNRAS.396.1709C,2012MNRAS.420.1462R}. Both
WDs with He-dominated  or thin H-envelopes cannot be explained by the
simple picture presented above and more complex evolutionary scenarios
have been developed to explain their existence. Besides the
binary-evolution channels to be discussed in the next section, some
single-evolution scenarios predict the formation of both He-dominated
atmospheres and thin H envelopes. 

\citet{1990ARA&A..28..139D} reviewed several evolutionary channels for
the formation of WDs with low H contents. In particular, thermal
pulses during the post-AGB phase lead in a  very natural way to the
formation of  WDs with low H contents. \cite{1984ApJ...277..333I}
showed that depending on the timing of the departure from the AGB
phase, a last thermal pulse can  happen during the post-AGB
evolution. If a late thermal pulse develops when the post-AGB star is
already entering the WD cooling track, then the H-rich envelope will
be ingested by the He-shell flash convective zone, where H will be
burnt in the hot interior of the star
\citep{1995LNP...443...48I,1999A&A...349L...5H}. This scenario  was
named a Very Late Thermal Pulse (VLTP) by \cite{2001Ap&SS.275....1B}.
Although numerical simulations never predict the burning of the
complete H content of the star, it has been argued that whatever the
traces of H that may be left by the VLTP, they will  very likely  be
peeled off by mass loss during  the subsequent  giant  phase
\citep{2006PASP..118..183W}. While this is a possibility, it has to be
noted that,  once the star is back on its giant phase, H is diluted in
the more massive convective envelope of the born again AGB star
($M_{\rm env}\sim 10^{-3} M_{\star}$).  The star would need to lose
all  that mass in order to get rid of its whole H content. If this is
the case, then this  scenario will produce a WD with a H-deficient
atmosphere.  It should be noted, however, that  the study of
\cite{2007MNRAS.380..763M} suggests that the amount of H burned might
depend on the total mass of H remaining in the star at the moment of
the VLTP, and VLTPs in low-mass stars might only burn a small fraction
of the total H content of the star.  Consequently, in this case a VLTP
will lead to the formation of DA WDs with very low H contents
\citep[$M_{\rm H}\lesssim 10^{-7} M_\odot$, see][and Miller Bertolami
  et al. 2020, in preparation]{2017ASPC..509..435M}.  If a thermal
pulse happens during the horizontal evolution of the post-AGB star in
the HR diagram, a scenario termed Late Thermal Pulse (LTP) by
\citet{2001Ap&SS.275....1B} occurs.  Then, the H-rich envelope is not
burned but diluted by the deepening of the convective  envelope once
the star evolves back to the AGB after the LTP. As shown  by
\citet{2005A&A...440L...1A}, the H diluted into the deeper parts of the
envelope  is later burned as the H-deficient central star contracts
again to the WD cooling track,  leading to the formation of WDs with a
low H-content ($M_{\rm H} \lesssim 10^{-6}-10^{-7} M_\odot$).
Finally, \citet{1990ARA&A..28..139D} mentioned that the
diffusion-induced nova studied by \citet{1986ApJ...301..164I} was also
a possible channel for the formation of DA  WDs with low-H contents,
but the later work by \citet{2011MNRAS.415.1396M} showed that this
scenario does not lead to a significant reduction in the H-content of
the future WD. 

A few final words of caution on some widespread misconceptions about
the relevance  of stellar winds for the formation of WDs with low-H
contents are in order. Although it has been  known for many decades
that the H content of the future WD cannot be arbitrarily reduced  by
winds during the AGB or post-AGB H-burning phases, \citep[see][and
  references therein]{1987ASSL..132..337S}, some confusion has arisen
in recent works. As already shown by \cite{1971AcA....21..417P},
H-burning post-AGB models have a very tight relationship between the
effective temperature and the envelope mass   of the post-AGB
object. As a consequence, and as long as the envelope can be
considered  in thermal equilibrium, the location of a H-burning
post-AGB remnant on the HR-diagram is independent of the mass-loss
history and only dependent on the value of the envelope mass. This
implies that an enhancement in the post-AGB winds does speed up the
post-AGB evolution but does not reduce the final H-content of  the WD;
see \citet{2015A&A...576A...9A}, \citet{2017ASPC..509..435M} 
and Miller Bertolami et al. (2019).
Note, however, that more intense winds in a He-burning post-AGB object
can indeed  reduce the final H-content of the WD significantly. Yet,
due to the relatively short  duration of the He-burning phase of AGB
and post-AGB models, the star has to undergo a  final He-flash at or
very close to the departure from the AGB, and this is only  relevant
within the late thermal pulse scenario discussed before.  A similar
situation holds for winds during the AGB and the departure from the
AGB phase.   An enhancement of AGB winds leads to a shortening of the
AGB phase. With less time for the H-free core to grow during the thermal
pulses, enhancing the winds on the AGB leads to a smaller final mass
for the same initial mass. Due to the tight $M_{\rm WD}$-$M_{\rm
  H}^{\rm WD}$ relation, this implies that more intense winds on the
AGB lead to the formation of WDs with higher H contents. This is true
also when looked at  the same value of the WD mass ($M_{\rm
  WD}$). Models of the same final mass but shorter AGB lifetimes are
less compact and luminous and have, consequently, larger post-AGB
envelope masses \citep{1995A&A...299..755B, 2016A&A...588A..25M}.
Consequently, more intense winds on the AGB usually lead to larger
H-envelopes for WDs. In closing, nowadays the LTP and VLTP scenarios
are the best explanations for the formation of WDs with low H-contents
in the context of single stellar evolution.

\begin{figure}
\includegraphics[width=1.0\textwidth]{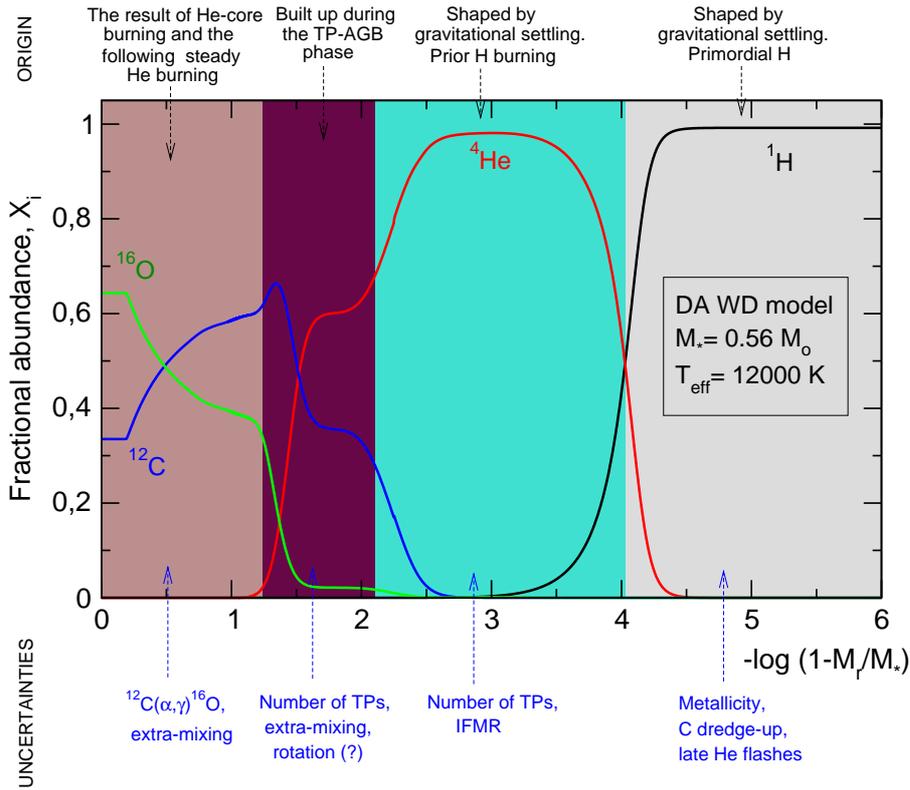}
\caption{Internal chemical structure of a typical DA WD model with
  $M_{\star}= 0.56 M_{\odot}$, $T_{\rm eff} \sim 12000$ K, and H
  envelope thickness of $\log(M_{\rm H}/M_{\star}) \sim -4$, resulting
  from the complete evolution of a single 1~$M_{\odot}$ progenitor from the
  ZAMS to the WD stage at the ZZ Ceti instability domain. 
  Plotted is the mass fraction ($X_i$) of
  $^{16}$O (green line), $^{12}$C (blue line), $^{4}$He (red line),
  and $^{1}$H (black line), in terms of the outer mass coordinate.  We
  include in the plot a  brief  explanation   about  the  origin (up)
  and the uncertainties (down) of  each  part  of  the  internal
  chemical structure, emphasized with different  colors.}
\label{fig02}
\end{figure}

The internal chemical constitution of WDs is a crucial issue for the
determination of the pulsational properties of these stars. In Fig. \ref{fig02}
we show the chemical structure of a template DA WD model with
$M_{\star}= 0.56 M_{\odot}$, $T_{\rm eff} \sim 12000$ K, and $\log(M_{\rm H}/M_{\star}) \sim -4$, resulting from the evolution of a 
single 1 $M_{\odot}$ progenitor from the ZAMS to the WD phase. We display
the mass fraction of $^{16}$O, $^{12}$C, $^{4}$He,
  and $^{1}$H, as a function of the outer mass coordinate. Note that 
  this coordinate strongly emphasizes the external part of the WD star. We
  include a short account of  the  origin and the uncertainties playing a 
  role  at  each  part  of  the  chemical structure. The extreme sensitivity of the pulsation properties of WDs to the details of the chemical structure links the asteroseismological inferences of pulsating WDs to the physical processes that take place during the progenitor's
evolution. However, it has not been until recently that the impact of
the current uncertainties in stellar evolution, both concerning the
modeling of physical processes and input physics of WD progenitors,
have begun to be assessed in asteroseismological fits.  In this
regard, \citet{2017A&A...599A..21D} explored for the first time the
impact of the occurrence of TPs on the AGB in WD
progenitors, the uncertainty in the $^{12}$C$(\alpha,\gamma)^{16}$O
cross section, and the occurrence of extra mixing episodes during core
He burning on the expected period spectrum of ZZ Ceti stars\footnote{Previous 
efforts to constrain the 
$^{12}$C$(\alpha,\gamma)^{16}$O reaction rate using WD 
asteroseismology have been done using DBVs \citep[see, e.g., ][and references therein]{2003ApJ...587L..43M}.}.  In this
connection, the  mixing and burning processes that take place along
the thermally pulsing AGB phase build  the chemical stratification of
the outer layers of the CO-core  of the emerging WD
\citep{2010ApJ...717..897A}, in particular  the He- and C-rich
inter-shell region that is formed during this stage as a result of the
short-lived He flash convection zone induced by the peak flash (see
Fig. \ref{fig02}).  The mass of this inter-shell, as well as  the mass
of the total He-content of the WD,  depend on both the occurrence of
overshooting (OV) during the He flash and the number of thermal  pulses. In
turn, the number of thermal pulses is determined by the initial mass,
chemical composition and by the poorly constrained efficiency of mass 
loss \citep{2014PASA...31...30K}.  As we will discuss, the shape
of the chemical profiles left by evolution during the thermally
pulsing AGB phase markedly impacts the $g$-mode pulsational periods, as
shown in \citet{2017A&A...599A..21D}, who also  concluded that the
occurrence or not of the TP-AGB phase during the evolution of the WD
progenitor constitutes a relevant issue that has to be taken into
account in seismological period fits of these stars. As mentioned,
despite the fact that the occurrence of the TP-AGB phase is expected
for most of single WD progenitors, it is not discarded that some WDs
could have evolved from progenitor stars that avoided this phase.  In
fact,  it is well known that low-mass He-burning stars located at the
extreme horizontal branch, and thus characterized by extremely thin H
envelopes \citep{1972ApJ...173..401F}, evolve directly to the WD stage,
avoiding the AGB \citep[the AGB-Manqu\'e and post early AGB
  stars; see][]{1989A&A...221...27C,1990ApJ...349..458B,1990ApJ...364...35G}. 
In line with this, recent evidence
suggests that most He rich stars of the globular cluster 
NGC2808 do not reach the AGB phase,
evolving directly to the WD state after the end of the He core burning
\citep{2017ApJ...843...66M}. In addition, departure from the AGB
before reaching the TP-AGB phase as a result of mass transfer by
binary interaction \citep{2000MNRAS.319..215H} or envelope ejection by
the swallowing of a planet or a very low mass companion
\citep{2002PASP..114..602D} is also possible. Accordingly, major
differences in the chemical structure of the outermost layers of WDs
should be expected depending on whether the progenitor stars evolved
through the  thermal pulses on the AGB or not, with consequences for
the expected  pulsational properties of pulsating WDs.

\subsection{Binary evolution}
\label{sec:2.2}

In a binary system, if the binary orbit is wide enough, the individual
stars are  not affected by the presence of a companion, so single
stellar evolution theory is enough   to describe their evolution (see
previous section). However, if the stars become close,  they can
interact, with severe consequences for their evolution. Interaction
can happen  by tidal forces, by stellar winds, or by mass transfer
and accretion. If either star fills  its Roche lobe, then gas flows
from the outer layers of that star into the Roche lobe of  the
companion star. Some or all of this gas may be captured by the
companion star so that mass transfer occurs. When the star filling the
Roche lobe is a giant  that has a convective envelope, and when the
donor star is significantly more massive  than its companion, the
transferred mass may not be accreted by the companion,  leading to the
formation of a common envelope (CE) surrounding both stars. The
outcome of CE evolution is still not fully understood
\citep{2013A&ARv..21...59I,2016MNRAS.460.3992N}, but possible outcomes
involve the formation of a closer binary or a stellar merger. If the
system is left in a very close binary configuration, then the
radiation of gravitational  waves will further shrink the orbits
leading to a stellar merger.

The stable Roche lobe overflow channel (RLOF) is of particular
interest for  the formation of He-core WDs. In fact, due to the very
long lifetimes on the main sequence,  low-mass He-core WDs cannot be
formed within single stellar evolution. For this reason,  the most
accepted channel for the formation of these stars involves one, or
more, phases  of mass transfer. In particular, if a low-mass star
fills its Roche lobe during the RGB,  the H-rich envelope can be
stripped before the He-core becomes massive (and hot)  enough to
ignite He. As a consequence, once all but a thin H-rich envelope is
stripped  from the star, it cannot support its giant configuration
anymore, and contracts to become  a He-core WD
\citep{2013A&A...557A..19A,2016A&A...595A..35I}. 

The coalescence of two stars within a common envelope evolution or in
a tight close binary system after a common envelope event has been
proposed to explain the properties of isolated WD stars. In particular, the
merger of two WDs may give rise to Type Ia Supernovae and to a variety
of objects such as Hot-subdwarf (sdO/sdB spectral types) stars and
R CrB stars that will finally evolve into WDs
\citep{2014MNRAS.438...14D}. The WDs formed by these channels might
harbor CO or ONe cores with either H-rich or H-deficient atmospheres.

\section{Asteroseismology of pulsating WDs and pre-WDs}
\label{sec:3}

\subsection{Recent observational achievements}
\label{sec:3.1}

\begin{table*}
\caption{ZZ Ceti stars and their effective temperatures, 
surface gravities and magnitudes. The $T_{\rm eff}$ and $\log g$ values have been 
corrected to 3D model atmosphere values. The letter in parentheses in the fourth column corresponds to the filter of the magnitude.  Specifically, ``g'' is SDSS g magnitude and ``GG'' is Gaia G magnitude.  
%The list does not include the variables found by \citet{2019MNRAS.486.4574R}, which do not have published spectral-temperature and gravity determinations.
}
\begin{tabular}{lccl}
\hline
\noalign{\smallskip}
Name &$T_{\rm eff}$ (K) & $\log g$ & Magnitude \\
\noalign{\smallskip}
\hline
\noalign{\smallskip}
SDSS J000006.75$-$004654.0    & 10620     & 8.18   &18.8 (g)  \\
SDSS J001836.11+003151.1      & 11530     & 8.04   &17.4 (g)  \\
MCT 0016$-$2553               & 11060     & 8.06   &15.9 (GG)  \\
HE 0031$-$5525                & 11662     & 7.71   &15.7 (g)  \\
G 132$-$12                    & 12480     & 8.00   &16.3 (g)  \\
SDSS J004345.78+005549.9      & 12130     & 8.14   &18.7 (g)  \\
LAMOST J004628.31+343319.9    & 11681     & 7.53   &16.3 (g)  \\%Guo et al. 2015, Su+17,ApJ,847,34 473s
SDSS J004855.17+152148.7      & 11280     & 8.17   &18.7 (g)  \\
SDSS J005208.42$-$005134.6      & 12300     & 8.46   &17.7 (g)  \\ %Pyrzas+15 DA+M
SDSS J010207.17$-$003259.4    & 10850     & 8.18   &18.0 (g) \\
EPIC 220274129                & 11810     & 8.03   &16.7 (g) \\ %Bell+17
LAMOST J010302.46+433756.2    & 11750     & 7.89   &18.3 (g)  \\%Gentile Fusillo et al. 2015, Su+17,ApJ,847,34 1174s
BPM 30551                     & 11240     & 8.16   &15.4 (g)    \\
SDSS J011100.63+001807.2      & 11490     & 8.08   &18.8 (g)  \\
SDSS J011123.89+000935.3      & 12321     & 7.50   &17.8 (g)     \\ %Pyrzas+15 DA+M
SDSS J012234.68+003025.8      & 11650     & 7.94   &16.8 (g)  \\
SDSS J012950.44$-$101842.0    & 12043     & 8.03   &18.4 (g)  \\
LAMOST J013033.90+273757.9    & 14127     & 7.69   &18.5 (g)  \\%Guo et al. 2015, Su+17,ApJ,847,34 310s
SDSS J013440.94$-$010902.3    & 10260     & 7.82   &18.1 (g)  \\
Ross 548                      & 12300     & 8.03   &14.3 (g) \\
MCT 0145$-$2211               & 11850     & 8.15   &14.9 (GG) \\
SDSS J020351.28+004025.1      & 10794     & 8.17   &19.4 (g) \\ %Pyrzas+15 DA+M
HS 0210+3302                  & 12176     & 7.38   &16.7 (g)  \\
SDSS J021406.78$-$082318.4    & 11580     & 7.86   &17.9 (g)  \\
HS 0235+0655                & 11008     & 7.56   &16.5 (g)  \\
SDSS J024922.35$-$010006.7  & 11030     & 8.19   &18.8 (g)  \\
KUV 02464+3239              & 11620     & 8.13   &16.0 (g)   \\
SDSS J030153.81+054020.0    & 11139     & 8.02   &18.1 (g)  \\
SDSS J030325.22$-$080834.9  & 11260     & 8.40   &18.8 (g)  \\
SDSS J031847.09+003029.9    & 11150     & 8.18   &17.8 (g)  \\
SDSS J033236.61$-$004918.   & 10940     & 8.05   &18.2 (g)  \\
BPM 31594                   & 11500     & 8.05   &15.1 (GG) \\
KUV 03442+0719              & 10870     & 7.78   &16.6 (g)  \\
HE 0344$-$1207              & 11497     & 7.91   &16.0 (g)  \\
SDSS J034939.35+103649.9    & 11896     & 8.21   &16.6 (g)  \\
EPIC 210377280              & 11590     & 7.94   &18.5 (g) \\%Bell+17
HL Tau76                    & 11470     & 7.92   &15.0 (g) \\
G 38$-$29                   & 11160     & 7.89   &15.6 (g) \\
G 191$-$16                  & 11440     & 8.04   &15.9 (g) \\
LP 119$-$10                 & 11342     & 8.085  &15.2 (g) \\
HS 0507+0435                & 12010     & 8.19   &15.3 (g) \\
GD 66                       & 12210     & 8.10   &15.5 (g)  \\
HE 0532$-$5605              & 11510     & 8.42   &15.9 (GG) \\
LAMOST J062159.49+252335.9  & 11728     & 8.25   & 17.5 (g) \\%Rebassa-Mansergas et al. 2015, Su+17,ApJ,847,34 830s
HS 0733+4119                & 11049     & 8.01   &15.8 (g)  \\
SDSS J075617.54+202010.2    & 11830     & 8.13   &18.3 (g)  \\
SDSS J081531.75+443710.3    & 11840     & 8.21   &19.3 (g)  \\
SDSS J081828.98+313153.0    & 11820     & 8.13   &17.4 (g)  \\
SDSS J082429.01+172345.4    & 11433     & 8.21   &18.3 (g)  \\ %Pyrzas+15 DA+M
SDSS J082518.86+032927.8    & 12120     & 8.15   &17.5 (g)  \\
SDSS J082547.00+411900.0    & 11510     & 8.37   &18.5 (g)  \\
SDSS J083203.98+142942.3    & 11643     & 7.99   &18.9 (g)  \\ %Hermes+17
KUV 08368+4026              & 12010     & 8.13   &15.6 (g)  \\
SDSS J084054.14+145709.0    & 10862     & 7.89   &18.3 (g)  \\
SDSS J084021.23+522217.4    & 12160     & 8.93   &18.2 (g)  \\
SDSS J084220.73+370701.7    & 11620     & 7.88   &18.8 (g)  \\
SDSS J084314.05+043131.6    & 11220     & 8.09   &17.8 (g)  \\
SDSS J084746.82+451006.3    & 11690     & 8.12   &18.3 (g)  \\
SDSS J085128.17+060551.1    & 11300     & 8.05   &16.8 (g)  \\
SDSS J085325.55+000514.2    & 11950     & 8.15   &18.2 (g)  \\
SDSS J085507.29+063540.9    & 10970     & 8.22   &17.2 (g)  \\
SDSS J085648.33+185804.9    & 11896     & 8.09   &18.9 (g)  \\
SDSS J090041.08+190714.3    & 11849     & 8.05   &17.6 (g)  \\
GD 99                       & 12110     & 8.20   &14.5 (g)  \\
\noalign{\smallskip}
\hline
   \end{tabular}
   \label{zzceti1}
   \end{table*}

\begin{table*}
\caption{ZZ Ceti stars (continuation of Table \ref{zzceti1}).}
\begin{tabular}{lccl}
\hline
\noalign{\smallskip}
Name &$T_{\rm eff}$ (K) & $\log g$ & Magnitude \\
\noalign{\smallskip}
\hline
\noalign{\smallskip}
SDSS J090231.76+183554.9    & 11191     & 7.89   &19.4 (g)  \\
SDSS J090624.26$-$002428.2  & 11260     & 8.07   &17.7 (g)  \\
SDSS J091118.42+031045.1    & 11630     & 8.14   &18.4 (g)  \\
SDSS J091312.74+403628.7    & 11850     & 8.09   &17.6 (g)  \\
SDSS J091635.07+385546.2    & 11320     & 8.04   &16.6 (g)  \\
SDSS J091731.00+092638.1    & 11340     & 8.09   &18.1 (g)  \\
SDSS J092329.81+012020.0    & 11190     & 8.38   &18.3 (g)  \\
G 117$-$B15A                & 12420     & 8.12   &15.5 (g) \\
SDSS J092511.60+050932.4    & 10830     & 8.21   &15.2 (g)  \\
SDSS J093944.89+560940.2    & 11690     & 8.29   &18.7 (g)  \\
SDSS J094000.27+005207.1    & 10590     & 8.34   &18.1 (g)  \\
SDSS J094213.13+573342.5    & 11360     & 8.12   &17.4 (g)  \\
SDSS J094917.04$-$000023.6  & 11130     & 8.21   &18.8 (g)  \\
HS 0951+1312                & 12010     & 8.05   &16.5 (g)  \\
HS 0952+1816                & 11390     & 8.11   &16.3 (g)  \\
SDSS J095833.13+013049.3    & 11730     & 8.08   &16.7 (g)  \\
SDSS J095936.96+023828.4    & 11830     & 8.06   &18.1 (g)  \\
SDSS J100238.58+581835.9    & 11440     & 8.11   &18.3 (g)  \\
SDSS J100718.26+524519.8    & 11390     & 8.12   &18.9 (g)  \\
SDSS J101519.65+595430.5    & 11440     & 8.06   &18.0 (g)  \\
SDSS J101540.14+234047.4    & 11320     & 8.44   &18.6 (g)  \\
SDSS J101548.01+030648.4    & 11630     & 8.12   &15.7 (g)  \\
HS 1039+4112                & 11730     & 8.12   &16.1 (g)  \\
SDSS J104358.59+060320.9    & 11173     & 8.19   &18.7 (g)  \\ %Pyrzas+15 DA+M
WD 1047+335                 & 11310     & 8.09   &16.5 (g)  \\
SDSS J105449.87+530759.1    & 10960     & 7.96   &17.9 (g)  \\
SDSS J105612.32$-$000621.7    & 11130     & 7.91   &17.5 (g)  \\
SDSS J110525.70$-$161328.3    & 11857     & 8.06   &17.5 (g)  \\
SDSS J110623.40+011520.8    & 10920     & 7.90   &18.4 (g)  \\
GD 133                      & 12430     & 8.10   &14.6 (g) \\
SDSS J112221.10+035822.4    & 11030     & 7.91   &18.2 (g)  \\
SDSS J112542.84+034506.3    & 11600     & 7.95   &18.1 (g)  \\
SDSS J111710.54-125540.9    & 11302     & 8.29   &19.6 (g)  \\ %Pyrzas+15 DA+M
EC 11266$-$2217               & 12010     & 8.08   &16.4 (g) \\
SDSS J113604.01$-$013658.1    & 11780     & 8.05   &17.8 (g)  \\
SDSS J113655.17+040952.6    & 12330     & 7.99   &17.0 (g)  \\ %DA+M Pyrzas+15 11699 \pm 152 7.99 \pm 0.08 PCEB
KUV 11370+4222              & 11940     & 8.17   &16.5 (g)  \\
PG 1149+057                 & 11060     & 8.06   &15.0 (g)  \\
EC 11507$-$1519             & 12440     & 8.20   &16.0 (g)  \\
SDSS J115707.43+055303.6    & 11040     & 8.04   &17.6 (g)  \\
SDSS J120054.55$-$025107.0    & 11970     & 8.24   &18.2 (g)  \\
G 255$-$2                     & 11440     & 8.14   &16.0 (g)\\
SDSS J121628.55+092246.4    & 11240     & 8.25   &18.6 (g)  \\
WD J1218+0042               & 11170     & 8.06   &18.5 (g)  \\
SDSS J122229.57$-$024332.5    & 11380     & 8.19   &16.7 (g)  \\
BPM 37093                   & 11620     & 8.69   &13.8 (GG) \\
SDSS 124949.36+304828.1     & 12127     & 8.299  &18.0 (g)  \\
HS 1249+0426                & 12160     & 8.21   &16.0 (g)  \\
SDSS J125535.41+021116.0    & 11580     & 8.15   &19.1 (g)  \\
SDSS J125710.50+012422.9    & 11490     & 8.30   &18.6 (g)  \\
HE 1258+0123                & 11420     & 8.02   &16.4 (g)  \\
GD 154                      & 11120     & 8.07   &15.3 (g)  \\
WD 1310$-$0159                & 10940     & 7.76   &17.7 (g)  \\
SDSS J132350.28+010304.2    & 11380     & 8.45   &18.5 (g)  \\
WD J1337+0104               & 11460     & 8.64   &18.6 (g)  \\
SDSS J133831.74$-$002328.0  & 11900     & 8.07   &17.1 (g)  \\
EPIC 229227292              & 11190     & 8.02   &16.7 (g) \\
SDSS J134550.93$-$005536.5    & 11760     & 8.10   &16.7 (g)  \\
LP 133$-$144                  & 12150     & 7.97   &15.7 (g)  \\
G 238$-$53                    & 12130     & 7.97   &15.5 (g)\\
SDSS J135459.89+010819.3    & 11650     & 8.03   &16.4 (g)  \\
SDSS J135531.03+545404.5    & 11480     & 7.93   &18.6 (g)  \\
EC 14012$-$1446               & 12020     & 8.18   &15.7 (g) \\
SDSS J140859.46+044554.7    & 10920     & 7.99   &17.9 (g)  \\
\noalign{\smallskip}
\hline
\end{tabular}
\label{zzceti2}
\end{table*}

\begin{table*}
\caption{ZZ Ceti stars (continuation of Table \ref{zzceti1}).}
\begin{tabular}{lccl}
\hline
\noalign{\smallskip}
Name &$T_{\rm eff}$ (K) & $\log g$ & Magnitude \\
\noalign{\smallskip}
\hline
\noalign{\smallskip}
GD 165                      & 12220     & 8.11   &14.3 (g)   \\
HE 1429$-$0343                & 11290     & 8.00   &16.0 (g) \\
L 19$-$2                      & 12070     & 8.13   &13.4 (GG) \\
SDSS J144330.93+013405.8    & 10450     & 7.85   &18.7 (g)  \\
SDSS J150207.02$-$000147.1    & 11090     & 7.75   &18.7 (g)  \\
WD 1526+558                 & 10860     & 7.73   &17.1 (g)  \\
HS 1531+7436                & 13270     & 8.49   &16.5 (g)  \\
SDSS J153332.96$-$020655.7  & 11390     & 8.04   &16.4 (g)  \\
PG 1541+650                 & 11560     & 8.12   &15.6 (g)  \\
SDSS J155438.35+241032.6    & 11470     & 8.49   &17.5 (g)  \\
Ross 808                    & 11120     & 7.98   &14.4 (g)  \\
WD 1607+205                 & 11140     & 7.81   &17.3 (g)  \\
SDSS J161218.08+083028.1    & 12250     & 8.29   &17.8 (g)  \\
SDSS J161737.63+432443.8    & 11070     & 8.07   &18.4 (g)  \\
SDSS J161837.25$-$002302.7  & 10292     & 7.97   &19.3 (g)  \\
HS 1625+1231                & 11690     & 8.06   &16.1 (g)  \\
SDSS J164115.61+352140.6    & 12025     & 8.34   &19.0 (g) \\
G 226$-$29                  & 12510     & 8.35   &12.2 (g) \\
SDSS J165020.53+301021.2    & 10830     & 8.43   &18.1 (g)  \\
GD 518                      & 11760     & 8.97   &17.3 (g)  \\
SDSS J170055.38+354951.1    & 11230     & 7.94   &17.3 (g)  \\
SDSS J171113.01+654158.3    & 11130     & 8.47   &16.9 (g)  \\
BPM 24754                   & 10840     & 7.93   &16.2 (GG) \\
SDSS J172428.42+583539.0    & 11640     & 7.88   &17.6 (g)  \\
SDSS J173235.19+590533.4    & 10770     & 7.97   &18.7 (g)  \\
HS 1824+6000                & 11520     & 7.73   &16.1 (g)  \\
G 207$-$9                   & 12080     & 8.37   &14.6 (g)  \\
KIC 7594781                 & 11730     & 8.11   &18.6 (g)  \\
KIC 10132702                & 11940     & 8.12   &18.8 (g)  \\
KIC 4552982                 & 10860     & 8.16   &17.8 (g)  \\
KIC 4357037                 & 10950     & 8.11   &18.2 (g)  \\
KIC 8293193                 & 12650     & 8.01   &18.4 (g)  \\
KIC 11911480                & 12160     & 7.94   &18.1 (g)  \\
KIC 4362927                 & 11140     & 7.84   &19.4 (g)  \\
G 185$-$32                  & 12470     & 8.10   &13.0 (g)  \\
KIC 9162396                 & 11070     & 8.06   &18.5 (g)  \\
KIC 7766212                 & 11890     & 8.01   &16.8 (g)  \\
KIS J1945+4455              & 11590     & 8.04   &17.2 (g)  \\
GD 385                      & 11820     & 8.07   &15.1 (g)  \\
GD 226                      & 10730     & 8.06   &16.4 (g) \\
SDSS J202857.52+771054.5    & 11940     & 8.38   &19.0 (g)  \\
WD 2102+233                 & 11712     & 8.28   &15.9 (g)  \\
SDSS J212808.49$-$000750.8    & 11420     & 8.24   &18.0 (g) \\
SDSS J213530.32$-$074330.7    & 10900     & 7.96   &18.7 (g) \\
SDSS J214723.73$-$001358.4    & 12098     & 7.93   &19.0 (g) \\
G 232$-$38                    & 11590     & 8.02   &16.8 (g) \\
WD 2148$-$291                 & 11490     & 8.06   &16.0 (g) \\
SDSS J215354.11$-$073121.9    & 11910     & 8.27   &18.7 (g) \\
SDSS J215628.26$-$004617.2    & 10680     & 8.01   &18.3 (g)  \\
SDSS J215905.52+132255.7    & 11370     & 8.69   &18.9 (g)  \\
SDSS J220830.02+065448.7    & 11147     & 8.25   &17.9 (g)  \\
 SDSS J220831.42+205909.66 &  11776 &  8.77 &  17.47 (g)  \\ %Barbara New3 Romero+13
SDSS J220915.84$-$091942.5    & 11630     & 8.49   &18.4 (g)  \\
SDSS J221458.37$-$002511.7    & 11650     & 8.30   &17.9 (g)  \\
SDSS J223135.71+134652.8    & 11060     & 7.89   &18.7 (g) \\
SDSS J223726.86$-$010110.9    & 11380     & 7.97   &18.9 (g)  \\
GD 244                      & 11760     & 8.09   &15.7 (g)    \\
PG 2303+242                 & 11500     & 8.07   &15.3 (g) \\ %23 06 18        24 32 08 
SDSS J230726.66$-$084700.3  & 10970     & 8.21   &18.9 (g) \\
SDSS J231934.52+515316.4 &  11435 &  8.45 &  19.37 (g)\\ %Barbara New3 Romero+13 
G 29$-$38                   & 11910     & 8.17   &13.3 (g) \\ %23 28 48        05 14 54
SDSS J233458.71+010303.1    & 11104     & 8.16   &19.2 (g)  \\
GD 1212                     & 10970     & 8.03   &13.3 (g)  \\
G 30$-$20                     & 11150     & 8.01   &16.1 (g)  \\
SDSS J235040.72$-$005430.9    & 10290     & 8.14   &18.1 (g)  \\
EC 23487$-$2424               & 11560     & 8.09   &15.3 (g)  \\
\noalign{\smallskip}
\hline
\end{tabular}
\label{zzceti3}
\end{table*}

\begin{table*}
\caption{ZZ Ceti stars (continuation of Table \ref{zzceti1}).}
\begin{tabular}{llcl}
\hline
\noalign{\smallskip}
Name &$T_{\rm eff}$ (K) & $\log g$ & Magnitude \\
\noalign{\smallskip}
\hline
\noalign{\smallskip}

LAMOST J004628.31+343319.9  & 11681     & 7.53   &16.3 (g) \\
LAMOST J062159.49+252335.9  & 11728     & 8.25   &17.6 (g) \\
LAMOST J010302.46+433756.2  & 11750     & 7.89   &18.3 (g) \\
LAMOST J013033.90+273757.9  & 14127     & 7.69   &18.6 (g) \\
KIC 4357037                 & 12650     & 8.01  &18.2 (g)  \\  %19 17 19.198 +39 27 19.11
KIC 4552982                 & 10950     & 8.11  &17.7 (g)  \\  %19 16 43.827 +39 38 49.69 
KIC 7594781                 & 11730     & 8.11  &18.1 (g)  \\  %19 08 35.882 +43 16 42.37
KIC 10132702                & 11940     & 8.12  &19.0 (g)  \\  %19 13 40.891 +47 09 31.28
KIC 11911480                & 11580     & 7.96  &18.0 (g)  \\  %19 20 24.900 +50 17 22.40
KIC 60017836                & 10980     & 8.00  &13.3 (g)  \\  %23 38 50.739 -07 41 19.90
EPIC 201355934              & 11770     & 7.97  &17.8 (g)  \\  %11 36 04.013 -01 36 58.09
EPIC 201719578              & 11070     & 7.94  &18.1 (g)  \\  %11 22 21.104 +03 58 22.41 
EPIC 201730811              & 12480     & 7.96  &17.1 (g)  \\  %11 36 55.180 +04 09 52.60
EPIC 201802933              & 12330     & 8.11  &17.6 (g)  \\  %11 51 26.148 +05 25 12.90
EPIC 201806008              & 10910     & 8.02  &14.9 (g)  \\  %11 51 54.200 +05 28 39.82
EPIC 206212611              & 10830     & 8.00  &17.3 (g)  \\  %22 20 24.230 -09 33 31.09
EPIC 210397465              & 11200     & 7.71  &17.6 (g)  \\  %03 58 24.233 +13 24 30.79
EPIC 211596649              & 11600     & 7.91  &18.9 (g)  \\  %08 32 03.984 +14 29 42.37
EPIC 211629697              & 10600     & 7.77  &18.3 (g)  \\  %08 40 54.144 +14 57 08.98
EPIC 211914185              & 13590     & 8.43  &18.8 (g)  \\  %08 37 02.160 +18 56 13.48 
EPIC 211916160              & 11510     & 7.96  &18.9 (g)  \\  %08 56 48.334 +18 58 04.92
EPIC 211926430              & 11420     & 7.98  &17.6 (g)  \\  %09 00 41.080 +19 07 14.40
EPIC 228682478              & 12070     & 8.18  &18.2 (g)  \\  %08 40 27.839 +13 20 09.96
EPIC 229227292              & 11210     & 8.03  &16.6 (g)  \\  %13 42 11.659 -07 35 38.30
EPIC 229228364              & 11030     & 8.03  &17.8 (g)  \\  %19 18 10.598 -26 21 05.00
EPIC 220204626              & 11620     & 8.17  &18.4 (g)  \\  %01 11 23.890 +00 09 35.20
EPIC 220258806              & 12800     & 8.09  &16.2 (g)  \\  %01 06 37.030 +01 45 03.00 
EPIC 220347759              & 12770     & 8.08  &17.6 (g)  \\  %00 51 24.245 +03 39 03.79
EPIC 220453225              & 11220     & 8.04  &17.9\\  %00 45 33.151 +05 44 46.96 
EPIC 229228478              & 12500     & 7.93  &16.9 (g)  \\  %01 22 34.682 +00 30 25.73 
EPIC 229228480              & 12450     & 8.18  &18.8 (g)  \\  %01 11 00.640 +00 18 07.04
 SDSS J002945.75+144214.9  &    11310     &   7.95  &     18.20 (GG) \\ %Rowan19   DAV
 SDSS J002959.14+145814.2  &    10589**   &   7.89  &     17.49 (GG) \\ %Rowan19   DAVa
  WD J003116.51+474828.39  &    10444**   &   8.04  &     18.36 (GG) \\ %Rowan19   DAVa
 SDSS J004154.66$-$030802.5  &  10936**   &   8.36  &     18.07 (GG) \\ %Rowan19   DAVa
 SDSS J010025.55+421840.9  &    $\cdots$    & $\cdots$ &  16.58 (GG) \\ %Rowan19   DAVa
 SDSS J010528.74+020501.1  &    $\cdots$   &  $\cdots$ &  16.73 (GG) \\ %Rowan19   DAVa
 SDSS J010539.14+321846.6  &    10737**    &  8.08  &     18.07 (GG) \\ %Rowan19   DAVa
           KUV 01595$-$1109  &  11062      &  8.14  &     16.89 (GG) \\ %Rowan19   DAV
 SDSS J022941.29$-$063842.7  &  9791**     &  7.24  &     18.23 (GG) \\ %Rowan19   DAVa
  WD J030648.49$-$172332.19  &  $\cdots$ & $\cdots$&      16.70 (GG) \\ %Rowan19   DAVa
 WD J053212.77$-$432006.05   &  $\cdots$ & $\cdots$ &     18.19 (GG) \\ %Rowan19   DAVa
 SDSS J080609.19+111231.4  &    10987**    &  7.90  &     18.11 (GG) \\ %Rowan19   DAVa
 SDSS J084055.71+130329.4  &    $\cdots$ & $\cdots$ &     17.22 (GG) \\ %Rowan19   DAVa
 SDSS J084652.93+442638.6  &    11565     &   8.06  &     18.19 (GG) \\ %Rowan19   DAV
 SDSS J093250.56+554315.4  &    $\cdots$ & $\cdots$ &     17.69 (GG) \\ %Rowan19   DAVa
 SDSS J094851.43+512448.0  &    10252     &   7.80  &     18.57 (GG) \\ %Rowan19   DAV
 SDSS J103642.25+211527.9  &    11329     &   8.27  &     17.60 (GG) \\ %Rowan19   DAV
                  CBS 130  &    11309     &   8.09  &     16.56 (GG) \\ %Rowan19   DAV
 SDSS J110505.94+583103.0  &    11256**   &   8.01  &     17.94 (GG) \\ %Rowan19   DAV
  WD J115057.43$-$055306.58  &   9611**   &   7.38  &     17.44 (GG) \\ %Rowan19   DAVa
 SDSS J120309.16+454520.3  &    11077     &   8.01  &     18.57 (GG) \\ %Rowan19   DAV
 SDSS J122155.79+050622.7  &    11720**   &   7.00  &     17.90 (GG) \\ %Rowan19   DAVa
 SDSS J124759.03+110703.0  &    $\cdots$ & $\cdots$ &     19.35 (GG) \\ %Rowan19   DAV
 SDSS J124804.03+282103.8  &    10987**   &   7.83&       18.04 (GG) \\ %Rowan19   DAVa
              WD 1452+600  &    10962**   &   7.91&       17.18 (GG) \\ %Rowan19   DAV
 SDSS J150626.18+063845.9  &    $\cdots$ & $\cdots$ &     16.60 (GG) \\ %Rowan19   DAVa
 SDSS J150739.34+074828.5  &    10540**    &  7.81&       18.21 (GG) \\ %R owan19   DAVa
 SDSS J162724.67+392026.3  &    $\cdots$ & $\cdots$ &     16.23 (GG) \\ %Rowan19   DAVa
 SDSS J173351.49+341012.5  &    $\cdots$ & $\cdots$ &     16.35 (GG) \\ %Rowan19   DAVa
  WD J202838.13$-$060842.11  &  $\cdots$ & $\cdots$ &     15.22 (GG) \\ %Rowan19   DAVa
  WD J204127.11$-$041724.22  &  11114**    &  8.34&       18.31 (GG) \\ %Rowan19   DAVa
  WD J212402.03$-$600100.05  &  $\cdots$ & $\cdots$ &     17.97  (GG) \\ %Rowan19   DAVa
 SDSS J215321.77+044020.0  &    10936**    &  8.06&       18.19 (GG) \\ %Rowan19   DAVa
 SDSS J231536.88+192449.0  &    10540**    &  7.85&       17.99 (GG) \\ %Rowan19   DAVa
  WD J231641.17$-$315352.74  &  11720**    &  8.08&       18.29 (GG) \\ %Rowan19   DAVa
 SDSS J235010.36+201914.0  &    10786**    &  7.86&       17.41 (GG) \\ %Rowan19   DAVa
% WD J085927.11-042916.34 &&&20.54 (g) \\%K. Dame+19 P=466s,318s
% WD J090051.51-044249.1  &&&20.60 (g) \\%K. Dame+19P=868s,757s

\noalign{\smallskip}
\hline
 **Photometric determination only, with large uncertainty.
\end{tabular}
\label{zzceti4}
\end{table*}

As shown in Table \ref{tabla-properties} and Fig. \ref{fig01}, at
present there are   eight families of pulsating WDs and pre-WDs,
although  two of those categories (hot DAVs and DQVs)  need to be confirmed as 
such. Here, we describe the new observational findings in  the field of these
pulsating degenerate stars. 

In the last decade, there have been numerous discoveries of pulsating
WD  stars, both from the ground and from space. Ground-based
observations, mainly with the spectral  observations of the Sloan
Digital Sky Survey \citep[SDSS;][]{2000AJ....120.1579Y},  have
increased the number of known WDs by a factor of $15$
\citep{2013ApJS..204....5K,2016MNRAS.455.3413K,2019MNRAS.486.2169K,2017EPJWC.15201011K,
2019MNRAS.482.4570G} and the number 
of pulsators by a 
factor of $4$, starting with
\cite{2004ApJ...607..982M,2005ApJ...625..966M,2006A&A...450..227C,2007ASPC..372..583V,
  2009ApJ...690..560N,2013MNRAS.430...50C}. In Tables \ref{zzceti1} to
\ref{zzceti4} we present the list of ZZ Ceti stars known at
the time of writing this review (March 2019), along with their  effective
temperatures, surface gravities, and magnitudes. 
  The determinations of $\log g$ and $T_{\rm eff}$ from the spectra, when available, were taken from the literature, 
mainly \cite{2011ApJ...743..138G} and \cite{2019MNRAS.486.2169K}, 
with 3D corrections \citep[following][]{2013A&A...559A.104T} applied. 

Very recently, 36 new DAVs have been discovered by \citet{2019MNRAS.486.4574R}, but for most of them their  spectroscopically determined gravities and effective temperatures are not available yet. When only colors were available, they were used as in  \cite{2019MNRAS.486.2169K}. %\citet{2019MNRAS.tmp..405D} discovered two new ZZ Cetis with the DECCam, again with no spectral determinations.
For ZZ Cetis, the
discovery of the rare ultra-massive pulsators, which started with
BPM~37093  with $M_{\star} \sim 1.1~M_\odot$
\citep{1992ApJ...390L..89K}, includes GD~518, with $M_{\star}\sim
1.24~M_\odot$ \citep{2013ApJ...771L...2H}, and SDSS~J084021.23+522217.4 \citep{2017MNRAS.468..239C}, with
$M_{\star}\sim 1.16~M_\odot$, opened the study of crystallized WDs. In addition, 
WD J212402.03$-$600100.0 is another possible 
ultra-massive DAV star \citep[$M_{\star} \sim 1.16 M_{\odot}$; see][]{2019MNRAS.486.4574R}. New theoretical  work about
ultra-massive ZZ Cetis will be introduced in Sect. \ref{sec:3.3}. On
the other hand, the 3D convection studies of
\citet{2013A&A...552A..13T,2015ApJ...809..148T} have converged the
determination of temperatures and gravities using different mixing
length models, which had plagued prior studies of the location of the ZZ
Ceti instability strip.

Despite numerous discoveries of pulsating WDs ---and notwithstanding
several successful  asteroseismological studies
carried out in the last decade, see Sect. \ref{sec:3.2}--- the
discovery of new  pulsators has not led to asteroseismological
solutions for most of them, because the available discovery  data in
general covers only a few hours of time series photometry, which
allows only for the determination of the dominant period, or worse, just
the beat period. The need of detecting a large number of pulsation
periods  to allow the determination of the internal structure of stars
can be traced back to  \citet{Legendre1805} and \citet{Gauss1809}, who
demonstrated that the best way to find an unknown parameter is to
minimize the sum of the square of the residuals.  \citet{Laplace1810}
presented  the generalization of least squares fits for multiple
parameters and defined the uncertainty as $\sigma^2 = \frac{S} {N-k} =
\frac{\sum_{i=1}^N (P^{\rm o}_i - P_i)^2} {N-k}$, where $N$ is the
number of measurements and $k$ the number of fit parameters, so if
$k\geq N$, the uncertainty is infinite.  $S$ decreases when $k$
increases, but when a term is added in a fit, it has to  be tested if
the decrease in $S$ is significant.  The confidence level at which we
can rule out the null hypothesis can be calculated   through the
Fisher $F$-distribution, as shown for example by
\citet{1975MNRAS.170..633P}. However, in some cases it is possible to constrain 
the structure of a pulsating WD with a few observed periods, i.e.,  when the number of 
observed periods is less than the number of parameters that define 
the stellar models \citep{2017A&A...598A.109G, 2017ApJ...834..136G}. This would be connected to the 
individual properties of the modes ---illustrated in 
particular by their weight functions, which reflect the 
shape of the eigenfunctions of the modes--- and the differential information 
contained in the distribution of these modes relative to each other. This is a 
very interesting aspect that deserves to be investigated in more detail.

With the aim of detecting a large number of pulsation periods,
\citet{1990ApJ...361..309N} established the Whole Earth Telescope
(WET). The studies by \cite{1991ApJ...378..326W} for PG1159$-$035 ---the prototypical GW Vir star--- and \cite{1994ApJ...430..839W} for GD~358 (V777 Her) ---the brightest ($m_V = 13.7)$ and best studied He-atmosphere WD
pulsator--- are excellent examples of the wealth of information that can be extracted from pulsating WDs 
from long, nearly uninterrupted data sets. As a more recent example of this, \citet{2019ApJ...871...13B} presented the
results of three decades of observations of GD~358, which allowed the detection of 
15 independent (13 consecutive) $\ell= 1$ pulsation modes. They also showed that the
frequencies and  amplitudes of these modes change with time.

The pulsation periods of WDs vary as a consequence of the evolution of these stars, giving place  to a detectable rate of period change. The estimate of the rate of period change for pulsating WDs, a direct
measurement of their evolutionary time scale
\citep{1983Natur.303..781W}, has progressed slowly because it demands
huge observational time
\citep{2005ApJ...634.1311K,2008A&A...489.1225C,
  2015ASPC..493..199S,2013ApJ...771...17M, 2013ApJ...766...42H}. Some
recent applications of the rate of  period change in ZZ Ceti stars to
derive constraints on fundamental particles is presented in
Sect. \ref{sec:3.9}. On the other hand, the first ZZ Ceti in a
detached WD with a main sequence star was discovered by
\citet{2015MNRAS.447..691P}. Cool ZZ Cetis, near the red edge of the
instability strip, show long pulsation periods, and those observed
---i.e., with significant amplitude--- in general change with time.
\citet{2018MNRAS.tmp.2745B} detected 14 frequencies for HS~0733+4119,
of which 8 are independent modes; for GD 154, 17 frequencies, with 4
independent modes, and for R808, the 28 frequencies discussed in
\citet{2009JPhCS.172a2067T} and 2 new ones.

For DBVs, the progress has been slower than for ZZ Cetis, first
because the number of He atmosphere WDs is smaller, and also
because the DBVs are hotter, and therefore evolve through the
instability strip faster, but also because the uncertainty in the
temperature estimate is much larger for DBs, in part caused by the H
contamination in the atmosphere. With the consistent determination of
the atmospheric parameters by \citet{2015A&A...583A..86K}, the group
led by \citet{zach18} increased the number of known DBV pulsators from
23 to 46, but their results are not yet published.
\citet{2017ApJ...835..277H} discovered pulsations in the hot ($T_{\rm
  eff}\sim 32\,000$~K) DB PG~0112+104, expanding the DBV instability
strip. In Table \ref{DBVs} we list the V777 Her stars known at the
time of writing this review, along with their $T_{\rm eff}$, $\log g$,
and magnitudes. There are 4 new additional DBV stars discovered by
\cite{2019MNRAS.486.4574R} for which the effective temperatures and 
gravities have not been assessed yet.
 
The {\it  Kepler} satellite observations, both main mission
\citep{2010Sci...327..977B} and K2 \citep[two-wheel
  operation,][]{2014PASP..126..398H}, increased the number of known
WD pulsators by a factor of 2 \citep{2017ApJS..232...23H}, until it
ran out of fuel by October 2018.  It had 42
2200$\times$1024 pixel charge coupled devices (CCD) to measure the
stellar brightness variations. Due to data storage and downloading
limitations, only 512 stars were observed in short cadence mode ---
$58.85$~s exposures --- at each quarter, while the remaining observations were co-added
on chip to $29.42$ minutes long-cadence exposures \citep{2010ApJ...713L.160G}.
%The short-cadence data shows instrumental patterns at the fundamental and up to the 14th harmonic
%of the inverse of the 29.42 minutes long-cadence period, i.e., multiples of
%0.566391~mHz. 
The extended time span of observations ---at least 3 months for
the main mission and around 80 days for each K2 cycle---  
allowed the determination of dozens of pulsation periods 
and the determination of rotation periods for
different modes.  
In addition, the nearly continuous light curves eliminated spectral leakage
in the Fourier Transforms (FTs), making period identification relatively easy.
When launched, there were no known pulsating WDs in
the {\it Kepler} field, and due to the telescope's relatively small 
1.4 meter primary mirror and fixed field of $10\deg\times 10\deg$, it took a
large effort from the WD community to find pulsating WD candidates 
that could be observed in the limited number of targets in short cadence mode
\citep{2010MNRAS.409.1470O,2011ApJ...736L..39O,2011MNRAS.414.2860O,
2011ApJ...741L..16H,2016MNRAS.457.2855G}.
An important aspect to be mentioned is the ability of {\it Kepler} to detect 
Nyquist-aliased pulsation signals, and  the possibility to recover accurate 
pulsation periods to a precision similar to that of the K2 mode with the help of 
ground-based data 
\citep{2017ApJ...851...24B}.

\begin{table*}
\caption{V777 Her stars and their effective temperatures, 
surface gravities, and magnitudes. The $T_{\rm eff}$ and $\log g$ values have been 
derived using 3D model atmospheres \citep{2018MNRAS.481.1522C}. The letter in parentheses in the fourth column corresponds
to the filter of the magnitude.   Specifically, ``g'' is SDSS g magnitude and ``GG'' is Gaia G magnitude.} 
%The list does not include the variables found by \citet{2019MNRAS.486.4574R} because those do not have spectral-temperature and gravity determinations.
\begin{tabular}{lccl}
\hline
\noalign{\smallskip}
Name & $T_{\rm eff}$ (K) & $\log g$ & Magnitude \\
\noalign{\smallskip}
\hline                
\noalign{\smallskip}
KUV 05134+2605            & 24680 & 8.21  & 16.70 (g)   \\
CBS 114                   & 26050 & 7.98  & 17.50 (g)   \\
PG 1115+158               & 23770 & 7.91  & 16.61 (g)   \\
PG 1351+489               & 26010 & 7.91  & 16.52 (g)   \\
PG 1456+103               & 24080 & 7.91  & 16.24 (g)   \\
GD 358                    & 24940 & 7.75  & 13.47 (g)   \\
PG 1654+160               & 29410 & 7.97  & 16.40 (g)   \\
PG 2246+121               & 27070 & 7.92  & 16.54 (g)   \\
EC 20058$-$5234             & 25500 & 8.01  & 15.58 (g) \\
SDSS J034153.03$-$054905.8  & 25087 & 8.02  & 18.25 (g)   \\
SDSS J085202.44+213036.5  & 25846 & 8.02  & 18.23 (g)    \\
SDSS J094749.40+015501.8  & 23453 & 8.13  & 19.95 (g)  \\
SDSS J104318.45+415412.5  & 26291 & 7.77  & 18.95 (g)  \\
SDSS J122314.25+435009.1  & 23442 & 7.84  & 18.98 (g) \\
SDSS J125759.03$-$021313.3  & 25820 & 7.57  & 19.16 (g) \\
SDSS J130516.51+405640.8  & 24080 & 8.14  & 17.46 (g) \\
SDSS J130742.43+622956.8  & 23841 & 8.14  & 18.82 (g) \\
SDSS J140814.63+003838.9  & 26073 & 7.98  & 19.19 (g)\\
PG 0112+104               & 31300 & 7.8   & 15.21 (g)   \\
EC 04207$-$4748             & 25970 & 7.79  & 15.25 (GG)\\
EC 01585$-$1600             & 25500 & 7.88  & 14.40 (g) \\
WD J192904.6+444708       & 28480 & 7.89  & 18.38 (g) \\
EC 05221$-$4725             & 27900 & 7.78  & 16.72 (GG)    \\
%SDSS J011607.92+330154.29 & 21040 & 8.03 & 18.87 (g) \\ Zach - not published
%SDSS J025352.96+332803.60 & 27557 & 7.72 & 18.82 (g)  \\Zach
%SDSS J073935.14+244505.27 & 21498 & 7.90 & 17.29 (g) \\Zach
%SDSS J080236.92+154813.58 & 21395 & 7.97 & 17.42 (g) \\Zach
%SDSS J081345.43+365140.50 & 27065 & 7.61 & 18.80 (g) \\Zach
%SDSS J081453.56+300734.90 & 22632 & 8.02 & 18.67 (g)\\Zach
%SDSS J083035.14+564459.40 & 23498 & 7.83 & 17.32 (g) \\Zach
%SDSS J083415.45+254819.90 & 22846 & 7.98 & 18.33 (g) \\Zach
%%DSS J084211.30+461819.00 & 24778 & 8.00 & 18.47 (g) \\Zach
%SDSS J101502.95+464835.30 & 23457 & 7.82 & 18.58 (g) \\Zach
%SDSS J110235.85+623416.10 & 23158 & 7.83 & 17.74 (g) \\Zach
%SDSS J131646.02+414639.00 & 21844 & 7.95 & 17.24 (g) \\Zach
  SDSS J102106.69+082724.8 &  21629   &  7.96 &      17.82 (GG) \\ %Rowan19   DBV
  WD J025121.71-125244.85  & 16567** &   7.63  &     18.23 (GG) \\ %Rowan19   DBVa
  SDSS J123654.96+170918.7 &  17673** &  7.51 &      18.17 (GG) \\ %Rowan19   DBV
  SDSS J132952.64+392150.5 &  20203** &  7.98 &      18.01 (GG) \\ %Rowan19   DBVa
\noalign{\smallskip}
\hline
**Photometric determination only, with large uncertainty.
\end{tabular}
\label{DBVs}
\end{table*}

The long quasi-continuous observations of the {\it Kepler} satellite
led to the unexpected discovery of outbursts in the WD pulsations by
\citet{2015ApJ...809...14B}, bringing another dimension to 
pulsation studies
\citep{2015ApJ...810L...5H, 2016ApJ...829...82B,
  2017ApJS..232...23H,2017ASPC..509..303B}, probably caused by
resonant mode coupling of the pulsations \citep[][see
  Sect. \ref{sec:3.5}]{2018ApJ...863...82L}. Still another benefit
from the long quasi-continuous data was the possibility to estimate,
for the first time, the natural width of the pulsation mode peaks
(related to the stability) and show that some modes are less stable than others.
This had been hinted at from the WET observations of
GD~358 \citep{2003A&A...401..639K} and PG~1159$-$035
itself \citep{2008A&A...477..627C}, but now shown in detail
\citep{2017ApJ...835..277H}. One of the major observational results from 
the {\it Kepler} mission and its K2 extension is the discovery of a  
dichotomy of mode widths: low-frequency modes with periods greater than 
roughly 800 s  are generally incoherent over the length of observations, 
while higher-frequency  modes are observed to be much more 
stable in phase and amplitude \citep{2017ApJS..232...23H}. 
Another example from space data is the
K2 Campaign 10 on the GW Vir star PG~1159$-$035, with almost 50 days of coverage,
showing at least 189 frequencies, and strong nonlinear effects
throughout the observations.  Finally, we mention the interesting 
finding of amplitude and frequency variations of the components of the triplets 
of frequencies caused by stellar rotation in the DBV star KIC 08626021 by \cite{2016A&A...585A..22Z}. 
The observed modulations of the frequencies of this star are the clearest 
hints of nonlinear resonant couplings occurring in WDs stars identified so far.  
Resonant mode coupling signals have been also detected in the pulsating sdB star 
KIC 10139564 observed with the {\it Kepler} satellite  \citep{2016A&A...594A..46Z}.
These findings open a new window on the study of non-linear theory of stellar pulsations.

The successor of {\it Kepler} is the Transiting Exoplanet Survey Satellite \citep[TESS,][]{2015JATIS...1a4003R},
composed of four 10.5~cm entrance pupil $24\deg\times 24\deg$ cameras,
with 2 minute cadence for guest observer mode, lasting at least 27
days. It will observe the 200\,000 brightest stars in 85\% of the
whole sky in two years. WD observations are limited because of their
faintness.  The main effort to get most of the brightest ($V < 16$)
known WDs (and hot subdwarfs) observed within the TESS Asteroseismic Science Consortium (TASC)
reserved 2-minute cadence slots has been coordinated by the 
TASC WG8\footnote{Working Group 8: Evolved compact stars with TESS ({\tt https://tasoc.dk/wg8/}).}.
Independent efforts to find new pulsators among the bright targets in the WG8 
lists have also been done. As an example of this, \citet{2018MNRAS.478.2676B} 
started looking for target WDs and there are a few WD pulsators down to $V= 16.3$ already
observed starting September 2018.  There are no exclusive use data
rights on TESS data, and the light curves are accessed through the
Mikulski Archive for Space Telescopes (MAST), similar to the {\it
  Kepler} and K2 mission data.

The discovery of pulsations in ELMs
\citep{2012ApJ...750L..28H} and pre-ELMs
\citep{2014MNRAS.444..208M} (see Sect. \ref{sec:3.6}), and even Blue
Large-Amplitude Pulsators  \citep[BLAPs,][see
  Sect. \ref{sec:3.7}]{2017NatAs...1E.166P}, is another chapter on the
asteroseismological study of the by-products of binary
interactions. DA WD stars with masses $M_\star \leq 0.45~M_\odot$ and
$T_{\rm eff} < 20\,000$~K are Low Mass (LM) WDs,  and if $M_{\star}
\lesssim 0.18-0.20 M_{\odot}$ they are  called Extremely Low Mass (ELM)
WDs. They have been found by \citet{2011ApJ...727....3K}, \citet{2010ApJ...723.1072B,2012ApJ...744..142B,2013ApJ...769...66B,
2016ApJ...818..155B}, and \citet{2014ApJ...794...35G,2015ApJ...812..167G}. 
\cite{2012ApJ...750L..28H, 2013ApJ...765..102H,2013MNRAS.436.3573H,
  2015MNRAS.446L..26K,2018MNRAS.479.1267K,2017ApJ...835..180B,
  2018A&A...617A...6B, 2018MNRAS.478..867P,2019MNRAS.482.3831P}
found pulsations in eleven of these ELMs, similar to the $g$-mode
pulsations seen in DAVs \citep{2013ApJ...762...57V}, although with
much longer periods. \citet{2014MNRAS.437.1681M} found 17 pre-ELMs,
i.e., He--core WD precursors, and
\cite{2013Natur.498..463M,2014MNRAS.444..208M} and \cite{2016ApJ...822L..27G}
 report  pulsations in
five of them. In Tables \ref{elmv} and \ref{pre-elmv} we show the
list of ELMVs and pre-ELMVs, respectively,  detected at the
time of writing this review. 

\begin{table}[!ht]
\caption{ELMV stars and their effective temperatures, 
surface gravities, magnitudes,  and period ranges. The $T_{\rm eff}$ and $\log g$ values have been derived using 3D model atmospheres. The letter in parentheses in the fourth column corresponds to the filter of the magnitude.}
\smallskip
\begin{center}
{\small
\begin{tabular}{lcccc}
\hline
\noalign{\smallskip}
Name & $T_{\rm eff}$  & $\log g$ &  Magnitude & Period range\\
       &          [K]             &    [cgs]           &         & [s]              \\
\noalign{\smallskip}
\hline
\noalign{\smallskip}
SDSS J184037.78+642312.3    & 9100 & 6.22  & 18.91 (g)&  $2094-4890$  \\%arXiv:1204.1338\\
SDSS J111215.82+111745.0    & 9590 & 6.36  & 16.35 (g)&  $ 108-2855$  \\%arXiv:1211.1022\\
SDSS J151826.68+065813.2    & 9900 & 6.80  & 17.54 (g)&  $1335-3848$  \\%arXiv:1211.1022\\
SDSS J161412.28+191219.4(*) & 8880 & 6.66  & 16.40 (g)&  $1184-1263$ \\%arXiv:1310.0013\\
SDSS J222859.93+362359.6(*) & 7870 & 6.03  & 16.83 (g)&  $3254-6235$ \\%arXiv:1310.0013\\
PSR J173853.96+033310.8     & 9130 & 6.55  & 21.3  (V)&  $1788-4980$  \\%rXiv:1410.4898\\
SDSS J161831.69+385415.2    & 9144 & 6.83  & 19.84 (g)&  $5000-6100$  \\%arXiv:1805.11129\\
SDSS J173521.69+213440.6(*) & 7940 & 5.76  & 16.12 (g)&  $3363-4961$ \\%arXiv:1612.06390\\
SDSS J213907.42+222708.9(*) & 7990 & 5.93  & 15.92 (g)&  $2119-3303$ \\%arXiv:1612.06390\\
SDSS J134336.44+082639.4    & 8100 & 5.97  & 16.27 (g) &  $3600$       \\%arXiv:1804.09059\\
SDSS J222009.74-092709.9(*) & 8230 & 6.10  & 15.84 (g)&  $2169-3591$ \\%arXiv:1804.09059\\
\noalign{\smallskip}
\hline
\label{elmv}
\end{tabular}
}
\end{center}
\begin{center}
{\footnotesize  (*) Probably not binary.}
\end{center}
\end{table}

\begin{table}[!ht]
\caption{Pre-ELMV stars and their effective temperatures, 
surface gravities,  magnitudes, and period ranges. The letter in parentheses in the fourth 
column corresponds to the filter of the magnitude.}
\smallskip
\begin{center}
{\small
\begin{tabular}{lcccc}
\hline
\noalign{\smallskip}
 Star &  $T_{\rm eff}$ & $\log g$    & Magnitude & Period range  \\ %&  Ref.\\
      &          [K] &    [cgs]      &  & [s]          \\ %&     \\
\noalign{\smallskip}
\hline
\noalign{\smallskip}
SDSS J115734.46+054645.6   & $11\,870$ & $4.81$      & 19.9 (g) & $364$          \\ %& (2) \\  
SDSS J075610.71+670424.7   & $11\,640$ & $4.90$      & 16.3 (g) & $521-587$     \\ %& (2) \\
WASP J024743.37$-$251549.2 & $11\,380$ & $4.576$     & 12.1 (g) & $380-420$      \\ %& (3) \\
SDSS J114155.56+385003.0   & $11\,290$ & $4.94$      & 19.1 (g) & $325-368$     \\ %& (2) \\
KIC  9164561(*)            & $10\,650$ & $4.86$      & 13.7 (g) & $3018-4668$   \\ %& (4) \\   
WASP J162842.31+101416.7   & $9200$    & $4.49$      & 13.0 (g) & $668-755$     \\ %& (5) \\
SDSS J173001.94+070600.25(*)  & $7972$ & $4.25$      & 16.4 (g) & $3367$       \\ %& (6) \\
SDSS J145847.02+070754.46(*)  & $7925$ & $4.25$      & 15.2 (g) & $1634-3279$  \\ %& (6) \\
SDSS J131011.61$-$014233.0(*)  & $8224$ & $5.33$     & 16.6 (g) & $2100-3100$ \\ %&  (1)\\        
SDSS J075738.94+144827.50(*)  & $8180$ & $4.75$      & 15.0 (g) & $803-2982$   \\ %&  (7) \\  
\noalign{\smallskip}
\hline
\label{pre-elmv}
\end{tabular}
}
\end{center}
\begin{center}
{\footnotesize  (*) Unconfirmed as a pre-ELM WD (see text)}.
\end{center}
%;
%(1) \cite{2018A&A...617A...6B};
%(2) \cite{2016ApJ...822L..27G}; 
%(3) \cite{2013Natur.498..463M}; 
%(4) \cite{2016ApJ...821L..32Z};  
%(5) \cite{2014MNRAS.444..208M}; 
%(6) \cite{2016A&A...587L...5C}; 
%(7) \cite{2018A&A...616A..80S} 
%\end{center}
\end{table}

The prototype  of  the DQV class, SDSS  J142625.71+575218.3,   was
discovered by \cite{2008ApJ...678L..51M}.  \citet{2008ApJ...688L..95B}
and \citet{2010ApJ...720L.159D} found three others,
SDSS~J220029.08-074121.5, SDSS~J234843.30-094245.3, and
SDSS~J133710.19-002643.6. \cite{2011ApJ...733L..19D} discovered
the fifth DQV star, SDSS J115305.55+005646.2. \citet{2013ApJ...769..123W} discuss the
variability in the strongly magnetic DQ WD SDSS J103655.39+652252.2,
but it is not evident if these are  {\it bona fide} pulsations or more
probably caused  by cool spots in rotating magnetic stars
\citep{2016ApJ...817...27W,2018MsT.........14D}. At present, 6 stars 
of this category exist (see Table \ref{tabla-DQV}).

\begin{table}[!ht]
\caption{DQV stars and their effective temperatures, 
surface gravities, stellar masses, magnitudes, period ranges, and remark. 
The letter in parentheses in the fifth 
column corresponds to the filter of the magnitude.}
\smallskip
\begin{center}
{\small
\begin{tabular}{lcccccc}
\hline
\noalign{\smallskip}
 Star &  $T_{\rm eff}$ & $\log g$  & $M_{\star}$  & Magnitude & Period range & Remark \\ 
      &          [K] &    [cgs]  & [$M_{\odot}$] &  & [s]                     & \\
\noalign{\smallskip}
\hline
\noalign{\smallskip}
SDSS J142625.70+575218.4   & $19\,800$ & $9$ & $\gtrsim 1.0$ & 19.16 (g) & $209-418$ & Magnetic \\   
SDSS J220029.08$-$074121.5 & $21\,240$ & $8$ & $0.60$ & 17.70 (g) & $327-653$ & Magnetic \\                  
SDSS J234843.30$-$094245.3 & $21\,550$ & $8$ & $0.59$ & 19.00 (g) & $417-1044$ & $-$\\
SDSS J133710.19$-$002643.6 &    $\cdots$       & $\cdots$    &    $\cdots$    & 18.70 (g) & $326-341$  & $-$\\
SDSS J103655.39+652252.2   & $15\,500$ & $9$ & $\gtrsim 1.0$ & 18.51 (g) & $1116$ & Magnetic \\
SDSS J115305.55+005646.2   &    $\cdots$       &     $\cdots$ &  $\cdots$     &   18.97 (g) & $159-375$  & $-$\\  
\noalign{\smallskip}
\hline
\end{tabular}
}
\label{tabla-DQV}
\end{center}
\end{table}

WDs in cataclysmic variables (CVs) also show pulsations,
as discovered by \citet{1998IAUS..185..321W} in the quiescent phase of
the dwarf nova GW Librae. Since then, pulsations have been
detected in nearly twenty dwarf novae \citep[see][]{2013ASPC..469...31S,2015ASPC..493..205S, 2016MNRAS.459.3929T}, that define the GW Lib class of pulsating WDs. 
The prototype star, GW Librae, showed nonradial pulsations during
quiescent phases before and after the 2007 nine magnitude outburst.
The location of the instability strip of the class of GW Lib stars is affected by accretion,
which changes the temperature structure of the envelope and therefore
the driving region. Theoretical models indicate excitation of $g$
modes due to the H/HeI ionization zone, but enhanced He abundance due to accretion 
also can drive pulsations as a result of the subsurface HeII partial ionization zone 
\citep{2006ApJ...643L.119A,2015A&A...575A.125V}. On the other hand, 
\citet{2019MNRAS.tmp.1346S} proposes that toroidal $r$ modes are also detected
---analyzing the substructure of the pulsation peaks--- and concludes the outburst in 
the star GW Librae increased the rotation rate of the H envelope significantly.

\begin{figure}
\includegraphics[width=1.0\textwidth]{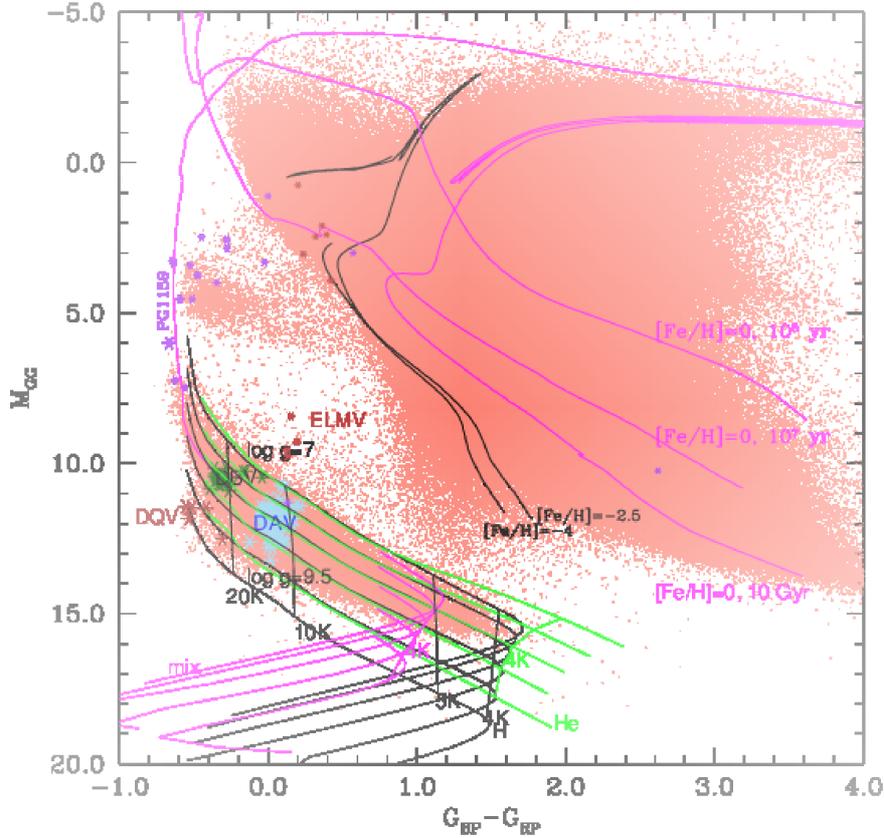}
\caption{Colour--magnitude diagram for the known variables DAVs, DBVs,
  pulsating PG1159 stars (including PNNVs), ELMVs, and variable DQVs,
  from Gaia DR2 distances and colours. The background is composed of
  the stars in Gaia DR2 with parallax/error $> 10$. The lines are MIST \citep[MESA Isochrones \& Stellar Tracks;][]{2018ApJ...863...65C}
  isochrones with  solar metallicity, [Fe/H]= 0, and also [Fe/H]= $-2.5$, and 
  [Fe/H]= $-4$, where [Fe/H]= [$\log({\rm Fe/H})_{\star}-\log({\rm Fe/H})_{\odot}$]. 
  The adopted age is for the low-metallicity isochrones. Three different ages are considered in the solar case. In the WD region, the lines are the pure H, pure He, and mixed models from
  \citet{2014A&A...565A..11C}. The Gaia DR2 data is less reliable for
  stars fainter than $M_{\rm G} > 19$.  This is because both, 
  the photometry and the parallax available on Gaia DR2 are less reliable 
  because of background noise in the Gaia data.}
\label{gaia}
\end{figure}

Gaia Data Release 2 \citep[DR2;][]{2018A&A...616A..10G, 2018A&A...616A...1G}
is a new source of detailed information for distances, and therefore
radii and masses, for the known WDs. It also allowed the discovery of
a few hundred thousand probable WDs
\citep{2019MNRAS.482.4570G,2019MNRAS.486.2169K}, and consequently of WD
 pulsator candidates, as pulsation appears to be a natural phenomenon 
during WD cooling.  Once Gaia make spectrophotometry and light curves available, it will be possible to confirm WD pulsators directly without the use of dedicated follow-ups, although it depends on the number of measurements and time scale 
(frequency) of the Gaia measurements. Fig.~\ref{gaia} shows the colour-magnitude diagram
for the stars detected with parallax/error $> 10\sigma$ in Gaia DR2,
and the location of the WD variables. Recently, by using observations with the Gaia 
satellite, \cite{2019Natur.565..202T} have confirmed that the cores of 
cooling WDs undergo crystallization, as predicted half a century ago \citep[][]{1968ApJ...151..227V}. 
 
In the case of pulsating PG~1159 stars, also called GW~Vir
stars, Table~3 in \cite{2008PASP..120.1043F} \citep[see also Table 5
  of][]{2010A&ARv..18..471A}, requires only minor revisions. 
We have to add two objects to the list. On one hand, the GW Vir
star  SDSS J075415.12+085232.18, discovered by
\cite{2014MNRAS.442.2278K} and characterized by $T_{\rm eff}=
120\,000 \pm 10\,000$ K, $\log g= 7.0 \pm 0.3$, $M_{\star}= 0.52 \pm
0.02 M_{\odot}$, and a dominant period of $\sim 525$ s. On the other hand,
the GW Vir star SDSS J0349$-$0059, discovered to be pulsating by  \cite{2012MNRAS.426.2137W}, and characterized by $T_{\rm eff}=
90\,000 \pm 900$ K, $\log g= 7.5 \pm 0.01$, $M_{\star}= 0.543 \pm
0.004 M_{\odot}$, and periods in the range $[301-964]$ s. 
This star was analyzed asteroseismologically by \cite{2016A&A...589A..40C}.
Finally, we have  to subtract VV47 ---a supposed PNNV star reported to be
variable by \cite{2006A&A...454..527G}--- from the list of GW Vir
stars. In fact, a re-analysis by \cite{2018MNRAS.479.2476S}  revealed
that this star does not show variability according to the new
photometric  data, but neither with the original data of its discovery!

Another issue that has experienced a remarkable development is the
derivation of the rotation rates of WDs through asteroseismology. 
Slow, solid-body rotation of a star produces a set of
equally spaced frequencies, with a separation between each component of the multiplet given by $\delta \sigma_{k, \ell, m}= m\ (1-C_{k\ell})\ \Omega_{\rm R}$. This rotational splitting of frequencies is found in a large number 
of pulsating WDs. This allows to estimate the rotation period of the star
($P_{\rm R}= 2 \pi\ /\Omega_{\rm R}$) as well as to identify the
harmonic degree $\ell$ and the azimuthal order $m$ of the modes
\citep[see][for details]{2010A&ARv..18..471A}. 
In general, the measured rotation period for single WDs ranges from 1~h
to 18~d, with a median  around 1~d \citep{2015ASPC..493...65K}. We
show in Table \ref{rot} a  compilation of pulsating WDs and pre-WDs
with rotation periods derived from asteroseismology. The fastest
single WD rotator from asteroseismological measurements
(Table~\ref{rot}) is the $0.79~M_\odot$ DAV SDSS~J161218.08+083028.1
discovered by \cite{2013MNRAS.430...50C}, assuming that the two
observed periods at 115.0~s and 117.0~s are two components of a
rotation triplet. This star seems to be rotating with a period of $0.93$ h! 
Differential rotation in WDs was studied by
\citet{1991ApJ...378..326W,1994ApJ...430..839W, 1999ApJ...516..349K, 2009Natur.461..501C,2011MNRAS.418.2519C,2013EPJWC..4305011F}
and \citet{2017ApJ...835..277H}, using the change in rotation
splitting of non-radial pulsations. In particular, 
\cite{1999ApJ...516..349K} \citep[see, also,][]{2011MNRAS.418.2519C} 
employed asteroseismological inversion methods for the inversion of the rotation 
profiles of GW Vir and V777 Her stars. \citet{2017ApJ...835..277H} have 
been able to assess, for the first time in a systematic way, WD rotation as a 
function of the stellar mass. In particular, they found that WDs with masses between 
$0.51$ and $0.74 M_{\odot}$ 
have a mean rotation period of 35 hr. The assessment of the rotation periods of WDs through 
asteroseismology provides final boundary conditions to the internal angular-moment evolution 
of isolated stars. 
The longest rotation period determined is for the hot polluted ZZ Ceti (DAV) GD~133
by \citet{2019MNRAS.486.3560F}.
In particular, the WD rotation rates as a function of mass can shed 
light on the unknown angular-momentum transport mechanism coupling red-giant cores to 
their envelopes \citep{2012ApJ...756...19D,2013ApJ...775L...1T,2014A&A...572L...5M,2014ApJ...788...93C,2019A&A...622A.187D}.
 The longest rotation period determined is for the hot polluted (DAZ) ZZ Ceti (DAV) GD~133
by \citet{2019MNRAS.486.3560F}. The metals imply ongoing accretion of planetesimals.

\begin{table*}
\caption{Rotation periods of WDs as determined via asteroseismology. The $T_{\rm eff}$  
values have been converted to 3D model atmospheres.}
{\begin{tabular}{lcccc}
\noalign{\smallskip}
\hline
\noalign{\smallskip}
Star & $P_{\rm rot}$ [h]  & $T_{\rm eff}$ [K] & Type & $M_{\star}$ [$M_{\odot}$]\\
\noalign{\smallskip}
\hline
\noalign{\smallskip}
RX J2117.1+3412  & 28  & 170\,000 & GW Vir & 0.72 \\
PG 1159$-$035   & 33  & 140\,000 & GW Vir & 0.54 \\
NGC 1501  & 28  & 134\,000 & [WCE] & 0.56 \\
PG 2131+066   &  5  &95\,000 & GW Vir & 0.55 \\
PG 1707+427   & 16  & 85\,000 & GW Vir & 0.53 \\
PG 0122+200   & 37  & 80\,000 & GW Vir & 0.53 \\
SDSS J0349$-$0059 & 9.8 & 90\,000 & GW Vir & 0.54\\
\noalign{\smallskip}
\hline
\noalign{\smallskip}
PG 0112+104 & 10.17 & 31\,040 & DBV & 0.58 \\
KIC 8626021 & 43  & 29\,700 & DBV & 0.56 \\
EC 20058$-$5234 &  2  & 25\,500 & DBV & 0.65 \\
GD 358    &   29  & 23\,740 & DBV & 0.54 \\
\noalign{\smallskip}
\hline
\noalign{\smallskip}
SDSS J083702.16+185613.4 & 1.13 & 13\,590 & ZZ Ceti & 0.88 \\
G 226$-$29                 &  9 & 12\,510 & ZZ Ceti & 0.83 \\
G 185$-$32                 & 15 & 12\,470 & ZZ Ceti & 0.67 \\
 GD 133                     & 169 & 12\,400 (*) & ZZ Ceti & 0.63 \\
%\footnote{Asteroseismological value of of $\Teff$\ and $\logg$\ because of %the observed metal pollution.}& ZZ Ceti & 0.63 \\ %Fu+19}
SDSS J113655.17+040952.6 & 2.6 & 12\,330 & ZZ Ceti & 0.55 \\
SDSS J161218.08+083028.1 & 0.93 & 12\,330 & ZZ Ceti & 0.79 \\
Ross 548                 & 37 & 12\,300 & ZZ Ceti & 0.63 \\
%HS 0507+0435             & 38   &  12\,290 & ZZ Ceti & 0.675 \\*** %Fu+13
GD 165                   & 50 & 12\,220 & ZZ Ceti & 0.68 \\
LP 133$-$144               & 41.8 & 12\,150 &ZZ Ceti & 0.59 \\
KIC 11911480             & 86.4 & 12\,160 &ZZ Ceti & 0.58 \\
L 19$-$2                   & 13 & 12\,070 & ZZ Ceti & 0.69 \\
HS 0507+0435             &  38   &  12\,010 & ZZ Ceti & 0.73 \\ %Fu+13 Prot
EC 14012$-$1446            & 14.4 & 12\,020 & ZZ Ceti & 0.72 \\
KUV 11370+4222           & 5.56 &  11\,940 & ZZ Ceti & 0.72 \\
G 29$-$38                  & 32 &   11\,910 & ZZ Ceti & 0.72 \\
EPIC 220274129             & 12.7 & 11\,810 & ZZ Ceti & 0.62 \\
KUV 02464+3239           & 90.7 & 11\,620 & ZZ Ceti & 0.70 \\ 
HL Tau 76                & 53   &  11\,470 & ZZ Ceti & 0.55 \\
SDSS J171113.01+654158.3 & 16.4 &  11\,130 & ZZ Ceti & 0.90\\
GD 154                   & 50.4 &  11\,120 & ZZ Ceti & 0.65 \\
KIC 4552982              & 15.0 &  10\,860 & ZZ Ceti & 0.71 \\
SDSS J094000.27+005207.1 & 11.8 & 10\,590 & ZZ Ceti & 0.82 \\
KIC  435703719   &  22.0  & 12\,650 & ZZ Ceti & 0.62 \\
KIC  455298219   &  18.4  & 10\,950 & ZZ Ceti & 0.67 \\
KIC  759478119   &  26.8  & 11\,730 & ZZ Ceti & 0.67 \\
KIC  1013270219  &  11.2  & 11\,940 & ZZ Ceti & 0.68 \\
KIC  1191148019  &  74.7  & 11\,580 & ZZ Ceti & 0.58 \\
EPIC 6001783623  &  6.9   & 10\,980 & ZZ Ceti & 0.57 \\
EPIC 20171957811 &  26.8  & 11\,070 & ZZ Ceti & 0.57 \\
EPIC 20173081111 &  2.6   & 12\,480 & ZZ Ceti & 0.58 \\
EPIC 20180293311 &  31.3  & 12\,330 & ZZ Ceti & 0.68 \\
EPIC 20180600811 &  31.3  & 10\,910 & ZZ Ceti & 0.61 \\
EPIC 21039746503 &  49.1  & 11\,200 & ZZ Ceti & 0.45 \\
EPIC 21159664908 &  81.8  & 11\,600 & ZZ Ceti & 0.56 \\
EPIC 21162969708 &  64.0  & 10\,600 & ZZ Ceti & 0.48 \\
EPIC 21191418508 &  1.1   & 13\,590 & ZZ Ceti & 0.88 \\
EPIC 21192643009 &  25.4  & 11\,420 & ZZ Ceti & 0.59 \\
EPIC 22868247808 &  109.1 & 12\,070 & ZZ Ceti & 0.72 \\
EPIC 22922729213 &  29.4  & 11\,210 & ZZ Ceti & 0.62 \\
EPIC 22020462601 &  24.3  & 11\,620 & ZZ Ceti & 0.71 \\
EPIC 22025880601 &  30.0  & 12\,800 & ZZ Ceti & 0.66 \\
EPIC 22034775900 &  31.7  & 12\,770 & ZZ Ceti & 0.66 \\
EPIC 220274129   &  12.7  & 11\,810 & ZZ Ceti & 0.62 \\
\noalign{\smallskip}
\hline
\end{tabular}
\label{rot}}
\begin{center}
(*) Asteroseismological value of $T_{\rm eff}$ and $\log g$ because of the observed metal polution.
\end{center}
\end{table*}

\subsection{Asteroseismic approaches}
\label{sec:3.2}

\begin{figure}
\includegraphics[width=1.0\textwidth]{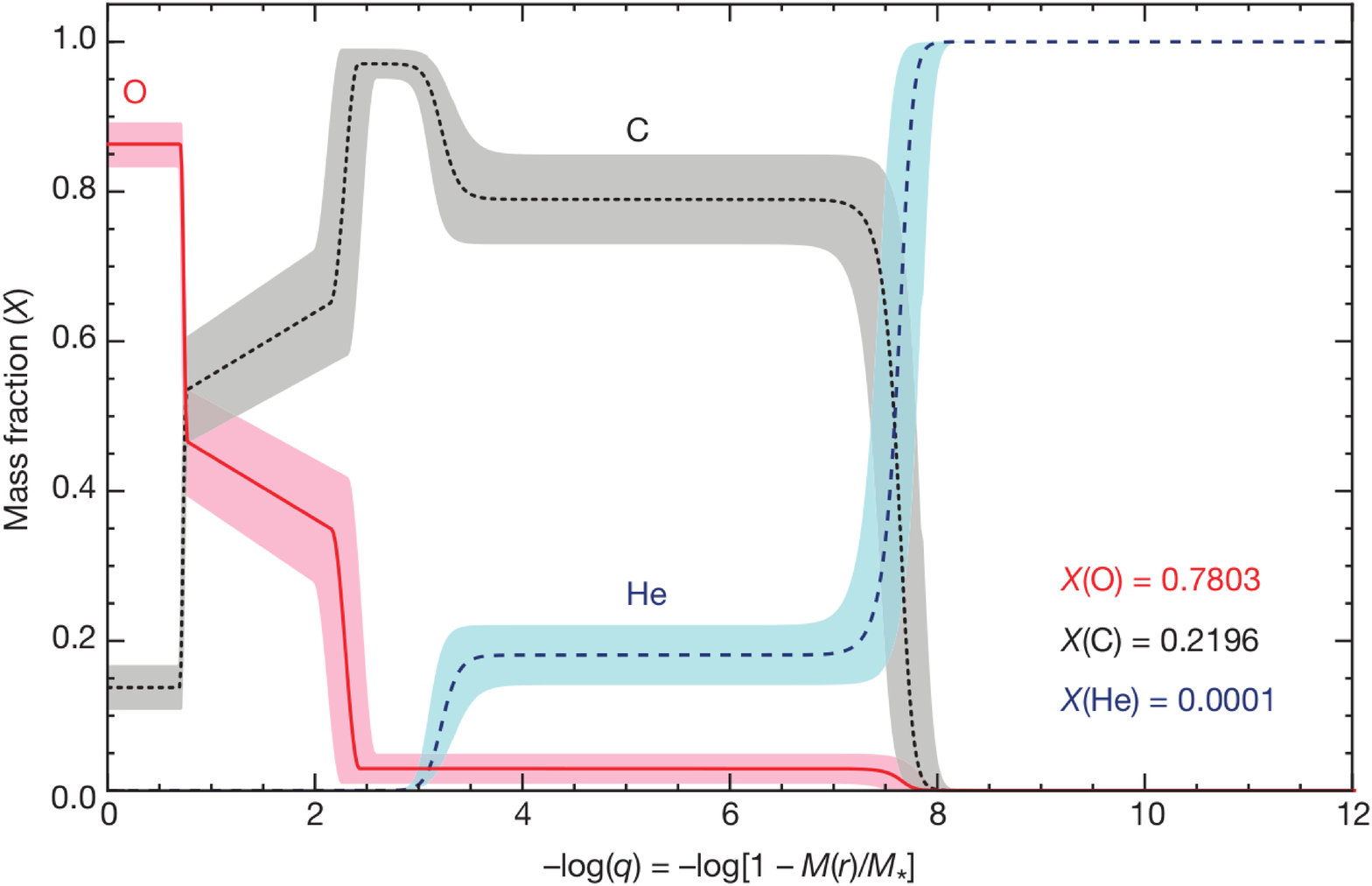}
\caption{Abundance by mass of $^{16}$O (in red), $^{12}$C (in gray)
  and $^{4}$He (in blue) in terms of the outer mass fraction,
  corresponding to the  asteroseismological model of KIC08626021. The
  shaded areas  around each curve are the  estimated $1 \sigma$
  errors. The total (integrated) mass fractions for  each chemical
  species are indicated at the bottom right of the plot
  \citep[extracted from][]{2018Natur.554...73G}.}
\label{giammichele-DBV}
\end{figure}

Up to now, two main standpoints have been adopted to perform
  forward asteroseismic modeling of pulsating WD stars.  One approach considers WD models characterized by
parameterized chemical composition profiles, while the other technique
involves fully evolutionary  models constructed with chemical profiles
resulting from all the processes experienced during the evolution of
the WD progenitors. The former approach constitutes a powerful forward
method with the flexibility of allowing a full exploration of the
parameter space (the total mass, the mass of the H and He envelopes,
the thickness of the chemical transition regions, the core chemical
structure and composition, etc) to find an  optimum
asteroseismological solution \citep[see][among
  others]{1998ApJS..116..307B,2001ApJ...552..326B,2008ApJ...675.1505B,2014ApJ...794...39B,2019ApJ...871...13B,
  2013MNRAS.429.1585F, 2016MNRAS.461.4059B,
  2017A&A...598A.109G,2017ApJ...834..136G}.
In particular, \cite{2017A&A...598A.109G,2017ApJ...834..136G} present
a new prescription for parameterizing the chemical profiles  in the
core of WDs which is based on Akima splines.  The method, in principle,
allows for period-to-period fits (theoretical periods minus observed periods)
with differences smaller than the uncertainties in the observed periods.
This technique has  been
applied by \cite{2018Natur.554...73G} to the DBV star KIC 08626021
monitored extensively  by the {\it Kepler} spacecraft. We describe
this specific application in some detail here due to the relevance of
this  asteroseismological technique. \cite{2018Natur.554...73G} have
derived the chemical stratification of $^{16}$O, $^{12}$C and $^{4}$He
in KIC08626021 using archival data.  They find an  asteroseismological
model for this star whose $^{16}$O  content and the extent of its core
exceed the predictions of current models of DB WDs derived from fully
evolutionary computations.  The chemical profiles of the
asteroseismological model are displayed  in
Fig. \ref{giammichele-DBV}. This model is characterized  by $T_{\rm
  eff}= 29\,968 \pm 200$~K and $\log g= 7.917 \pm 0.009$,  which
closely match the independent measurements obtained from  spectroscopy
($T_{\rm eff}= 29\,360 \pm 780$~K and $\log g= 7.89 \pm 0.05$).  The
central homogeneous part of the core of the asteroseismological model
has a mass of $0.45 M_{\odot}$, and is composed of $\sim 86 \%$
$^{16}$O by mass. These values are respectively $40 \%$ and $15 \%$
larger than those predicted for typical DB WD models.  The total
$^{16}$O content of the WD core reaches $78.0\pm 4.2 \%$,  much higher
than the expected value of around $64 \%$ for a standard evolutionary
DB WD model of the same mass. 

A  disturbing feature of the method described above  is that it can
lead to asteroseismological models characterized by chemical
structures that  cannot be reconciled with the predictions of the
currently accepted scenarios of WD formation.  For instance, the
derived asteroseismological models  may have a pure (or nearly pure)
$^{12}$C buffer ---which is difficult 
to  predict from the  existing
channels of WD formation; see De Ger\'onimo et al. (2019),  submitted--- or central abundances of $^{12}$C and $^{16}$O that are
at variance with the current uncertainty of the
$\rm{}^{12}C(\alpha,\gamma)^{16}O$   reaction rate (see
Fig. \ref{giammichele-DBV}). In addition, as  precise as they have
become, parameterized approaches rely on an  educated guess of the
internal composition profiles, due to the large  number of parameters
involved and the small number of periods typically
available.
%\footnote{The method seems not to follow   the
%  statistical-method limitations of number of measurements larger than
% the number of fit parameters \citep{Laplace1810}.}.  
%In particular,
%\cite{2018ApJ...867L..30T} and De Ger\'onimo et al. (2019) showed that
%asteroseismological fits to observed pulsation frequencies below the
%level of tens of $\mu$Hz by means of models with simplified thermal
%and/or chemical  structures would not be physically meaningful.

\begin{figure}
\includegraphics[width=0.9\textwidth]{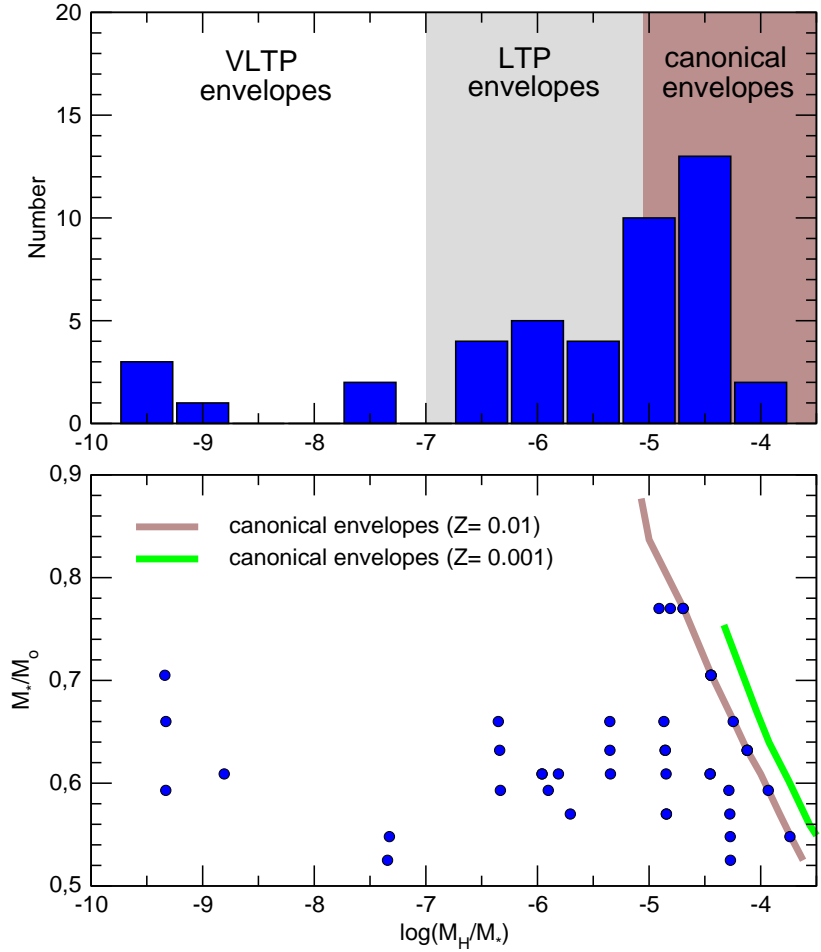}
\caption{Upper panel: histogram of the H envelope thickness
  distribution corresponding  to  the  asteroseismological  models
  of  the sample of 44  ZZ Ceti  stars analyzed in
  \cite{2012MNRAS.420.1462R}. The regions marked with different
  colors (brown, gray and white) refer to three possible  evolutionary
  origins  (single progenitors) of the WDs  according to the value of
  the  thickness of their H envelopes. The acronyms LTP and VLTP stand for 
  Late Thermal Pulse and Very Late Thermal Pulse, respectively 
  (see Section \ref{sec:2.1}). Lower panel:  stellar  mass in
  terms of the H envelope  mass  for the same sample of ZZ Ceti stars.
  The thick brown (green) line corresponds to the canonical envelopes
  for  post-AGB WDs at the beginning of the cooling track coming from
  progenitor stars with $Z= 0.01$ ($Z= 0.001$) at the ZAMS.} 
\label{fig_envelopes_H}
\end{figure}

The second asteroseismological approach for WDs  was developed at La
Plata Observatory and  is based on non-static WD evolutionary models
that result from the  complete evolution of the progenitor stars. This
method has been already applied to GW Virginis stars \citep[see, e.g.,][and references therein]{2007A&A...461.1095C,2007A&A...475..619C,
2008A&A...478..869C,2009A&A...499..257C,2016A&A...589A..40C}, to DBV WDs
\citep[see, e.g.,][]{2012A&A...541A..42C,2014JCAP...08..054C,2014A&A...570A.116B}, and 
recently to  ELMV stars
\citep{2017A&A...607A..33C,2018A&A...620A.196C}.  Regarding ZZ Ceti
stars, this approach has been employed  by
\cite{2012MNRAS.420.1462R,2013ApJ...779...58R,2017ApJ...851...60R}.
In the context of this approach, the chemical structure of WD models
is consistent with the pre-WD evolution. This is a crucial aspect
because WD models with  consistent and detailed chemical profiles from
the center to the surface are needed to correctly assess the adiabatic
pulsation periods and also  the mode-trapping properties of ZZ Ceti
stars. These chemical  profiles are computed  from the complete
evolution  of  the  progenitor  stars  from  the ZAMS,  through  the
thermally  pulsing and mass-loss phases on the AGB, and from time-dependent 
element diffusion predictions during
the WD stage.  For realistic asteroseismic inferences, this method
requires a quantification  of the uncertainties inherent to WD
progenitor evolution and an assessment of their impact on the
pulsational expectations.

A notable example of application of this second approach to DAV stars
is the asteroseismological study of 44 ZZ Ceti  stars extracted from a
sample of bright stars --- for which the surface parameters are
accurately  known --- performed by \cite{2012MNRAS.420.1462R}. These
authors employed a large grid of fully evolutionary models
characterized by consistent chemical profiles covering a wide range
of stellar masses, thicknesses of the H envelope and effective
temperatures.  The different thicknesses of the H envelope were
artificially obtained from the canonical (thickest) envelopes. One of
the main results of \cite{2012MNRAS.420.1462R} is a strong suggestion of 
the existence
of  a  {\it range}  of  thicknesses  of  the  H  envelope  in the
studied ZZ Ceti stars, with a distribution  characterized  by  a
strong  peak  at  thick  envelopes $[\log (M_{\rm H}/M_{\star}) \sim
  -4.5]$ and another much less pronounced peak at very thin envelopes
$[\log (M_{\rm H}/M_{\star}) \sim -9.5]$, with an evident paucity for
intermediate thicknesses; see upper panel of
Fig. \ref{fig_envelopes_H}.  The expected range of H-envelope
thickness, that is in agreement with the seismological  results of
\cite{2009MNRAS.396.1709C}, is extremely relevant and is in  line with
the spectroscopic finding of \cite{2008ApJ...672.1144T}, that about 15
\% of  the DA WDs with $10\,000\ {\rm K} \lesssim T_{\rm eff} \lesssim
15\,000$ K  should have H envelopes in the range $-10 \leq \log
(M_{\rm H}/M_{\star}) \leq -8$. These two pieces of evidence reinforce
the idea that a non-negligible fraction of DA WDs could harbor thin H
envelopes. Because the standard theory of stellar evolution predicts
the formation of WDs with thick H envelopes, then the following
question arises: which are the possible channels of   evolution that
can lead to the formation of DA WDs with a range of envelopes of H? In
the case of  single progenitors, the expected theoretical predictions
are displayed in Fig. \ref{fig_envelopes_H}. In the lower panel of
this figure, the stellar mass in terms of the H envelope thickness
corresponding to the asteroseismological models derived by
\cite{2012MNRAS.420.1462R} is shown. The thick brown line corresponds
to the  thickest H envelopes possible (canonical envelopes)
corresponding to progenitor stars with nearly solar metallicity, $Z=
0.01$. The thick green line, in turn, corresponds to the canonical H
envelopes of DA WD models coming from progenitor stars with  sub-solar
metallicity ($Z= 0.001$). There is a widespread belief that progenitor
metallicity  and/or enhanced winds on the AGB and H-burning post-AGB
phases  could be at the root of accounting for the existence of WDs
with very thin H envelopes \cite[see][for a thorough
  discussion]{2015A&A...576A...9A}.  From inspection of
Fig. \ref{fig_envelopes_H}, it  becomes clear that, according to fully
evolutionary computations,  the metallicity of  the WD progenitors 
cannot be the explanation for the formation of WDs  with H envelopes
thinner than the canonical ones; see  \cite{2013ApJ...775L..22M, 2015A&A...576A...9A} and
\cite{2017ASPC..509..435M}. These authors also show that neither the
action of  enhanced winds during the AGB and post-AGB phases can be
responsible for the thin H envelopes in DA WDs. Rather, the
explanation for the existence of WDs with thinner H envelopes than the
canonical ones can be found in connection with the occurrence of late
thermal pulses experienced by progenitors after departing from the
TP-AGB phase; specifically LTP and VLTP events,   as proposed by
\cite{2005A&A...440L...1A} and \cite{2017ASPC..509..435M} (see also
Section \ref{sec:2} for details). The range of H envelope mass
expected from these scenarios is schematically illustrated in the
upper panel of Fig. \ref{fig_envelopes_H}.

The sample of ZZ Ceti stars studied in \cite{2012MNRAS.420.1462R}
includes the archetypal ZZ Ceti star G117$-$B15A.   For this star, a
unique asteroseismological solution, characterized by  $T_{\rm eff}=
11\,985 \pm 200$ K, $\log g= 8.00 \pm 0.09$ and $M_{\star}= 0.593 \pm
0.007 M_{\odot}$ ---in excellent agreement with the spectroscopic
determinations--- was found. For  the  first  time,   the  degeneracy
of  the  asteroseismological solutions for this star reported by
previous studies \citep[e.g.,][]{1998ApJS..116..307B} regarding  the
thickness  of  the  H  envelope was broken. The best-fit model has a H
envelope with $M_{\rm H}= (1.25 \pm 0.7) \times 10^{-6} M_{\star}$,
about  two  orders  of  magnitude  thinner  than  the value predicted
by canonical evolutionary computations, of $M_{\rm H} \sim 10^{-4}
M_{\star}$ at this stellar-mass value. 

As stated before, reliable asteroseismological inferences of WD
parameters require an assessment of the uncertainties of the
predictions of WD progenitor evolution. In this sense,
\cite{2017A&A...599A..21D, 2018A&A...613A..46D} have
explored for the first time the impact of the occurrence of TPs in WD
progenitors,   the  uncertainty  in  the
$^{12}$C$(\alpha,\gamma)^{16}$O  cross  section,  and  the  occurrence
of extra mixing  (referred to as OV later on) during core He burning on the expected period spectrum
of ZZ Ceti stars, and  also on the asteroseismologically-derived
parameters of ZZ Ceti stars.

\begin{figure}
\includegraphics[width=1.0\textwidth]{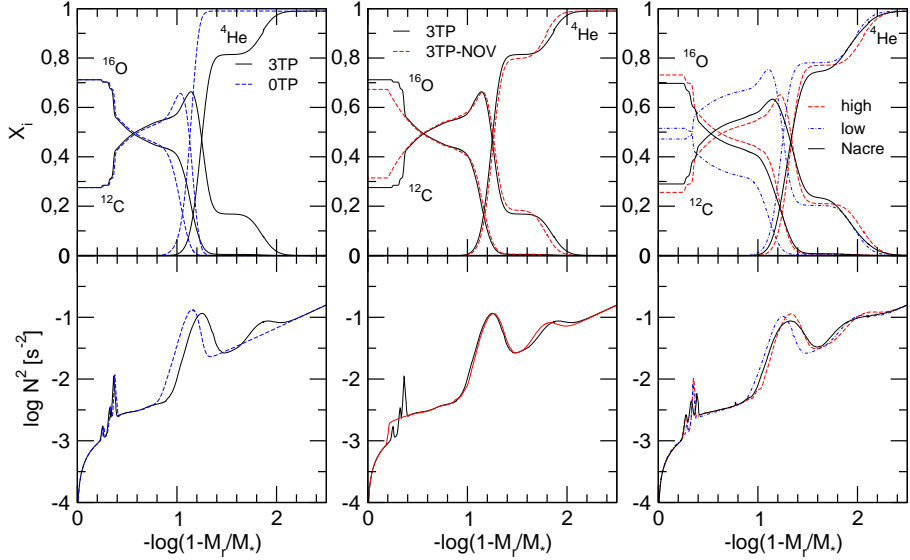}
\caption{Left panels: inner $^{16}$O, $^{12}$C and $^{4}$He abundance
  distribution (upper panel) and the logarithm of the squared
  Brunt-V\"ais\"al\"a frequency (lower panel)  in terms of the outer
  mass fraction for two DA WD models at $T_{\rm eff} \sim 12\,000$ K. 0TP
  and 3TP refer to the expected chemical profiles  when the progenitor
  departed from the AGB before the occurrence of the first TP  and at
  the end  of the third TP, blue dashed and black solid lines,
  respectively. Middle panels:  same as left panels, but for models at
  the third TP calculated with  OV (3TP; black solid line) and without
  OV (3TP-NOV; red dashed line)  during core He burning. Right panels:
  same as left panels, but for models computed with three different
  $^{12}$C$(\alpha,\gamma)^{16}$O reaction rates:
  \cite{1999NuPhA.656....3A} (Nacre; black solid line), the higher
  (high; red dashed line), and the lower (low; blue dot-dashed line)
  reaction rates from \cite{2002ApJ...567..643K}.} 
\label{fig_uncertainties}
\end{figure}

In particular, the number of TPs experienced by the WD progenitor star
through the AGB phase is quite uncertain and depends on the rate at
which mass is lost during the TP-AGB phase, on the initial metallicity
of the progenitor star,  and the occurrence of extra mixing in the
pulse-driven convection zone.  To quantify the impact of all these
uncertainties on the pulsational properties of ZZ Ceti stars,
\cite{2017A&A...599A..21D} explored  the situation in which the
progenitor is forced to abandon the AGB phase before the occurrence of
the first TP (0TP case) and also the  case in which it experiences
three TPs (3TP case). The results are shown in the left panels of
Fig. \ref{fig_uncertainties},  in which the  chemical  profiles
(upper  panel)  for $^{16}$O, $^{12}$C, and $^{4}$He,  and the
logarithm of the squared Brunt-V\"ais\"al\"a frequency  (lower  panel)
in  terms  of  the outer mass, for a DA WD model with $M_{\star}=
0.548 M_{\odot}$,  $T_{\rm eff}\sim 12\ 000$ K, and $M_{\rm H}\sim  4
\times 10^{-6} M_{\star}$, are displayed. In the 3TP case,  there is
an  inter-shell  region   rich  in  $^{4}$He  and  $^{12}$C  at  the
bottom  of  the  $^{4}$He  buffer (see Fig. \ref{fig02}). At variance
with  this, in the 0TP situation no inter-shell region is
expected. The presence or  absence of an inter-shell region has a
non-negligible impact on the  Brunt-V\"ais\"al\"a frequency, and
consequently on the period spectrum.  Specifically, the resulting
changes in the periods (at fixed radial order, $k$)  are on  average
between  5  and  10  s. In the case of more massive models,  the
average differences reduce to $\sim 2-3$ s. On the other hand,  the
amount of OV  (central panel of Fig. \ref{fig_uncertainties}) and
different assumptions about the  $^{12}$C$(\alpha,\gamma)^{16}$O
reaction rate during core He burning (right panels of
Fig. \ref{fig_uncertainties}) impact to a lesser extent the pulsation
periods, with average period differences of about $5$ s when account
is made of the current uncertainties in OV prescription and
$^{12}$C$(\alpha,\gamma)^{16}$O reaction rate
\citep{2002ApJ...567..643K}.

The impact of the chemical structure  built up during the TP-AGB
evolution and the $^{12}$C$(\alpha,\gamma)^{16}$O reaction rate on the
stellar parameters inferred from asteroseismological period-to-period
fits of ZZ Ceti stars was studied by \cite{2018A&A...613A..46D}. These
authors found that the occurrence or not of TPs  during AGB evolution
implies an average deviation in the asteroseismological  effective
temperature ($\sigma_{T_{\rm eff}}$) of ZZ Ceti stars of at most 8 \%,
and of the order of 5 \% in the stellar mass
($\sigma_{M_{\star}}$). For the mass  of the H envelope, they find
deviations up to 2 orders of magnitude ---although  generally much
lower. These trends remain even when a sample of real ZZ Ceti stars
is considered in the analysis. Noteworthy, the mean deviations in
$T_{\rm eff}$,  $M_{\star}$ and $M_{\rm H}$ inflicted by the
uncertainties in the  $^{12}$C$(\alpha,\gamma)^{16}$O reaction rate
and  OV  are smaller than those produced
by the uncertainties in the number of TPs.

In summary,  the uncertainties in the parameters of the
asteroseismological models due to uncertainties in the chemical
profiles of WDs during the evolution of the  progenitor stars are
admittedly not negligible. However,  the impact of the uncertainties 
in the chemical structure over the asteroseismological determinations 
are quantifiable and bounded, with average values close to the observational errors of the effective temperature and surface gravity. 
These results add confidence to the  use  of  fully  evolutionary  models  with
consistent  chemical profiles in WD asteroseismology, and render this
approach as a robust way  to peer  into the internal structure of WD
stars.

As a corollary of this Section, the two asteroseismological approaches
commonly used in pulsating WDs described above are very different in
nature, but complementary to each other. On the one hand, the parametric
approach needs as a starting point certain constraints about the shape
of the chemical profiles, the physically plausible chemical abundances
in each part of the star ($X_i$), and the specific chemical species
that make up a WD ($^{1}$H, $^{4}$He, $^{12}$C, $^{16}$O, etc). This
information is provided by the approach that uses detailed
evolutionary models of WDs because in this treatment, the chemical
profiles are consistent with the previous evolutionary history. At the
same time, the parametric approach leads to seismological models with
characteristics that differ somewhat from the canonical models of WDs,
this way providing certain hints that can help to improve the
evolutionary calculations. In other words, the parametric approach is
useless if it does not take into account some robust constraints
predicted by stellar evolution for the structure of a WD, and on the
other hand, the approach using evolutionary models cannot ignore the
apparently discordant results that the parametric approach predicts
when it is applied to real stars. We conclude that it would be very
important that efforts were combined in both methods in order to make
progress in the detailed knowledge of the evolutionary origin and
internal structure of WDs.

\subsection{Massive and ultra-massive WDs}
\label{sec:3.3}

Another relevant progress in the field of WD asteroseismology in the last decade concerns the massive and ultra-massive WDs, which are key to understand physical processes
in the AGB phase, the theory of crystallization, high-field magnetic
WDs, and type Ia supernova explosions. These stars, which cluster
around well-defined peaks in the WD mass distribution at $M_{\star}
\sim 0.8 M_{\odot}$ and $M_{\star} \sim 1.10 M_{\odot}$, 
have been reported in several studies \citep[see,
  e.g.,][]{2016MNRAS.455.3413K}. Massive WDs ($0.8 \lesssim M_{\star}/M_{\odot}\lesssim 1.0$) are expected to have C/O
cores, with progenitor stars of $\sim 3-7 M_{\odot}$, while
ultra-massive WDs ($M_{\star}/M_{\odot}\gtrsim 1.0$) would result either from the evolution of single
progenitors with masses in the range $\sim 7-8.5 M_{\odot}$, and
characterized by O/Ne cores, or  from the merger of two $\sim 0.5-0.6
M_{\odot}$ WDs, and hence harboring  C/O cores
\citep{2012ApJ...749...25G}.  The internal structure and chemical
stratification of massive and ultra-massive  WDs can be, in principle,
probed through asteroseismology. Fortunately, in recent years a
significant number of massive and ultra-massive ZZ Ceti stars have been
uncovered
\citep{2005A&A...432..219K,2010MNRAS.405.2561C,2013MNRAS.430...50C,
  2013MNRAS.436.3573H,2017MNRAS.468..239C}. The ultra-massive ZZ Ceti
star BPM 37093 \citep{1992ApJ...390L..89K,2005A&A...432..219K} was the
first object of this kind to be studied in detail, opening the
opportunity to test crystallization theory
\citep[e.g.,][]{1968ApJ...151..227V} through asteroseismology
\citep{2004ApJ...605L.133M,2005ApJ...622..572B}.

\begin{figure}
\includegraphics[width=1.0\textwidth]{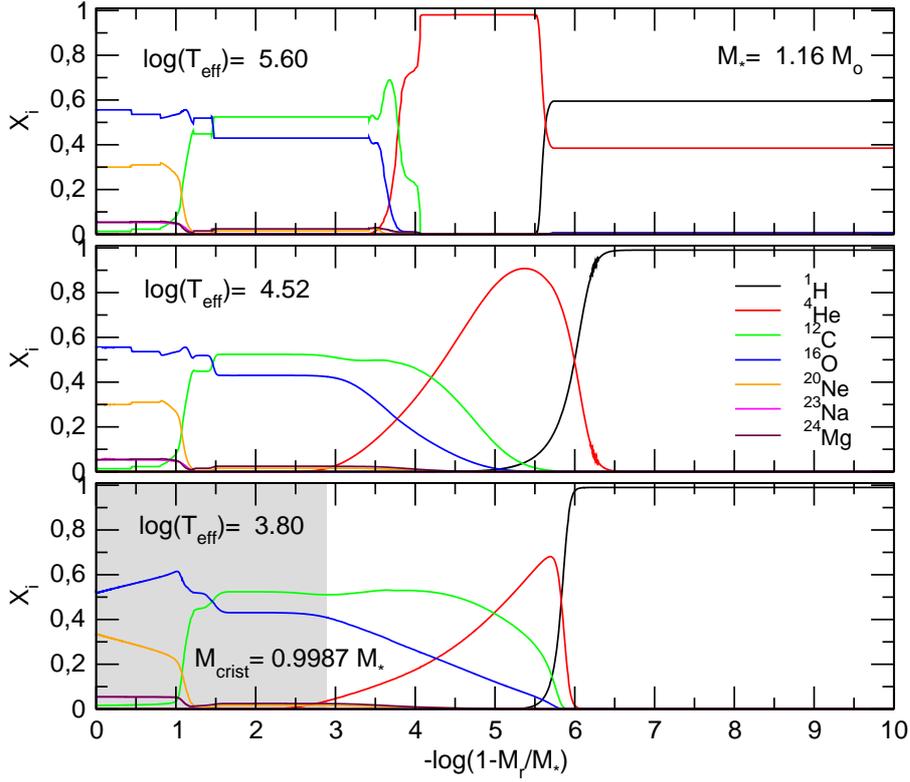}
\caption{Internal chemical profiles for $1.16 M_{\odot}$
DA WD models at three specific values of $\log(T_{\rm eff})$, 
as indicated at the left of each panel. 
The gray area in the lower panel
emphasizes the crystallized region of the model, that comprises 
$99.87 \%$ of the total mass.} 
\label{fig_profiles-122}
\end{figure}

Asteroseismological studies based on fully evolutionary models (see
Section \ref{sec:3.1})  have been done for samples of massive
and ultra-massive ZZ Ceti stars. For massive WDs,   \cite{2013ApJ...779...58R} have been able to
place constraints on the process of crystallization and phase
separation of C/O mixtures by using the phase diagrams of
\cite{1993A&A...271L..13S} and  \cite{2010PhRvL.104w1101H}. In line
with this,   \cite{2012A&A...537A..33A} found  that,  for   a  given
value of $M_{\star}$,  the  amount  of matter   redistributed  by
phase  separation  is  smaller  when  the \cite{2010PhRvL.104w1101H}
phase  diagram  is  considered  instead of the
\cite{1993A&A...271L..13S} one, leading to a smaller energy release
from the process of C/O separation.  Regarding ultra-massive WDs,
\cite{2019A&A...625A..87C} have presented  detailed evolutionary
sequences of ONe-core DA and DB WD models with stellar masses in the
range $1.10-1.29 M_{\odot}$ extracted from  the full evolution of
9-10.5 $M_{\odot}$ single progenitors from the ZAMS  through core
H and He stable burning and semi-degenerate C burning during the
thermally-pulsing Super-AGB (SAGB) phase \citep{2010A&A...512A..10S}.  In this
way,  realistic and consistent chemical profiles for the resulting WDs
can be obtained, including the  O/Ne inner profiles and the outer
chemical stratification, in particular the mass of the He inter-shell
built up during the SAGB,  a key issue as far as  the assessment of
cooling times at low luminosities is concerned. But the most novel
aspect of  these ultra-massive WD models, relevant for both
evolutionary and pulsational inferences,   is that the release of
energy and the ensuing core  chemical redistribution resulting from
the phase separation of $^{16}$O and $^{20}$Ne upon crystallization
\citep{2010PhRvE..81c6107M}   is included for the first time,  thus
substantially improving previous  attempts at modelling these stars
\citep{2004A&A...427..923C,2007A&A...465..249A}. 

\begin{figure}
\includegraphics[width=1.0\textwidth]{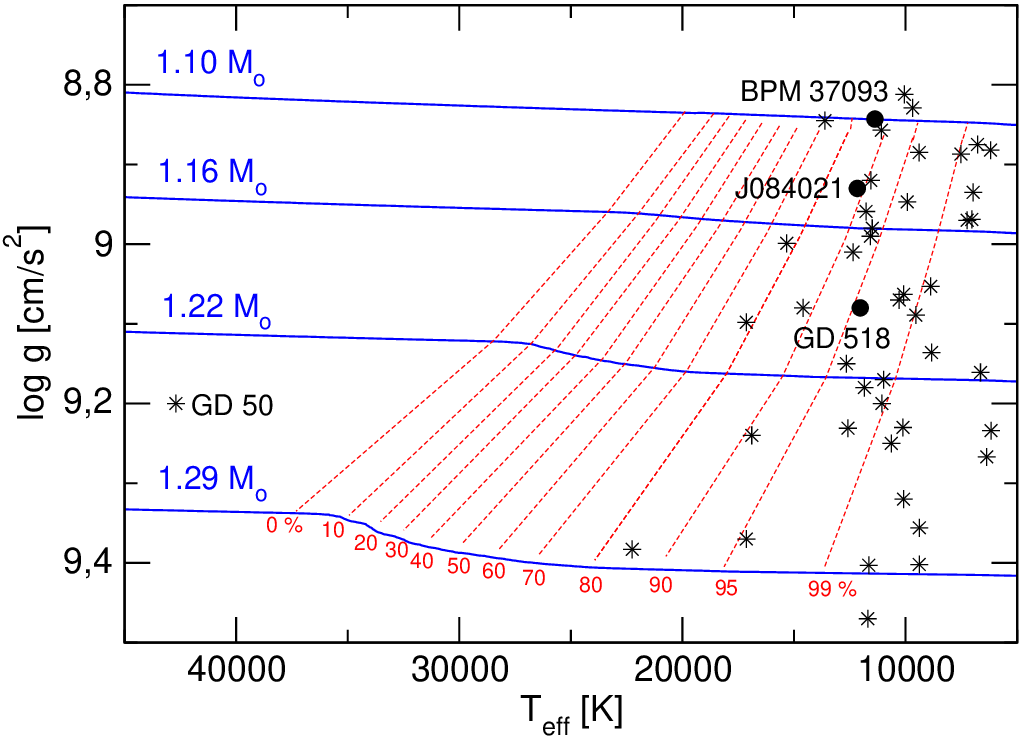}
\caption{Evolutionary tracks of the ultra-massive DA WD models 
computed by \cite{2019A&A...625A..87C} in the
  plane  $T_{\rm eff}-\log g$. Red dashed  lines indicate $0, 10, 20,
  30, 40, 50, 60, 70, 80, 90, 95$ and $99 \%$ of crystallized
  mass. The  location of ultra-massive DA WD stars
  \citep{2013ApJS..204....5K,2016MNRAS.455.3413K,2017MNRAS.468..239C}
  are indicated with black star  symbols. The black circles indicate
  the location of the known  ultra-massive ZZ Ceti stars: BPM 37093
  \citep{2016IAUFM..29B.493N}, J084021 \citep{2017MNRAS.468..239C},
  and GD 518 \citep{2013ApJ...771L...2H}.}
\label{fig-tracks-masivos}
\end{figure}

As shown by  \cite{2019A&A...625A..87C},  element diffusion strongly
modifies the inner chemical distribution of ultra-massive WDs from the
very early stages of WD evolution, as illustrated in
Fig. \ref{fig_profiles-122} for selected  $1.16 M_{\odot}$
models. Note that the heavy species are depleted from the outer layers
as a result of gravitational settling and that the initial $^{4}$He
and  $^{12}$C distribution in the  deep envelope results drastically
modified. This is in contrast to the situation of average-mass
WDs. These changes in the $^{4}$He and $^{12}$C profiles affect the
radiative opacity in the envelope and thus the cooling times at late
stages. The core chemical distribution is also strongly modified
during the WD cooling  due to phase separation during crystallization.
It can be appreciated from the bottom panel of the figure, that
corresponds to a WD model with $T_{\rm eff} \sim 6\,300$ K. Note 
that most of the star should be crystallized ($99.87 \%$,  gray zone). 

The phase diagram for $^{16}$O-$^{20}$Ne plasmas predicts that all
ultra-massive  ONe-core DAVs reported in the literature with masses
higher than $1.10 M_{\odot}$ should have more than $80 \%$ of their
mass crystallized \citep{2019A&A...625A..87C}. This is borne out in
Fig. \ref{fig-tracks-masivos}, in which  the evolutionary tracks for
ultra-massive WDs with $M_{\star}= 1.10, 1.16, 1.22$, and $1.29
M_{\odot}$  in the $T_{\rm eff}-\log g$ diagram is displayed. The
increase in the surface gravity for all of the models at the onset of
crystallization is due to a decrease in the stellar radius caused by
the changes in the abundance distribution of $^{16}$O and $^{20}$Ne at
the core, induced by phase separation during crystallization. Also
plotted in Fig. \ref{fig-tracks-masivos} is the location of 
spectroscopically identified ultra-massive WDs (asterisk symbols). It is apparent that a large
fraction   of stars in this sample should have more than $99 \%$ of
their mass crystallized. We also include three known ultra-massive ZZ
Ceti stars: BPM 37093, J084021, and GD 518. Note that, in particular,
BPM 37093 should have $\sim 88 \%$ of its mass in a crystalline state,
whereas the more massive stars, J084021 and GD 518, should be $\sim 90
\%$ and $\sim 96 \%$ crystallized. 

\begin{figure}
\includegraphics[width=1.0\textwidth]{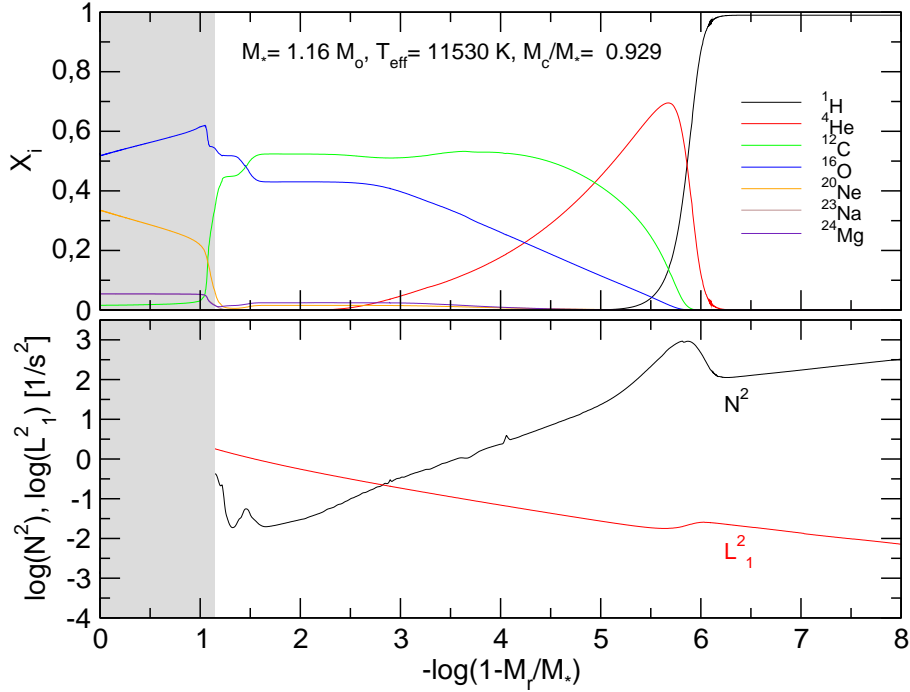}
\caption{Abundances by mass of $^{1}$H, $^{4}$He, $^{12}$C, $^{16}$O,
  $^{20}$Ne,$^{23}$Na, and $^{24}$Mg (upper panel) and the logarithm
  of the squared Brunt-V\"is\"al\"a and Lamb (with $\ell= 1$)
  frequencies (lower panel) as a function of the fractional mass,
  corresponding to an ONe-core WD model with $M_{\star}= 1.16M_{\odot}$
  and $T_{\rm eff} \sim 11\,500$ K. The model has been computed taking
  into account latent heat release and chemical redistribution due to
  phase separation during crystallization. The gray area marks the
  domain of crystallization.  $M_{\rm c}/M_{\star}$ is the fraction of
  the crystallized mass of the model.}
\label{X-BVF-116-11500}
\end{figure}

\begin{figure}
\includegraphics[width=1.0\textwidth]{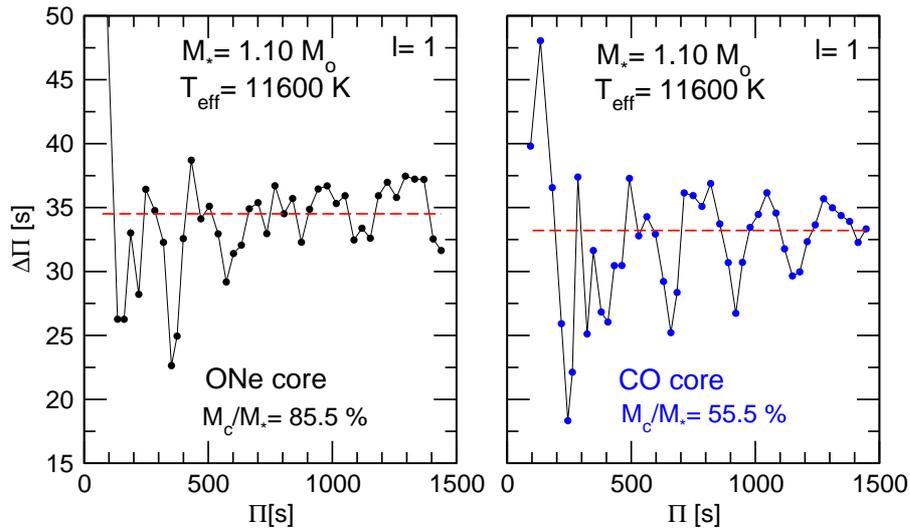}
\caption{Forward period spacing ($\Delta \Pi$) in terms of the periods
  of $\ell= 1$ pulsation $g$ modes for $1.10 M_{\odot}$  WD models at
  $T_{\rm eff} \sim 11600$ K with an  ONe core (left panel) and a CO
  core (right panel). In both models, latent-heat release and chemical
  redistribution caused by phase separation have  been taken into
  account during crystallization. The percentages of the
  crystallized mass are indicated. The horizontal red-dashed line is
  the asymptotic period spacing.}
\label{delp-110}
\end{figure}

Asteroseismology constitutes a very promising way to infer and test
the occurrence of crystallization in WD interiors as well as physical
processes related with dense plasmas. A first step in this direction
has been given by \cite{2019A&A...621A.100D}, who explored the
pulsational properties ---mainly the period spectrum and the
distribution of the period spacings--- of the ultra-massive DA WDs
models of \cite{2019A&A...625A..87C}, taking into account the impact
of crystallization on the $g$-mode period spectrum through the
hard-sphere boundary condition \citep{1999ApJ...526..976M}.  In
Fig. \ref{X-BVF-116-11500} we plot the chemical profiles and the
logarithm of the squared Brunt-V\"ais\"al\"a and Lamb frequencies  as
a function of the fractional mass for an ONe-core WD model with
$M_{\star}= 1.16 M_{\odot}$, $T_{\rm eff} \sim 11\,500$~K, and a
percentage of  crystallized mass of $\sim 93 \%$. For this model,
which is located at the middle  of the ZZ Ceti instability strip,
phase separation due to crystallization ---which  shapes the $^{16}$O
and $^{20}$Ne chemical profiles--- has already finished.
\cite{2019A&A...621A.100D} discuss the possibility of discerning
whether an ultra-massive ZZ Ceti star has a core made of $^{12}$C and
$^{16}$O or a nucleus of $^{16}$O and $^{20}$Ne on the basis of
period-spacing diagrams. They find that period spacing departures due
to mode trapping effects are weaker for ONe-core WD models than for
CO-core WD models with  the same effective temperature and stellar
mass (Fig. \ref{delp-110}). These authors conclude that the features
found in the period-spacing diagrams could be used as a  seismological
tool to discern the core composition of ultra-massive ZZ Ceti stars,
something that should be complemented with detailed asteroseismic
analysis using the individual observed periods.

Independent research on the evolution of ultra-massive WDs has been
performed by \cite{2018MNRAS.480.1547L}, who used {\tt MESA} \citep{2011ApJS..192....3P,2013ApJS..208....4P,2015ApJS..220...15P,
2018ApJS..234...34P,2019arXiv190301426P} to
compute evolutionary sequences of H- and He-rich atmosphere WD models
with masses in the range $1.012$ and $1.307 M_{\odot}$. These authors
present chemical profiles for the whole mass range considered,
covering different core chemical compositions (i.e. CO, ONe and NeOMg)
and its  dependence on the stellar mass. In addition, they  present
the initial-to-final mass relation, the mass-radius relation and
cooling times considering the effects of the atmosphere and the core
composition. At variance with \cite{2019A&A...625A..87C}, however, the
models of  \cite{2018MNRAS.480.1547L} do not take into account the
phase separation and the consequent core chemical rehomogeneization
during crystallization. Also, the predicted compositions of the WD
cores differ from those of \cite{1996ApJ...460..489R} and
\cite{2010A&A...512A..10S}. Differences can be traced back to the
extreme choice of macrophysics parameters done in
\cite{2018MNRAS.480.1547L},  particularly the choice of a large
convective boundary mixing $f$ value below  the C-burning flame ($f=
0.1$), which is at variance with  expectations from hydrodynamical
simulations \citep{2016ApJ...832...71L}. Also, another source of
discrepancy that may have an important impact on the composition and
properties  of the WD models is  the extreme mass-loss rates on the
SAGB phase assumed in \cite{2018MNRAS.480.1547L} \citep[rates of][with
  $\eta= 10$]{1995A&A...297..727B}.  Regarding this last issue, due to
the extreme SAGB winds adopted in their modeling, stellar models
avoided the thermally pulsing SAGB phase that shapes the outer regions
of the ultra-massive WD cores. These are particularly important points
with regard to the pulsational properties of ultra-massive  DA WDs.

\subsection{Excitation mechanisms}
\label{sec:3.4}

\begin{table*}
\scriptsize
\caption{Observed and theoretical effective temperature (in K) of the edges of the ZZ Ceti and  
V777 Her instability strip. The theoretical values are obtained for $0.6 M_{\odot}$ CO-core WD 
models.}
\begin{tabular}{lrrrr}
\hline
\noalign{\smallskip}
 & Blue edge & Red edge  & Blue edge & Red edge \\
 & ZZ Ceti   & ZZ Ceti   & V777 Her & V777 Her \\  
\noalign{\smallskip}
\hline
\noalign{\smallskip}
Observed & $\sim 12\,500$ & $\sim 10\,800$ & $\sim 31\,100$  & $\sim 23\,000$ \\
Theoretical, TDC & $11\,970$ & $5\,600$ & $30\,000$ & $13\,000$ \\
Theoretical, FC  & $11\,750$ & $5\,520$ & $28\,000$ & $13\,000$\\
\noalign{\smallskip}
\hline
\end{tabular}
\label{tabla-edges}
\end{table*}

The early non-adiabatic works on the excitation of ZZ Ceti pulsations
\citep{1981A&A...102..375D,1981A&A....97...16D,1982ApJ...252L..65W} 
assumed that  $g$-mode
periods are much shorter than the turn-over  time of convection of a
ZZ Ceti star ($\tau_{\rm conv} \gg \Pi_g$) and neglected the
perturbation of the convective flux ($\delta \mathbf F_c$)   in the
pulsation equations ---the so-called ``frozen-convection''
approximation \citep[FC;][]{1989nos..book.....U}. Under this
assumption, these authors found  that the physical agent responsible
for the driving of $g$-mode   pulsations at the hot boundary (blue
edge) of the  ZZ Ceti instability strip should be the $\kappa-\gamma$
mechanism  acting at the H partial ionization zone\footnote{Actually, 
\cite{1981A&A....97...16D} found $g$-mode instability due to the partial 
ionization of He.}. However,
\cite{1991MNRAS.251..673B}  realized that the turn-over timescale in
the convection zone of a ZZ Ceti star, at  least at the blue edge, is
shorter than the pulsation periods of  interest ($\tau_{\rm conv} \ll
\Pi_g$), that is, the exact opposite of  the FC approximation.
\cite{1991MNRAS.251..673B} \citep[see also][]{1999ApJ...511..904G}
proposed the ``convective-driving'' mechanism as the responsible for
pulsations in ZZ Ceti stars. Under this hypothesis, the convective
flux should react instantaneously to the macroscopic motions of
pulsations \citep{1997ASSL..214..451B,1999ASPC..173..329B}. 

In an effort to include consistently the impact of possible
interactions  between convection  and  pulsations on the precise
location of the blue and red edges of the ZZ Ceti instability strip,
\cite{2012A&A...539A..87V} applied a time-dependent convection (TDC)
treatment in the framework of the mixing-length theory
\citep{2005A&A...434.1055G} for the first time in WDs, thus avoiding
in their stability computations the extreme assumptions adopted in
previous approaches. These authors also performed nonadiabatic
computations assuming the FC approximation for comparison. The results
of this work ---summarized in Table \ref{tabla-edges}--- are a bit
surprising.  Indeed, the predicted  boundaries of the ZZ Ceti
instability strip according to the TDC treatment are not much
different from those obtained  with the FC approach. Specifically, the
effective temperature  of the blue edge derived from the TDC treatment
is only  $\sim 240$~K higher than that obtained with the FC
approximation, and in good agreement with the instantaneous-convection
adaptation results of
\cite{1997ASSL..214..451B,1999ASPC..173..329B}. In the case of the red
edge, the TDC approach predicts an effective temperature barely
higher ($\sim 80$~K) than the FC approach does,  but both fail to
match the  observed red edge of the ZZ Ceti instability strip.
Similar results to those described for the ZZ Ceti stars using the TDC
and FC  treatments were obtained for V777 Her stars \citep[][see Table
  \ref{tabla-edges}]{2008JPhCS.118a2051D,2008ASPC..391..183Q,2017ASPC..509..321V}. All
these results indicate that the interaction between pulsations and
convection does not seem to be relevant in ZZ Ceti and V777 Her
pulsating stars in terms of the location of  the edges of their
instability strips. 

Recently, \cite{2018ApJ...863...82L} showed that $g$ modes of
pulsating DA WDs with angular frequency $\omega < L_{\ell, b}$
($L_{\ell, b}$ being the Lamb frequency at the base  of the surface
convection zone)  suffer  enhanced  radiative  damping  that  exceeds
convective  driving  rendering them damped, thus giving origin to the
red edge of the ZZ Ceti instability strip. These authors also give an
explanation for the sporadic outbursts exhibited by  some ZZ Ceti
stars observed by the {\it  Kepler} mission that  are located near the
red edge of the instability strip (See Sect. \ref{sec:3.4}).

A potentially efficient mechanism of mode excitation in DA WD  stars
is  the $\epsilon$ mechanism, in which  pulsational instability is
induced by thermonuclear reactions  \citep{1989nos..book.....U}. In
this mechanism, the driving is due to the strong dependence of nuclear
burning on temperature. Usually,  WDs are considered to lack nuclear
burning. In the case of DA WDs this is true in general,  but in some
circumstances these stars may sustain substantial nuclear burning even
at low luminosities, and in particular, during the ZZ Ceti stage. This
has been explored by  \cite{2015A&A...576A...9A} in the case of DA WDs
coming from low-metallicity progenitors. These authors found that for
progenitor metallicities $Z$ in the range $0.00003-0.001$,  and in the
absence of third dredge-up during the TP-AGB phase (and the ensuing
carbon enrichment of the envelope), the resulting H envelope of the
average-mass DA WDs is thick enough to make stable H burning the most
important energy source even at low  luminosities. This finding has
been exploited by \cite{2016A&A...595A..45C} who have investigated the
possible excitation of $g$ modes in ZZ Ceti stars due to the
$\epsilon$ mechanism by H burning. They demonstrated that, for WDs
with masses smaller than $\sim 0.70 M_{\odot}$  and effective
temperatures lower than $11\,600$~K that evolved  from low-metallicity
progenitors, the dipole and  quadrupole $g$ modes with radial order
$k= 1$ ($g_1$) are excited mostly as a result of the H-burning shell
through the $\epsilon$ mechanism. However, the $\epsilon$ mechanism is
insufficient to drive these modes in WDs descended from
solar-metallicity progenitors.  These results  encourage the
possibility of placing constraints on H shell burning in cool WDs and
the efficiency of third dredge-up episodes during the preceding AGB
phase if the $g_1$ mode were detected in DA WDs associated with
low-metallicity environments, such as globular clusters and/or the
Galactic halo. 

\begin{figure}
\includegraphics[width=1.0\textwidth]{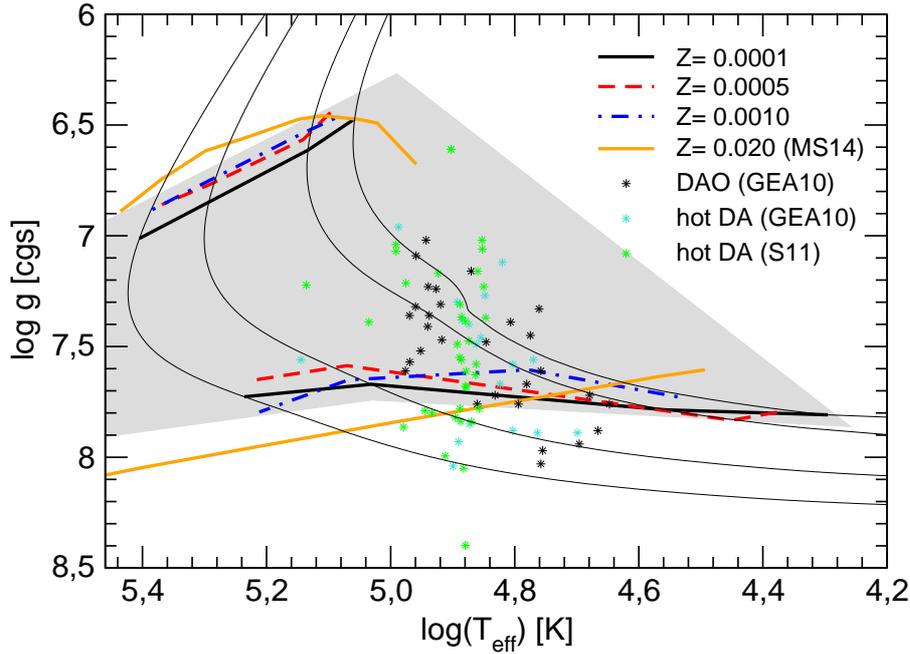}
\caption{Instability domain (shaded area) of hot DA pre-WDs with
  low-order $g$ modes excited by the $\epsilon$ mechanism on the $\log
  T_{\rm eff}-\log g$ plane.  Thick lines show the edges of the
  instability region for different progenitor metallicities. Also
  included are the results of  \cite{2014PASJ...66...76M} (MS14)  for
  solar metallicity. Star symbols of different colors show the DAO and
  hot  DA stars from \cite{2010ApJ...720..581G} (GEA10) and hot DA
  stars  from \cite{2011wdac.book....1S} (S11).  Black lines depict
  evolutionary tracks corresponding to  $M_{\star}=  0.535, 0.568,
  0.666, 0.738 M_{\odot}$ and $Z= 0.0001$.}
\label{HR-blue-red-stars}
\end{figure}

Finally, it is worth mentioning the possible existence of very hot
pulsating DA pre-WDs with  $g$ modes excited by the $\epsilon$
mechanism\footnote{Pulsational excitation of $g$ modes due to the $\epsilon$ 
mechanism in H-deficient pre-WD models was investigated 
by \cite{1986ApJ...306L..41K} and \cite{2009ApJ...701.1008C} 
\citep[see, e.g.,]
[for a review of this topic]{2010A&ARv..18..471A,2015pust.book.....C}.}. 
Early work on this issue by \cite{1988ApJ...334..220K} revealed 
pulsation instability in models of H shell-burning planetary nebula nuclei.
Later, \cite{1997ApJ...489L.149C} predicted $g$-mode instability promoted by H burning through 
the $\epsilon$ mechanism in post-EHB stars. These pioneering researches were 
extended by  \cite{2014PASJ...66...76M}, who found that nuclear
reactions in the H burning shell can drive low-degree $g$ modes with
periods in the  range $\sim 40-200$ s for DA pre-WDs with $40\,000
\leq T_{\rm eff} \leq 300\,000$~K coming from  solar-metallicity
progenitors, the instability domain in the HR diagram being sensitive
to the H content. Similar results, although on the basis of DA pre-WD
models coming from low-metallicity progenitors
\citep{2015A&A...576A...9A}, have been found by
\cite{2017EPJWC.15206012C}.  In Fig. \ref{HR-blue-red-stars} we show
the instability domain  of hot DA pre-WDs with low-order $g$ modes
excited by the $\epsilon$  mechanism on the $\log T_{\rm eff}-\log g$
plane.  The boundaries  of the instability domain for different
sub-solar metallicities  computed by \cite{2017EPJWC.15206012C} are
depicted with thick lines of different colors and styles. We also
include the results of  \cite{2014PASJ...66...76M} for solar
metallicity. Star symbols show the location of DAO and hot DA stars
that could be candidate targets to be scrutinized to search for
pulsations. The finding of any of these stars to be pulsating with
short-period $g$ modes could allow to place constraints on the H thickness of DA WDs \citep{2014PASJ...66...76M}. 

\begin{table*}
\scriptsize
\caption{Properties of the known Hot DAV stars \citep[adapted from][]{2016IBVS.6184....1B}.}
\begin{tabular}{lccccc}
\hline
\noalign{\smallskip}
 Star & $T_{\rm eff}$ & $\log g$  & $V$ & $\Pi$  & Reference\\
      &   [K]         &   [cgs]   &  [mag]  &  [s]  &            \\ 
\noalign{\smallskip}
\hline
\noalign{\smallskip}
WD 0101+145 & $29\,980$  & 7.38 & 18.8 (g) & 159   & \cite{2008MNRAS.389.1771K} \\
WD 0232-097A & $30\,110$ & 7.30 & 17.8 (g) & 705  & \cite{2008MNRAS.389.1771K} \\ 
WD 1017-138 &  $32\,600$ & 7.80 & 14.6     & 624   & \cite{2013MNRAS.432.1632K} \\
\noalign{\smallskip}
\hline
\end{tabular}
\label{table-hotDAV}
\end{table*}

An alternative possibility for driving pulsations in WDs 
could be tidal excitation of $g$ modes in compact WD binary systems
due to resonances between the orbital frequency and the discrete 
spectrum of eigenfrequencies of the WDs. 
These tidally-forced  $g$-mode oscillations have been thoroughly investigated by \cite{2011MNRAS.412.1331F,2012MNRAS.421..426F}. 
They showed that the excited $g$ modes can reach very large 
amplitudes at the regions close to the stellar surface, 
where  they  are  likely  dissipated  through  a  combination  of
non-linear processes and radiative damping, thus probably 
preventing the excitation of discrete  $g$ modes. 
In any case, no tidally excited $g$ modes have  been detected in 
any WD so far \citep[see][]{2018MNRAS.479.1267K}. 

We close this section by describing the possible existence of pulsating 
DA WDs somewhat hotter than ZZ Ceti stars, with effective temperatures 
comparable to those of the DBV WDs. \cite{1982ApJ...252L..65W} 
\citep[see, also,][]{1982PhDT........27W} were the first to find $g$-mode pulsational 
instability due to the partial ionisation of He in models of DA WDs harboring 
very thin H envelopes ($M_{\rm H}/M_{\star} \lesssim 10^{-10}$), for effective 
temperatures of $\sim 19\,000$~K. On the other hand, \cite{2013MNRAS.432.1632K} 
announced the existence of a new class of DA WD pulsators, the so-called 
pulsating hot DAV WDs, characterized by $T_{\rm eff}\sim  30\,000$~K, that show $g$ modes 
probably excited by a driving mechanism very different than those described 
above. The driving mechanism for hot DAVs was formulated by
\cite{2005EAS....17..143S,2007AIPC..948...35S} \citep[see,
  also,][]{2013EAS....63..185S} who theoretically found that in DA WDs
with thin H envelopes at the cool edge of the $45\,000-30\,000$~K
effective temperature range (the ``DB gap''), a super-adiabatic,
chemically-inhomogeneous ($\mu$-gradient) zone drives pulsation $g$
modes  with high $\ell$ values, but even with some $\ell < 3$ modes
excited, which  could be observable. Three objects  of this predicted
class are known, the characteristics of which are summarized in Table
\ref{table-hotDAV},  although the observations should be confirmed
with additional monitoring. Further investigation of these stars could
open the chance of a direct test of the explanation/existence of the
DB gap through asteroseismology.

\subsection{Outbursting pulsating DAV stars}
\label{sec:3.5}

\begin{figure}
\includegraphics[width=1.0\textwidth]{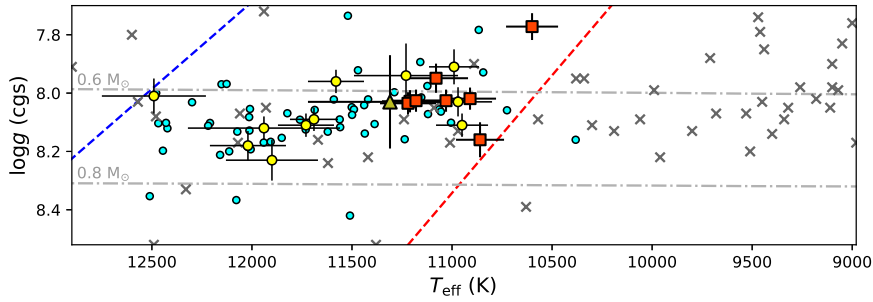}
\caption{$T_{\rm eff}-\log g$ diagram showing the location of
  presently known outbursting DAVs (8 objects) marked with red
  squares. The dark-yellow triangle corresponds to the object EPIC
  211891315 from \cite{2016ApJ...829...82B}, that exhibited a single
  outburst-type event. The small yellow circles and gray  crosses
  correspond to the DAVs and long-cadence WD targets that do not show
  outbursts in K2 data through Campaign 6 \citep{2016ApJ...829...82B}.
  The dashed lines are the observational boundaries of the ZZ Ceti
  instability  strip, according to
  \cite{2015ApJ...809..148T}. Dash-dotted lines are  WD cooling tracks
  for $M_{\star}= 0.6 M_{\odot}$ and $M_{\star}= 0.8 M_{\odot}$  WDs
  from \cite{2001PASP..113..409F}. The cyan circles correspond to  the
  location of DAVs known from the ground \citep{2011ApJ...743..138G},
  including SDSS J2350$-$0054
  \citep{2004ApJ...607..982M,2016MNRAS.455.3413K}  that is cooler than
  the red edge of the instability strip   \cite[Figure adapted
    from][]{2017PhDT........14C}.}
\label{DAV-outbursts}
\end{figure}

{\it Kepler} spacecraft observations of ZZ Ceti stars revealed a
new type of phenomena never before observed from the ground:
outburst-like events in DAV stars with effective temperatures near the
red edge of the instability strip \citep{2017ASPC..509..303B}\footnote{A phenomenon reminiscent of these outburst-like events
was the {\it sforzando} event detected in 1996 for the 
DBV GD358, in which the star dramatically altered its 
pulsation characteristics on a 
timescale of hours \citep{2009ApJ...693..564P}.}.  The
first ZZ Ceti star exhibiting outbursts was WD J1916+3938
(KIC4552982). This star was the first ZZ Ceti identified in the
original {\it Kepler} mission field \citep{2011ApJ...741L..16H}. After 1.5 years of observations
which provided the longest pseudo-continuous light curve ever recorded
for a ZZ Ceti, \cite{2015ApJ...809...14B} found that  the star shows a
rich period spectrum typical of DAVs,  with at least 20 modes detected,  along with a total of 178
enhancements of brightness typical of outburst phenomena, with peaks
of up to $17 \%$ above the quiescent level and involving very
energetic events ($\sim 10^{33}$ erg)  with a mean recurrence period
of about 2.7 days. Remarkably, KIC4552982 is located at the cool
boundary of the ZZ Ceti instability strip ($T_{\rm eff}= 10\,860$~K,
$\log g= 8.16$). The second ZZ Ceti star showing outbursts was
PG1149+057 \citep{2015ApJ...810L...5H}, discovered with the {\it K2}
mission. This star exhibits flux enhancements of up to $45 \%$. Being
relatively bright, it was possible to determine for this ZZ Ceti that
the outbursts actually affect the normal pulsations, thus
demonstrating that  outbursts are an intrinsic phenomenon of the
star. At present,  a total of eight outbursting ZZ Ceti stars have
been  discovered
\citep[][]{2016ApJ...829...82B,2017ASPC..509..303B,2017PhDT........14C}.
In Fig. \ref{DAV-outbursts} we depict the ZZ Ceti instability strip in
the $T_{\rm eff}-\log g$ diagram with the location of the known DAVs,
emphasizing the outbursting objects with red square symbols. All of
the outbursting ZZ Ceti stars share some common properties: (i) the
outbursts can enhance the  mean flux of the star over timescales of
hours, with an irregular  recurrence on timescales of days; (ii) the
normal ZZ Ceti-like $g$-mode pulsation spectrum changes in amplitude
and frequency during outbursts,  relative to those in quiescence; and
(iii) the effective temperatures of outbursting ZZ Cetis locate them
close to the cool edge of the  instability strip, suggesting that
outbursts could be related  to the cessation of $g$-mode pulsations
\citep{2017ASPC..509..303B}.

A physical explanation for the occurrence of outbursts in cool ZZ Ceti
stars  was suggested by \cite{2015ApJ...810L...5H} and formally drawn
up by  \cite{2018ApJ...863...82L}, in terms of parametric  instability
via mode coupling of WD pulsations
\citep{1982AcA....32..147D,2001ApJ...546..469W}.  In this scenario, it
is possible that the oscillation kinetic energy of  a parent mode (or
multiple parent modes) grows linearly until it reaches a critical
value. When this value is exceeded, the mode enters a nonlinear regime
and rapidly transfers its energy to resonant daughter modes, or a
cascade of resonant daughter modes, which may be quickly damped by
turbulence in the convection zone.  The sum of the frequencies of the
daughter modes must be close to  the parent frequency. The energy of
these modes is deposited at the basis of  the outer convection zone,
producing the sudden enhancement of the stellar brightness
characterizing the outbursts. \cite{2018ApJ...863...82L}   attribute
ZZ Ceti outbursts  to  limit  cycles  arising  from   sufficiently
resonant  3-mode  couplings  between overstable parent  modes and
pairs of radiatively damped daughter modes.   Limit cycles account for
the durations and energies of outbursts ($\sim 10^{33}-10^{34}$ erg)
and their prevalence near the cool edge of the ZZ Ceti instability
strip.

\subsection{ELMV and pre-ELMV stars} 
\label{sec:3.6}

\begin{figure}
\includegraphics[width=1.05\textwidth]{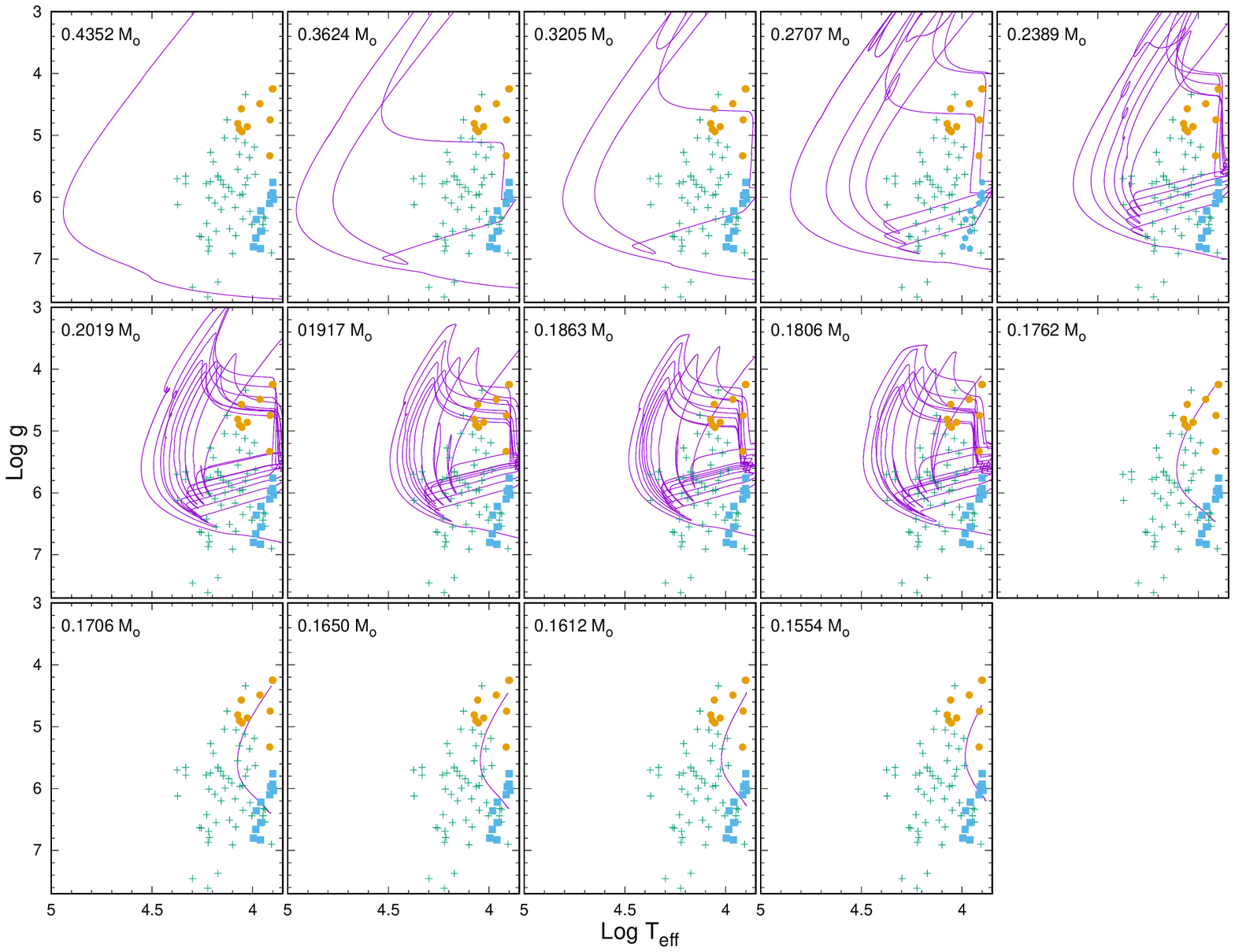}
\caption{$\log T_{\rm eff} - \log g$ diagrams showing the He-core WD
  evolutionary tracks computed in \cite{2013A&A...557A..19A} 
  (purple curves; see text for details). Green plus
  symbols correspond to the observed LM and ELM WDs from
  \cite{2012MNRAS.424.1752S} and \cite{2013ApJ...769...66B}. Filled light blue
  circles correspond to  pulsating LM and ELM WDs (ELMV variables) and
  filled orange circles  correspond to pulsating LM and ELM pre-WDs (pre-ELMV
  variables). Numbers in the left  upper corner of each panel
  correspond to the stellar mass at the WD stage.}
\label{tracks-elms}
\end{figure}

One of the most important findings of recent years in the field of
variable WDs is the discovery of pulsations in DA low-mass (LM)
and extremely low-mass (ELM) WDs and pre-WDs.   DA WDs have a mass
distribution that peaks at $0.59 M_{\odot}$, but it also shows a peak at
low masses: $M_{\star}/M_{\odot} \lesssim 0.45$
\citep{2013ApJS..204....5K,2015MNRAS.446.4078K}. LM  WDs can be
the result of enhanced mass loss  before the occurrence of the He flash
during the red giant branch phase of low-mass stars.  At variance with
average WDs with C/O cores, these stars are expected to  harbor cores
made of He, since He ignition is avoided. In particular,  
strong mass loss resulting
from interactive binary evolution  is {\it needed} to explain the origin of the so-called ELM WDs, which have masses below $\sim 0.18-0.20
M_{\odot}$\footnote{The definition of an ELM WD is still under debate.
  In the context of the ELM Survey \citep{2010ApJ...723.1072B}, an 
  ELM WD is defined as an object
  with surface  gravity of $5 \lesssim \log g \lesssim  7$ and
  effective temperature in the range of  $8000 \lesssim T_{\rm eff}
  \lesssim 22\,000$~K \citep[see, e.g.,][]{2016ApJ...818..155B}.  Here
  \citep[see also][]{2014A&A...569A.106C}, we propose to define an ELM
  WD as a WD that does not undergo H shell flashes, because in this
  way the pulsational  properties are quite different as compared with
  the systems that experience flashes, although this mass limit
  depends on the metallicity of the progenitor stars
  \citep{2002MNRAS.337.1091S,2016A&A...595A..35I}.} \citep[see  Sect. \ref{sec:2.2}
  and][for details]{2013A&A...557A..19A,2016A&A...595A..35I}. In
Fig. \ref{tracks-elms} we show the evolutionary tracks of
\cite{2013A&A...557A..19A} for He-core LM and ELM WDs in the $T_{\rm
  eff}-\log g$ diagram. Sequences with masses in the range $0.18-0.20
\lesssim M_{\star}/M_{\odot} \lesssim 0.4$ experience multiple CNO-cycle
flashes during the early-cooling phase, which leads to the intricate
loops in the diagrams. In the last decade, numerous LM WDs, including
ELM WDs, have been detected with the ELM Survey and the SPY and WASP
surveys \citep[see][]{2009A&A...505..441K,  2010ApJ...723.1072B,
  2012ApJ...744..142B,2013ApJ...769...66B,2016ApJ...818..155B,
  2017ApJ...847...10B,  2011MNRAS.418.1156M, 2011ApJ...727....3K,
  2012ApJ...751..141K, 2015MNRAS.446L..26K, 2014ApJ...794...35G,
   2015ApJ...812..167G}. We include in Fig. \ref{tracks-elms} the
location of a sample of LM and ELM WDs and pre-WDs with the  aim of
illustrating the region in the $\log T_{\rm eff}- \log g$ plane  that
is populated by these stars. 
  
The possible existence of pulsations  in ELM WDs was predicted by
\cite{2010ApJ...718..441S} for the first time, by scaling of the
thermal  timescale at the basis of the outer   convective zone of
their ELM WD models with surface gravity. They found that,  contrary
to what happens in the case of ZZ Ceti stars ($M_{\star} \gtrsim 0.50
M_{\odot}$), in pulsating ELM WDs the $g$-mode pulsations 
should probe the core regions of the  WDs, by virtue that the
eigenfunctions are not excluded from those regions due to the low
degeneracy of ELM WDs. The long time  (several Gyr) that ELM WDs
should spend at low effective temperatures  ($T_{\rm eff} \lesssim
14\,000$~K) due to vigorous stable H burning via the $pp$ chain,
motivated \cite{2012PASP..124....1S} to  carry out a careful search
for pulsating objects with  masses  $M_{\star} \lesssim 0.20
M_{\odot}$, but with null results. Shortly after,
\cite{2012ApJ...750L..28H} reported the exciting discovery of the
first  pulsating ELM WD, SDSS J184037.78$+$642312.3, the coolest and
the  lowest-mass pulsating WD at that time, with $T_{\rm eff}= 9100
\pm 170$~K, $\log g = 6.22 \pm 0.06$, and $M_{\star} \sim 0.17
M_{\odot}$. This pulsating ELM WD ---which is in a 4.6 hr binary
system with another WD star--- exhibits high-amplitude, multi-periodic
variability with periods in the range $2000-4900$ s, compatible with
intermediate- and high-order $g$ modes. The increasing interest in
pulsating LM and ELM WDs led to the discovery of more objects of
this type. At present, eleven pulsating WDs of this kind are known
\citep{2012ApJ...750L..28H, 2013ApJ...765..102H,2013MNRAS.436.3573H,
  2015MNRAS.446L..26K,2018MNRAS.479.1267K,2017ApJ...835..180B,
  2018A&A...617A...6B, 2018MNRAS.478..867P}. 
  This number includes all
the known and suspected ELMV stars, that is, objects that show
radial velocity (RV) variations  confirming the binary nature
expected for He-core WDs, and objects that do  not\footnote{According
  to \cite{2018MNRAS.479.1267K}, there  are only four confirmed
  pulsating ELM WDs in short-period binaries (which are the four that show RV variations),  that occupy  a  similar
  parameter   space  and  there  is  no question about their nature as
  WDs. These are: SDSS J1112+1117, SDSS J1518+0658, SDSS J1840+6423,
  and PSR J1738+0333. We have to add SDSSJ1618+3854 to that list, based on \cite{2018A&A...617A...6B}.}. This new class of variable stars has been
designed as ELMV. Note that both the pulsating LM WDs
($0.18-0.20 \lesssim M_{\star}/M_{\odot} \lesssim 0.45$) and the
pulsating ELM WDs ($M_{\star}/M_{\odot} \lesssim 0.18-0.20$) are
designated by the common name of ELMVs.  They are listed in  Table
\ref{elmv},  along with updated effective temperatures and
gravities \citep[3D corrected;][]{2015ApJ...809..148T}, magnitudes,  
and period ranges.  We emphasize that
Table \ref{elmv} includes all ---the known and suspected--- ELMV
stars.

\begin{figure}
\includegraphics[width=1.0\textwidth]{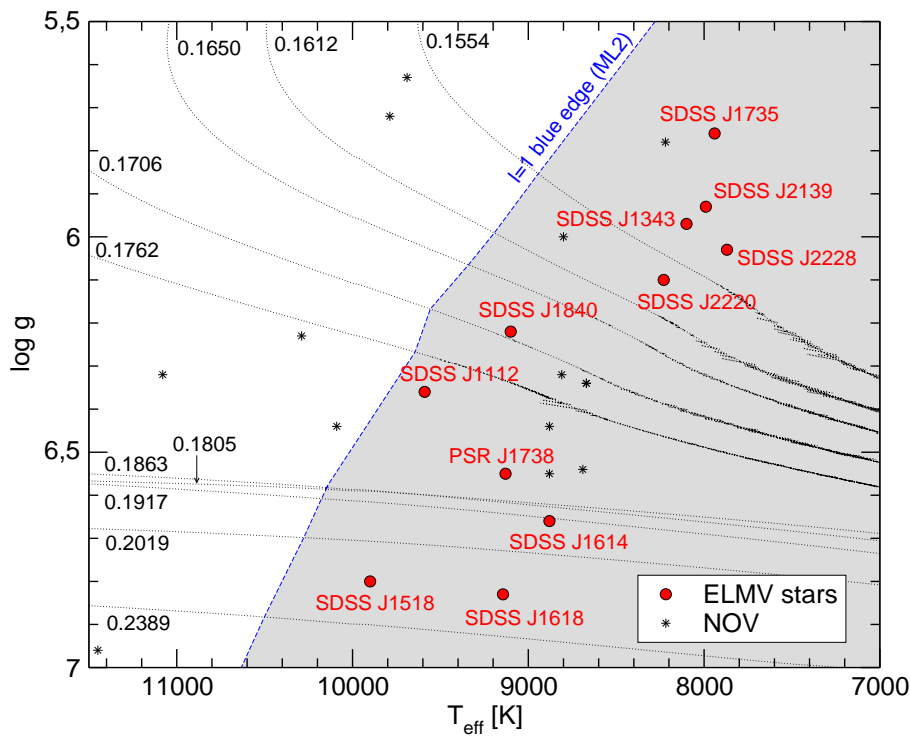}
\caption{$T_{\rm eff}-\log g$ diagram showing the location of the eleven
  known ELMVs,  marked with red circles ($T_{\rm eff}$ and $\log g$
  computed with 1D model atmospheres after 3D corrections). Black-star symbols
  are objects not observed to vary (NOV).  Dotted lines correspond to
  the low-mass He-core WD evolutionary  tracks (final cooling
  branches) of \cite{2013A&A...557A..19A}. The ``spikes'' exhibited 
 by the tracks with $M_{\star} \leq 0.1762 M_{\odot}$ are the result 
 of intense H residual burning 
 ($pp$ chain) at low $T_{\rm eff}$ values. The gray region limited by
  the dashed blue line corresponds to the instability domain of $\ell=
  1$ $g$ modes due to the $\kappa-\gamma$ mechanism
  acting at the H$-$H$^{+}$ partial ionization region, according  to
  \cite{2016A&A...585A...1C}.}
\label{HR-ELMVs}
\end{figure}  

\begin{figure}
\includegraphics[width=1.0\textwidth]{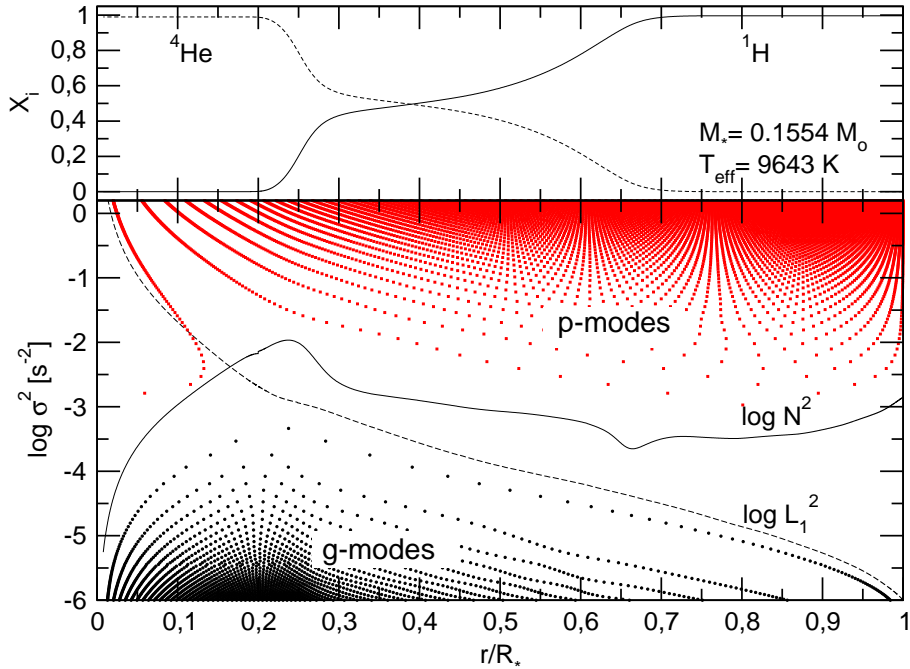}
\caption{Fractional abundances of He  and H as a function of the 
normalized radius (upper panel), and a
propagation diagram (lower panel) corresponding 
to a representative ELM WD model with $M_{\star}= 0.1554 M_{\odot}$
and $T_{\rm eff} \sim  9600$~K. Small squares (in
red) correspond to the spatial location of the nodes of the radial eigenfunction 
of $\ell= 1$ $p$ modes, whereas small circles (in black) represent the
location of the nodes of dipole $g$ modes. Solid (dashed) black curve corresponds 
to the logarithm of the squared Brunt-V\"ais\"al\"a (Lamb) frequency.}
\label{x-bvf-elm}
\end{figure}

\begin{figure}
\includegraphics[width=1.0\textwidth]{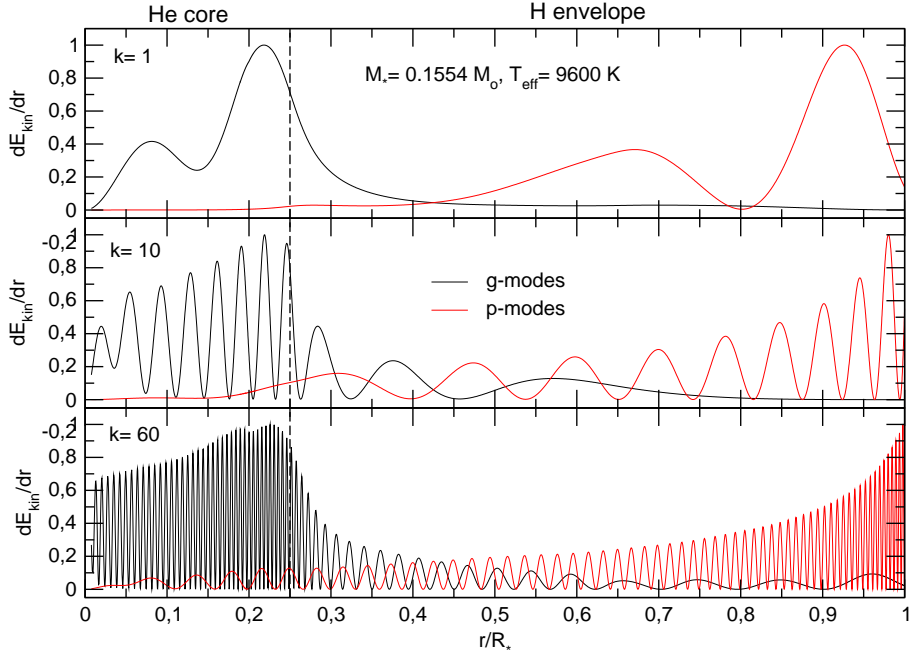}
\caption{Run of the density of oscillation kinetic energy for dipole
$p$ modes (red) and $g$ modes (black) with $k= 1$ (upper panel), $k=10$  
(middle panel),  and $k= 60$ (lower panel), in terms of the normalized radius, 
corresponding to the same ELM WD model shown in Fig. \ref{x-bvf-elm}. 
The vertical dashed line marks the approximate location of the He/H chemical 
transition region.}
\label{ekin-elm}
\end{figure}

The existence of ELMV stars (red circles in
Fig. \ref{HR-ELMVs}) constitutes an  unprecedented opportunity  for
probing their subsurface layers and ultimately to place constraints on
the currently accepted formation scenarios  by means  of
asteroseismology. Indeed, the importance of these stars has motivated
intensive theoretical work to explore the basic nature of their
pulsations
\citep{2010ApJ...718..441S,2012A&A...547A..96C,2013ApJ...762...57V,2014A&A...569A.106C,
  2014ApJ...793L..17C,2016A&A...585A...1C,2017A&A...600A..73C,
  2017A&A...607A..33C,2018A&A...620A.196C}. 
  In line with the pioneering work by \cite{2010ApJ...718..441S}, \cite{2014A&A...569A.106C} 
  concluded that for ELM WDs, $g$ modes mainly probe the core  regions, while $p$
and radial  modes (if excited) sound the stellar envelope. Hence,
pulsations in ELMVs offer the rare opportunity ---in the context of 
pulsating WDs--- to put constraints to both the
core and envelope chemical  structure of these stars by means of
asteroseismology. In  Fig. \ref{x-bvf-elm} we display the chemical
profiles (upper panel) and the propagation diagram (lower panel),  for
a representative ELM WD model characterized by $M_{\star}= 0.1554 M_{\odot}$ and
$T_{\rm eff} \sim 9600$~K.  Fig. \ref{ekin-elm} shows the  density
of kinetic energy of oscillation ($dE_{\rm kin}/dr$, normalized to 1)
for dipole ($\ell= 1$) $g$ and $p$ modes with  different values of the
radial order $k$, for the same ELM WD model. The fact that the
Brunt-V\"ais\"al\"a frequency adopts large values in the core
(Fig. \ref{x-bvf-elm}) allows eigenfunctions of $g$ modes to penetrate
to those deep regions of the star, where most of the kinetic  energy
of oscillation is concentrated (Fig. \ref{ekin-elm}). On the other
hand, in LM WDs, $g$ modes are very sensitive to the He/H compositional gradient \citep[see Fig. 8 of][]{2014A&A...569A.106C} and
then, they could be employed as a diagnostic tool for constraining the
H envelope thickness, similar to ZZ Ceti stars.

Another important finding in this area is that
time-dependent   element    diffusion markedly modifies the shape of
the He/H chemical transition region, and in turn, has a strong impact
on the $g$-mode pulsation spectrum of ELM WDs \citep[][]{2012A&A...547A..96C,2014A&A...569A.106C}. The effects are weaker ---but still
non-negligible--- for LM WDs. Therefore, it is apparent that
time-dependent element  diffusion must be taken into account in any
pulsational study of ELMV stars.

A detailed linear pulsation-stability analysis of ELMV stars has been carried out by  \cite{2013ApJ...762...57V}. These authors employed a sophisticated treatment  of TDC in their nonadiabatic
pulsation  computations, similar to that applied by the  same authors
to ZZ Ceti and V777 Her stars \citep[see][and Section
  \ref{sec:3.4}]{2012A&A...539A..87V,2017ASPC..509..321V}. The
excitation mechanism is the  convective-driving mechanism. They obtained
numerous excited $g$ modes with  periods in a range that depends on
the stellar mass. For instance, for a $0.2 M_{\odot}$ model, the range
of periods of the excited modes is $\sim 500-8000$~s.  Their
computations, based on envelope models, predict a blue edge  of the
ELMV instability domain in good agreement with the
observations. However, as  for ZZ Ceti and V777 Her stars, their
calculations are unable to account for  the cool edge, which according
to their predictions should be located at $T_{\rm eff} \sim 5\,500$~K,
i.e., much cooler than the observed one  ($T_{\rm eff} \sim
7\,900$~K). Instead, \cite{2013ApJ...762...57V} estimate the location
of the red  edge by requiring that the  thermal timescale in the
driving region at the base of the H convection zone be  equal to the
critical period beyond which $\ell= 1$ $g$ modes cease to exist.  This
estimation is based on the atmosphere energy-leakage argument
elaborated by \cite{1985ApJ...297..544H}. This estimate indicates a
theoretical red edge close to the observed one; however, it is not
completely  satisfactory, because several ELMVs are cooler than the
red edge predicted in  that way. Another relevant result, that is in
line with the  previous findings of these authors for the ZZ Ceti and
V777 Her stars, is that the  blue edge of the instability domain of
ELMVs does not change substantially  if the FC approach is used
instead of the more physically sound TDC treatment.  Again, it seems
that the interaction between pulsation and convection does not
dramatically affect the position in the $\log T_{\rm eff}-\log g$
diagram at  which the LM and ELM WDs begin to pulsate.

%\begin{figure}
%\includegraphics[width=1.0\textwidth]{per-teff-eft-1518-review.eps}
%\caption{Periods  of  unstable $\ell= 1$  $g$ modes ($\Pi \gtrsim 250$ s) 
%and $p$ modes ($\Pi \lesssim 40$ s) 
%in  terms  of  the  effective temperature, corresponding to the $0.1917$ and 
%$0.2019 M_{\odot}$ ELM WD sequences \citep{2016A&A...585A...1C}.
%Also shown are the seven pulsation periods of the 
%ELMV star SDSS J1518+0658. The $T_{\rm eff}$ adopted for the star is its 1D 
%value corrected by 3D model atmosphere effects.}
%\label{per-teff-eft-1518}
%\end{figure} 

\cite{2014ApJ...793L..17C,2016A&A...585A...1C} have
independently  explored in detail the instability strip of ELMV stars,
based on the fully  evolutionary low-mass He-core WD models with
H-pure envelopes of  \cite{2013A&A...557A..19A}, and assuming the FC
approximation for their nonadiabatic computations. In particular, \cite{2016A&A...585A...1C} took into
account  three different prescriptions for the  MLT theory of
convection  and covered a wide range of effective   temperatures and
stellar masses. For each model, the pulsation stability  of radial
($\ell= 0$) and nonradial ($\ell= 1, 2$) $p$ and $g$ modes was
assessed.  Their main predictions are illustrated in
Fig. \ref{HR-ELMVs},  which shows  the instability domain of $\ell= 1$
$g$ modes in the spectroscopic HR diagram  due to the $\kappa-\gamma$
mechanism acting at the H-H$^{+}$ partial ionization region, together
with some selected low-mass He-core WD evolutionary tracks (final
cooling branches),  and the location of ELMVs (red circles). The
instability domain is emphasized  by a  gray region bounded by a
dashed blue line corresponding  to the blue  edge of the  instability
domain.   Notably, the FC results of  \cite{2016A&A...585A...1C} are
in good agreement  with the TDC results of \cite{2013ApJ...762...57V}
in terms of the location of the blue edge of the ELMV instability
strip. Similar to \cite{2013ApJ...762...57V},
\cite{2016A&A...585A...1C} did not find a red edge of the instability
strip consistent with the observed one. It is found that the blue edge
of the instability domain in the $T_{\rm eff}-\log g$ plane is hotter
for higher stellar mass and larger convective efficiency and that the
blue edge  of $p$ modes does not depend on the harmonic
degree. However, in the case of $g$ modes, a weak sensitivity of the
blue edge with $\ell$ is found. In addition,  the blue edges
corresponding to radial and nonradial $p$ modes are almost coincident
with each other, and  somewhat hotter ($\sim 200$ K) than the blue edges of
$g$ modes.  Finally, we emphasize 
that some short-period $g$ modes can be destabilized mainly  by the $\varepsilon$ mechanism due to stable nuclear burning at the basis of the  H
envelope, particularly for model  sequences with $M_{\star} \lesssim
0.18 M_{\odot}$ \citep[see][for details]{2014ApJ...793L..17C,
2016A&A...585A...1C,2017ASPC..509..347F}. 

It is worth mentioning that the ranges of unstable-mode
periods predicted by current stability  analyses are in excellent
agreement with  the  ranges of periods observed in the ELMV stars
\citep[see, for instance, Figs. 10 to 17 of][]{2016A&A...585A...1C}. 
However,  for all  the  analyzed ELMVs, the number of periods  detected is alarmingly low in comparison with
the rich spectrum of periods of  radial and nonradial $p$ and $g$
unstable modes  expected from theoretical stability analyses. Similar to
other kinds of pulsating WDs,  it is suspected that an unknown
mechanism should be  at work in real stars that  favors only a few
modes (out of the available dense spectrum of eigenmodes)  to reach
observable amplitudes. Finding that missing piece of physics in the
pulsation models is a challenge for future research.

\begin{figure}
\includegraphics[width=1.0\textwidth]{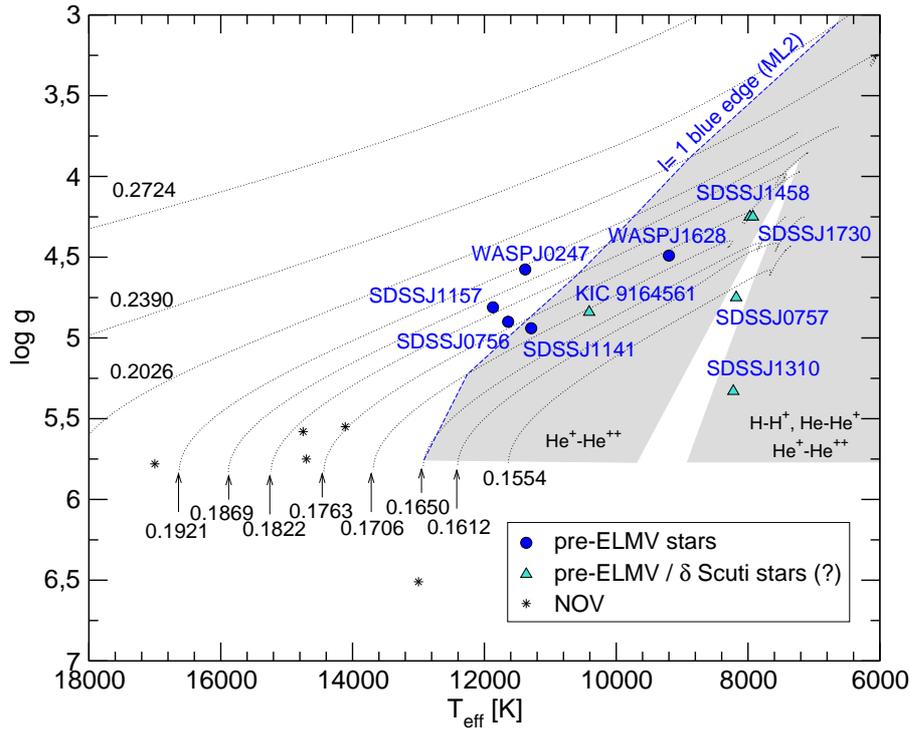}
\caption{$T_{\rm eff}-\log g$ diagram showing low-mass He-core pre-WD
  evolutionary tracks (dotted curves) computed neglecting element
  diffusion. Numbers  correspond to the stellar masses of each
  sequence. Blue circles correspond to the known
  pre-ELMV stars with a secure identification, turquoise triangles stand
  for  objects that could be pre-ELMVs or $\delta$ Scuti/SX Phoenicis
  stars as well,  and black-star symbols depict the location of stars not
  observed to vary (NOV). The dashed blue line indicates the nonradial
  dipole ($\ell= 1$) blue edge of the pre-ELMV instability domain
  (emphasized as a gray area) due to the $\kappa-\gamma$ mechanism
  acting at  the He$^{+}-$He$^{++}$ partial ionization region, as
  obtained by  \cite{2016A&A...588A..74C}.}
\label{HR-preELMVs}
\end{figure}

Apart from ELMVs, multi-periodic pulsations in five  objects that are
supposed to be precursors of LM and ELM WDs have been detected  in the
last few years \citep{2013Natur.498..463M,2014MNRAS.444..208M,2016ApJ...822L..27G}.  These stars, that typically have $8000 \lesssim
T_{\rm eff} \lesssim 13\,000$ K  and $4 \lesssim \log g \lesssim 5.5$,
are represented in Fig. \ref{HR-preELMVs} as   blue circles. They show
a surface composition made of  a mixture of H and He, suggesting that some
mechanism must be delaying  (or even inhibiting) the effects of
chemical diffusion.  They are called   pre-ELMV stars and constitute a
new class  of pulsating stars.  Also, the discovery of long-period
($\Pi \sim 800-4700$ s) pulsations in five additional objects located
at the same region of the $T_{\rm eff}-\log g$ diagram has been
reported.  These stars are represented with turquoise triangles in
Fig. \ref{HR-preELMVs}.  The exact nature of these pulsating stars is
uncertain since   they could be identified as pre-ELMVs or as SX Phe
and/or $\delta$ Scuti   pulsating stars as well. In Table
\ref{pre-elmv} we  include an updated compilation of the effective
temperature, gravity,  and range of observed periods for
all the known pre-ELMV stars, including those with an indefinite
identification. The data were extracted from \cite{2013Natur.498..463M,2014MNRAS.444..208M} and \cite{,2016ApJ...822L..27G} for the confirmed pre-ELMVs, and from \cite{2016ApJ...821L..32Z,2016A&A...587L...5C,2018A&A...617A...6B} and \cite{2018MNRAS.478..867P} for stars with ambiguous 
identification (pre-ELMVs or $\delta$ Scuti/SX Phoenicis stars).

The first theoretical study exploring the pulsation stability
properties  of radial modes of static low-mass He-core pre-WD models
was that of \cite{2013MNRAS.435..885J}.  They were successful in
identifying the instability boundaries associated with radial modes
characterized by low-to-high radial orders, and showed that they are
very sensitive to the chemical composition at the driving region. In
particular, these authors  found that the modes are excited by
the $\kappa-\gamma$ mechanism operating mainly in the second He
ionization zone (He$^{+}-$He$^{++}$),  provided that the driving
region is depleted in H. \cite{2016A&A...588A..74C} extended  the work
of \cite{2013MNRAS.435..885J} by analyzing the pulsational stability
of  radial and nonradial $p$ and $g$ modes with periods in the range
$10\ {\rm s} \lesssim \Pi  \lesssim 20\,000$ s, on the basis of
He-core, low-mass pre-WD evolutionary models.  In
Fig. \ref{HR-preELMVs} we show with a  dashed blue line the nonradial
$\ell= 1$ blue edge of the pre-ELMV instability domain  (gray area)
due to the $\kappa-\gamma$ mechanism acting at the  He$^+$-He$^{++}$
partial ionization region ($\log T \sim 4.7$). Non-diffusion  low-mass
He-core pre-WD evolutionary tracks are displayed with dotted curves.
The results of \cite{2016A&A...588A..74C} for radial modes (not shown)
are in good agreement with the predictions of
\cite{2013MNRAS.435..885J}.    At $T_{\rm eff} \lesssim 7800$ K there
is also a non-negligible contribution to mode driving due to the
He-He$^{+}$ and H-H$^{+}$ partial ionization zones  ($\log T \sim
4.42$ and $\log T \sim 4.15$, respectively).  The blue edge for $\ell=
2$ modes (not shown) is slightly ($\sim 10-30$ K) hotter than the
$\ell= 1$ blue edge. The location of the blue edges does  not depend
on the prescription for the MLT theory of convection adopted in the
equilibrium models, at variance with what happens in the case of
ELMVs. The blue edge of radial modes is substantially cooler ($\sim
1000$ K) than for nonradial modes.  The nature of the pulsation modes
excited in pre-ELMV stars has been  established as being
high-frequency $p$ modes and  intermediate-frequency ``mixed'' $p$-$g$
modes \citep{1974A&A....36..107S}. These  modes  behave  like $g$
modes  in  the inner parts  of  the  star  and like $p$ modes in  the
outer parts. This is due to the very peculiar shape of the
Brunt-V\"ais\"al\"a frequency profile in the inner regions of the star
\citep[see Fig. 2 and 3 of][]{2016A&A...588A..74C}.

\begin{figure}
\includegraphics[width=1.0\textwidth]{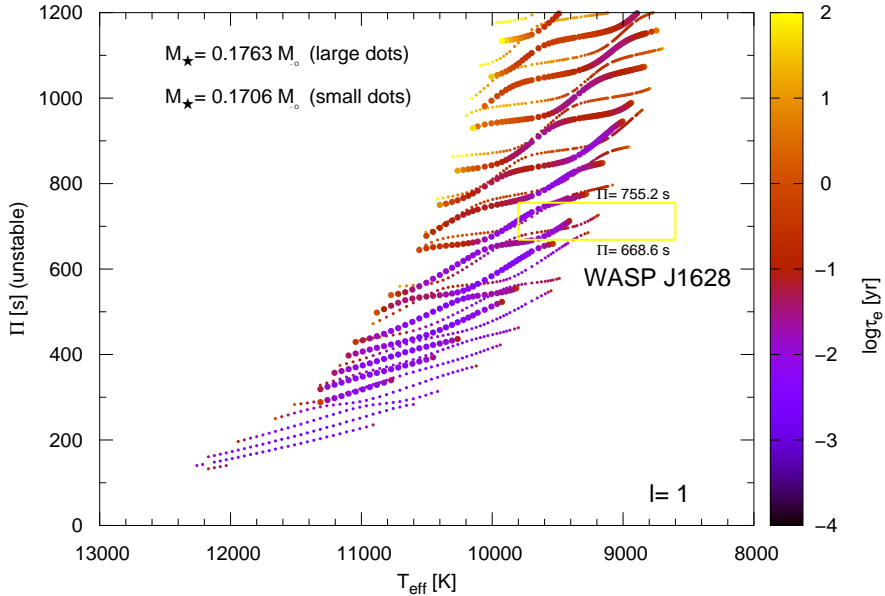}
\caption{Periods of unstable $\ell= 1$ modes in terms of the
  effective temperature, with the palette of colors (right scale)
  indicating the value of the logarithm of the $e$-folding time (in
  years),  corresponding to the sequences with $M_{\star}= 0.1706
  M_{\odot}$ (small dots) and $M_{\star}= 0.1763 M_{\odot}$ (large
  dots)  neglecting element diffusion. Also shown are the pulsation
  periods of the pre-ELM WD star WASP J1628+10B.}
\label{per-teff-eft-wasp}
\end{figure}

The theoretical  computations of \cite{2016A&A...588A..74C} account
for  the existence of some of  the known pre-ELMVs, including the
observed ranges  of  excited periods. As an example,  in
Fig. \ref{per-teff-eft-wasp} we show  the case of WASP J1628+10B. In
this figure, we depict the theoretical periods of  unstable dipole
modes in terms of $T_{\rm eff}$ with the values of the logarithm of
the $e$-folding time, corresponding to sequences with $M_{\star}=
0.1706 M_{\odot}$ and $M_{\star}= 0.1763 M_{\odot}$.  The most
unstable modes (smaller $e$-folding times) have the shorter periods.
The pulsation periods of WASP J1628+10B are marked with horizontal
segments.  Note that the two periods of this star are well accounted
for by the theoretical  predictions at the right effective
temperature. The computations of \cite{2016A&A...588A..74C}, however,
fail to predict  pulsations in three pulsating objects: WASP0247,
SDSSJ1157, and SDSS0756   (the hottest ones,  see
Fig. \ref{HR-preELMVs}).  This is intimately linked to the abundance
of He at the driving zone;  i.e.,  larger He abundances at the
envelope  of the models are required in order for the blue edge to be
hotter, so as to include these three stars within the  instability
domain. This could be achieved by adopting different  masses for the
initial donor star in the original binary system.  That could lead to
low-mass pre-WD models with different He abundances at  their
envelopes. 

The results described above are derived from computations in which
element diffusion is not allowed to operate. When element diffusion
is considered in the  pre-ELM WD models,  the driving region is
quickly depleted of He by  virtue of gravitational settling, and hence
the instability region in the $T_{\rm eff} - \log g$ plane shrinks. In
this case, none  of the known pre-ELMVs is within the instability
region. This very important result strongly suggests that element
diffusion  should not be operative in the low-mass pre-WD stage. This
is a necessary  condition also for pre-ELMVs to have envelopes with
mixed H and He,  as observed \citep{2016ApJ...822L..27G}. In order to prevent (or diminish) the
effects of element diffusion,   two main  factors could be playing a
role, namely stellar winds and/or stellar  rotation. This interesting
issue has been recently addressed by  \cite{2016A&A...595L..12I}, who
demonstrated  that rotational mixing, which counteracts gravitational
settling,  can maintain a sufficient amount of He within the driving
region of the  pre-ELM WDs. In this way, He can drive pulsational
instabilities  with periods that are in line  with the observed
periods. These authors point out that, by the time such a star enters its
cooling track, gravitational settling overcomes rotational mixing,
allowing the formation  of a pure H envelope. At this point, the star
is able to develop  pulsation $g$ modes due to the partial ionization
of H, becoming an  ELMV variable star. According to this scenario, it
is expected that the  ELMVs are rotating, and that rotation should be
evidenced in splittings  of the pulsation frequencies. However, no
rotational-splitting signals  have yet been detected in the pulsation
spectrum of any ELMV star. 

Very interesting research about the evolutionary and  pulsational
properties of WASP 0247$-$25B, a pre-ELMV star  component of the
double-lined eclipsing system WASP 0247$-$25,  was presented by
\cite{2017ApJ...847..130I}. \cite{2013Natur.498..463M} provided
fundamental parameters of this star at a unique level of precision. By
employing state-of-the-art evolutionary models   that take into
account rotational mixing and diffusion processes along with a
comprehensive stability pulsation analysis, \cite{2017ApJ...847..130I}
were able  to find a stellar model that closely reproduces observed
properties of the star such as stellar mass, orbital period, surface
gravity, effective temperature,  surface chemical composition, and
also the pulsation periods detected  in the star. This reinforces the
validity of the pre-ELM WD models of \cite{2016A&A...595L..12I} that
incorporate rotational mixing. 

\begin{figure*}[t]
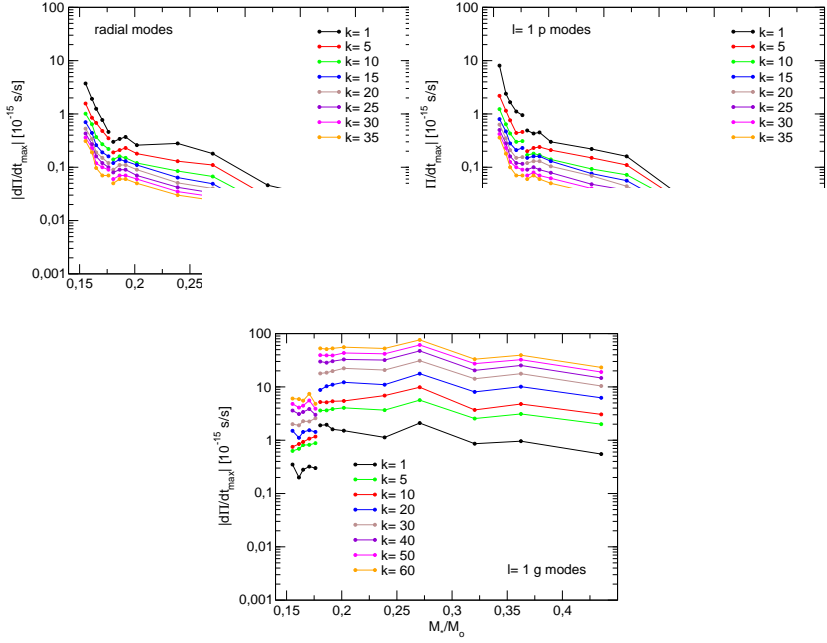

  \begin{center}
\subfigure{\includegraphics[width=.45\textwidth]{max-r.eps}}
\subfigure{\includegraphics[width=.45\textwidth]{max-p.eps}}
\subfigure{\includegraphics[width=.45\textwidth]{max-g.eps}}
  \end{center}
  \caption{Upper-left panel: absolute value of the maximum $\dot{\Pi}$
  value in terms of the stellar mass corresponding
  to radial modes ($\ell= 0$) with selected radial orders $k$,
  for low-mass WD models with effective temperatures in the range
  $8000 \lesssim T_{\rm eff} \lesssim 10000$ K.
  Upper-right panel: same as in the upper left panel, 
  but for $\ell= 1$ $p$ modes.
  Lower panel: same as in the upper left panel, 
 but for $\ell= 1$ $g$ modes.}
\label{max_rate_g_p_r}
\end{figure*}

An aspect of crucial importance in the investigation of low-mass WDs
is  the precise identification of the individual stars, either as
precursor objects of WDs (pre-WDs) or as stars that have  already
entered their WD cooling tracks. According to theoretical models
\citep{2013A&A...557A..19A,2016A&A...595A..35I}, low-mass WDs cool
more slowly than low-mass pre-WDs. On these grounds, it is expected that
ELMVs should exhibit smaller rates of period  change, $\dot{\Pi}$
($\equiv d\Pi/dt$),  than  pre-ELMVs.   Hence,  the  eventual
measurement of the rate of period change  for a given pulsating star
in the region of the  $\log T_{\rm eff}-\log g$ diagram populated by
ELMVs and pre-ELMVs could  be potentially  useful  to  know its
evolutionary  stage. Also, a measurement of $\dot{\Pi}$ could help to
distinguish  ELM  WDs  $(M_{\star} \lesssim 0.18-0.20 M_{\odot})$,
that  have  thick  H  envelopes  and  long  cooling timescales,  from
LM WDs ($M_{\star} \gtrsim 0.18-0.20 M_{\odot}$), characterized by
thinner H envelopes and shorter cooling timescales.  On the other
hand, some of these stars may be headed towards a CNO-cycle flash, 
and thus have much larger ---and more easily  detectable--- $\dot{\Pi}$ values.  Also, in view of the systematic  difficulties in the
spectroscopic classification of stars of the ELM  Survey
\citep{2017ApJ...835..180B},  an eventual measurement of $\dot{\Pi}$
could help to confirm that a  given pulsating star is an authentic
low-mass WD and not a star at a different evolutionary stage. 

\cite{2017A&A...600A..73C} have carried  out a
comprehensive analysis of the secular rates of period change
theoretically expected in ELMV and pre-ELMV stars, as well as WD
precursors that are evolving at stages prior to the  development of
CNO-cycle flashes during the early cooling phase.  For the case of ELMVs, we
show in Fig. \ref{max_rate_g_p_r} the  maximum absolute value expected
for the rate of period change,  $|\dot{\Pi}_{\rm max}|$, as a function
of $M_{\star}$, for $\ell= 1$  $g$ and $p$ modes,  and also for radial
modes ($\ell= 0$), for selected values of the radial order $k$, covering the range
of periods typically  observed in ELMV stars.  In the case of $g$
modes, which are the modes that have been detected so far in these
stars \footnote{\cite{2013ApJ...765..102H} reported the discovery
  of short-period pulsations  compatible with $p$ modes or radial
  modes in an  ELMV WD (SDSS  J111215.82+111745.0), but this needs to
  be confirmed  with further observations.}, there is a clear
distinction in  the magnitude of $|\dot{\Pi}_{\rm max}|$ depending on
whether $M_{\star} \lesssim 0.18 M_{\odot}$  or $M_{\star} \gtrsim
0.18 M_{\odot}$.  Specifically, the rates of period change  for ELM
WDs  are about ten times smaller
than for more massive models. In contrast, for radial modes and $p$
modes there is no clear differentiation in  $|\dot{\Pi}_{\rm max}|$
between models, no matter than the mass satisfies   $M_{\star} \lesssim 0.18 M_{\odot}$  or $M_{\star} \gtrsim 0.18 M_{\odot}$. Also, the  $|\dot{\Pi}_{\rm max}|$
values are  systematically larger for $g$ modes as compared with
radial and $p$ modes. 

\begin{table*}
\centering
\caption{Absolute value of the {\it maximum} expected rates of period change,
  $|\dot{\Pi}_{\rm max}|$ (s/s), and the sign --- that indicates that the periods are decreasing or increasing with time--- for nonradial $\ell= 1$ $g$ and $p$ modes (and $p-g$ mixed modes for pre-ELMVs) corresponding to low-mass WD and pre-WD models and also objects evolving just before the CNO-cycle flashes.}
\begin{tabular}{l|cc|cc|cc} 
\hline
\hline
\noalign{\smallskip}
Evolutionary phase & $|\dot{\Pi}_{\rm max}|$ &   &  $|\dot{\Pi}_{\rm max}|$ &  & $|\dot{\Pi}_{\rm max}|$  &  \\
 & $g$ modes & & $p-g$ modes & & $p$ modes &  \\
\noalign{\smallskip}
\hline
\noalign{\smallskip}
pre-WD (pre-ELMVs) & $\sim 3 \times 10^{-13}$ & $(<0)$ & $\sim 5 \times 10^{-12}$ & $(<0)$ & $\sim 3 \times 10^{-12}$ & $(<0)$ \\
pre-CNO-cycle flashes & $\sim 3 \times 10^{-11}$ & $(<0)$ & $\cdots$ & &$\sim 2 \times 10^{-13}$ & $(>0)$ \\
WD (ELMVs)      & $\sim 8 \times 10^{-14}$ & $(>0)$  & $\cdots$ & & $\sim 8 \times 10^{-15}$ & $(<0)$\\
\noalign{\smallskip}
\hline
\hline
\end{tabular}
\label{max-pdot}
\end{table*}

In Table \ref{max-pdot}, adapted from  \cite{2017A&A...600A..73C}, we
show the maximum expected rates of period change (absolute value)
for  dipole $g$, $p$, and $p-g$ mixed modes (in the case of pre-ELMVs)
for low-mass WD and pre-WD models and also objects that are expected
to  be briefly residing in stages prior to flashes. We envisage that
any  future  measurement  of  $\dot{\Pi}$ for a given pulsating
low-mass pre-WD or WD star could help to establish  its evolutionary
status. For instance,  it could be possible to distinguish a star that
is in its pre-WD phase,  if it is evolving in stages just prior to a H
flash, or if it is already settled on its final cooling stage as a
WD. Although less likely, it would also be possible to discriminate
whether or not an ELMV star is an ELM or an LM WD.  Finally, a measured
value of $\dot{\Pi}$ larger than $\sim 10^{-10}$ s/s would mean that
the object is still evolving  quickly between CNO-cycle flashes.

We close this section by briefly describing the first attempts of
asteroseismology of  ELMV stars. A detailed asteroseismological study
of all the known and suspected  ELMV stars based on the low-mass
He-core WD models of  \cite{2013A&A...557A..19A} was presented by
\cite{2017A&A...607A..33C}.  Despite  ELMVs exhibiting very few
periods and the period-to-period  fits showing multiple
solutions, they found that it is still possible to find
asteroseismological models with $M_{\star}$ and $T_{\rm eff}$
compatible with the values derived by spectroscopy for most 
cases. Due to the scarcity of periods generally   exhibited by these
stars, it is not feasible to assess the mean period spacing to
constrain the stellar masses. A step forward in asteroseismology of
ELMVs  was given by \cite{2018A&A...620A.196C} who redid the
asteroseismological analysis of these stars but this time expanding
the parameter space, i.e., adopting the thickness of the H envelope
($M_{\rm H}/M_{\star}$) as a free parameter, in addition to
$M_{\star}$ and $T_{\rm eff}$.  They found again multiple
asteroseismological solutions in all the cases,  something that could
be due to the few periods exhibited by these stars.  Only with the
inclusion of external constraints, that is, the effective temperature
and surface gravity derived from spectroscopy, was it possible to
adopt an asteroseismological model or a family of solutions for each
star.  Interestingly enough, some of the stars analyzed are
better represented by asteroseismological models that harbor thin H
envelopes.  In connection with this finding, it is predicted that
stable mass transfer  during the binary evolution of the ELM WD
progenitors leads to the formation  of ELM WDs with thick H
envelopes. However, the formation of ELM WDs with thinner H envelopes
from unstable mass loss cannot be discarded. The results of
\cite{2018A&A...620A.196C} seem to reinforce the idea that this
scenario of formation could work in practice, and that the existence 
of ELM WDs ($M_{\star} \lesssim 0.18-0.20 M_{\odot}$) with thin H envelopes is
possible. Such WDs would not have residual H burning and thus
should cool very quickly. Hence, they should  have had time enough to
cool down to very low effective temperatures \citep[up to $T_{\rm eff}
  \sim 2500-3000$~K; see ][]{2018A&A...614A..49C}.

Going back to the issue of asteroseismology of ELMV stars, we conclude
that with the current sets of observed periods of ELMVs, it is not
possible to find a unique asteroseismological solution for each
star  without invoking the spectroscopic determination of $T_{\rm
  eff}$ and $\log g$.  The situation gets worse if one or more periods
are affected by large  uncertainties. Since  a complete set of state-of-the-art 
evolutionary models representative of
He-core ELM WDs with different H-envelope thicknesses is employed,  
it is likely that a limit regarding the possibility of the
asteroseismological approach to find a representative model in order
to  infer the internal structure of these stars has been reached. 
This implies the urgent necessity  of detecting more pulsation modes in these
stars in order to make more robust asteroseismological
inferences. On the other hand,  the discovery of additional ELMV stars
is also a pressing need in order to better determine their 
internal structure and the nature of their progenitors. 

\subsection{Blue Large-Amplitude Pulsators (BLAPs)}
\label{sec:3.7}

\begin{figure}
\includegraphics[width=1.0\textwidth]{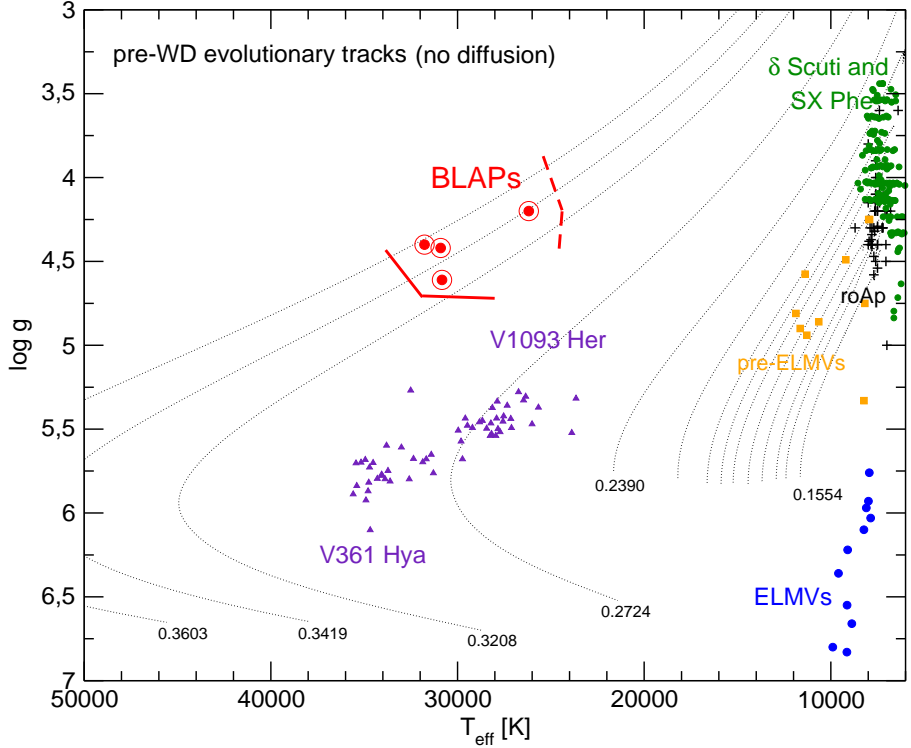}
\caption{$T_{\rm eff} - \log g$ diagram showing the location of four
  BLAP stars with measured atmospheric parameters (red dots surrounded by circles), along
  with other families of already known pulsating stars: ELMVs (blue
  dots), pre-ELMVs (orange squares), pulsating sdBs (V361 Hya and V1093
  Her; violet triangles), $\delta$ Sct/SX Phe stars (green
  dots), and roAp stars (black plus symbols). Dotted lines 
  correspond to low-mass  He-core pre-WD
  evolutionary tracks for $Z= 0.01$ computed neglecting element
  diffusion. Numbers correspond to the stellar mass of some
  sequences. Also included are the boundaries of the theoretical
  instability domain of BLAPs (thick solid and dashed red lines) of
  $\ell= 1, 2$ $g$ modes and radial fundamental modes ($\ell= 0, k= 0$), 
  according to \cite{2018MNRAS.477L..30R} (see the text).}
\label{Teff-logg}
\end{figure}

Recently, a new class of pulsating stars, the Blue Large-Amplitude
Pulsators \citep[BLAPs;][]{2017NatAs...1E.166P}, was discovered in the
Optical Gravitational Lensing Experiment \citep[OGLE; ][]
{2015AcA....65....1U}. The object named OGLE-BLAP-001 was the  first BLAP star discovered, becoming the prototypical member of the class. It was originally
misclassified as a $\delta$ Scuti variable star
(OGLE-GD-DSCT-0058). At the time of writing this review, 14 BLAPs are
known.  BLAPs have been discovered only in the Galactic
disk  and bulge (high-metallicity environments), but not in
the low-metallicity  environment of the Magellanic Clouds
\citep{2018pas6.conf..258P}.  These variables have not been observed
in globular clusters nor in  the Galactic halo.  It seems that high
metallicity is a crucial ingredient in these pulsating stars, and this
is confirmed by the theoretical models.  BLAPs are very hot stars,
with an average effective temperature of $T_{\rm eff}\sim 30\,000$
K. Their effective temperature and colour  change over a complete
pulsation cycle, confirming that their  variability is due to
pulsations. Their lightcurves are  similar to those of
classical pulsators like Cepheid- and RR Lyrae-type  stars that
exhibit just the radial fundamental mode ($\ell= 0, k= 0$),  show
large amplitudes ($0.2-0.4$ mag) and short periods  ($\sim 1200-2400$
sec). The pulsation periods of BLAPs show a  secular drift with
typical values of $\dot{\Pi}/\Pi=d(\log \Pi)/dt= 10^{-7}$ yr$^{-1}$,
both positive (increasing periods) and negative (decreasing
periods). The magnitudes of the rate of period change suggest that
BLAPs are stars that are evolving on nuclear timescales.
Finally, BLAPs exhibit envelopes made of a mixture of H and He.  In
Fig. \ref{Teff-logg} we show the location of some BLAPs with available
atmospheric parameters in the $T_{\rm eff} - \log g$ diagram. For
comparison,  we also plot the location of other classes of
known pulsating stars.  Notably, BLAPs populate a region of the
diagram not occupied by any category  of pulsating stars previously
studied. Indeed, they have effective temperatures that may be approached by the 
hottest $\beta$ Cep stars (not included in Fig. \ref{Teff-logg}), but 
are also much fainter; BLAPs are much hotter than pre-ELMVs, $\delta$
Scuti/SX Phe stars, and roAp stars, although they share similar surface gravities;
BLAPs have similar $T_{\rm eff}$ as pulsating sdB stars, but are
much less compact; finally, BLAPs are much hotter and less compact than
ELMVs.  Gaia Data Release 2 \citep[DR2;][]{2018A&A...616A...1G} has
measured the parallax of 10 BLAPs, 6 of which have  absolute
magnitudes and intrinsic colours consistent with the temperature
derived with  optical  spectra  and theoretical predictions, whereas 4
stars have properties which appear different and may correspond to other types of
pulsating variables \citep{2018A&A...620L...9R}.

The peculiar characteristics of these pulsating stars ---they show 
very high amplitudes that are unusual in very hot pulsating stars, they exhibit 
short periods and small rates of period change--- pose a challenge 
to the theory of stellar evolution and  pulsations \citep{2017NatAs...1E.166P}. 
No evolutionary/pulsational model has been proposed up to now that 
explains entirely the existence and
the pulsational properties of  BLAPs. At the outset, the evolution of
single isolated low-mass stars has to be discarded to explain BLAPs,
because the evolutionary timescales involved in such a scenario should
be much longer  than the Hubble time. Instead,
\cite{2017NatAs...1E.166P} have proposed that binary-star evolution
through stable mass transfer and/or common  envelope ejection could be
a plausible evolutionary channel for these intriguing pulsating
stars. \cite{2017NatAs...1E.166P} examined two possibilities: {\it
  (i)} BLAPs are He-core,  H shell burning low-mass stars ($\sim 0.30
M_{\odot}$), or  {\it (ii)} BLAPs are core He-burning stars ($\sim 1.0
M_{\odot}$). 

The first detailed evolutionary and pulsational study  focused on the
evolutionary origin and the nature of BLAPs was that of
\cite{2018MNRAS.477L..30R}. They examined in detail the possibility
that BLAPs are hot  He-core,  H shell burning low-mass pre-WD stars
with masses $\sim 0.30 M_{\odot}$ coming from binary star
evolution. In Fig. \ref{Teff-logg} we display  the evolutionary tracks
of \cite{2013A&A...557A..19A}  (element diffusion not considered)
corresponding to $Z= 0.01$ for  $M_{\star}=
0.3208, 0.3419$ and $0.3603 M_{\odot}$ evolving models.  The location
of the four BLAPs  with known atmospheric parameters ($T_{\rm eff}$
and $g$) is  nicely accounted for by the evolutionary  tracks,
suggesting that the scenario studied by \cite{2018MNRAS.477L..30R} is
a plausible one. The next step  taken by these authors was to try to
identify the type of pulsation modes  responsible for the pulsations
observed in BLAPs. Specifically, they examined the  possibility that
periods of radial ($\ell= 0$) modes and/or nonradial ($\ell > 0$) $p$
and $g$ modes match the periods of oscillation of BLAPs. They found
that the periods of the fundamental radial mode or the first overtone
($k= 0, 1$), and the periods of high radial order ($25 \lesssim k
\lesssim 50$ for $\ell= 1$) $g$ modes of solar-metallicity  low-mass
He-core pre-WD models ($M_{\star} \sim  0.32 M_{\odot}$, $T_{\rm eff}
\sim 31000$ K, $\log g \sim 4.8$), are compatible with the
periodicities detected in BLAPs. When tested for pulsational
stability, these stellar models proved inadequate  since all the
pulsation modes with periods compatible with those of the BLAPs  are
globally stable, although substantial driving due to the $\kappa$
mechanism  acting at the location of the $Z$ bump of the Rosseland
opacity was found. This  driving guided the authors to investigate
template models with higher metallicity, finding that for
representative models with $Z= 0.05$,  the radial fundamental mode and
nonradial $g$ modes with the right periods  are unstable, confirming
the need for high metallicity (see below) as a necessary 
condition to explain the pulsations of BLAPs.  The limits of the 
theoretical instability domain
of BLAPs associated to $\ell= 1, 2$ $g$ modes and the radial
fundamental mode corresponding to stellar models with $Z= 0.05$ are
displayed with thick red lines in   Fig. \ref{Teff-logg}. 

\begin{figure*}[t]
  \begin{center}
\subfigure{\includegraphics[width=.49\textwidth]{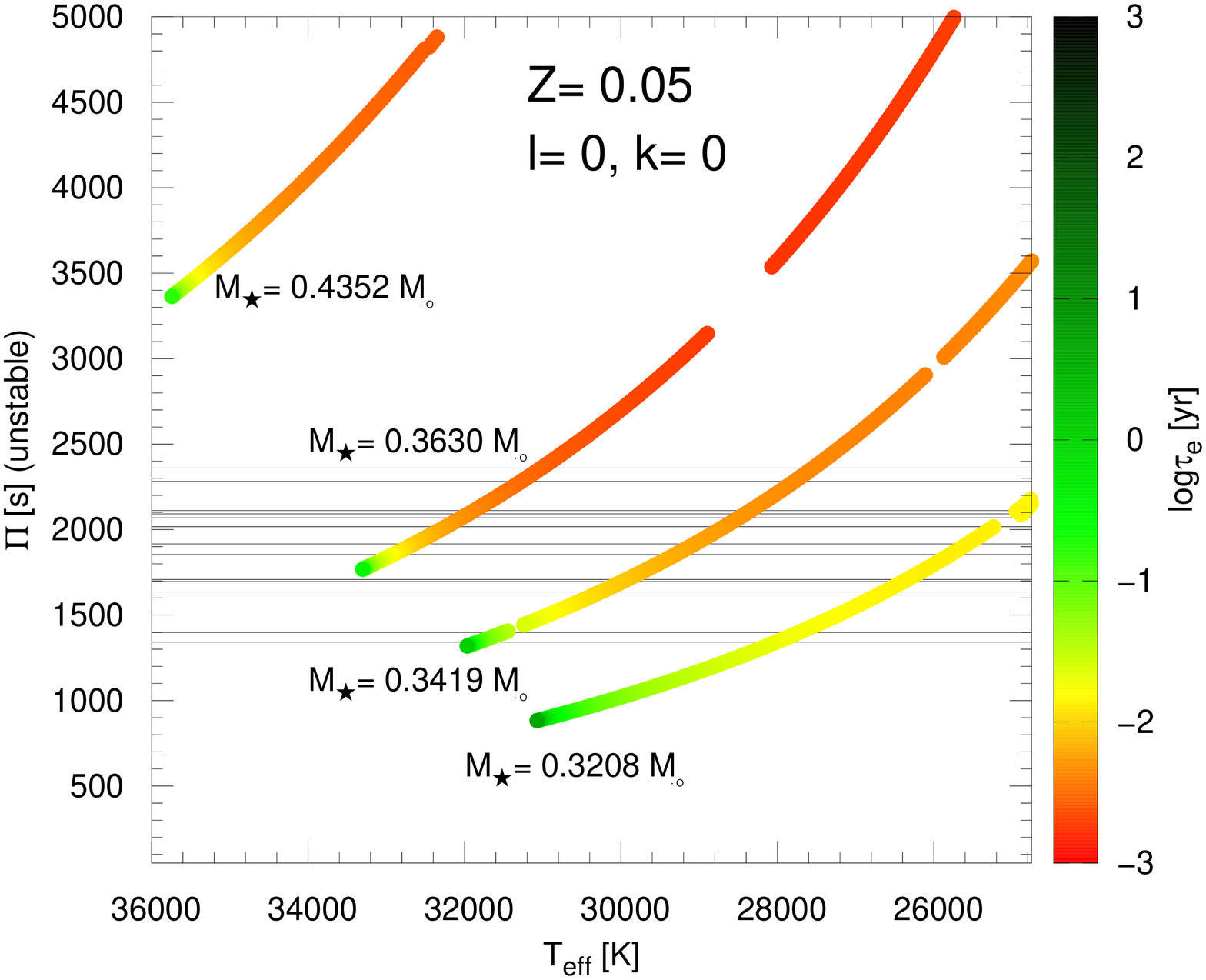}}
\subfigure{\includegraphics[width=.49\textwidth]{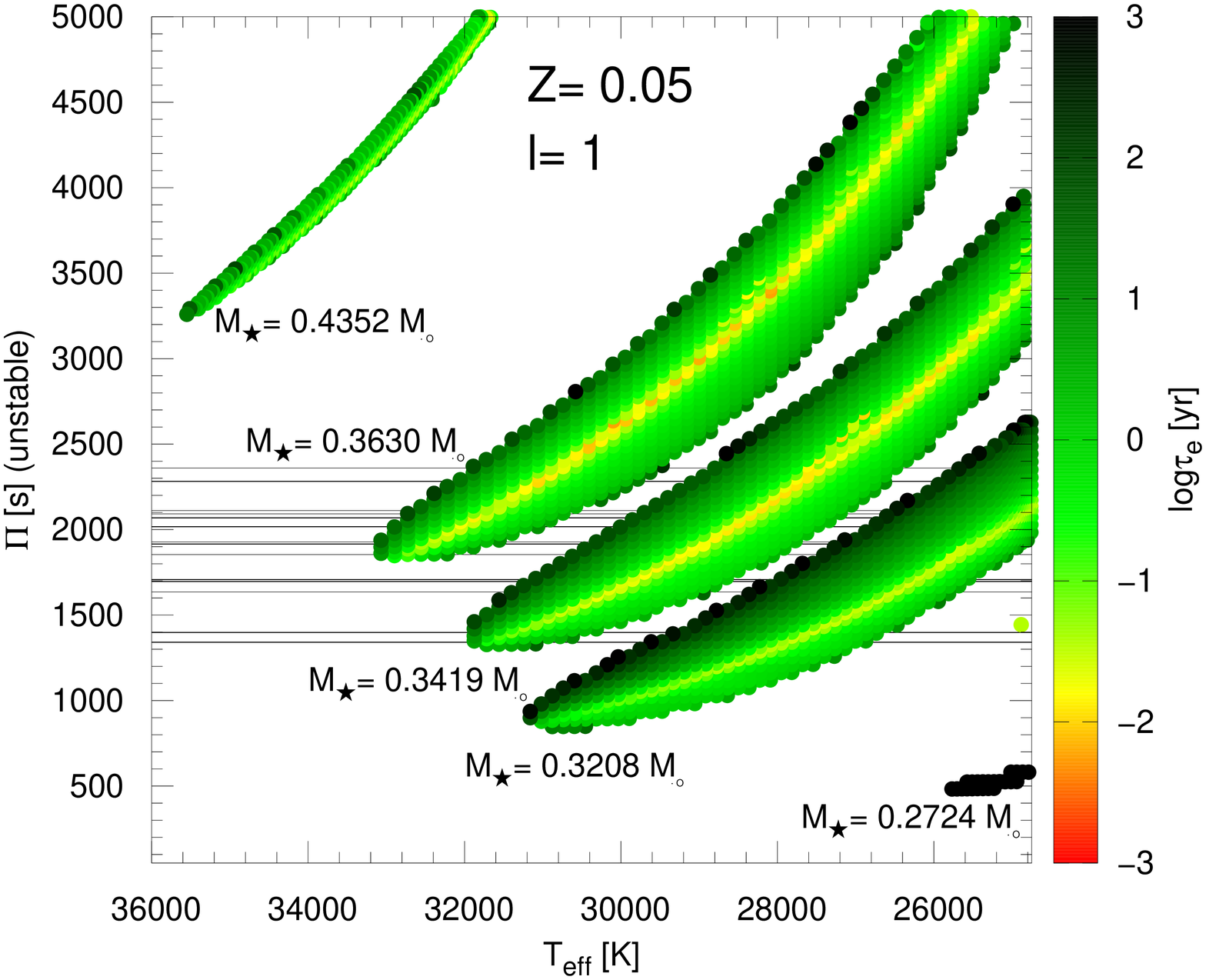}}
  \end{center}
  \caption{Left panel: periods of the unstable fundamental  ($k= 0$)
    radial mode ($\ell= 0$) in terms of $T_{\rm eff}$ for pre-WD
    models  of the indicated masses and $Z= 0.05$. The palette of
    colors at the right scale indicates the value of the logarithm of
    the $e$-folding time (in yrs). The $e$-folding times range from
    $\sim 10^{-3}$ to $\sim 10^3$ yr, much shorter than the typical
    evolutionary timescales at that stage of evolution. The horizontal
    lines correspond to the periods detected in BLAPs. Right panel:
    same as in the left  panel, but for the case of nonradial $g$
    modes with $\ell= 1$ and a range of radial  orders $k$.}
\label{per-teff-eft-blaps}
\end{figure*}

In Fig. \ref{per-teff-eft-blaps} we show the domains of instability of
BLAPs in the $T_{\rm eff}-\Pi$ diagram  for radial and nonradial
dipole modes corresponding to different values of the stellar
mass. Similar results are found for quadrupole modes (not shown).  In
the case of  radial modes (left panel), only the fundamental mode is
unstable.  The observed periodicities $(1200 \leq P \leq 2400)$ s are
well accounted for by the theoretical computations considering a range
of stellar masses ($0.33 \leq M_{\star}/M_{\odot} \leq 0.36$). The
radial fundamental modes are the most  unstable ones among the studied
cases ($\ell= 0, 1$). In fact, they are destabilized during very short
times ($e$-folding times) as compared with the evolutionary
timescales at that stage of evolution. 

The pulsation analysis of \cite{2018MNRAS.477L..30R} suggests  that
the pulsations of the BLAPs are better explained by the excitation of
the fundamental radial mode in these stars. Indeed, the fundamental
radial mode has the correct period, it is pulsationally unstable in
the range of effective temperatures of interest, and it is more
unstable than the nonradial $g$ modes. This conclusion  is reinforced
by the fact that BLAPs exhibit a {\it single} mode with large
amplitude in the lightcurves, reminiscent of typical radial
fundamental-mode pulsations.  It cannot be discarded, however, that the pulsations
of BLAPs correspond to nonradial  $g$ modes, and that the fact that only one
period is detected is due to just an insufficient observing time. 
On the other hand,  evidence in  favor of the
interpretation of $g$ modes comes from the comparison of  the rates of
period change of BLAPs ---measured by  \cite{2017NatAs...1E.166P}---
with the theoretical expectations for  radial and nonradial
modes. Indeed, \cite{2018arXiv180907451C} have  shown that the
$\dot{\Pi}/\Pi$ values   of nonradial $g$ modes with high radial order
$k$ are in much better  agreement  with the values measured in BLAPs
than the  fundamental radial mode.  Clearly, the exact nature of the
pulsation  modes responsible for the variability of BLAPs
remains a matter of  debate.

In closing this section, we mention the issue of the puzzling
super-solar metallicity required to drive pulsation modes through the
$\kappa$ mechanism.  In the models of \cite{2018MNRAS.477L..30R}, the
metallicity is {\it globally} augmented to have an enhanced $Z$ bump
in the Rosseland opacity. However,  it would be possible to find
instability with a {\it local} enhancement of the opacity  at the
location of the $Z$ bump in the star. This issue has been addressed
by \cite{2018MNRAS.481.3810B}, who examined the pulsation instability
of radial  modes of single-star evolution models of post-RGB stars
that have  undergone a common envelope ejection in the form of a high
mass-loss rate. These authors have included the effects of radiative
levitation  which leads to mode excitation by the $\kappa$ mechanism
at $T_{\rm eff}$  values comparable to those of BLAPs and the right
period interval.  By comparing with the observations,
\cite{2018MNRAS.481.3810B}  favor models with $\sim 0.31 M_{\odot}$ as
the more likely candidates for BLAPs,  in very good agreement with the scenario
proposed by \cite{2018MNRAS.477L..30R}. Furthermore, and more
importantly, the proposal of  \cite{2018MNRAS.477L..30R} that the $Z$
bump (iron and nickel) in the opacity is responsible for the
excitation of  pulsations in BLAPs is confirmed  by the more detailed
radiative-levitation  calculations carried out by
\cite{2018MNRAS.481.3810B}. 

\subsection{The sdA problem}
\label{sec:3.8}

In the course of the analysis of the Sloan Digital Sky Survey
(SDSS)  Data Release 12 (DR12), thousands of objects showing H-rich
spectra and sub-MS surface gravities ($4.75 \lesssim \log g  \lesssim
6.5$),  but effective temperatures lower than the ZAHB ($T_{\rm
  eff}\sim 10\,000$~K), were identified and classified with the
spectral class of subdwarf A stars
\citep[``sdAs'';][]{2016MNRAS.455.3413K}. The evolutionary origin of
the sdAs has  been elusive since their
discovery. \cite{2017ApJ...839...23B} suggested that sdAs could be
mainly metal-poor A/F stars in the halo with an overestimated $\log g$
value,  due to the use of pure H atmospheres to fit the spectra of
these objects.  However, \cite{2018MNRAS.475.2480P} showed that the
addition of metals  to the H atmosphere models does not necessarily
lower the estimated $\log g$ value.  At present, it is understood that
sdAs consist of multiple  populations, including  byproducts of binary
evolution  (blue-stragglers, ELM WDs, and their  precursors, pre-ELM
WDs),  metal-poor A/F  dwarfs,  or  even stars
accreted from dwarf galaxies. Preliminary constraints on sdA radii
using parallaxes from Gaia DR2 suggest that at least dozens of sdAs
are ELM WDs or  their precursors \citep{2019MNRAS.482.3831P}.

A promising avenue to study the origin and nature of sdAs is to  look
for pulsational variability in objects of this category. 
Indeed,  the pulsation properties exhibited by any sdA
star could help to understand their internal structure and
evolutionary phase. \cite{2018A&A...617A...6B} carried  out
photometric campaigns targeting 24 sdA stars classified from SDSS
spectra.  They found 7 new pulsating stars, which show pulsation
characteristics of ELMVs, pre-ELMVs, $\delta$ Scuti and even ``RR Lyrae-like'' 
pulsators\footnote{They have been called ``binary evolution pulsators'' by 
\cite{2013MNRAS.428.3034S}.}. Also,  \cite{2018MNRAS.478..867P} obtained time-series
photometry for 21 sdAs,  and found 7 new photometrically variable
stars, one of which with  pulsation  characteristics of ELMV
stars. The  diversity  in   pulsation  properties  of  the  variable
sdAs  supports  the idea that this population comes from a mixture of
formation and  evolution scenarios. Also, it seems that the fraction
of variable  stars among the sdAs is large, something that could be
exploited by asteroseismology to shed some light on their internal
structure and  evolutionary status. 
%Theoretical computations of
%pulsation stability  at the CNO-flashes stage of low-mass pre-WDs
%indicate that numerous $g$  modes are excited through the $\epsilon$
%mechanism for effective  temperatures and gravities compatible with
%pulsating  sdA stars (Calcaferro 2019, PhD Thesis).
%However, the evolutionary  timescales at the phase of interest is so
%short (of the order of 1-10 yr)  that it would be extremely difficult
%to  detect a star at that stage. 

\subsection{WD pulsators as cosmic laboratories for fundamental physics}
\label{sec:3.9}

WD asteroseismology constitutes a novel tool for applications of WDs
to fields beyond stellar astrophysics. A vivid example of this is the 
application of pulsating WDs to constrain properties 
of elementary particles and to test  
the possible variation of fundamental constants. In this Section, we summarize 
the use of pulsating WDs to assess stringent constraints on the mass of the axions, 
the  magnetic dipole moment of the neutrino, and the secular rate of variation of the gravitational constant, $G$. For a background about the application of stars to the study of weakly interacting particles and the possible drift of fundamental constants, we recommend the reader consult the excellent review articles by \cite{1990PhR...198....1R} and \cite{2007A&ARv..14..113G}, respectively.  The application of 
 pulsating WDs to study theories with large extra dimensions, WIMPs (weakly interacting massive particles), axions
 and the possible time variation of $G$ has been discussed by \cite{2013ASPC..469...21M} \citep[see, also,][]{2002PhRvD..65d3008B,2003Ap&SS.283..601B}.

%These  applications of pulsating WDs are feasible  thanks to the so-called ``stellar energy loss argument'', according to which, {\it novel,  low-mass particles  would be produced in the interior of stars and,   because of their weak interaction with matter and radiation, would   escape almost freely, draining the star of energy and changing the   course of stellar  evolution} \citep{1990PhR...198....1R}.

\subsubsection{Upper bounds on the axions mass}
\label{sec:3.9.1}

Axions are hypothetical pseudo Nambu-Goldstone  bosons \citep[that is,
  bosons with  a   tiny mass;][]{1996slfp.book.....R} that were
postulated by \cite{1977PhRvL..38.1440P}, \cite{1978PhRvL..40..223W},
and \cite{1978PhRvL..40..279W} to solve the long-standing problem in
particle physics known as the  ``strong CP problem''
\citep{2010RvMP...82..557K}.  The strong CP problem consists of
non-violation of the charge-parity (CP)  symmetry\footnote{The CP
  symmetry establishes that the laws of physics  should be the same if
  particles were replaced with their antiparticles  (C symmetry) and
  their spatial coordinates were inverted (P symmetry)
  \citep{Luders54,1955nbdp.book.....P}.} in strong interactions. The
breaking of this symmetry is  predicted by  quantum chromodynamics
(QCD)\footnote{The name of axion  comes from the Axion laundry detergent,
  and was introduced by Frank Wilczek ``to clean QCD from the CP
  problem''.}. At present, axions are the focus of a plethora 
  of theoretical and experimental investigations  aimed at proving their existence
\citep{2018PrPNP.102...89I}. Despite much effort, this elusive
particle has not yet been detected.
\cite{2017NatPh..13..530G}, \cite{2017JCAP...10..010G}, and \cite{2019JHEP...03..191H}
present updated accounts of observational  hints from astrophysics
pointing to the existence of stellar energy losses beyond the ones
accounted for by neutrino emission, and future  experiments aimed
at detecting axions. A property of utmost  importance of axions is the
axion mass ($m_{\rm a}$). The importance of knowing the  mass of the
axion lies in the fact that,  depending on its value, they could
contribute substantially or not  to the cold dark matter of the
Universe. Axions are electrically  neutral, and they interact very
weakly with normal matter and radiation.   There exist several axion
models,  the most important one in the WD context being the DFSZ
 model \citep[Dine-Fischler-Srednicki-Zhitnitsky;][]{Zhitnitsky:1980tq,1981PhLB..104..199D}, where axions couple to charged  leptons
like electrons with a strength defined by the dimensionless coupling
constant, $g_{\rm ae}$, being $g_{\rm ae}=  2.8 \times
10^{-14}\ m_{\rm a} \cos^2 \beta$, where $\cos \beta$ is
undetermined. 

Since the axion mass is not predicted by the theory that postulates
its existence \citep{2007JPhA...40.6607R}, it must be derived from
either terrestrial  experiments  \citep{2016arXiv160100578R},  or
indirectly  by  using  well-studied  properties  of stars
\citep{Vysotsky:1978dc,1996slfp.book.....R}.  Here, we describe how pulsating WDs can
be used to constrain $m_{\rm a}$. The degenerate  cores of WDs contain
plenty of free electrons \citep{2010A&ARv..18..471A}, therefore
axions would be abundantly produced in their interiors
\citep{1986PhLB..166..402R}. In stars, the energy drain by weakly
interacting  particles (such as axions) is equivalent to a local
energy sink.  Indeed, the stellar energy generation rate can be
written $\varepsilon_{\rm eff}= \varepsilon_{\rm nuc} -
\varepsilon_{\nu}- \varepsilon_{\rm a}$, where $\varepsilon_{\rm nuc}$
is the nuclear burning rate, $\varepsilon_{\nu}$  is the neutrino
loss, and $\varepsilon_{\rm a}$ is the axion loss. Since WDs are
strongly  degenerate  and  generally do  not  have  relevant  nuclear energy
sources ($\varepsilon_{\rm nuc}= 0$)\footnote{Exceptions to this assertion 
are WDs coming from low-metallicity progenitors \citep{2013ApJ...775L..22M,2015A&A...576A...9A} 
and ELM WDs \citep{2013A&A...557A..19A}.}, 
their evolution consists of a
slow cooling process in which the gravothermal energy release is  the
main energy source regulating their evolution
\citep{1952MNRAS.112..583M}.  Thus, in WDs the emission of axions
means speeding up the cooling, with observable consequences. The axion
emission rate in WDs is $\varepsilon_{\rm a} \propto g_{\rm
  ae}^2$, with Bremsstrahlung processes being the dominant
mechanism. Therefore, the more massive the axions are, the larger the
axion emission is.   Since axions can freely escape from the interior
of WDs, their emission would accelerate cooling, with more massive
axions producing larger cooling rates. In the case of pulsating WDs,
the cooling of the star is reflected by a secular change of the
pulsation periods. The rate of period change ($\dot{\Pi}/\Pi$) is
connected to the rate of change of the core temperature ($\dot{T}/T$)
and the rate of variation of the stellar radius
($\dot{R_{\star}}/R_{\star}$) through the order-of-magnitude relation:
$\dot{\Pi}/\Pi \approx -(\dot{T}/T) + (\dot{R_{\star}}/R_{\star})$
\citep{1983Natur.303..781W}. For ZZ Ceti and V777 Her  stars, $\dot{T}
< 0$ and $\dot{R_{\star}} \approx 0$, so that   $\dot{\Pi} > 0$ and
the periods should lengthen with cooling,  something that is confirmed
by observations. 

\begin{figure}
\includegraphics[width=1.0\textwidth]{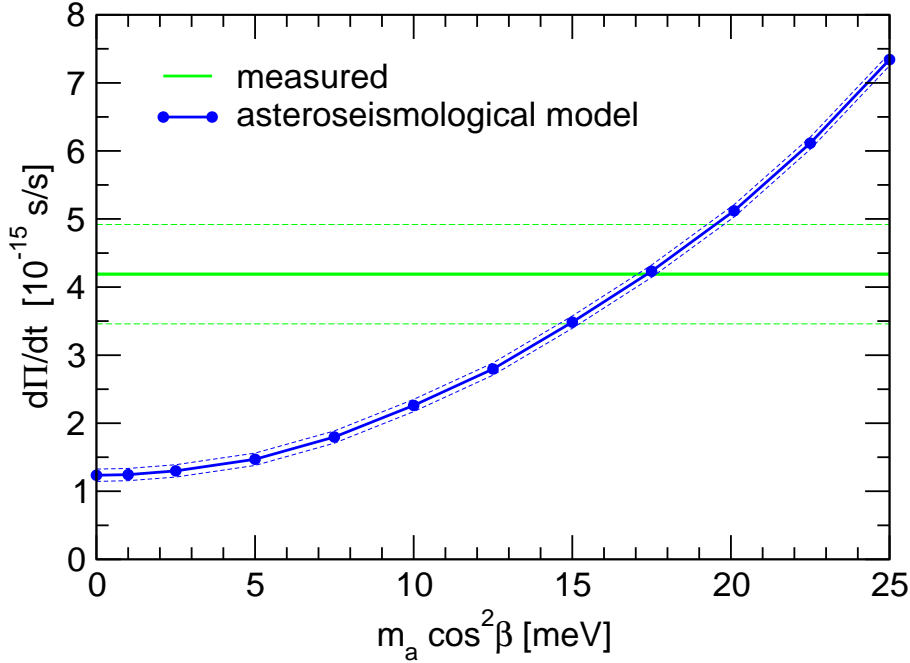}
\caption{Rate of period change for the mode with $\ell=1, k= 2$ of
  the asteroseismological model of G117$-$B15A (solid blue curve with
  dots) in terms of the axion mass. Dashed curves represent the errors
  in $\dot{\Pi}$ due to {\it internal} uncertainties in the
  asteroseismological procedure. The horizontal green lines indicate
  the observed value with its corresponding uncertainties.}
\label{dpdt-vs-max}
\end{figure}

If a pulsating WD emits axions, then it should cool faster than
otherwise expected, with the acceleration  of cooling being
proportional to the mass of the axion.  The enhancement of cooling
of the WD should be reflected in  a larger rate of
period change. This is the principle  by which it is possible to infer
constraints on the mass of the axion  through pulsating WDs. This
approach was first employed by  \cite{1992ApJ...392L..23I} considering
the archetype  ZZ Ceti star G117$-$B15A. They assessed the rate of
period change  of the largest amplitude mode ($\Pi \sim 215$ s) of
this star by means of a semi-analytic treatment considering the
evolution of DA WDs with and without axion emission, and compared the
theoretical values of $\dot{\Pi}$ for increasing values of $m_{\rm a}$
to the observed rate of change  of period with time of
G117$-$B15A. They found that in order to match the observed
$\dot{\Pi}$, the axion mass should be $m_{\rm a} \cos^2 \beta\leq 8.7$
meV  ($g_{\rm ae}\leq 2.4 \times 10^{-13}$).  Later, the approach was
refined by \cite{2001NewA....6..197C} and \cite{2008ApJ...675.1512B}
considering detailed asteroseismological analyses of  G117$-$B15A, and
obtaining  $m_{\rm a} \cos^2 \beta\leq 4.4$ meV  ($g_{\rm ae}\leq 1.2
\times 10^{-13}$) and $m_{\rm a} \cos^2 \beta\leq 26.5$ meV  ($g_{\rm
  ae}\leq 7.4 \times 10^{-13}$), respectively. The asteroseismological
models for G117$-$B15A employed in those works, however, were not
robust enough, and the observed values of the rate of period change
had  large uncertainties. The situation strongly improved with a new
measurement of the observed value of $\dot{\Pi}$ for this star
\citep[($4.19 \pm 0.73) \times 10^{-15}$ s/s;][]{2012ASPC..462..322K},
and a new asteroseismological model for G117$-$B15A \citep{2012MNRAS.420.1462R}. Armed with these new determinations,
\cite{2012MNRAS.424.2792C} obtained a new upper bound of the axion mass.  In
Fig. \ref{dpdt-vs-max} we show the rate of period change for the mode
with  period of $\sim 215$ s ($\ell= 1, k= 2$) of the
asteroseismological model for G117$-$B15A derived by
\cite{2012MNRAS.420.1462R} in terms of the axion mass, where the
errors in $\dot{\Pi}$ due to only internal uncertainties in the
asteroseismological procedure are included.  The true uncertainties of
the theoretical $\dot{\Pi}$ could be  larger if the uncertainties in
the WD previous evolution were taken into account
\citep[see][]{2017A&A...599A..21D,2018A&A...613A..46D}.  The observed
value with its uncertainties is also plotted. The comparison of the
theoretical rate  of period change associated to the
asteroseismological model with the observed  one suggests the
existence of an additional cooling mechanism in  this pulsating WD,
consistent with axions of mass of $m_{\rm a} \cos^2 \beta \leq 17.4$
meV ($g_{\rm ae}\leq 4.9 \times 10^{-13}$). This is similar to the constraint 
obtained from RGB stars in globular clusters \citep[$g_{\rm ae}\leq 4.3 \times 10^{-13}$, ][]{2013PhRvL.111w1301V}.

Similar analyses have been carried out considering other ZZ Ceti stars
for which the rate of period change has been measured. Specifically,
\cite{2012JCAP...12..010C} obtained  an upper limit of $m_{\rm a}
\cos^2 \beta \leq 17.1$ meV ($g_{\rm ae} \leq 4.8 \times 10^{-13}$)
considering the prototype ZZ Ceti star  R548 from the rate of period
change for the mode with period $\sim 212$ s measured by
\cite{2013ApJ...771...17M}, $\dot{\Pi}= (3.3 \pm 1.1) \times 10^{-15}$
s/s. A third ZZ Ceti star, L19$-$2, was  employed to infer an
upper bound for the axion mass. Indeed, \cite{2016JCAP...07..036C}
derived $m_{\rm a} \cos^2 \beta\leq 25$ meV ($g_{\rm ae}\leq 7
\times 10^{-13}$) from the value of the rate of  period change of the
modes with  periods $113$ s and $192$ s of $\dot{\Pi}= (3.0 \pm 0.6)
\times 10^{-15}$ s/s measured by \cite{2015ASPC..493..199S}. A fourth
pulsating WD, the V777 Her star PG 1351+489, has been scrutinized for
a   rate of period change. \cite{2011MNRAS.415.1220R} have  measured a preliminary value of 
the rate of period change of $\dot{\Pi}= (2.0 \pm 0.9) \times 10^{-13}$
s/s for the period at $\sim 490$ s of this star, and this value has
has been employed by \cite{2016JCAP...08..062B} to constrain the axion mass:
$m_{\rm a} \cos^2 \beta \leq 19.5$ meV $(g_{\rm ae} \leq 5.5 \times 10^{-13})$.
An alternative way  to present the results of the axion analyses from WD pulsations is in terms of
$\alpha_{26}= (g_{\rm ae}/10^{-13})^2/4\pi$, rather than in terms 
of $g_{\rm ae}$ or $m_{\rm a}$. In fact, the emission rate is proportional to
$g_{\rm ae}^2$. In Table \ref{axions-alpha26} (M. Giannotti, private communication) 
we depict the 1$\sigma$ and 2$\sigma$ intervals of $\alpha_{26}$ for G117$-$B15A \citep{2012MNRAS.420.1462R}, R548 \citep{2012JCAP...12..010C}, L19$-$2 \citep {2016JCAP...07..036C}, and  PG1351+489 \citep{2016JCAP...08..062B}.

\begin{table*}
\centering
\caption{$1\sigma$ and $2\sigma$ intervals in $\alpha_{26}$ as inferred from the WD pulsation analyses.}
\begin{tabular}{lcccc} 
\hline
\hline
\noalign{\smallskip}
Star & $\Pi$ (s) & $\alpha_{26}$ & 1$\sigma$ & $2\sigma$\\
\noalign{\smallskip}
\hline
\noalign{\smallskip}
G117$-$B15A       &    215  & 1.89  & 0.48  & 0.95  \\
R548              &    212  & 1.84  & 0.93  & 1.85  \\ 
L19$-$2           &    113  & 2.08  & 1.35  & 2.70  \\ 
L19$-$2           &    192  & 0.50 & 1.21   & 2.43  \\ 
PG1351$+$489      &    490  & 0.36 & 0.38   & 0.76  \\ 
\noalign{\smallskip}
\hline
\hline
\end{tabular}
\label{axions-alpha26}
\end{table*}

\begin{figure}
\includegraphics[width=1.0\textwidth]{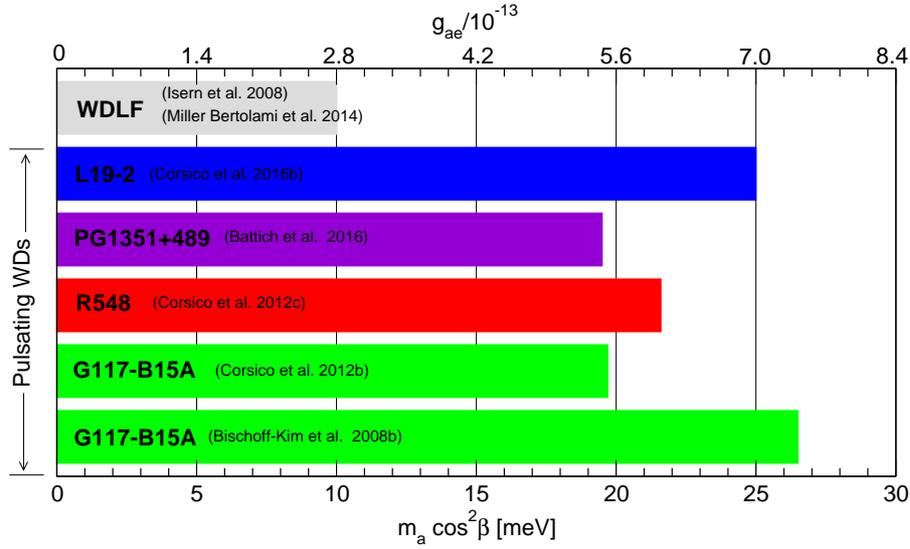}
\caption{Upper limits for the axion mass from pulsating WDs  
and from the WDLF.}
\label{cotas-axiones}
\end{figure}

Another method to determine the mass of the axion employs the WD
luminosity function (WDLF), which is defined as the number of WDs per
unit bolometric magnitude and unit volume.   The shape of the WDLF is
sensitive to the characteristics of the cooling of WDs. This was  exploited
by \cite{2008ApJ...682L.109I} who found that when axion emission is
included in the WD models, the agreement between the theoretical and
the observed WDLFs significantly improves
\citep{2008ApJ...682L.109I,2009JPhCS.172a2005I}.  More recent work
employs new  WDLFs  of  the  Galactic  disk and state-of-the-art
theoretical WD  cooling sequences \citep{2014JCAP...10..069M}. 
A tight constraint for the axion mass has been obtained by \cite{2015ApJ...809..141H} 
from the WDLF of the globular cluster 47 Tucanae, $m_{\rm a}\leq 4.1$ meV ($g_{\rm ae} \sim  8.4 \times 10^{-14}$) to a 95 \% confidence limit.
On the other hand, \cite{2018MNRAS.478.2569I} have  computed the theoretical
WDLF of the thin and thick disk, and of the stellar  halo including
axion emission, and compared them with the existing observed
WDLFs. All these studies do not discard  the existence of
extra-cooling in WDs,  compatible with axions with an upper limit for
the mass of  $\sim 10$ meV ($g_{\rm ae} \sim  2.4 \times
10^{-13}$). We summarize the   most recent DFSZ axion-mass limits
derived from pulsating  WDs as well as employing the WDLF in
Fig. \ref{cotas-axiones}. The WDLF approach seems to discard axion
masses larger than $\sim 10$ meV (in agreement with the SN1987A
determination), so, this method seems to be more stringent than that  
using pulsating WDs (axion masses $\lesssim 25$ meV). However, there exist
important  uncertainties in the determination of the
asteroseismological models,  connected with the prior evolution of WDs
and the microphysics  (EoS, opacities), and also large uncertainties
in the theoretical and observed WDLFs. Therefore, taking into account
the uncertainties affecting both approaches, the results obtained from
the WDLF and  asteroseismological models of pulsating WDs are
compatible. We conclude from both
independent methods that the existence of
extra-cooling in WDs, compatible with the emission of axions with
masses lower than  $\sim 10-25$ meV cannot be discarded. Both methods are expected to
become more precise with more pulsating WDs having $\dot{\Pi}$ measured
(for instance, through the TESS mission), and also substantial
improvements in the determining of the observed WDLF from the Gaia collaboration.

\subsubsection{Constraints on the neutrino magnetic dipole moment}
\label{sec:3.9.2}

As often happens in physics, the conservation of a given fundamental quantity
requires the existence of a particle hitherto unknown. Neutrinos have 
not been the exception.  The existence of neutrinos was first 
postulated by
W. Pauli in 1930  to explain the conservation of energy, momentum, and
angular momentum in $\beta$-decays \citep{1996slfp.book.....R}. In the
frame of the Standard Model (SM) of particle physics, neutrinos are
massless, electrically neutral, have zero decay rate, and, in
particular, they have zero dipole moment. This simple characterization of
neutrinos fails when the  observed neutrino mixing and oscillation
have to be explained. In this case, it is necessary to go beyond the
SM, allowing a finite neutrino mass, neutrino decays, and in
particular, a non-zero magnetic dipole moment, $\mu_{\nu}$
\citep{1996slfp.book.....R}. The neutrino magnetic dipole moment was
computed for the first time by \cite{1980PhRvL..45..963F}.

Neutrino emission represents an efficient energy-loss mechanism in a
variety of stars,  from low-mass red giants and horizontal-branch
stars to WDs, neutron stars and core-collapse supernovae. In the case
of WDs ---and other objects at advanced stages of stellar evolution
like the cores of red giant stars--- neutrinos are produced through thermal
effects, without nuclear reactions being  involved\footnote{This is at
  variance with solar neutrinos, which are the result of  nuclear
  fusion. For the Sun, thermal neutrino emission is negligible
  \citep{1996slfp.book.....R}.}. In particular, in the case of pre-WDs
and very hot WDs, neutrinos are produced mainly as a result of {\it
  plasmon decay} processes $[\gamma \rightarrow \overline{\nu} \nu]$,
in which a neutrino ($\nu$) and anti-neutrino  ($\overline{\nu}$) pair
is  generated due to an indirect coupling between neutrinos and
photons ($\gamma$) through electrons in a plasma. If neutrinos have a
non-zero magnetic dipole moment,  then a direct coupling between
neutrinos and the photons (electromagnetic field)  is allowed. In this
case, plasmon emission processes have to be much more efficient.

The fact that plasma neutrino emission is the dominant cooling
mechanism in pre-WDs and  hot WDs
\citep{1996slfp.book.....R,2004ApJ...602L.109W}  implies that the
evolutionary timescale of these stars is sensitive to this
mechanism. In other words, the larger the neutrino plasma emission,
the faster the evolution of these stars. This fact offers the
possibility of employing the observed properties of pre-WD and hot WD
stars to constrain the possible amount of anomalous energy loss, if it
exists, and then the magnitude of the neutrino magnetic dipole moment
\citep{1963PhRv..132.1227B,2012arXiv1201.1637R}. 

\begin{figure}
\includegraphics[width=1.0\textwidth]{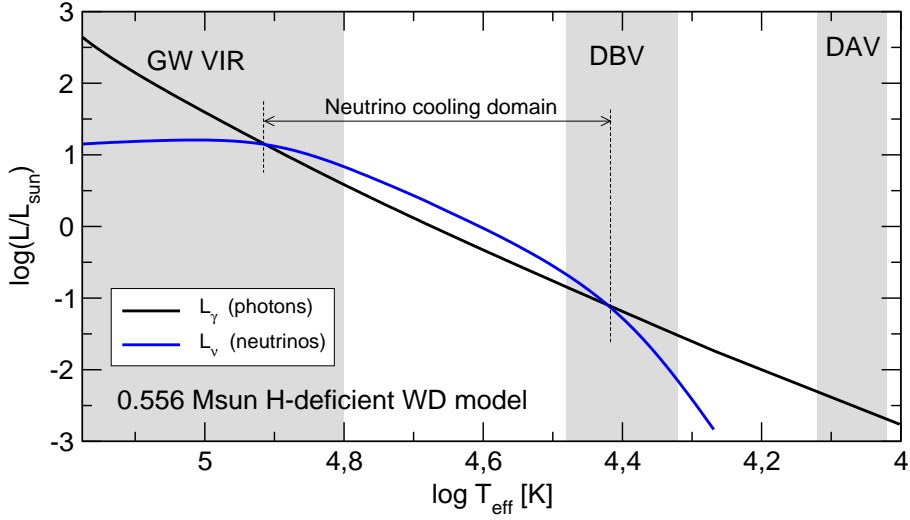}
\caption{Neutrino (blue line) and photon (black line) luminosities for a 
H-deficient WD model with $M_{\star}= 0.556 M_{\odot}$. The shaded areas 
emphasize the location of the instability domains of GW Vir, V777 Her (DBV) and 
ZZ Ceti (DAV) stars.}
\label{neutri-normal}
\end{figure}

A way to constrain the neutrino magnetic dipole moment from pre-WDs
and hot WDs is provided by asteroseismology. In particular, the DBV
(V777 Her) instability  strip ($22\,400\ {\rm K} \lesssim  T_{\rm eff} \lesssim
32\,000\ {\rm K}$) and the GW Vir instability domain  ($80\,000\ {\rm K} \lesssim
T_{\rm eff} \lesssim 180\,000\ {\rm K}$) partly overlap with the $T_{\rm eff}$
interval in which plasmon neutrino emission controls the evolution of
H-deficient WDs.  This is borne out in Fig. \ref{neutri-normal}, where
we plot the neutrino and photon luminosities of a
0.556~$M_{\odot}$ H-deficient WD model in terms of the effective
temperature. From $\log T_{\rm eff} \sim 4.9$ down to $\log T_{\rm eff}
\sim 4.4$, neutrino luminosity exceeds photon luminosity. The
evolutionary timescale can be derived, in principle, by measuring the
rates of period changes. Therefore, in the overlapping  regions, the
magnitude of the rates of period change of GW Vir and V777 Her stars
is expected to be influenced by plasmon neutrino emission. This means
that, if a non-vanishing  neutrino magnetic dipole moment exists, the
value of $\dot{\Pi}$ should be  enhanced as compared with the case in
which $\mu_{\nu}= 0$, and this  excess in the period drift provides a
constraint to $\mu_{\nu}$.  This  approach is exactly the same as that
adopted  to derive upper  bounds to the axion mass (see
Sect. \ref{sec:3.9.1}).

In the GW Vir regime, the prospect of using a measured $\dot{\Pi}$ to
constrain plasmon neutrino emission was explored by
\cite{2000ApJ...539..372O},  and in the V777 Her (DBV) domain by
\cite{2004ApJ...602L.109W}. In the case of GW Vir stars, the pulsating
star PG 0122+200 could, in principle, be  an appropriate candidate to
place constraints on the plasmon neutrino emission rate on the basis of an
observed  value of $\dot{\Pi}$. This is because, at the effective
temperature of this pulsating star ($T_{\rm eff}= 80\,000 \pm 4000$
K), the neutrino luminosity should be comparable to the photon
luminosity.  However, \cite{2011A&A...528A...5V} found $\dot{\Pi}$ values  
for the periods of this star that are $100-1000$ times larger than the
value expected from the asteroseismological model derived by
\cite{2007A&A...475..619C}. This period drift cannot be explained by
neutrino cooling only, so that probably, another mechanism is at
work. \cite{2011A&A...528A...5V} suggest that the resonant mode coupling
induced within triplets by  rotation could be such a mechanism. In
short, PG 0122+200 is not useful for constraining neutrino
emission. In the case of V777 Her stars, the pulsating  star EC
20058$-$5234 \citep[$T_{\rm eff} = 25\,000-27\,000$
  K;][]{2013MNRAS.431..520C} could be a good candidate for exploring
the efficiency of plasmon neutrino emission. From theoretical grounds,
neutrino luminosity should be a factor of $\sim 3$ larger than the photon
luminosity in this star. Unfortunately, no measurement of $\dot{\Pi}$
is possible because the long-term period stability is not good enough
\citep{2017ASPC..509..315S}. Another V777 Her star that could  
place constraints on neutrino emission  is PG 0112+104
\citep{2017ApJ...835..277H}, the hottest DBV of the class ($T_{\rm
  eff} > 30\,000$ K), although a measurement of the rate of change of
its periods is not available yet. Fortunately, it has been possible to
determine a preliminary value of $\dot{\Pi} \sim (2.0 \pm 0.9) \times
10^{-13}$ s/s  for the largest amplitude mode ($\Pi \sim 489$ s) of
another V777 Her star,  the pulsating star PG 1351+489
\citep{2011MNRAS.415.1220R}. Regrettably,  with a $T_{\rm eff}\sim
22\,000-26\,000$ K, this star could be  too cool to allow
measurement of the {\it normal} plasmon emission rate, but it can be still
employed for detecting {\it anomalous} neutrino emission.

\begin{figure}
\includegraphics[width=1.0\textwidth]{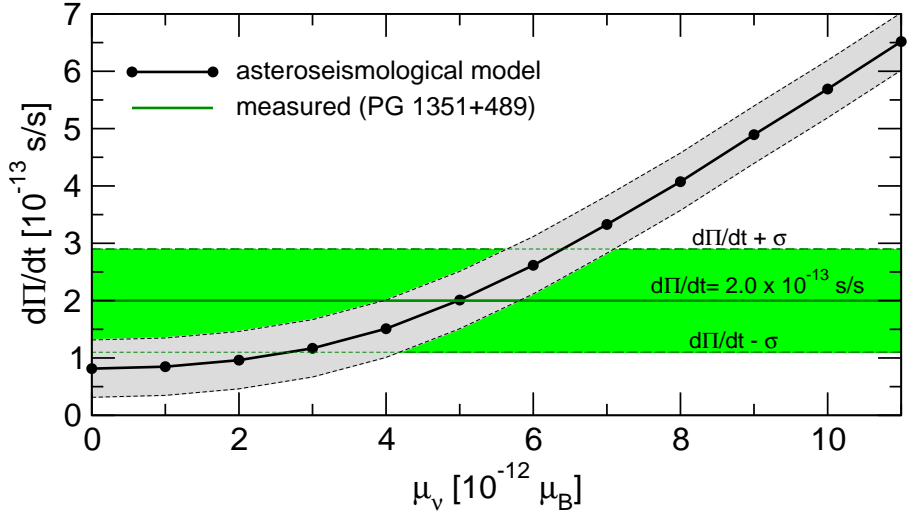}
\caption{Theoretical $\dot{\Pi}$ value corresponding to the period
of 489 s of the DBV star PG 1351+489 in  terms of the neutrino magnetic dipole moment
(black solid curve and dots), and the measured value
(horizontal green solid line) with its corresponding $1\sigma$
uncertainties $(\pm 0.9 \times 10^{-13}$ s/s, dashed lines).}
\label{dpdt-mu-0664}
\end{figure}

\cite{2014JCAP...08..054C} employed the estimate of the rate of period
change for PG 1351+489 to  derive an upper limit to the neutrino
magnetic dipole moment. Specifically, these authors computed the
anomalous energy loss due to the existence of a magnetic dipole
moment, $\varepsilon_{\nu}^{\rm mdm}$, from the plasmon neutrino
emission, $\varepsilon_{\nu}^{\rm p}$, employing the scaling relation
of \cite{1994ApJ...425..222H}: $\varepsilon_{\nu}^{\rm mdm}=
0.318\ \mu_{12}^2 \left(10 {\rm keV}/\hbar \omega_{\rm p}\right)^2
(Q_2/Q_3)\ \varepsilon_{\nu}^{\rm p}$, where $\mu_{12}=
\mu_{\nu}/10^{12}\ \mu_{\rm B}$ ($\mu_{\rm B}$  is the Bohr
magneton). \cite{2014JCAP...08..054C} compared the theoretical
$\dot{\Pi}$ value for asteroseismological models considering
increasing values of  $\mu_{\nu}$ with the observed rate of period
change of PG 1351+489,  and assessed the possible existence of
additional cooling by  neutrinos with magnetic dipole moment.
Fig. \ref{dpdt-mu-0664} displays the  rate of period change of the
period $\Pi= 489.3$ s (for $\mu_{\nu}= 0)$, in terms of increasing
values of $\mu_{\nu}$,  corresponding to the asteroseismological model
for PG 1351+489.   The estimate of the rate of period change of the
489 s period  of  PG 1351+489 and its uncertainties are also
shown. These results  suggest the existence of additional cooling in
this V777 Her star,  consistent with a non-zero magnetic dipole moment
with an upper limit  of $\mu_{\nu} \lesssim 7 \times 10^{-12} \mu_{\rm
  B}$. A similar analysis,  but without considering an
asteroseismological model for PG 1351+489,  but instead the
spectroscopically derived effective temperature of this  star,
indicates an upper limit of $\mu_{\nu} \lesssim 9  \times 10^{-12}
\mu_{\rm B}$. In summary, a conservative limit of $\mu_{\nu} \lesssim
10^{-11} \mu_{\rm B}$ can be adopted as the main result of the
analysis. This constraint is a bit less restrictive than  the upper
bound derived from the Galactic WDLF, of  $\mu_{\nu} \lesssim 5 \times 10^{-12}
\mu_{\rm B}$ \citep{2014A&A...562A.123M}, the upper bound for
$\mu_{\nu}$ derived from the analysis of red giants from the
color-magnitude diagram of the Galactic globular cluster M5,
$\mu_{\nu} \lesssim 4.5 \times 10^{-12} \mu_{\rm B}$
corresponding to the $2\sigma$ limit \citep{2013PhRvL.111w1301V,2013A&A...558A..12V},
and the 95 \% confidence  limit  of $\mu_{\nu} \lesssim 3.4 \times 10^{-12} \mu_{\rm B}$ derived by \cite{2015ApJ...809..141H} from the WDLF of the globular cluster 47 Tucanae.

\subsubsection{Limits on the secular rate of change of the Gravitational Constant}
\label{sec:3.9.3}

According to the General Relativity theory, the gravitational constant, $G$, does not vary with time or location. However, some  alternative theories of gravity predict that the gravitational  constant is both time and space dependent
\citep{2003RvMP...75..403U,2007A&ARv..14..113G}. If these theories are
correct, it is expected that the gravitational  constant would vary
slowly over long timescales.  A large variety of methods aimed to
place limits on a  possible variation of $G$ exists. The most
restrictive upper  bounds are derived using Lunar Laser Ranging
\citep[$\dot{G}/G= (0.2 \pm 0.7) \times 10^{-12}$
  yr$^{-1}$,][]{2010A&A...522L...5H} and Big Bang nucleosynthesis
\citep[$-0.3 \times 10^{-12}  \lesssim  \dot{G}/G \lesssim 0.4 \times
  10^{-12}$ yr$^{-1}$,][]{2004PhRvL..92q1301C}. Other constraints,
somewhat less restrictive,  come from  the WDLF in clusters
\citep[$\dot{G}/G \lesssim 1.8 \times 10^{-12}$
  yr$^{-1}$,][]{2011JCAP...05..021G} and also from the Hubble diagram of
Type Ia supernovae \citep[$\dot{G}/G \lesssim  1 \times 10^{-11}$
  yr$^{-1}$  at $z\sim 0.5$,][]{2006IJMPD..15.1163G}, see also
\cite{2014PASA...31...15M} ($-3 \times 10^{-11}  \lesssim  \dot{G}/G
\lesssim 7.3 \times 10^{-11}$ yr$^{-1}$).

WDs are ideal targets to detect a hypothetical secular change of $G$
because they have very  long evolutionary timescales, so that the
cumulative effect of a slightly changing $G$ should lead sizable
imprints in their evolutionary properties. This has been investigated
by \cite{2011A&A...527A..72A}, who showed  that the mechanical
structure and the energy balance of WDs are strongly modified by a
varying the gravitational constant, and the impact is more pronounced for more
massive WDs. \cite{2013JCAP...06..032C} have taken advantage of the
sensitivity of the evolutionary timescale of WDs to $G$ to put
limits on its rate of variation by comparing  the theoretical
$\dot{\Pi}$ obtained taking into account the effects of a running $G$
with the measured rates of period change of the ZZ Ceti stars
G117$-$B15A and R548. The rates of period change  measured for the
$\sim 215$ s mode (G117$-$B15A) and the $\sim 212$ s mode (R548) are
larger than those predicted by the asteroseismological models  for
these stars, derived by assuming  $\dot{G}= 0$. Hypothesizing that
this discrepancy can be attributed exclusively to a variable $G$,
assuming for simplicity that $\dot{G}/G$ is constant, considering for
simplicity that  $\dot{G}<0$,  and adopting different values  for this
quantity, these authors derived upper limits for the rate of variation  
of  the gravitation constant of $\dot{G}/G= -1.8 \times 10^{-10}$ yr$^{-1}$ from
G117$-$B15A and $\dot{G}/G= -1.3 \times 10^{-10}$ yr$^{-1}$ from R548.
These values are completely compatible with each other, and are currently less
restrictive than those derived using other methods such as Lunar Laser
Ranging and Big Bang nucleosynthesis. They are comparable to the upper
limits  derived from Hubble diagram of Type Ia supernovae and the
WDLF.  These bounds could be improved if the $\dot{\Pi}$ value were 
measured for a massive pulsating WD, since the effects of a varying $G$ are
stronger for the more massive WDs \citep{2011A&A...527A..72A}. 

\section{Summary}
\label{sec:5}

The study of WD stars exceeds by far the scope of  the theory of
stellar evolution. Examples abound of WD applications to  various
areas of astrophysics, such as cosmochronology, and also as
laboratories to study exotic physics. Fortunately,  WDs go through
certain stages during which the pulsations they  experience  allow to
``see'' their interiors, thus enabling to elucidate their inner
structure and evolutionary stage through asteroseismology. The
progress in  the study of pulsating WDs has been remarkable in the
last decade. This progress has been partly boosted with both  the
advent of observations  from space such as the {\it Kepler} and K2 missions,
and the  generation of new detailed evolutionary models, along with the
development of powerful asteroseismological techniques. Obviously, the
refinement of the  observations has led to new challenges for the
theoreticians, being  perhaps the best example the exciting detection
of outbursts in cool  ZZ Ceti stars. 

With the availability of parallax measurements from the Gaia
collaboration, and  with the  arrival of new data from the TESS
mission, the field of WD asteroseismology will surely experience an
even  greater revolution in a few years. These circumstances force
the members  of the WD community to be ready to interpret new and
unexpected data, which will ultimately lead to correcting the current
theoretical models. Fields such as the dating of stellar populations
through cosmochronology, the  theory of accretion in WDs,  the physics
of matter  under extreme conditions (such as crystallization), as well
as  the study of fundamental particles (both hypothetical and real),
will ultimately be strongly benefited.

\begin{acknowledgements}

We thank our referees (Pier-Emmanuel Tremblay and an anonymous referee), for their very relevant comments and suggestions that largely improved the content of the paper. We warmly thank Tiara Battich, Leila M. Calcaferro, and Francisco C. De Ger\'onimo,  
members of the La Plata Stellar Evolution and Pulsation 
Research Group\footnote{{\tt http://evolgroup.fcaglp.unlp.edu.ar/}}, for their valuable 
suggestions concerning the presentation of the material of this review. We also thank Keaton J. Bell, Paul A. Bradley, M\'arcio Catelan, St\'ephane Charpinet, 
Maurizio Giannotti, Steve D. Kawaler, and Don E. Winget
for the invaluable effort of reading the manuscript critically and 
suggesting changes that profoundly enriched the content of this article.
We particularly thank Keaton J. Bell for being so kind in preparing Figure 12 for this 
review. Special thanks also to Paul A. Bradley, M\'arcio Catelan, and Leila M. Calcaferro for editing the text and greatly improving the presentation. Finally, we warmly appreciate the analysis kindly provided by Maurizio Giannotti on axion constraints. This paper is dedicated to the memory of Enrique Garc\'ia-Berro, our great friend and mentor. Part of this 
work was supported by AGENCIA through the Programa de
Modernizaci\'on Tecnol\'ogica BID 1728/OC-AR, and by the PIP 112-200801-00940 grant
from CONICET. M3B is partially supported by  ANPCyT through  grant  PICT 2016-0053, and
by  the MinCyT-DAAD  bilateral cooperation program through grant DA/16/07. This research has
made intensive use of NASA's Astrophysics Data System.
\end{acknowledgements}

\bibliographystyle{spbasic}      % basic style, author-year citations
\bibliography{paper_bibliografia}   % name your BibTeX data base

\begin{thebibliography}{395}
\providecommand{\natexlab}[1]{#1}
\providecommand{\url}[1]{{#1}}
\providecommand{\urlprefix}{URL }
\expandafter\ifx\csname urlstyle\endcsname\relax
  \providecommand{\doi}[1]{DOI~\discretionary{}{}{}#1}\else
  \providecommand{\doi}{DOI~\discretionary{}{}{}\begingroup
  \urlstyle{rm}\Url}\fi
\providecommand{\eprint}[2][]{\url{#2}}

\bibitem[{{Aerts} et~al.(2010){Aerts}, {Christensen-Dalsgaard}, and
  {Kurtz}}]{2010aste.book.....A}
{Aerts} C, {Christensen-Dalsgaard} J, {Kurtz} DW (2010) {Asteroseismology}

\bibitem[{{Althaus} et~al.(2005){Althaus}, {Miller Bertolami}, {C{\'o}rsico},
  {Garc{\'{\i}}a-Berro}, and {Gil-Pons}}]{2005A&A...440L...1A}
{Althaus} LG, {Miller Bertolami} MM, {C{\'o}rsico} AH, {Garc{\'{\i}}a-Berro} E,
  {Gil-Pons} P (2005) {The formation of DA white dwarfs with thin hydrogen
  envelopes}. \aap 440:L1--L4, \doi{10.1051/0004-6361:200500159},
  \eprint{astro-ph/0507415}

\bibitem[{{Althaus} et~al.(2007){Althaus}, {Garc{\'{\i}}a-Berro}, {Isern},
  {C{\'o}rsico}, and {Rohrmann}}]{2007A&A...465..249A}
{Althaus} LG, {Garc{\'{\i}}a-Berro} E, {Isern} J, {C{\'o}rsico} AH, {Rohrmann}
  RD (2007) {The age and colors of massive white dwarf stars}. \aap
  465:249--255, \doi{10.1051/0004-6361:20066059}, \eprint{astro-ph/0702024}

\bibitem[{{Althaus} et~al.(2010{\natexlab{a}}){Althaus}, {C{\'o}rsico},
  {Bischoff-Kim}, {Romero}, {Renedo}, {Garc{\'{\i}}a-Berro}, and {Miller
  Bertolami}}]{2010ApJ...717..897A}
{Althaus} LG, {C{\'o}rsico} AH, {Bischoff-Kim} A, {Romero} AD, {Renedo} I,
  {Garc{\'{\i}}a-Berro} E, {Miller Bertolami} MM (2010{\natexlab{a}}) {New
  Chemical Profiles for the Asteroseismology of ZZ Ceti Stars}. \apj
  717:897--907, \doi{10.1088/0004-637X/717/2/897}, \eprint{1005.2612}

\bibitem[{{Althaus} et~al.(2010{\natexlab{b}}){Althaus}, {C{\'o}rsico},
  {Isern}, and {Garc{\'{\i}}a-Berro}}]{2010A&ARv..18..471A}
{Althaus} LG, {C{\'o}rsico} AH, {Isern} J, {Garc{\'{\i}}a-Berro} E
  (2010{\natexlab{b}}) {Evolutionary and pulsational properties of white dwarf
  stars}. \aapr 18:471--566, \doi{10.1007/s00159-010-0033-1},
  \eprint{1007.2659}

\bibitem[{{Althaus} et~al.(2011){Althaus}, {C{\'o}rsico}, {Torres},
  {Lor{\'e}n-Aguilar}, {Isern}, and
  {Garc{\'{\i}}a-Berro}}]{2011A&A...527A..72A}
{Althaus} LG, {C{\'o}rsico} AH, {Torres} S, {Lor{\'e}n-Aguilar} P, {Isern} J,
  {Garc{\'{\i}}a-Berro} E (2011) {The evolution of white dwarfs with a varying
  gravitational constant}. \aap 527:A72, \doi{10.1051/0004-6361/201015849},
  \eprint{1101.0986}

\bibitem[{{Althaus} et~al.(2012){Althaus}, {Garc{\'{\i}}a-Berro}, {Isern},
  {C{\'o}rsico}, and {Miller Bertolami}}]{2012A&A...537A..33A}
{Althaus} LG, {Garc{\'{\i}}a-Berro} E, {Isern} J, {C{\'o}rsico} AH, {Miller
  Bertolami} MM (2012) {New phase diagrams for dense carbon-oxygen mixtures and
  white dwarf evolution}. \aap 537:A33, \doi{10.1051/0004-6361/201117902},
  \eprint{1110.5665}

\bibitem[{{Althaus} et~al.(2013){Althaus}, {Miller Bertolami}, and
  {C{\'o}rsico}}]{2013A&A...557A..19A}
{Althaus} LG, {Miller Bertolami} MM, {C{\'o}rsico} AH (2013) {New evolutionary
  sequences for extremely low-mass white dwarfs. Homogeneous mass and age
  determinations and asteroseismic prospects}. \aap 557:A19,
  \doi{10.1051/0004-6361/201321868}, \eprint{1307.1882}

\bibitem[{{Althaus} et~al.(2015){Althaus}, {Camisassa}, {Miller Bertolami},
  {C{\'o}rsico}, and {Garc{\'\i}a-Berro}}]{2015A&A...576A...9A}
{Althaus} LG, {Camisassa} ME, {Miller Bertolami} MM, {C{\'o}rsico} AH,
  {Garc{\'\i}a-Berro} E (2015) {White dwarf evolutionary sequences for
  low-metallicity progenitors: The impact of third dredge-up}. \aap 576:A9,
  \doi{10.1051/0004-6361/201424922}, \eprint{1502.03882}

\bibitem[{{Althaus} et~al.(2017){Althaus}, {De Ger{\'o}nimo}, {C{\'o}rsico},
  {Torres}, and {Garc{\'{\i}}a-Berro}}]{2017A&A...597A..67A}
{Althaus} LG, {De Ger{\'o}nimo} F, {C{\'o}rsico} A, {Torres} S,
  {Garc{\'{\i}}a-Berro} E (2017) {The evolution of white dwarfs resulting from
  helium-enhanced, low-metallicity progenitor stars}. \aap 597:A67,
  \doi{10.1051/0004-6361/201629909}, \eprint{1611.06191}

\bibitem[{{Angulo} et~al.(1999){Angulo}, {Arnould}, {Rayet}, {Descouvemont},
  {Baye}, {Leclercq-Willain}, {Coc}, {Barhoumi}, {Aguer}, {Rolfs}, {Kunz},
  {Hammer}, {Mayer}, {Paradellis}, {Kossionides}, {Chronidou}, {Spyrou},
  {degl'Innocenti}, {Fiorentini}, {Ricci}, {Zavatarelli}, {Providencia},
  {Wolters}, {Soares}, {Grama}, {Rahighi}, {Shotter}, and {Lamehi
  Rachti}}]{1999NuPhA.656....3A}
{Angulo} C, {Arnould} M, {Rayet} M, {Descouvemont} P, {Baye} D,
  {Leclercq-Willain} C, {Coc} A, {Barhoumi} S, {Aguer} P, {Rolfs} C, {Kunz} R,
  {Hammer} JW, {Mayer} A, {Paradellis} T, {Kossionides} S, {Chronidou} C,
  {Spyrou} K, {degl'Innocenti} S, {Fiorentini} G, {Ricci} B, {Zavatarelli} S,
  {Providencia} C, {Wolters} H, {Soares} J, {Grama} C, {Rahighi} J, {Shotter}
  A, {Lamehi Rachti} M (1999) {A compilation of charged-particle induced
  thermonuclear reaction rates}. Nuclear Physics A 656:3--183,
  \doi{10.1016/S0375-9474(99)00030-5}

\bibitem[{{Arras} et~al.(2006){Arras}, {Townsley}, and
  {Bildsten}}]{2006ApJ...643L.119A}
{Arras} P, {Townsley} DM, {Bildsten} L (2006) {Pulsational Instabilities in
  Accreting White Dwarfs}. \apjl 643:L119--L122, \doi{10.1086/505178},
  \eprint{astro-ph/0604319}

\bibitem[{{Baade}(1992)}]{1992RvMA....5..125B}
{Baade} D (1992) {Observational Aspects of Stellar Seismology.} In: {Klare} G
  (ed) Reviews in Modern Astronomy, Reviews in Modern Astronomy, vol~5, pp
  125--142, \doi{10.1007/978-3-642-77543-7$_9$}

\bibitem[{{Balona}(2010)}]{2010csp..book.....B}
{Balona} LA (2010) {Challenges In Stellar Pulsation}

\bibitem[{{Barlow} et~al.(2008){Barlow}, {Dunlap}, {Rosen}, and
  {Clemens}}]{2008ApJ...688L..95B}
{Barlow} BN, {Dunlap} BH, {Rosen} R, {Clemens} JC (2008) {Two New Variable Hot
  DQ Stars}. \apjl 688:L95, \doi{10.1086/595584}, \eprint{0810.2140}

\bibitem[{{Battich} et~al.(2016){Battich}, {C{\'o}rsico}, {Althaus}, and
  {Miller Bertolami}}]{2016JCAP...08..062B}
{Battich} T, {C{\'o}rsico} AH, {Althaus} LG, {Miller Bertolami} MM (2016)
  {First axion bounds from a pulsating helium-rich white dwarf star}. \jcap
  8:062, \doi{10.1088/1475-7516/2016/08/062}, \eprint{1605.07668}

\bibitem[{{Bedin} et~al.(2009){Bedin}, {Salaris}, {Piotto}, {Anderson}, {King},
  and {Cassisi}}]{2009ApJ...697..965B}
{Bedin} LR, {Salaris} M, {Piotto} G, {Anderson} J, {King} IR, {Cassisi} S
  (2009) {The End of the White Dwarf Cooling Sequence in M4: An Efficient
  Approach}. apj 697:965--979, \doi{10.1088/0004-637X/697/2/965},
  \eprint{0903.2839}

\bibitem[{{Bedin} et~al.(2015){Bedin}, {Salaris}, {Anderson}, {Cassisi},
  {Milone}, {Piotto}, {King}, and {Bergeron}}]{2015MNRAS.448.1779B}
{Bedin} LR, {Salaris} M, {Anderson} J, {Cassisi} S, {Milone} AP, {Piotto} G,
  {King} IR, {Bergeron} P (2015) {Hubble Space Telescope observations of the
  Kepler-field cluster NGC 6819 - I. The bottom of the white dwarf cooling
  sequence}. mnras 448:1779--1788, \doi{10.1093/mnras/stv069},
  \eprint{1501.02953}

\bibitem[{Bell(2017)}]{2017PhDT........14C}
Bell KJ (2017) {Pulsational oddities at the extremes of the DA white dwarf
  instability strip}. PhD thesis, University of Texas

\bibitem[{{Bell} et~al.(2015){Bell}, {Hermes}, {Bischoff-Kim}, {Moorhead},
  {Montgomery}, {{\O}stensen}, {Castanheira}, and
  {Winget}}]{2015ApJ...809...14B}
{Bell} KJ, {Hermes} JJ, {Bischoff-Kim} A, {Moorhead} S, {Montgomery} MH,
  {{\O}stensen} R, {Castanheira} BG, {Winget} DE (2015) {KIC 4552982: Outbursts
  and Asteroseismology from the Longest Pseudo-continuous Light Curve of a ZZ
  Ceti}. \apj 809:14, \doi{10.1088/0004-637X/809/1/14}, \eprint{1506.07878}

\bibitem[{{Bell} et~al.(2016){Bell}, {Hermes}, {Montgomery}, {Gentile Fusillo},
  {Raddi}, {G{\"a}nsicke}, {Winget}, {Dennihy}, {Gianninas}, {Tremblay},
  {Chote}, and {Winget}}]{2016ApJ...829...82B}
{Bell} KJ, {Hermes} JJ, {Montgomery} MH, {Gentile Fusillo} NP, {Raddi} R,
  {G{\"a}nsicke} BT, {Winget} DE, {Dennihy} E, {Gianninas} A, {Tremblay} PE,
  {Chote} P, {Winget} KI (2016) {Outbursts in Two New Cool Pulsating DA White
  Dwarfs}. \apj 829:82, \doi{10.3847/0004-637X/829/2/82}, \eprint{1607.01392}

\bibitem[{{Bell} et~al.(2017{\natexlab{a}}){Bell}, {Gianninas}, {Hermes},
  {Winget}, {Kilic}, {Montgomery}, {Castanheira}, {Vanderbosch}, {Winget}, and
  {Brown}}]{2017ApJ...835..180B}
{Bell} KJ, {Gianninas} A, {Hermes} JJ, {Winget} DE, {Kilic} M, {Montgomery} MH,
  {Castanheira} BG, {Vanderbosch} Z, {Winget} KI, {Brown} WR
  (2017{\natexlab{a}}) {Pruning The ELM Survey: Characterizing Candidate
  Low-mass White Dwarfs through Photometric Variability}. \apj 835:180,
  \doi{10.3847/1538-4357/835/2/180}, \eprint{1612.06390}

\bibitem[{{Bell} et~al.(2017{\natexlab{b}}){Bell}, {Hermes}, {Montgomery},
  {Winget}, {Gentile Fusillo}, {Raddi}, and
  {G{\"a}nsicke}}]{2017ASPC..509..303B}
{Bell} KJ, {Hermes} JJ, {Montgomery} MH, {Winget} DE, {Gentile Fusillo} NP,
  {Raddi} R, {G{\"a}nsicke} BT (2017{\natexlab{b}}) {The First Six Outbursting
  Cool DA White Dwarf Pulsators}. In: {Tremblay} PE, {Gaensicke} B, {Marsh} T
  (eds) 20th European White Dwarf Workshop, Astronomical Society of the Pacific
  Conference Series, vol 509, p 303, \eprint{1609.09097}

\bibitem[{{Bell} et~al.(2017{\natexlab{c}}){Bell}, {Hermes}, {Vanderbosch},
  {Montgomery}, {Winget}, {Dennihy}, {Fuchs}, and
  {Tremblay}}]{2017ApJ...851...24B}
{Bell} KJ, {Hermes} JJ, {Vanderbosch} Z, {Montgomery} MH, {Winget} DE,
  {Dennihy} E, {Fuchs} JT, {Tremblay} PE (2017{\natexlab{c}}) {Destroying
  Aliases from the Ground and Space: Super-Nyquist ZZ Cetis in K2 Long Cadence
  Data}. \apj 851:24, \doi{10.3847/1538-4357/aa9702}, \eprint{1710.10273}

\bibitem[{{Bell} et~al.(2018){Bell}, {Pelisoli}, {Kepler}, {Brown}, {Winget},
  {Winget}, {Vanderbosch}, {Castanheira}, {Hermes}, {Montgomery}, and
  {Koester}}]{2018A&A...617A...6B}
{Bell} KJ, {Pelisoli} I, {Kepler} SO, {Brown} WR, {Winget} DE, {Winget} KI,
  {Vanderbosch} Z, {Castanheira} BG, {Hermes} JJ, {Montgomery} MH, {Koester} D
  (2018) {The McDonald Observatory search for pulsating sdA stars.
  Asteroseismic support for multiple populations}. \aap 617:A6,
  \doi{10.1051/0004-6361/201833279}, \eprint{1805.11129}

\bibitem[{{Bellini} et~al.(2013){Bellini}, {Anderson}, {Salaris}, {Cassisi},
  {Bedin}, {Piotto}, and {Bergeron}}]{2013ApJ...769L..32B}
{Bellini} A, {Anderson} J, {Salaris} M, {Cassisi} S, {Bedin} LR, {Piotto} G,
  {Bergeron} P (2013) {A Double White-dwarf Cooling Sequence in {$\omega$}
  Centauri}. \apjl 769:L32, \doi{10.1088/2041-8205/769/2/L32},
  \eprint{1305.0265}

\bibitem[{{Bernstein} et~al.(1963){Bernstein}, {Ruderman}, and
  {Feinberg}}]{1963PhRv..132.1227B}
{Bernstein} J, {Ruderman} M, {Feinberg} G (1963) {Electromagnetic Properties of
  the Neutrino}. Physical Review 132:1227--1233, \doi{10.1103/PhysRev.132.1227}

\bibitem[{{Biesiada} and {Malec}(2002)}]{2002PhRvD..65d3008B}
{Biesiada} M, {Malec} B (2002) {White dwarf cooling and large extra
  dimensions}. \prd 65(4):043008, \doi{10.1103/PhysRevD.65.043008},
  \eprint{astro-ph/0109545}

\bibitem[{{Biesiada} and {Malec}(2003)}]{2003Ap&SS.283..601B}
{Biesiada} M, {Malec} B (2003) {White Dwarf Constraints On Exotic Physics}.
  \apss 283:601--606, \doi{10.1023/A:1022582802326}

\bibitem[{{Bischoff-Kim} et~al.(2008{\natexlab{a}}){Bischoff-Kim},
  {Montgomery}, and {Winget}}]{2008ApJ...675.1505B}
{Bischoff-Kim} A, {Montgomery} MH, {Winget} DE (2008{\natexlab{a}}) {Fine Grid
  Asteroseismology of G117-B15A and R548}. \apj 675:1505--1511,
  \doi{10.1086/527287}, \eprint{0711.2039}

\bibitem[{{Bischoff-Kim} et~al.(2008{\natexlab{b}}){Bischoff-Kim},
  {Montgomery}, and {Winget}}]{2008ApJ...675.1512B}
{Bischoff-Kim} A, {Montgomery} MH, {Winget} DE (2008{\natexlab{b}}) {Strong
  Limits on the DFSZ Axion Mass with G117-B15A}. \apj 675:1512--1517,
  \doi{10.1086/526398}, \eprint{0711.2041}

\bibitem[{{Bischoff-Kim} et~al.(2014){Bischoff-Kim}, {{\O}stensen}, {Hermes},
  and {Provencal}}]{2014ApJ...794...39B}
{Bischoff-Kim} A, {{\O}stensen} RH, {Hermes} JJ, {Provencal} JL (2014)
  {Seven-period Asteroseismic Fit of the Kepler DBV}. \apj 794:39,
  \doi{10.1088/0004-637X/794/1/39}

\bibitem[{{Bischoff-Kim} et~al.(2019){Bischoff-Kim}, {Provencal}, {Bradley},
  {Montgomery}, {Shipman}, {Harrold}, {Howard}, {Strickland}, {Chandler},
  {Campbell}, {Arredondo}, {Linn}, {Russell}, {Doyle}, {Brickhouse}, {Peters},
  {Kim}, {Jiang}, {Mao}, {Kusakin}, {Sergeev}, {Andreev}, {Velichko},
  {Janulis}, {Pakstiene}, {Ali{\c c}avu{\c s}}, {Horoz}, {Zola}, {Og{\l}oza},
  {Koziel-Wierzbowska}, {Kundera}, {Jableka}, {Debski}, {Baran}, {Meingast},
  {Nagel}, {Loebling}, {Heinitz}, {Hoyer}, {Bogn{\'a}r}, {Castanheira}, and
  {Erdem}}]{2019ApJ...871...13B}
{Bischoff-Kim} A, {Provencal} JL, {Bradley} PA, {Montgomery} MH, {Shipman} HL,
  {Harrold} ST, {Howard} B, {Strickland} W, {Chandler} D, {Campbell} D,
  {Arredondo} A, {Linn} R, {Russell} DP, {Doyle} D, {Brickhouse} A, {Peters} D,
  {Kim} SL, {Jiang} XJ, {Mao} YN, {Kusakin} AV, {Sergeev} AV, {Andreev} M,
  {Velichko} S, {Janulis} R, {Pakstiene} E, {Ali{\c c}avu{\c s}} F, {Horoz} N,
  {Zola} S, {Og{\l}oza} W, {Koziel-Wierzbowska} D, {Kundera} T, {Jableka} D,
  {Debski} B, {Baran} A, {Meingast} S, {Nagel} T, {Loebling} L, {Heinitz} C,
  {Hoyer} D, {Bogn{\'a}r} Z, {Castanheira} BG, {Erdem} A (2019) {GD358: Three
  Decades of Observations for the In-depth Asteroseismology of a DBV Star}.
  \apj 871:13, \doi{10.3847/1538-4357/aae2b1}, \eprint{1810.11708}

\bibitem[{{Bl\"ocker}(1995)}]{1995A&A...299..755B}
{Bl\"ocker} T (1995) {Stellar evolution of low- and intermediate-mass stars.
  II. Post-AGB evolution.} \aap 299:755

\bibitem[{{Bl{\"o}cker}(2001)}]{2001Ap&SS.275....1B}
{Bl{\"o}cker} T (2001) {Evolution on the AGB and beyond: on the formation of
  H-deficient post-AGB stars}. \apss 275:1--14, \eprint{astro-ph/0102135}

\bibitem[{{Bloecker}(1995)}]{1995A&A...297..727B}
{Bloecker} T (1995) {Stellar evolution of low and intermediate-mass stars. I.
  Mass loss on the AGB and its consequences for stellar evolution.} \aap
  297:727

\bibitem[{{Bognar} and {Sodor}(2016)}]{2016IBVS.6184....1B}
{Bognar} Z, {Sodor} A (2016) {White Dwarf Period Tables I. Pulsators with
  hydrogen-dominated atmospheres}. Information Bulletin on Variable Stars 6184,
  \eprint{1610.07470}

\bibitem[{{Bogn{\'a}r} et~al.(2014){Bogn{\'a}r}, {Papar{\'o}}, {C{\'o}rsico},
  {Kepler}, and {Gy{\H o}rffy}}]{2014A&A...570A.116B}
{Bogn{\'a}r} Z, {Papar{\'o}} M, {C{\'o}rsico} AH, {Kepler} SO, {Gy{\H o}rffy}
  {\'A} (2014) {Revealing the pulsational properties of the V777 Herculis star
  KUV 05134+2605 by its long-term monitoring}. \aap 570:A116,
  \doi{10.1051/0004-6361/201423757}, \eprint{1408.3569}

\bibitem[{{Bogn{\'a}r} et~al.(2016){Bogn{\'a}r}, {Papar{\'o}}, {Moln{\'a}r},
  {P{\'a}pics}, {Plachy}, {Vereb{\'e}lyi}, and
  {S{\'o}dor}}]{2016MNRAS.461.4059B}
{Bogn{\'a}r} Z, {Papar{\'o}} M, {Moln{\'a}r} L, {P{\'a}pics} PI, {Plachy} E,
  {Vereb{\'e}lyi} E, {S{\'o}dor} {\'A} (2016) {G 207-9 and LP 133-144:
  light-curve analysis and asteroseismology of two ZZ Ceti stars}. \mnras
  461:4059--4070, \doi{10.1093/mnras/stw1597}, \eprint{1606.09506}

\bibitem[{{Bogn{\'a}r} et~al.(2018{\natexlab{a}}){Bogn{\'a}r}, {Kalup},
  {S{\'o}dor}, {Charpinet}, and {Hermes}}]{2018MNRAS.478.2676B}
{Bogn{\'a}r} Z, {Kalup} C, {S{\'o}dor} {\'A}, {Charpinet} S, {Hermes} JJ
  (2018{\natexlab{a}}) {Searching for new white dwarf pulsators for TESS
  observations at Konkoly Observatory}. \mnras 478:2676--2685,
  \doi{10.1093/mnras/sty1393}, \eprint{1805.10165}

\bibitem[{{Bogn{\'a}r} et~al.(2018{\natexlab{b}}){Bogn{\'a}r}, {Papar{\'o}},
  {S{\'o}dor}, {Jenei}, {Kalup}, {Bertone}, {Chavez-Dagostino}, {Montgomery},
  {Gy{\H o}rffy}, {Moln{\'a}r}, {Oll{\'e}}, {P{\'a}pics}, {Plachy}, and
  {Vereb{\'e}lyi}}]{2018MNRAS.tmp.2745B}
{Bogn{\'a}r} Z, {Papar{\'o}} M, {S{\'o}dor} {\'A}, {Jenei} DI, {Kalup} C,
  {Bertone} E, {Chavez-Dagostino} M, {Montgomery} MH, {Gy{\H o}rffy} {\'A},
  {Moln{\'a}r} L, {Oll{\'e}} H, {P{\'a}pics} PI, {Plachy} E, {Vereb{\'e}lyi} E
  (2018{\natexlab{b}}) {Wandering near the red edge: photometric observations
  of three cool ZZ Ceti stars}. \mnras \doi{10.1093/mnras/sty2884},
  \eprint{1810.09711}

\bibitem[{{Borucki} et~al.(2010){Borucki}, {Koch}, {Basri}, {Batalha}, {Brown},
  {Caldwell}, {Caldwell}, {Christensen-Dalsgaard}, {Cochran}, {DeVore},
  {Dunham}, {Dupree}, {Gautier}, {Geary}, {Gilliland}, {Gould}, {Howell},
  {Jenkins}, {Kondo}, {Latham}, {Marcy}, {Meibom}, {Kjeldsen}, {Lissauer},
  {Monet}, {Morrison}, {Sasselov}, {Tarter}, {Boss}, {Brownlee}, {Owen},
  {Buzasi}, {Charbonneau}, {Doyle}, {Fortney}, {Ford}, {Holman}, {Seager},
  {Steffen}, {Welsh}, {Rowe}, {Anderson}, {Buchhave}, {Ciardi}, {Walkowicz},
  {Sherry}, {Horch}, {Isaacson}, {Everett}, {Fischer}, {Torres}, {Johnson},
  {Endl}, {MacQueen}, {Bryson}, {Dotson}, {Haas}, {Kolodziejczak}, {Van Cleve},
  {Chandrasekaran}, {Twicken}, {Quintana}, {Clarke}, {Allen}, {Li}, {Wu},
  {Tenenbaum}, {Verner}, {Bruhweiler}, {Barnes}, and
  {Prsa}}]{2010Sci...327..977B}
{Borucki} WJ, {Koch} D, {Basri} G, {Batalha} N, {Brown} T, {Caldwell} D,
  {Caldwell} J, {Christensen-Dalsgaard} J, {Cochran} WD, {DeVore} E, {Dunham}
  EW, {Dupree} AK, {Gautier} TN, {Geary} JC, {Gilliland} R, {Gould} A, {Howell}
  SB, {Jenkins} JM, {Kondo} Y, {Latham} DW, {Marcy} GW, {Meibom} S, {Kjeldsen}
  H, {Lissauer} JJ, {Monet} DG, {Morrison} D, {Sasselov} D, {Tarter} J, {Boss}
  A, {Brownlee} D, {Owen} T, {Buzasi} D, {Charbonneau} D, {Doyle} L, {Fortney}
  J, {Ford} EB, {Holman} MJ, {Seager} S, {Steffen} JH, {Welsh} WF, {Rowe} J,
  {Anderson} H, {Buchhave} L, {Ciardi} D, {Walkowicz} L, {Sherry} W, {Horch} E,
  {Isaacson} H, {Everett} ME, {Fischer} D, {Torres} G, {Johnson} JA, {Endl} M,
  {MacQueen} P, {Bryson} ST, {Dotson} J, {Haas} M, {Kolodziejczak} J, {Van
  Cleve} J, {Chandrasekaran} H, {Twicken} JD, {Quintana} EV, {Clarke} BD,
  {Allen} C, {Li} J, {Wu} H, {Tenenbaum} P, {Verner} E, {Bruhweiler} F,
  {Barnes} J, {Prsa} A (2010) {Kepler Planet-Detection Mission: Introduction
  and First Results}. Science 327:977, \doi{10.1126/science.1185402}

\bibitem[{{Bradley}(1998)}]{1998ApJS..116..307B}
{Bradley} PA (1998) {Asteroseismological Constraints on the Structure of the ZZ
  Ceti Stars G117-B15A and R548}. ApJs 116:307--319, \doi{10.1086/313102}

\bibitem[{{Bradley}(2001)}]{2001ApJ...552..326B}
{Bradley} PA (2001) {Asteroseismological Constraints on the Structure of the ZZ
  Ceti Stars L19-2 and GD 165}. \apj 552:326--339, \doi{10.1086/320454}

\bibitem[{{Brassard} and {Fontaine}(1997)}]{1997ASSL..214..451B}
{Brassard} P, {Fontaine} G (1997) {Recent advances in the theoretical
  determination of the ZZ Ceti instability strip}. In: {Isern} J, {Hernanz} M,
  {Garcia-Berro} E (eds) White dwarfs, Astrophysics and Space Science Library,
  vol 214, p 451, \doi{10.1007/978-94-011-5542-7-65}

\bibitem[{{Brassard} and {Fontaine}(1999)}]{1999ASPC..173..329B}
{Brassard} P, {Fontaine} G (1999) {Convective Efficiency and the Blue Edge in
  ZZ Ceti Stars}. In: {Gimenez} A, {Guinan} EF, {Montesinos} B (eds) Stellar
  Structure: Theory and Test of Connective Energy Transport, Astronomical
  Society of the Pacific Conference Series, vol 173, p 329

\bibitem[{{Brassard} and {Fontaine}(2005)}]{2005ApJ...622..572B}
{Brassard} P, {Fontaine} G (2005) {Asteroseismology of the Crystallized ZZ Ceti
  Star BPM 37093: A Different View}. \apj 622:572--576, \doi{10.1086/428116}

\bibitem[{{Brickhill}(1991)}]{1991MNRAS.251..673B}
{Brickhill} AJ (1991) {The pulsations of ZZ Ceti stars. III - The driving
  mechanism}. \mnras 251:673--680, \doi{10.1093/mnras/251.4.673}

\bibitem[{{Brocato} et~al.(1990){Brocato}, {Matteucci}, {Mazzitelli}, and
  {Tornambe}}]{1990ApJ...349..458B}
{Brocato} E, {Matteucci} F, {Mazzitelli} I, {Tornambe} A (1990) {Synthetic
  colors and the chemical evolution of elliptical galaxies}. \apj 349:458--470,
  \doi{10.1086/168330}

\bibitem[{{Brown} et~al.(2010){Brown}, {Kilic}, {Allende Prieto}, and
  {Kenyon}}]{2010ApJ...723.1072B}
{Brown} WR, {Kilic} M, {Allende Prieto} C, {Kenyon} SJ (2010) {The ELM Survey.
  I. A Complete Sample of Extremely Low-mass White Dwarfs}. \apj
  723:1072--1081, \doi{10.1088/0004-637X/723/2/1072}, \eprint{1011.3050}

\bibitem[{{Brown} et~al.(2012){Brown}, {Kilic}, {Allende Prieto}, and
  {Kenyon}}]{2012ApJ...744..142B}
{Brown} WR, {Kilic} M, {Allende Prieto} C, {Kenyon} SJ (2012) {The ELM Survey.
  III. A Successful Targeted Survey for Extremely Low Mass White Dwarfs}. \apj
  744:142, \doi{10.1088/0004-637X/744/2/142}, \eprint{1111.6588}

\bibitem[{{Brown} et~al.(2013){Brown}, {Kilic}, {Allende Prieto}, {Gianninas},
  and {Kenyon}}]{2013ApJ...769...66B}
{Brown} WR, {Kilic} M, {Allende Prieto} C, {Gianninas} A, {Kenyon} SJ (2013)
  {The ELM Survey. V. Merging Massive White Dwarf Binaries}. \apj 769:66,
  \doi{10.1088/0004-637X/769/1/66}, \eprint{1304.4248}

\bibitem[{{Brown} et~al.(2016){Brown}, {Gianninas}, {Kilic}, {Kenyon}, and
  {Allende Prieto}}]{2016ApJ...818..155B}
{Brown} WR, {Gianninas} A, {Kilic} M, {Kenyon} SJ, {Allende Prieto} C (2016)
  {The ELM Survey. VII. Orbital Properties of Low-Mass White Dwarf Binaries}.
  \apj 818:155, \doi{10.3847/0004-637X/818/2/155}, \eprint{1604.04268}

\bibitem[{{Brown} et~al.(2017{\natexlab{a}}){Brown}, {Kilic}, and
  {Gianninas}}]{2017ApJ...839...23B}
{Brown} WR, {Kilic} M, {Gianninas} A (2017{\natexlab{a}}) {The Physical Nature
  of Subdwarf A Stars: White Dwarf Impostors}. \apj 839:23,
  \doi{10.3847/1538-4357/aa67e4}, \eprint{1703.07799}

\bibitem[{{Brown} et~al.(2017{\natexlab{b}}){Brown}, {Kilic}, {Kosakowski}, and
  {Gianninas}}]{2017ApJ...847...10B}
{Brown} WR, {Kilic} M, {Kosakowski} A, {Gianninas} A (2017{\natexlab{b}})
  {Discovery of a Detached, Eclipsing 40 Minute Period Double White Dwarf
  Binary and a Friend: Implications for He+CO White Dwarf Mergers}. \apj
  847:10, \doi{10.3847/1538-4357/aa8724}, \eprint{1708.05287}

\bibitem[{{Byrne} and {Jeffery}(2018)}]{2018MNRAS.481.3810B}
{Byrne} CM, {Jeffery} CS (2018) {Post-common envelope binary stars, radiative
  levitation, and blue large-amplitude pulsators}. \mnras 481:3810--3820,
  \doi{10.1093/mnras/sty2545}, \eprint{1809.04183}

\bibitem[{{Calamida} et~al.(2008){Calamida}, {Corsi}, {Bono}, {Stetson}, {Prada
  Moroni}, {Degl'Innocenti}, {Ferraro}, {Iannicola}, {Koester}, {Pulone},
  {Monelli}, {Amico}, {Buonanno}, {Caputo}, {D'Odorico}, {Freyhammer},
  {Marchetti}, {Nonino}, and {Romaniello}}]{2008ApJ...673L..29C}
{Calamida} A, {Corsi} CE, {Bono} G, {Stetson} PB, {Prada Moroni} P,
  {Degl'Innocenti} S, {Ferraro} I, {Iannicola} G, {Koester} D, {Pulone} L,
  {Monelli} M, {Amico} P, {Buonanno} R, {Caputo} F, {D'Odorico} S, {Freyhammer}
  LM, {Marchetti} E, {Nonino} M, {Romaniello} M (2008) {On the White Dwarf
  Cooling Sequence of the Globular Cluster {$\omega$} Centauri}. \apjl 673:L29,
  \doi{10.1086/527436}, \eprint{0712.0603}

\bibitem[{{Calcaferro} et~al.(2016){Calcaferro}, {C{\'o}rsico}, and
  {Althaus}}]{2016A&A...589A..40C}
{Calcaferro} LM, {C{\'o}rsico} AH, {Althaus} LG (2016) {Asteroseismology of the
  GW Virginis stars SDSS J0349-0059 and VV 47}. \aap 589:A40,
  \doi{10.1051/0004-6361/201527996}, \eprint{1602.06355}

\bibitem[{{Calcaferro} et~al.(2017{\natexlab{a}}){Calcaferro}, {C{\'o}rsico},
  and {Althaus}}]{2017A&A...600A..73C}
{Calcaferro} LM, {C{\'o}rsico} AH, {Althaus} LG (2017{\natexlab{a}}) {Pulsating
  low-mass white dwarfs in the frame of new evolutionary sequences. IV. The
  secular rate of period change}. \aap 600:A73,
  \doi{10.1051/0004-6361/201630376}, \eprint{1701.08880}

\bibitem[{{Calcaferro} et~al.(2017{\natexlab{b}}){Calcaferro}, {C{\'o}rsico},
  and {Althaus}}]{2017A&A...607A..33C}
{Calcaferro} LM, {C{\'o}rsico} AH, {Althaus} LG (2017{\natexlab{b}}) {Pulsating
  low-mass white dwarfs in the frame of new evolutionary sequences. V.
  Asteroseismology of ELMV white dwarf stars}. \aap 607:A33,
  \doi{10.1051/0004-6361/201731230}, \eprint{1708.00482}

\bibitem[{{Calcaferro} et~al.(2017{\natexlab{c}}){Calcaferro}, {C{\'o}rsico},
  {Camisassa}, {Althaus}, and {Shibahashi}}]{2017EPJWC.15206012C}
{Calcaferro} LM, {C{\'o}rsico} AH, {Camisassa} ME, {Althaus} LG, {Shibahashi} H
  (2017{\natexlab{c}}) {Pulsational instability of high-luminosity H-rich
  pre-white dwarf star}. In: European Physical Journal Web of Conferences,
  European Physical Journal Web of Conferences, vol 152, p 06012,
  \doi{10.1051/epjconf/201715206012}

\bibitem[{{Calcaferro} et~al.(2018{\natexlab{a}}){Calcaferro}, {Althaus}, and
  {C{\'o}rsico}}]{2018A&A...614A..49C}
{Calcaferro} LM, {Althaus} LG, {C{\'o}rsico} AH (2018{\natexlab{a}}) {The
  coolest extremely low-mass white dwarfs}. \aap 614:A49,
  \doi{10.1051/0004-6361/201732551}, \eprint{1802.06753}

\bibitem[{{Calcaferro} et~al.(2018{\natexlab{b}}){Calcaferro}, {C{\'o}rsico},
  {Althaus}, {Romero}, and {Kepler}}]{2018A&A...620A.196C}
{Calcaferro} LM, {C{\'o}rsico} AH, {Althaus} LG, {Romero} AD, {Kepler} SO
  (2018{\natexlab{b}}) {Pulsating low-mass white dwarfs in the frame of new
  evolutionary sequences. VI. Thin H-envelope sequences and asteroseismology of
  ELMV stars revisited}. \aap 620:A196, \doi{10.1051/0004-6361/201833781},
  \eprint{1810.11502}

\bibitem[{{Caloi}(1989)}]{1989A&A...221...27C}
{Caloi} V (1989) {Evolution of extreme horizontal branch stars}. \aap
  221:27--35

\bibitem[{{Camisassa} et~al.(2016){Camisassa}, {C{\'o}rsico}, {Althaus}, and
  {Shibahashi}}]{2016A&A...595A..45C}
{Camisassa} ME, {C{\'o}rsico} AH, {Althaus} LG, {Shibahashi} H (2016)
  {Pulsations powered by hydrogen shell burning in white dwarfs}. \aap 595:A45,
  \doi{10.1051/0004-6361/201628857}, \eprint{1606.04367}

\bibitem[{{Camisassa} et~al.(2019){Camisassa}, {Althaus}, {C{\'o}rsico}, {De
  Ger{\'o}nimo}, {Miller Bertolami}, {Novarino}, {Rohrmann}, {Wachlin}, and
  {Garc{\'{\i}}a-Berro}}]{2019A&A...625A..87C}
{Camisassa} ME, {Althaus} LG, {C{\'o}rsico} AH, {De Ger{\'o}nimo} FC, {Miller
  Bertolami} MM, {Novarino} ML, {Rohrmann} RD, {Wachlin} FC,
  {Garc{\'{\i}}a-Berro} E (2019) {The evolution of ultra-massive white dwarfs}.
  \aap 625:A87, \doi{10.1051/0004-6361/201833822}, \eprint{1807.03894}

\bibitem[{{Campos} et~al.(2013){Campos}, {Kepler}, {Bonatto}, and
  {Ducati}}]{2013MNRAS.433..243C}
{Campos} F, {Kepler} SO, {Bonatto} C, {Ducati} JR (2013) {Multichromatic
  colour-magnitude diagrams of the globular cluster NGC 6366}. \mnras
  433:243--250, \doi{10.1093/mnras/stt719}, \eprint{1307.4499}

\bibitem[{{Campos} et~al.(2016){Campos}, {Bergeron}, {Romero}, {Kepler},
  {Ourique}, {Costa}, {Bonatto}, {Winget}, {Montgomery}, {Pacheco}, and
  {Bedin}}]{2016MNRAS.456.3729C}
{Campos} F, {Bergeron} P, {Romero} AD, {Kepler} SO, {Ourique} G, {Costa} JES,
  {Bonatto} CJ, {Winget} DE, {Montgomery} MH, {Pacheco} TA, {Bedin} LR (2016)
  {A comparative analysis of the observed white dwarf cooling sequence from
  globular clusters}. \mnras 456:3729--3742, \doi{10.1093/mnras/stv2911},
  \eprint{1512.03114}

\bibitem[{{Cantiello} et~al.(2014){Cantiello}, {Mankovich}, {Bildsten},
  {Christensen-Dalsgaard}, and {Paxton}}]{2014ApJ...788...93C}
{Cantiello} M, {Mankovich} C, {Bildsten} L, {Christensen-Dalsgaard} J, {Paxton}
  B (2014) {Angular Momentum Transport within Evolved Low-mass Stars}. \apj
  788:93, \doi{10.1088/0004-637X/788/1/93}, \eprint{1405.1419}

\bibitem[{{Carrasco} et~al.(2014){Carrasco}, {Catal{\'a}n}, {Jordi},
  {Tremblay}, {Napiwotzki}, {Luri}, {Robin}, and
  {Kowalski}}]{2014A&A...565A..11C}
{Carrasco} JM, {Catal{\'a}n} S, {Jordi} C, {Tremblay} PE, {Napiwotzki} R,
  {Luri} X, {Robin} AC, {Kowalski} PM (2014) {Gaia photometry for white
  dwarfs}. \aap 565:A11, \doi{10.1051/0004-6361/201220596}, \eprint{1403.6045}

\bibitem[{{Casewell} et~al.(2009){Casewell}, {Dobbie}, {Napiwotzki},
  {Burleigh}, {Barstow}, and {Jameson}}]{2009MNRAS.395.1795C}
{Casewell} SL, {Dobbie} PD, {Napiwotzki} R, {Burleigh} MR, {Barstow} MA,
  {Jameson} RF (2009) {High-resolution optical spectroscopy of Praesepe white
  dwarfs}. \mnras 395:1795--1804, \doi{10.1111/j.1365-2966.2009.14593.x},
  \eprint{0901.4464}

\bibitem[{{Castanheira} and {Kepler}(2009)}]{2009MNRAS.396.1709C}
{Castanheira} BG, {Kepler} SO (2009) {Seismological studies of ZZ Ceti stars -
  II. Application to the ZZ Ceti class}. \mnras 396:1709--1731,
  \doi{10.1111/j.1365-2966.2009.14855.x}

\bibitem[{{Castanheira} et~al.(2006){Castanheira}, {Kepler}, {Mullally},
  {Winget}, {Koester}, {Voss}, {Kleinman}, {Nitta}, {Eisenstein}, {Napiwotzki},
  and {Reimers}}]{2006A&A...450..227C}
{Castanheira} BG, {Kepler} SO, {Mullally} F, {Winget} DE, {Koester} D, {Voss}
  B, {Kleinman} SJ, {Nitta} A, {Eisenstein} DJ, {Napiwotzki} R, {Reimers} D
  (2006) {Discovery of eleven new ZZ Ceti stars}. \aap 450:227--231,
  \doi{10.1051/0004-6361:20053500}, \eprint{astro-ph/0511804}

\bibitem[{{Castanheira} et~al.(2010){Castanheira}, {Kepler}, {Kleinman},
  {Nitta}, and {Fraga}}]{2010MNRAS.405.2561C}
{Castanheira} BG, {Kepler} SO, {Kleinman} SJ, {Nitta} A, {Fraga} L (2010) {New
  developments of the ZZ Ceti instability strip: the discovery of 11 new
  variables}. \mnras 405:2561--2569, \doi{10.1111/j.1365-2966.2010.16633.x},
  \eprint{1007.0524}

\bibitem[{{Castanheira} et~al.(2013){Castanheira}, {Kepler}, {Kleinman},
  {Nitta}, and {Fraga}}]{2013MNRAS.430...50C}
{Castanheira} BG, {Kepler} SO, {Kleinman} SJ, {Nitta} A, {Fraga} L (2013)
  {Discovery of five new massive pulsating white dwarf stars}. \mnras
  430:50--59, \doi{10.1093/mnras/sts474}

\bibitem[{{Catelan}(2018)}]{2018IAUS..334...11C}
{Catelan} M (2018) {The ages of (the oldest) stars}. In: {Chiappini} C,
  {Minchev} I, {Starkenburg} E, {Valentini} M (eds) Rediscovering Our Galaxy,
  IAU Symposium, vol 334, pp 11--20, \doi{10.1017/S1743921318000868},
  \eprint{1709.08656}

\bibitem[{{Catelan} and {Smith}(2015)}]{2015pust.book.....C}
{Catelan} M, {Smith} HA (2015) {Pulsating Stars}

\bibitem[{{Charpinet} et~al.(1997){Charpinet}, {Fontaine}, {Brassard}, and
  {Dorman}}]{1997ApJ...489L.149C}
{Charpinet} S, {Fontaine} G, {Brassard} P, {Dorman} B (1997) {Gravity-Mode
  Instabilities in Models of Post--Extreme Horizontal Branch Stars: Another
  Class of Pulsating Stars?} \apjl 489:L149, \doi{10.1086/316792},
  \eprint{astro-ph/9707319}

\bibitem[{{Charpinet} et~al.(2009){Charpinet}, {Fontaine}, and
  {Brassard}}]{2009Natur.461..501C}
{Charpinet} S, {Fontaine} G, {Brassard} P (2009) {Seismic evidence for the loss
  of stellar angular momentum before the white-dwarf stage}. \nat 461:501--503,
  \doi{10.1038/nature08307}

\bibitem[{{Charpinet} et~al.(2011){Charpinet}, {Van Grootel}, {Fontaine},
  {Green}, {Brassard}, {Randall}, {Silvotti}, {{\O}stensen}, {Kjeldsen},
  {Christensen-Dalsgaard}, {Kawaler}, {Clarke}, {Li}, and
  {Wohler}}]{2011A&A...530A...3C}
{Charpinet} S, {Van Grootel} V, {Fontaine} G, {Green} EM, {Brassard} P,
  {Randall} SK, {Silvotti} R, {{\O}stensen} RH, {Kjeldsen} H,
  {Christensen-Dalsgaard} J, {Kawaler} SD, {Clarke} BD, {Li} J, {Wohler} B
  (2011) {Deep asteroseismic sounding of the compact hot B subdwarf pulsator
  KIC02697388 from Kepler time series photometry}. \aap 530:A3,
  \doi{10.1051/0004-6361/201016412}

\bibitem[{{Choi} et~al.(2018){Choi}, {Conroy}, {Ting}, {Cargile}, {Dotter}, and
  {Johnson}}]{2018ApJ...863...65C}
{Choi} J, {Conroy} C, {Ting} YS, {Cargile} PA, {Dotter} A, {Johnson} BD (2018)
  {Star Cluster Ages in the Gaia Era}. \apj 863:65,
  \doi{10.3847/1538-4357/aad18c}, \eprint{1807.03789}

\bibitem[{{Chote} et~al.(2013){Chote}, {Sullivan}, {Montgomery}, and
  {Provencal}}]{2013MNRAS.431..520C}
{Chote} P, {Sullivan} DJ, {Montgomery} MH, {Provencal} JL (2013) {Time series
  photometry of the helium atmosphere pulsating white dwarf EC 04207-4748}.
  \mnras 431:520--527, \doi{10.1093/mnras/stt180}, \eprint{1412.5683}

\bibitem[{{Constantino} et~al.(2015){Constantino}, {Campbell},
  {Christensen-Dalsgaard}, {Lattanzio}, and {Stello}}]{2015MNRAS.452..123C}
{Constantino} T, {Campbell} SW, {Christensen-Dalsgaard} J, {Lattanzio} JC,
  {Stello} D (2015) {The treatment of mixing in core helium burning models - I.
  Implications for asteroseismology}. \mnras 452:123--145,
  \doi{10.1093/mnras/stv1264}, \eprint{1506.01209}

\bibitem[{{Copi} et~al.(2004){Copi}, {Davis}, and
  {Krauss}}]{2004PhRvL..92q1301C}
{Copi} CJ, {Davis} AN, {Krauss} LM (2004) {New Nucleosynthesis Constraint on
  the Variation of G}. Physical Review Letters 92(17):171301,
  \doi{10.1103/PhysRevLett.92.171301}, \eprint{astro-ph/0311334}

\bibitem[{{C{\'o}rsico}(2018)}]{2018tjae.conf...13C}
{C{\'o}rsico} AH (2018) {Pulsating white dwarf stars and asteroseismology}. In:
  Terceras Jornadas de Astrof{\'{\i}}sica Estelar, pp 13--22,
  \eprint{1703.00934}

\bibitem[{{C{\'o}rsico} and
  {Althaus}(2014{\natexlab{a}})}]{2014A&A...569A.106C}
{C{\'o}rsico} AH, {Althaus} LG (2014{\natexlab{a}}) {Pulsating low-mass white
  dwarfs in the frame of new evolutionary sequences. I. Adiabatic properties}.
  \aap 569:A106, \doi{10.1051/0004-6361/201424352}, \eprint{1408.6708}

\bibitem[{{C{\'o}rsico} and
  {Althaus}(2014{\natexlab{b}})}]{2014ApJ...793L..17C}
{C{\'o}rsico} AH, {Althaus} LG (2014{\natexlab{b}}) {Short-period g-mode
  Pulsations in Low-mass White Dwarfs Triggered by H-shell Burning}. \apjl
  793:L17, \doi{10.1088/2041-8205/793/1/L17}, \eprint{1408.6724}

\bibitem[{{C{\'o}rsico} and {Althaus}(2016)}]{2016A&A...585A...1C}
{C{\'o}rsico} AH, {Althaus} LG (2016) {Pulsating low-mass white dwarfs in the
  frame of new evolutionary sequences. II. Nonadiabatic analysis}. \aap 585:A1,
  \doi{10.1051/0004-6361/201527162}, \eprint{1510.00645}

\bibitem[{{C{\'o}rsico} et~al.(2001){C{\'o}rsico}, {Benvenuto}, {Althaus},
  {Isern}, and {Garc{\'{\i}}a-Berro}}]{2001NewA....6..197C}
{C{\'o}rsico} AH, {Benvenuto} OG, {Althaus} LG, {Isern} J,
  {Garc{\'{\i}}a-Berro} E (2001) {The potential of the variable DA white dwarf
  G117-B15A as a tool for fundamental physics}. \nat 6:197--213,
  \doi{10.1016/S1384-1076(01)00055-0}, \eprint{astro-ph/0104103}

\bibitem[{{C{\'o}rsico} et~al.(2004){C{\'o}rsico}, {Garc{\'{\i}}a-Berro},
  {Althaus}, and {Isern}}]{2004A&A...427..923C}
{C{\'o}rsico} AH, {Garc{\'{\i}}a-Berro} E, {Althaus} LG, {Isern} J (2004)
  {Pulsations of massive ZZ Ceti stars with carbon/oxygen and oxygen/neon
  cores}. \aap 427:923--932, \doi{10.1051/0004-6361:20040416},
  \eprint{astro-ph/0408238}

\bibitem[{{C{\'o}rsico} et~al.(2007{\natexlab{a}}){C{\'o}rsico}, {Althaus},
  {Miller Bertolami}, and {Werner}}]{2007A&A...461.1095C}
{C{\'o}rsico} AH, {Althaus} LG, {Miller Bertolami} MM, {Werner} K
  (2007{\natexlab{a}}) {Asteroseismological constraints on the pulsating
  planetary nebula nucleus (PG 1159-type) RX J2117.1+3412}. \aap
  461:1095--1102, \doi{10.1051/0004-6361:20066452}, \eprint{astro-ph/0610420}

\bibitem[{{C{\'o}rsico} et~al.(2007{\natexlab{b}}){C{\'o}rsico}, {Miller
  Bertolami}, {Althaus}, {Vauclair}, and {Werner}}]{2007A&A...475..619C}
{C{\'o}rsico} AH, {Miller Bertolami} MM, {Althaus} LG, {Vauclair} G, {Werner} K
  (2007{\natexlab{b}}) {Asteroseismological constraints on the coolest GW
  Virginis variable star (PG 1159-type) PG 0122+200}. \aap 475:619--627,
  \doi{10.1051/0004-6361:20078145}, \eprint{0709.0280}

\bibitem[{{C{\'o}rsico} et~al.(2008){C{\'o}rsico}, {Althaus}, {Kepler},
  {Costa}, and {Miller Bertolami}}]{2008A&A...478..869C}
{C{\'o}rsico} AH, {Althaus} LG, {Kepler} SO, {Costa} JES, {Miller Bertolami} MM
  (2008) {Asteroseismological measurements on PG 1159-035, the prototype of the
  GW Virginis variable stars}. \aap 478:869--881,
  \doi{10.1051/0004-6361:20078646}, \eprint{0712.0795}

\bibitem[{{C{\'o}rsico} et~al.(2009{\natexlab{a}}){C{\'o}rsico}, {Althaus},
  {Miller Bertolami}, and {Garc{\'{\i}}a-Berro}}]{2009A&A...499..257C}
{C{\'o}rsico} AH, {Althaus} LG, {Miller Bertolami} MM, {Garc{\'{\i}}a-Berro} E
  (2009{\natexlab{a}}) {Asteroseismology of hot pre-white dwarf stars: the case
  of the DOV stars PG 2131+066 and PG 1707+427, and the PNNV star NGC 1501}.
  \aap 499:257--266, \doi{10.1051/0004-6361/200810727}, \eprint{0903.3628}

\bibitem[{{C{\'o}rsico} et~al.(2009{\natexlab{b}}){C{\'o}rsico}, {Althaus},
  {Miller Bertolami}, {Gonz{\'a}lez P{\'e}rez}, and
  {Kepler}}]{2009ApJ...701.1008C}
{C{\'o}rsico} AH, {Althaus} LG, {Miller Bertolami} MM, {Gonz{\'a}lez P{\'e}rez}
  JM, {Kepler} SO (2009{\natexlab{b}}) {On the Possible Existence of
  Short-Period g-Mode Instabilities Powered by Nuclear-Burning Shells in
  Post-Asymptotic Giant Branch H-Deficient (PG1159-Type) Stars}. \apj
  701:1008--1014, \doi{10.1088/0004-637X/701/2/1008}, \eprint{0906.2387}

\bibitem[{{C{\'o}rsico} et~al.(2011){C{\'o}rsico}, {Althaus}, {Kawaler},
  {Miller Bertolami}, {Garc{\'{\i}}a-Berro}, and
  {Kepler}}]{2011MNRAS.418.2519C}
{C{\'o}rsico} AH, {Althaus} LG, {Kawaler} SD, {Miller Bertolami} MM,
  {Garc{\'{\i}}a-Berro} E, {Kepler} SO (2011) {Probing the internal rotation of
  pre-white dwarf stars with asteroseismology: the case of PG 0122+200}. \mnras
  418:2519--2526, \doi{10.1111/j.1365-2966.2011.19642.x}, \eprint{1108.3359}

\bibitem[{{C{\'o}rsico} et~al.(2012{\natexlab{a}}){C{\'o}rsico}, {Althaus},
  {Miller Bertolami}, and {Bischoff-Kim}}]{2012A&A...541A..42C}
{C{\'o}rsico} AH, {Althaus} LG, {Miller Bertolami} MM, {Bischoff-Kim} A
  (2012{\natexlab{a}}) {Asteroseismology of the Kepler V777 Herculis variable
  white dwarf with fully evolutionary models}. \aap 541:A42,
  \doi{10.1051/0004-6361/201118736}, \eprint{1112.5882}

\bibitem[{{C{\'o}rsico} et~al.(2012{\natexlab{b}}){C{\'o}rsico}, {Althaus},
  {Miller Bertolami}, {Romero}, {Garc{\'{\i}}a-Berro}, {Isern}, and
  {Kepler}}]{2012MNRAS.424.2792C}
{C{\'o}rsico} AH, {Althaus} LG, {Miller Bertolami} MM, {Romero} AD,
  {Garc{\'{\i}}a-Berro} E, {Isern} J, {Kepler} SO (2012{\natexlab{b}}) {The
  rate of cooling of the pulsating white dwarf star G117-B15A: a new
  asteroseismological inference of the axion mass}. \mnras 424:2792--2799,
  \doi{10.1111/j.1365-2966.2012.21401.x}

\bibitem[{{C{\'o}rsico} et~al.(2012{\natexlab{c}}){C{\'o}rsico}, {Althaus},
  {Romero}, {Mukadam}, {Garc{\'{\i}}a-Berro}, {Isern}, {Kepler}, and
  {Corti}}]{2012JCAP...12..010C}
{C{\'o}rsico} AH, {Althaus} LG, {Romero} AD, {Mukadam} AS,
  {Garc{\'{\i}}a-Berro} E, {Isern} J, {Kepler} SO, {Corti} MA
  (2012{\natexlab{c}}) {An independent limit on the axion mass from the
  variable white dwarf star R548}. \jcap 12:010,
  \doi{10.1088/1475-7516/2012/12/010}, \eprint{1211.3389}

\bibitem[{{C{\'o}rsico} et~al.(2012{\natexlab{d}}){C{\'o}rsico}, {Romero},
  {Althaus}, and {Hermes}}]{2012A&A...547A..96C}
{C{\'o}rsico} AH, {Romero} AD, {Althaus} LG, {Hermes} JJ (2012{\natexlab{d}})
  {The seismic properties of low-mass He-core white dwarf stars}. \aap 547:A96,
  \doi{10.1051/0004-6361/201220114}, \eprint{1209.5107}

\bibitem[{{C{\'o}rsico} et~al.(2013){C{\'o}rsico}, {Althaus},
  {Garc{\'{\i}}a-Berro}, and {Romero}}]{2013JCAP...06..032C}
{C{\'o}rsico} AH, {Althaus} LG, {Garc{\'{\i}}a-Berro} E, {Romero} AD (2013) {An
  independent constraint on the secular rate of variation of the gravitational
  constant from pulsating white dwarfs}. JCAP 6:032,
  \doi{10.1088/1475-7516/2013/06/032}, \eprint{1306.1864}

\bibitem[{{C{\'o}rsico} et~al.(2014){C{\'o}rsico}, {Althaus}, {Miller
  Bertolami}, {Kepler}, and {Garc{\'{\i}}a-Berro}}]{2014JCAP...08..054C}
{C{\'o}rsico} AH, {Althaus} LG, {Miller Bertolami} MM, {Kepler} SO,
  {Garc{\'{\i}}a-Berro} E (2014) {Constraining the neutrino magnetic dipole
  moment from white dwarf pulsations}. \jcap 8:054,
  \doi{10.1088/1475-7516/2014/08/054}, \eprint{1406.6034}

\bibitem[{{C{\'o}rsico} et~al.(2016{\natexlab{a}}){C{\'o}rsico}, {Althaus},
  {Serenelli}, {Kepler}, {Jeffery}, and {Corti}}]{2016A&A...588A..74C}
{C{\'o}rsico} AH, {Althaus} LG, {Serenelli} AM, {Kepler} SO, {Jeffery} CS,
  {Corti} MA (2016{\natexlab{a}}) {Pulsating low-mass white dwarfs in the frame
  of new evolutionary sequences. III. The pre-ELM white dwarf instability
  strip}. \aap 588:A74, \doi{10.1051/0004-6361/201528032}, \eprint{1602.03195}

\bibitem[{{C{\'o}rsico} et~al.(2016{\natexlab{b}}){C{\'o}rsico}, {Romero},
  {Althaus}, {Garc{\'{\i}}a-Berro}, {Isern}, {Kepler}, {Miller Bertolami},
  {Sullivan}, and {Chote}}]{2016JCAP...07..036C}
{C{\'o}rsico} AH, {Romero} AD, {Althaus} LG, {Garc{\'{\i}}a-Berro} E, {Isern}
  J, {Kepler} SO, {Miller Bertolami} MM, {Sullivan} DJ, {Chote} P
  (2016{\natexlab{b}}) {An asteroseismic constraint on the mass of the axion
  from the period drift of the pulsating DA white dwarf star L19-2}. \jcap
  7:036, \doi{10.1088/1475-7516/2016/07/036}, \eprint{1605.06458}

\bibitem[{{C{\'o}rsico} et~al.(2018){C{\'o}rsico}, {Romero}, {Althaus},
  {Pelisoli}, and {Kepler}}]{2018arXiv180907451C}
{C{\'o}rsico} AH, {Romero} AD, {Althaus} LG, {Pelisoli} I, {Kepler} SO (2018)
  {Blue Large-Amplitude Pulsators (BLAPs): possible origin, evolutionary
  status, and nature of their pulsations}. ArXiv e-prints \eprint{1809.07451}

\bibitem[{{Corti} et~al.(2016){Corti}, {Kanaan}, {C{\'o}rsico}, {Kepler},
  {Althaus}, {Koester}, and {S{\'a}nchez Arias}}]{2016A&A...587L...5C}
{Corti} MA, {Kanaan} A, {C{\'o}rsico} AH, {Kepler} SO, {Althaus} LG, {Koester}
  D, {S{\'a}nchez Arias} JP (2016) {Two new pulsating low-mass pre-white dwarfs
  or SX Phoenicis stars?} \aap 587:L5, \doi{10.1051/0004-6361/201527458},
  \eprint{1602.00142}

\bibitem[{{Costa} and {Kepler}(2008)}]{2008A&A...489.1225C}
{Costa} JES, {Kepler} SO (2008) {The temporal changes of the pulsational
  periods of the pre-white dwarf PG 1159-035}. \aap 489:1225--1232,
  \doi{10.1051/0004-6361:20079118}, \eprint{0807.5137}

\bibitem[{{Costa} et~al.(2008){Costa}, {Kepler}, {Winget}, {O'Brien},
  {Kawaler}, {Costa}, {Giovannini}, {Kanaan}, {Mukadam}, {Mullally}, {Nitta},
  {Proven{\c c}al}, {Shipman}, {Wood}, {Ahrens}, {Grauer}, {Kilic}, {Bradley},
  {Sekiguchi}, {Crowe}, {Jiang}, {Sullivan}, {Sullivan}, {Rosen}, {Clemens},
  {Janulis}, {O'Donoghue}, {Ogloza}, {Baran}, {Silvotti}, {Marinoni},
  {Vauclair}, {Dolez}, {Chevreton}, {Dreizler}, {Schuh}, {Deetjen}, {Nagel},
  {Solheim}, {Gonzalez Perez}, {Ulla}, {Barstow}, {Burleigh}, {Good},
  {Metcalfe}, {Kim}, {Lee}, {Sergeev}, {Akan}, {{\c C}ak{\i}rl{\i}}, {Paparo},
  {Viraghalmy}, {Ashoka}, {Handler}, {H{\"u}rkal}, {Johannessen}, {Kleinman},
  {Kalytis}, {Krzesinski}, {Klumpe}, {Larrison}, {Lawrence}, {Meistas},
  {Martinez}, {Nather}, {Fu}, {Pakstien{e}}, {Rosen}, {Romero-Colmenero},
  {Riddle}, {Seetha}, {Silvestri}, {Vuckovi{\'c}}, {Warner}, {Zola}, {Althaus},
  {C{\'o}rsico}, and {Montgomery}}]{2008A&A...477..627C}
{Costa} JES, {Kepler} SO, {Winget} DE, {O'Brien} MS, {Kawaler} SD, {Costa} AFM,
  {Giovannini} O, {Kanaan} A, {Mukadam} AS, {Mullally} F, {Nitta} A, {Proven{\c
  c}al} JL, {Shipman} H, {Wood} MA, {Ahrens} TJ, {Grauer} A, {Kilic} M,
  {Bradley} PA, {Sekiguchi} K, {Crowe} R, {Jiang} XJ, {Sullivan} D, {Sullivan}
  T, {Rosen} R, {Clemens} JC, {Janulis} R, {O'Donoghue} D, {Ogloza} W, {Baran}
  A, {Silvotti} R, {Marinoni} S, {Vauclair} G, {Dolez} N, {Chevreton} M,
  {Dreizler} S, {Schuh} S, {Deetjen} J, {Nagel} T, {Solheim} JE, {Gonzalez
  Perez} JM, {Ulla} A, {Barstow} M, {Burleigh} M, {Good} S, {Metcalfe} TS,
  {Kim} SL, {Lee} H, {Sergeev} A, {Akan} MC, {{\c C}ak{\i}rl{\i}} {\"O},
  {Paparo} M, {Viraghalmy} G, {Ashoka} BN, {Handler} G, {H{\"u}rkal} {\"O},
  {Johannessen} F, {Kleinman} SJ, {Kalytis} R, {Krzesinski} J, {Klumpe} E,
  {Larrison} J, {Lawrence} T, {Meistas} E, {Martinez} P, {Nather} RE, {Fu} JN,
  {Pakstien{e}} E, {Rosen} R, {Romero-Colmenero} E, {Riddle} R, {Seetha} S,
  {Silvestri} NM, {Vuckovi{\'c}} M, {Warner} B, {Zola} S, {Althaus} LG,
  {C{\'o}rsico} AH, {Montgomery} MH (2008) {The pulsation modes of the
  pre-white dwarf PG 1159-035}. \aap 477:627--640,
  \doi{10.1051/0004-6361:20053470}, \eprint{0711.2244}

\bibitem[{{Cox}(1980)}]{1980tsp..book.....C}
{Cox} JP (1980) {Theory of stellar pulsation}

\bibitem[{{Cukanovaite} et~al.(2018){Cukanovaite}, {Tremblay}, {Freytag},
  {Ludwig}, and {Bergeron}}]{2018MNRAS.481.1522C}
{Cukanovaite} E, {Tremblay} PE, {Freytag} B, {Ludwig} HG, {Bergeron} P (2018)
  {Pure-helium 3D model atmospheres of white dwarfs}. \mnras 481:1522--1537,
  \doi{10.1093/mnras/sty2383}, \eprint{1809.00590}

\bibitem[{{Cummings} et~al.(2018){Cummings}, {Kalirai}, {Tremblay},
  {Ramirez-Ruiz}, and {Choi}}]{2018ApJ...866...21C}
{Cummings} JD, {Kalirai} JS, {Tremblay} PE, {Ramirez-Ruiz} E, {Choi} J (2018)
  {The White Dwarf Initial-Final Mass Relation for Progenitor Stars from 0.85
  to 7.5 M $_{\odot}$}. \apj 866:21, \doi{10.3847/1538-4357/aadfd6},
  \eprint{1809.01673}

\bibitem[{{Curd} et~al.(2017){Curd}, {Gianninas}, {Bell}, {Kilic}, {Romero},
  {Allende Prieto}, {Winget}, and {Winget}}]{2017MNRAS.468..239C}
{Curd} B, {Gianninas} A, {Bell} KJ, {Kilic} M, {Romero} AD, {Allende Prieto} C,
  {Winget} DE, {Winget} KI (2017) {Four new massive pulsating white dwarfs
  including an ultramassive DAV}. \mnras 468:239--249,
  \doi{10.1093/mnras/stx320}, \eprint{1702.03343}

\bibitem[{{Dan} et~al.(2014){Dan}, {Rosswog}, {Br{\"u}ggen}, and
  {Podsiadlowski}}]{2014MNRAS.438...14D}
{Dan} M, {Rosswog} S, {Br{\"u}ggen} M, {Podsiadlowski} P (2014) {The structure
  and fate of white dwarf merger remnants}. \mnras 438:14--34,
  \doi{10.1093/mnras/stt1766}, \eprint{1308.1667}

\bibitem[{{D'Antona} and {Mazzitelli}(1990)}]{1990ARA&A..28..139D}
{D'Antona} F, {Mazzitelli} I (1990) {Cooling of white dwarfs}. Annual Review
  Astronomy and Astrophyscs 28:139--181,
  \doi{10.1146/annurev.aa.28.090190.001035}

\bibitem[{{De Ger{\'o}nimo} et~al.(2017){De Ger{\'o}nimo}, {Althaus},
  {C{\'o}rsico}, {Romero}, and {Kepler}}]{2017A&A...599A..21D}
{De Ger{\'o}nimo} FC, {Althaus} LG, {C{\'o}rsico} AH, {Romero} AD, {Kepler} SO
  (2017) {Asteroseismology of ZZ Ceti stars with fully evolutionary white dwarf
  models. I. The impact of the uncertainties from prior evolution on the period
  spectrum}. \aap 599:A21, \doi{10.1051/0004-6361/201629806},
  \eprint{1611.10298}

\bibitem[{{De Ger{\'o}nimo} et~al.(2018){De Ger{\'o}nimo}, {Althaus},
  {C{\'o}rsico}, {Romero}, and {Kepler}}]{2018A&A...613A..46D}
{De Ger{\'o}nimo} FC, {Althaus} LG, {C{\'o}rsico} AH, {Romero} AD, {Kepler} SO
  (2018) {Asteroseismology of ZZ Ceti stars with full evolutionary white dwarf
  models. II. The impact of AGB thermal pulses on the asteroseismic inferences
  of ZZ Ceti stars}. \aap 613:A46, \doi{10.1051/0004-6361/201731982},
  \eprint{1801.10589}

\bibitem[{{De Ger{\'o}nimo} et~al.(2019){De Ger{\'o}nimo}, {C{\'o}rsico},
  {Althaus}, {Wachlin}, and {Camisassa}}]{2019A&A...621A.100D}
{De Ger{\'o}nimo} FC, {C{\'o}rsico} AH, {Althaus} LG, {Wachlin} FC, {Camisassa}
  ME (2019) {Pulsation properties of ultra-massive DA white dwarf stars with
  ONe cores}. \aap 621:A100, \doi{10.1051/0004-6361/201833789},
  \eprint{1807.03810}

\bibitem[{{De Marco} and {Soker}(2002)}]{2002PASP..114..602D}
{De Marco} O, {Soker} N (2002) {A New Look at the Evolution of Wolf-Rayet
  Central Stars of Planetary Nebulae}. PASP 114:602--611, \doi{10.1086/341691},
  \eprint{astro-ph/0204230}

\bibitem[{{Deheuvels} et~al.(2012){Deheuvels}, {Garc{\'{\i}}a}, {Chaplin},
  {Basu}, {Antia}, {Appourchaux}, {Benomar}, {Davies}, {Elsworth}, {Gizon},
  {Goupil}, {Reese}, {Regulo}, {Schou}, {Stahn}, {Casagrande},
  {Christensen-Dalsgaard}, {Fischer}, {Hekker}, {Kjeldsen}, {Mathur}, {Mosser},
  {Pinsonneault}, {Valenti}, {Christiansen}, {Kinemuchi}, and
  {Mullally}}]{2012ApJ...756...19D}
{Deheuvels} S, {Garc{\'{\i}}a} RA, {Chaplin} WJ, {Basu} S, {Antia} HM,
  {Appourchaux} T, {Benomar} O, {Davies} GR, {Elsworth} Y, {Gizon} L, {Goupil}
  MJ, {Reese} DR, {Regulo} C, {Schou} J, {Stahn} T, {Casagrande} L,
  {Christensen-Dalsgaard} J, {Fischer} D, {Hekker} S, {Kjeldsen} H, {Mathur} S,
  {Mosser} B, {Pinsonneault} M, {Valenti} J, {Christiansen} JL, {Kinemuchi} K,
  {Mullally} F (2012) {Seismic Evidence for a Rapidly Rotating Core in a
  Lower-giant-branch Star Observed with Kepler}. \apj 756:19,
  \doi{10.1088/0004-637X/756/1/19}, \eprint{1206.3312}

\bibitem[{{den Hartogh} et~al.(2019){den Hartogh}, {Eggenberger}, and
  {Hirschi}}]{2019A&A...622A.187D}
{den Hartogh} JW, {Eggenberger} P, {Hirschi} R (2019) {Constraining transport
  of angular momentum in stars. Combining asteroseismic observations of core
  helium burning stars and white dwarfs}. \aap 622:A187,
  \doi{10.1051/0004-6361/201834330}, \eprint{1902.04293}

\bibitem[{{Dine} et~al.(1981){Dine}, {Fischler}, and
  {Srednicki}}]{1981PhLB..104..199D}
{Dine} M, {Fischler} W, {Srednicki} M (1981) {A simple solution to the strong
  CP problem with a harmless axion}. Physics Letters B 104:199--202,
  \doi{10.1016/0370-2693(81)90590-6}

\bibitem[{{Dolez} and {Vauclair}(1981)}]{1981A&A...102..375D}
{Dolez} N, {Vauclair} G (1981) {Gravity modes instability in DA white dwarfs}.
  \aap 102:375--385

\bibitem[{{Dufour} et~al.(2011){Dufour}, {B{\'e}land}, {Fontaine}, {Chayer},
  and {Bergeron}}]{2011ApJ...733L..19D}
{Dufour} P, {B{\'e}land} S, {Fontaine} G, {Chayer} P, {Bergeron} P (2011)
  {Taking Advantage of the COS Time-tag Capability: Observations of Pulsating
  Hot DQ White Dwarfs and Discovery of a New One}. \apjl 733:L19,
  \doi{10.1088/2041-8205/733/2/L19}, \eprint{1104.2543}

\bibitem[{{Dunlap} et~al.(2010){Dunlap}, {Barlow}, and
  {Clemens}}]{2010ApJ...720L.159D}
{Dunlap} BH, {Barlow} BN, {Clemens} JC (2010) {A New Small-amplitude Variable
  Hot DQ White Dwarf}. \apjl 720:L159--L163,
  \doi{10.1088/2041-8205/720/2/L159}, \eprint{1007.5293}

\bibitem[{{Dupret} et~al.(2008){Dupret}, {Quirion}, {Fontaine}, {Brassard}, and
  {Grigahc{\`e}ne}}]{2008JPhCS.118a2051D}
{Dupret} MA, {Quirion} PO, {Fontaine} G, {Brassard} P, {Grigahc{\`e}ne} A
  (2008) {Time-dependent convection study of the driving mechanism in the DBV
  white dwarfs}. In: Journal of Physics Conference Series, Journal of Physics
  Conference Series, vol 118, p 012051, \doi{10.1088/1742-6596/118/1/012051}

\bibitem[{{Dupuis}(2018)}]{2018MsT.........14D}
{Dupuis} CM (2018) {A Search for Variability in C-rich DQ White Dwarfs}.
  Master's thesis, Texas A{\&}M University - Commerce

\bibitem[{{Dziembowski}(1982)}]{1982AcA....32..147D}
{Dziembowski} W (1982) {Nonlinear mode coupling in oscillating stars. I -
  Second order theory of the coherent mode coupling}. \actaa 32:147--171

\bibitem[{{Dziembowski} and {Koester}(1981)}]{1981A&A....97...16D}
{Dziembowski} W, {Koester} D (1981) {Excitation of gravity modes in white
  dwarfs with chemically stratified envelopes}. \aap 97:16--26

\bibitem[{{Ekstr{\"o}m} et~al.(2012){Ekstr{\"o}m}, {Georgy}, {Eggenberger},
  {Meynet}, {Mowlavi}, {Wyttenbach}, {Granada}, {Decressin}, {Hirschi},
  {Frischknecht}, {Charbonnel}, and {Maeder}}]{2012A&A...537A.146E}
{Ekstr{\"o}m} S, {Georgy} C, {Eggenberger} P, {Meynet} G, {Mowlavi} N,
  {Wyttenbach} A, {Granada} A, {Decressin} T, {Hirschi} R, {Frischknecht} U,
  {Charbonnel} C, {Maeder} A (2012) {Grids of stellar models with rotation. I.
  Models from 0.8 to 120 M$_{\odot}$ at solar metallicity (Z = 0.014)}. \aap
  537:A146, \doi{10.1051/0004-6361/201117751}, \eprint{1110.5049}

\bibitem[{{Farihi}(2016)}]{2016NewAR..71....9F}
{Farihi} J (2016) {Circumstellar debris and pollution at white dwarf stars}.
  \nar 71:9--34, \doi{10.1016/j.newar.2016.03.001}, \eprint{1604.03092}

\bibitem[{{Faulkner}(1972)}]{1972ApJ...173..401F}
{Faulkner} J (1972) {On the Nature of the Horizontal Branch. II. Extremely Blue
  Halo Stars: a Theoretical Viewpoint}. \apj 173:401, \doi{10.1086/151430}

\bibitem[{{Fields} et~al.(2016){Fields}, {Farmer}, {Petermann}, {Iliadis}, and
  {Timmes}}]{2016ApJ...823...46F}
{Fields} CE, {Farmer} R, {Petermann} I, {Iliadis} C, {Timmes} FX (2016)
  {Properties of Carbon-Oxygen White Dwarfs From Monte Carlo Stellar Models}.
  apj 823:46, \doi{10.3847/0004-637X/823/1/46}, \eprint{1603.06666}

\bibitem[{{Fontaine} and {Brassard}(2008)}]{2008PASP..120.1043F}
{Fontaine} G, {Brassard} P (2008) {The Pulsating White Dwarf Stars}. PASP
  120:1043--1096, \doi{10.1086/592788}

\bibitem[{{Fontaine} et~al.(2001){Fontaine}, {Brassard}, and
  {Bergeron}}]{2001PASP..113..409F}
{Fontaine} G, {Brassard} P, {Bergeron} P (2001) {The Potential of White Dwarf
  Cosmochronology}. \pasp 113:409--435, \doi{10.1086/319535}

\bibitem[{{Fontaine} et~al.(2013){Fontaine}, {Brassard}, and
  {Charpinet}}]{2013EPJWC..4305011F}
{Fontaine} G, {Brassard} P, {Charpinet} S (2013) {The angular momentum of
  isolated white dwarfs}. In: European Physical Journal Web of Conferences,
  European Physical Journal Web of Conferences, vol~43, p 05011,
  \doi{10.1051/epjconf/20134305011}

\bibitem[{{Fontaine} et~al.(2017){Fontaine}, {Istrate}, {Gianninas},
  {Brassard}, and {Van Grootel}}]{2017ASPC..509..347F}
{Fontaine} G, {Istrate} A, {Gianninas} A, {Brassard} P, {Van Grootel} V (2017)
  {Making Sense Out of Pulsating Pre-ELM and ELM White Dwarfs}. In: {Tremblay}
  PE, {Gaensicke} B, {Marsh} T (eds) 20th European White Dwarf Workshop,
  Astronomical Society of the Pacific Conference Series, vol 509, p 347

\bibitem[{{Fu} et~al.(2013){Fu}, {Dolez}, {Vauclair}, {Fox-Machado}, {Kim},
  {Li}, {Chen}, {Alvarez}, {Su}, {Charpinet}, {Chevreton}, {Michel}, {Yang},
  {Li}, {Zhang}, {Molnar}, and {Plachy}}]{2013MNRAS.429.1585F}
{Fu} JN, {Dolez} N, {Vauclair} G, {Fox-Machado} L, {Kim} SL, {Li} C, {Chen} L,
  {Alvarez} M, {Su} J, {Charpinet} S, {Chevreton} M, {Michel} R, {Yang} XH,
  {Li} Y, {Zhang} YP, {Molnar} L, {Plachy} E (2013) {Asteroseismology of the ZZ
  Ceti star HS 0507+0434B}. \mnras 429:1585--1595, \doi{10.1093/mnras/sts438},
  \eprint{1211.6226}

\bibitem[{{Fu} et~al.(2019){Fu}, {Vauclair}, {Su}, {Fox Machado}, {Colas},
  {Kim}, {Cang}, {Li}, {Niu}, {Xue}, {Li}, {Jiang}, {Michel}, {Alvarez},
  {Dolez}, {Ma}, {Esamdin}, and {Liu}}]{2019MNRAS.486.3560F}
{Fu} JN, {Vauclair} G, {Su} J, {Fox Machado} L, {Colas} F, {Kim} SL, {Cang} TQ,
  {Li} C, {Niu} HB, {Xue} HF, {Li} Y, {Jiang} XJ, {Michel} R, {Alvarez} M,
  {Dolez} N, {Ma} L, {Esamdin} A, {Liu} JZ (2019) {Precise determination of
  stellar parameters of the ZZ Ceti and DAZ white dwarf GD 133 through
  asteroseismology}. \mnras 486(3):3560--3568, \doi{10.1093/mnras/stz1088},
  \eprint{1904.07586}

\bibitem[{{Fujikawa} and {Shrock}(1980)}]{1980PhRvL..45..963F}
{Fujikawa} K, {Shrock} RE (1980) {Magnetic Moment of a Massive Neutrino and
  Neutrino-Spin Rotation}. Physical Review Letters 45:963--966,
  \doi{10.1103/PhysRevLett.45.963}

\bibitem[{{Fuller} and {Lai}(2011)}]{2011MNRAS.412.1331F}
{Fuller} J, {Lai} D (2011) {Tidal excitations of oscillation modes in compact
  white dwarf binaries - I. Linear theory}. \mnras 412:1331--1340,
  \doi{10.1111/j.1365-2966.2010.18017.x}, \eprint{1009.3316}

\bibitem[{{Fuller} and {Lai}(2012)}]{2012MNRAS.421..426F}
{Fuller} J, {Lai} D (2012) {Dynamical tides in compact white dwarf binaries:
  tidal synchronization and dissipation}. \mnras 421:426--445,
  \doi{10.1111/j.1365-2966.2011.20320.x}, \eprint{1108.4910}

\bibitem[{{Gaia Collaboration} et~al.(2018{\natexlab{a}}){Gaia Collaboration},
  {Babusiaux}, {van Leeuwen}, {Barstow}, {Jordi}, {Vallenari}, {Bossini},
  {Bressan}, {Cantat-Gaudin}, {van Leeuwen}, and et~al.}]{2018A&A...616A..10G}
{Gaia Collaboration}, {Babusiaux} C, {van Leeuwen} F, {Barstow} MA, {Jordi} C,
  {Vallenari} A, {Bossini} D, {Bressan} A, {Cantat-Gaudin} T, {van Leeuwen} M,
  et~al (2018{\natexlab{a}}) {Gaia Data Release 2. Observational
  Hertzsprung-Russell diagrams}. \aap 616:A10,
  \doi{10.1051/0004-6361/201832843}, \eprint{1804.09378}

\bibitem[{{Gaia Collaboration} et~al.(2018{\natexlab{b}}){Gaia Collaboration},
  {Brown}, {Vallenari}, {Prusti}, {de Bruijne}, {Babusiaux}, {Bailer-Jones},
  {Biermann}, {Evans}, {Eyer}, and et~al.}]{2018A&A...616A...1G}
{Gaia Collaboration}, {Brown} AGA, {Vallenari} A, {Prusti} T, {de Bruijne} JHJ,
  {Babusiaux} C, {Bailer-Jones} CAL, {Biermann} M, {Evans} DW, {Eyer} L, et~al
  (2018{\natexlab{b}}) {Gaia Data Release 2. Summary of the contents and survey
  properties}. \aap 616:A1, \doi{10.1051/0004-6361/201833051},
  \eprint{1804.09365}

\bibitem[{{G{\"a}nsicke} et~al.(2010){G{\"a}nsicke}, {Koester}, {Girven},
  {Marsh}, and {Steeghs}}]{2010Sci...327..188G}
{G{\"a}nsicke} BT, {Koester} D, {Girven} J, {Marsh} TR, {Steeghs} D (2010) {Two
  White Dwarfs with Oxygen-Rich Atmospheres}. Science 327:188,
  \doi{10.1126/science.1180228}, \eprint{0911.2246}

\bibitem[{{G{\"a}nsicke} et~al.(2012){G{\"a}nsicke}, {Koester}, {Farihi},
  {Girven}, {Parsons}, and {Breedt}}]{2012MNRAS.424..333G}
{G{\"a}nsicke} BT, {Koester} D, {Farihi} J, {Girven} J, {Parsons} SG, {Breedt}
  E (2012) {The chemical diversity of exo-terrestrial planetary debris around
  white dwarfs}. \mnras 424:333--347, \doi{10.1111/j.1365-2966.2012.21201.x},
  \eprint{1205.0167}

\bibitem[{{Garc{\'{\i}}a-Berro} and {Oswalt}(2016)}]{2016NewAR..72....1G}
{Garc{\'{\i}}a-Berro} E, {Oswalt} TD (2016) {The white dwarf luminosity
  function}. New Astronomy Reviews 72:1--22, \doi{10.1016/j.newar.2016.08.001},
  \eprint{1608.02631}

\bibitem[{{Garcia-Berro} et~al.(2006){Garcia-Berro}, {Kubyshin},
  {Loren-Aguilar}, and {Isern}}]{2006IJMPD..15.1163G}
{Garcia-Berro} E, {Kubyshin} Y, {Loren-Aguilar} P, {Isern} J (2006) {The
  Variation of the Gravitational Constant Inferred from the Hubble Diagram of
  Type ia Supernovae}. International Journal of Modern Physics D 15:1163--1174,
  \doi{10.1142/S0218271806008772}, \eprint{gr-qc/0512164}

\bibitem[{{Garc{\'{\i}}a-Berro} et~al.(2007){Garc{\'{\i}}a-Berro}, {Isern}, and
  {Kubyshin}}]{2007A&ARv..14..113G}
{Garc{\'{\i}}a-Berro} E, {Isern} J, {Kubyshin} YA (2007) {Astronomical
  measurements and constraints on the variability of fundamental constants}.
  \aapr 14:113--170, \doi{10.1007/s00159-006-0004-8}

\bibitem[{{Garc{\'{\i}}a-Berro} et~al.(2010){Garc{\'{\i}}a-Berro}, {Torres},
  {Althaus}, {Renedo}, {Lor{\'e}n-Aguilar}, {C{\'o}rsico}, {Rohrmann},
  {Salaris}, and {Isern}}]{2010Natur.465..194G}
{Garc{\'{\i}}a-Berro} E, {Torres} S, {Althaus} LG, {Renedo} I,
  {Lor{\'e}n-Aguilar} P, {C{\'o}rsico} AH, {Rohrmann} RD, {Salaris} M, {Isern}
  J (2010) {A white dwarf cooling age of 8Gyr for NGC 6791 from physical
  separation processes}. \nat 465:194--196, \doi{10.1038/nature09045},
  \eprint{1005.2272}

\bibitem[{{Garc{\'{\i}}a-Berro} et~al.(2011){Garc{\'{\i}}a-Berro},
  {Lor{\'e}n-Aguilar}, {Torres}, {Althaus}, and {Isern}}]{2011JCAP...05..021G}
{Garc{\'{\i}}a-Berro} E, {Lor{\'e}n-Aguilar} P, {Torres} S, {Althaus} LG,
  {Isern} J (2011) {An upper limit to the secular variation of the
  gravitational constant from white dwarf stars}. \jcap 5:021,
  \doi{10.1088/1475-7516/2011/05/021}, \eprint{1105.1992}

\bibitem[{{Garc{\'{\i}}a-Berro} et~al.(2012){Garc{\'{\i}}a-Berro},
  {Lor{\'e}n-Aguilar}, {Aznar-Sigu{\'a}n}, {Torres}, {Camacho}, {Althaus},
  {C{\'o}rsico}, {K{\"u}lebi}, and {Isern}}]{2012ApJ...749...25G}
{Garc{\'{\i}}a-Berro} E, {Lor{\'e}n-Aguilar} P, {Aznar-Sigu{\'a}n} G, {Torres}
  S, {Camacho} J, {Althaus} LG, {C{\'o}rsico} AH, {K{\"u}lebi} B, {Isern} J
  (2012) {Double Degenerate Mergers as Progenitors of High-field Magnetic White
  Dwarfs}. \apj 749:25, \doi{10.1088/0004-637X/749/1/25}, \eprint{1202.0461}

\bibitem[{{Gauss}(1809)}]{Gauss1809}
{Gauss} KF (1809) {Theoria motvs corporvm coelestivm in sectionibvs conicis
  solem ambientivm.}

\bibitem[{{Gautschy} et~al.(2005){Gautschy}, {Althaus}, and
  {Saio}}]{2005A&A...438.1013G}
{Gautschy} A, {Althaus} LG, {Saio} H (2005) {On the excitation of PG 1159-type
  pulsations}. \aap 438:1013--1020, \doi{10.1051/0004-6361:20042486},
  \eprint{astro-ph/0504495}

\bibitem[{{Gentile Fusillo} et~al.(2019){Gentile Fusillo}, {Tremblay},
  {G{\"a}nsicke}, {Manser}, {Cunningham}, {Cukanovaite}, {Hollands}, {Marsh},
  {Raddi}, {Jordan}, {Toonen}, {Geier}, {Barstow}, and
  {Cummings}}]{2019MNRAS.482.4570G}
{Gentile Fusillo} NP, {Tremblay} PE, {G{\"a}nsicke} BT, {Manser} CJ,
  {Cunningham} T, {Cukanovaite} E, {Hollands} M, {Marsh} T, {Raddi} R, {Jordan}
  S, {Toonen} S, {Geier} S, {Barstow} M, {Cummings} JD (2019) {A Gaia Data
  Release 2 catalogue of white dwarfs and a comparison with SDSS}. \mnras
  482:4570--4591, \doi{10.1093/mnras/sty3016}, \eprint{1807.03315}

\bibitem[{{Giammichele} et~al.(2017{\natexlab{a}}){Giammichele}, {Charpinet},
  {Brassard}, and {Fontaine}}]{2017A&A...598A.109G}
{Giammichele} N, {Charpinet} S, {Brassard} P, {Fontaine} G (2017{\natexlab{a}})
  {The potential of asteroseismology for probing the core chemical
  stratification in white dwarf stars}. \aap 598:A109,
  \doi{10.1051/0004-6361/201629935}, \eprint{1611.05071}

\bibitem[{{Giammichele} et~al.(2017{\natexlab{b}}){Giammichele}, {Charpinet},
  {Fontaine}, and {Brassard}}]{2017ApJ...834..136G}
{Giammichele} N, {Charpinet} S, {Fontaine} G, {Brassard} P (2017{\natexlab{b}})
  {Toward High-precision Seismic Studies of White Dwarf Stars: Parametrization
  of the Core and Tests of Accuracy}. \apj 834:136,
  \doi{10.3847/1538-4357/834/2/136}, \eprint{1610.06036}

\bibitem[{{Giammichele} et~al.(2018){Giammichele}, {Charpinet}, {Fontaine},
  {Brassard}, {Green}, {Van Grootel}, {Bergeron}, {Zong}, and
  {Dupret}}]{2018Natur.554...73G}
{Giammichele} N, {Charpinet} S, {Fontaine} G, {Brassard} P, {Green} EM, {Van
  Grootel} V, {Bergeron} P, {Zong} W, {Dupret} MA (2018) {A large
  oxygen-dominated core from the seismic cartography of a pulsating white
  dwarf}. \nat 554:73--76, \doi{10.1038/nature25136}

\bibitem[{{Gianninas} et~al.(2010){Gianninas}, {Bergeron}, {Dupuis}, and
  {Ruiz}}]{2010ApJ...720..581G}
{Gianninas} A, {Bergeron} P, {Dupuis} J, {Ruiz} MT (2010) {Spectroscopic
  Analysis of Hot, Hydrogen-rich White Dwarfs: The Presence of Metals and the
  Balmer-line Problem}. \apj 720:581--602, \doi{10.1088/0004-637X/720/1/581}

\bibitem[{{Gianninas} et~al.(2011){Gianninas}, {Bergeron}, and
  {Ruiz}}]{2011ApJ...743..138G}
{Gianninas} A, {Bergeron} P, {Ruiz} MT (2011) {A Spectroscopic Survey and
  Analysis of Bright, Hydrogen-rich White Dwarfs}. \apj 743:138,
  \doi{10.1088/0004-637X/743/2/138}, \eprint{1109.3171}

\bibitem[{{Gianninas} et~al.(2014){Gianninas}, {Dufour}, {Kilic}, {Brown},
  {Bergeron}, and {Hermes}}]{2014ApJ...794...35G}
{Gianninas} A, {Dufour} P, {Kilic} M, {Brown} WR, {Bergeron} P, {Hermes} JJ
  (2014) {Precise Atmospheric Parameters for the Shortest-period Binary White
  Dwarfs: Gravitational Waves, Metals, and Pulsations}. \apj 794:35,
  \doi{10.1088/0004-637X/794/1/35}, \eprint{1408.3118}

\bibitem[{{Gianninas} et~al.(2015){Gianninas}, {Kilic}, {Brown}, {Canton}, and
  {Kenyon}}]{2015ApJ...812..167G}
{Gianninas} A, {Kilic} M, {Brown} WR, {Canton} P, {Kenyon} SJ (2015) {The ELM
  Survey. VI. Eleven New Double Degenerates}. \apj 812:167,
  \doi{10.1088/0004-637X/812/2/167}, \eprint{1509.07134}

\bibitem[{{Gianninas} et~al.(2016){Gianninas}, {Curd}, {Fontaine}, {Brown}, and
  {Kilic}}]{2016ApJ...822L..27G}
{Gianninas} A, {Curd} B, {Fontaine} G, {Brown} WR, {Kilic} M (2016) {Discovery
  of Three Pulsating, Mixed-atmosphere, Extremely Low-mass White Dwarf
  Precursors}. \apjl 822:L27, \doi{10.3847/2041-8205/822/2/L27},
  \eprint{1604.04621}

\bibitem[{{Giannotti}(2017)}]{2017NatPh..13..530G}
{Giannotti} M (2017) {Axion searches: Exciting times}. Nature Physics
  13:530--531, \doi{10.1038/nphys4139}

\bibitem[{{Giannotti} et~al.(2017){Giannotti}, {Irastorza}, {Redondo},
  {Ringwald}, and {Saikawa}}]{2017JCAP...10..010G}
{Giannotti} M, {Irastorza} IG, {Redondo} J, {Ringwald} A, {Saikawa} K (2017)
  {Stellar recipes for axion hunters}. \jcap 10:010,
  \doi{10.1088/1475-7516/2017/10/010}, \eprint{1708.02111}

\bibitem[{{Gilliland} et~al.(2010){Gilliland}, {Jenkins}, {Borucki}, {Bryson},
  {Caldwell}, {Clarke}, {Dotson}, {Haas}, {Hall}, {Klaus}, {Koch}, {McCauliff},
  {Quintana}, {Twicken}, and {van Cleve}}]{2010ApJ...713L.160G}
{Gilliland} RL, {Jenkins} JM, {Borucki} WJ, {Bryson} ST, {Caldwell} DA,
  {Clarke} BD, {Dotson} JL, {Haas} MR, {Hall} J, {Klaus} T, {Koch} D,
  {McCauliff} S, {Quintana} EV, {Twicken} JD, {van Cleve} JE (2010) {Initial
  Characteristics of Kepler Short Cadence Data}. \apjl 713:L160--L163,
  \doi{10.1088/2041-8205/713/2/L160}, \eprint{1001.0142}

\bibitem[{{Goldreich} and {Wu}(1999)}]{1999ApJ...511..904G}
{Goldreich} P, {Wu} Y (1999) {Gravity Modes in ZZ Ceti Stars. I.
  Quasi-adiabatic Analysis of Overstability}. \apj 511:904--915,
  \doi{10.1086/306705}, \eprint{astro-ph/9804305}

\bibitem[{{Gonz{\'a}lez P{\'e}rez} et~al.(2006){Gonz{\'a}lez P{\'e}rez},
  {Solheim}, and {Kamben}}]{2006A&A...454..527G}
{Gonz{\'a}lez P{\'e}rez} JM, {Solheim} JE, {Kamben} R (2006) {A search for
  photometric variability of hydrogen-deficient planetary-nebula nuclei}. \aap
  454:527--536, \doi{10.1051/0004-6361:20053468}

\bibitem[{{Greggio} and {Renzini}(1990)}]{1990ApJ...364...35G}
{Greggio} L, {Renzini} A (1990) {Clues on the hot star content and the
  ultraviolet output of elliptical galaxies}. \apj 364:35--64,
  \doi{10.1086/169384}

\bibitem[{{Greiss} et~al.(2016){Greiss}, {Hermes}, {G{\"a}nsicke}, {Steeghs},
  {Bell}, {Raddi}, {Tremblay}, {Breedt}, {Ramsay}, {Koester}, {Carter},
  {Vanderbosch}, {Winget}, and {Winget}}]{2016MNRAS.457.2855G}
{Greiss} S, {Hermes} JJ, {G{\"a}nsicke} BT, {Steeghs} DTH, {Bell} KJ, {Raddi}
  R, {Tremblay} PE, {Breedt} E, {Ramsay} G, {Koester} D, {Carter} PJ,
  {Vanderbosch} Z, {Winget} DE, {Winget} KI (2016) {The search for ZZ Ceti
  stars in the original Kepler mission}. \mnras 457:2855--2863,
  \doi{10.1093/mnras/stw053}, \eprint{1601.01316}

\bibitem[{{Grigahc{\`e}ne} et~al.(2005){Grigahc{\`e}ne}, {Dupret}, {Gabriel},
  {Garrido}, and {Scuflaire}}]{2005A&A...434.1055G}
{Grigahc{\`e}ne} A, {Dupret} MA, {Gabriel} M, {Garrido} R, {Scuflaire} R (2005)
  {Convection-pulsation coupling. I. A mixing-length perturbative theory}. \aap
  434:1055--1062, \doi{10.1051/0004-6361:20041816}

\bibitem[{{Haft} et~al.(1994){Haft}, {Raffelt}, and
  {Weiss}}]{1994ApJ...425..222H}
{Haft} M, {Raffelt} G, {Weiss} A (1994) {Standard and nonstandard plasma
  neutrino emission revisited}. \apj 425:222--230, \doi{10.1086/173978},
  \eprint{astro-ph/9309014}

\bibitem[{{Han} et~al.(2000){Han}, {Tout}, and
  {Eggleton}}]{2000MNRAS.319..215H}
{Han} Z, {Tout} CA, {Eggleton} PP (2000) {Low- and intermediate-mass close
  binary evolution and the initial-final mass relation}. \mnras 319:215--222,
  \doi{10.1046/j.1365-8711.2000.03839.x}, \eprint{astro-ph/0010269}

\bibitem[{{Hansen}(2004)}]{2004PhR...399....1H}
{Hansen} B (2004) {The astrophysics of cool white dwarfs}. \physrep 399:1--70,
  \doi{10.1016/j.physrep.2004.07.001}

\bibitem[{{Hansen} and {Liebert}(2003)}]{2003ARA&A..41..465H}
{Hansen} BMS, {Liebert} J (2003) {Cool White Dwarfs}. \araa 41:465--515,
  \doi{10.1146/annurev.astro.41.081401.155117}

\bibitem[{{Hansen} et~al.(2015){Hansen}, {Richer}, {Kalirai}, {Goldsbury},
  {Frewen}, and {Heyl}}]{2015ApJ...809..141H}
{Hansen} BMS, {Richer} H, {Kalirai} J, {Goldsbury} R, {Frewen} S, {Heyl} J
  (2015) {Constraining Neutrino Cooling Using the Hot White Dwarf Luminosity
  Function in the Globular Cluster 47 Tucanae}. \apj 809(2):141,
  \doi{10.1088/0004-637X/809/2/141}, \eprint{1507.05665}

\bibitem[{{Hansen} et~al.(1985){Hansen}, {Winget}, and
  {Kawaler}}]{1985ApJ...297..544H}
{Hansen} CJ, {Winget} DE, {Kawaler} SD (1985) {Upper and lower bounds of
  periods in variable white dwarfs}. \apj 297:544--547, \doi{10.1086/163549}

\bibitem[{{Hermes} et~al.(2011){Hermes}, {Mullally}, {{\O}stensen}, {Williams},
  {Telting}, {Southworth}, {Bloemen}, {Howell}, {Everett}, and
  {Winget}}]{2011ApJ...741L..16H}
{Hermes} JJ, {Mullally} F, {{\O}stensen} RH, {Williams} KA, {Telting} J,
  {Southworth} J, {Bloemen} S, {Howell} SB, {Everett} M, {Winget} DE (2011)
  {Discovery of a ZZ Ceti in the Kepler Mission Field}. \apjl 741:L16,
  \doi{10.1088/2041-8205/741/1/L16}, \eprint{1109.6023}

\bibitem[{{Hermes} et~al.(2012){Hermes}, {Montgomery}, {Winget}, {Brown},
  {Kilic}, and {Kenyon}}]{2012ApJ...750L..28H}
{Hermes} JJ, {Montgomery} MH, {Winget} DE, {Brown} WR, {Kilic} M, {Kenyon} SJ
  (2012) {SDSS J184037.78+642312.3: The First Pulsating Extremely Low Mass
  White Dwarf}. ApJl 750:L28, \doi{10.1088/2041-8205/750/2/L28},
  \eprint{1204.1338}

\bibitem[{{Hermes} et~al.(2013{\natexlab{a}}){Hermes}, {Kepler}, {Castanheira},
  {Gianninas}, {Winget}, {Montgomery}, {Brown}, and
  {Harrold}}]{2013ApJ...771L...2H}
{Hermes} JJ, {Kepler} SO, {Castanheira} BG, {Gianninas} A, {Winget} DE,
  {Montgomery} MH, {Brown} WR, {Harrold} ST (2013{\natexlab{a}}) {Discovery of
  an Ultramassive Pulsating White Dwarf}. \apjl 771:L2,
  \doi{10.1088/2041-8205/771/1/L2}, \eprint{1306.4024}

\bibitem[{{Hermes} et~al.(2013{\natexlab{b}}){Hermes}, {Montgomery},
  {Gianninas}, {Winget}, {Brown}, {Harrold}, {Bell}, {Kenyon}, {Kilic}, and
  {Castanheira}}]{2013MNRAS.436.3573H}
{Hermes} JJ, {Montgomery} MH, {Gianninas} A, {Winget} DE, {Brown} WR, {Harrold}
  ST, {Bell} KJ, {Kenyon} SJ, {Kilic} M, {Castanheira} BG (2013{\natexlab{b}})
  {A new class of pulsating white dwarf of extremely low mass: the fourth and
  fifth members}. \mnras 436:3573--3580, \doi{10.1093/mnras/stt1835},
  \eprint{1310.0013}

\bibitem[{{Hermes} et~al.(2013{\natexlab{c}}){Hermes}, {Montgomery},
  {Mullally}, {Winget}, and {Bischoff-Kim}}]{2013ApJ...766...42H}
{Hermes} JJ, {Montgomery} MH, {Mullally} F, {Winget} DE, {Bischoff-Kim} A
  (2013{\natexlab{c}}) {A New Timescale for Period Change in the Pulsating DA
  White Dwarf WD 0111+0018}. \apj 766:42, \doi{10.1088/0004-637X/766/1/42},
  \eprint{1302.1875}

\bibitem[{{Hermes} et~al.(2013{\natexlab{d}}){Hermes}, {Montgomery}, {Winget},
  {Brown}, {Gianninas}, {Kilic}, {Kenyon}, {Bell}, and
  {Harrold}}]{2013ApJ...765..102H}
{Hermes} JJ, {Montgomery} MH, {Winget} DE, {Brown} WR, {Gianninas} A, {Kilic}
  M, {Kenyon} SJ, {Bell} KJ, {Harrold} ST (2013{\natexlab{d}}) {Discovery of
  Pulsations, Including Possible Pressure Modes, in Two New Extremely Low Mass,
  He-core White Dwarfs}. \apj 765:102, \doi{10.1088/0004-637X/765/2/102},
  \eprint{1211.1022}

\bibitem[{{Hermes} et~al.(2015){Hermes}, {Montgomery}, {Bell}, {Chote},
  {G{\"a}nsicke}, {Kawaler}, {Clemens}, {Dunlap}, {Winget}, and
  {Armstrong}}]{2015ApJ...810L...5H}
{Hermes} JJ, {Montgomery} MH, {Bell} KJ, {Chote} P, {G{\"a}nsicke} BT,
  {Kawaler} SD, {Clemens} JC, {Dunlap} BH, {Winget} DE, {Armstrong} DJ (2015)
  {A Second Case of Outbursts in a Pulsating White Dwarf Observed by Kepler}.
  \apjl 810:L5, \doi{10.1088/2041-8205/810/1/L5}, \eprint{1507.06319}

\bibitem[{{Hermes} et~al.(2017{\natexlab{a}}){Hermes}, {G{\"a}nsicke},
  {Kawaler}, {Greiss}, {Tremblay}, {Gentile Fusillo}, {Raddi}, {Fanale},
  {Bell}, {Dennihy}, {Fuchs}, {Dunlap}, {Clemens}, {Montgomery}, {Winget},
  {Chote}, {Marsh}, and {Redfield}}]{2017ApJS..232...23H}
{Hermes} JJ, {G{\"a}nsicke} BT, {Kawaler} SD, {Greiss} S, {Tremblay} PE,
  {Gentile Fusillo} NP, {Raddi} R, {Fanale} SM, {Bell} KJ, {Dennihy} E, {Fuchs}
  JT, {Dunlap} BH, {Clemens} JC, {Montgomery} MH, {Winget} DE, {Chote} P,
  {Marsh} TR, {Redfield} S (2017{\natexlab{a}}) {White Dwarf Rotation as a
  Function of Mass and a Dichotomy of Mode Line Widths: Kepler Observations of
  27 Pulsating DA White Dwarfs through K2 Campaign 8}. \apjs 232:23,
  \doi{10.3847/1538-4365/aa8bb5}, \eprint{1709.07004}

\bibitem[{{Hermes} et~al.(2017{\natexlab{b}}){Hermes}, {Kawaler},
  {Bischoff-Kim}, {Provencal}, {Dunlap}, and {Clemens}}]{2017ApJ...835..277H}
{Hermes} JJ, {Kawaler} SD, {Bischoff-Kim} A, {Provencal} JL, {Dunlap} BH,
  {Clemens} JC (2017{\natexlab{b}}) {A Deep Test of Radial Differential
  Rotation in a Helium-atmosphere White Dwarf. I. Discovery of Pulsations in PG
  0112+104}. \apj 835:277, \doi{10.3847/1538-4357/835/2/277},
  \eprint{1612.07807}

\bibitem[{{Herwig}(2005)}]{2005ARA&A..43..435H}
{Herwig} F (2005) {Evolution of Asymptotic Giant Branch Stars}. \araa
  43:435--479, \doi{10.1146/annurev.astro.43.072103.150600}

\bibitem[{{Herwig} et~al.(1999){Herwig}, {Bl{\"o}cker}, {Langer}, and
  {Driebe}}]{1999A&A...349L...5H}
{Herwig} F, {Bl{\"o}cker} T, {Langer} N, {Driebe} T (1999) {On the formation of
  hydrogen-deficient post-AGB stars}. \aap 349:L5--L8,
  \eprint{astro-ph/9908108}

\bibitem[{{Hofmann} et~al.(2010){Hofmann}, {M{\"u}ller}, and
  {Biskupek}}]{2010A&A...522L...5H}
{Hofmann} F, {M{\"u}ller} J, {Biskupek} L (2010) {Lunar laser ranging test of
  the Nordtvedt parameter and a possible variation in the gravitational
  constant}. \aap 522:L5, \doi{10.1051/0004-6361/201015659}

\bibitem[{{H{\"o}fner} and {Olofsson}(2018)}]{2018A&ARv..26....1H}
{H{\"o}fner} S, {Olofsson} H (2018) {Mass loss of stars on the asymptotic giant
  branch. Mechanisms, models and measurements}. \aapr 26:1,
  \doi{10.1007/s00159-017-0106-5}

\bibitem[{{Hollands} et~al.(2018){Hollands}, {G{\"a}nsicke}, and
  {Koester}}]{2018MNRAS.477...93H}
{Hollands} MA, {G{\"a}nsicke} BT, {Koester} D (2018) {Cool DZ white dwarfs II:
  compositions and evolution of old remnant planetary systems}. \mnras
  477:93--111, \doi{10.1093/mnras/sty592}, \eprint{1801.07714}

\bibitem[{{Hoof} et~al.(2019){Hoof}, {Kahlhoefer}, {Scott}, {Weniger}, and
  {White}}]{2019JHEP...03..191H}
{Hoof} S, {Kahlhoefer} F, {Scott} P, {Weniger} C, {White} M (2019) {Axion
  global fits with Peccei-Quinn symmetry breaking before inflation using
  GAMBIT}. Journal of High Energy Physics 3:191, \doi{10.1007/JHEP03(2019)191},
  \eprint{1810.07192}

\bibitem[{{Horowitz} et~al.(2010){Horowitz}, {Schneider}, and
  {Berry}}]{2010PhRvL.104w1101H}
{Horowitz} CJ, {Schneider} AS, {Berry} DK (2010) {Crystallization of
  Carbon-Oxygen Mixtures in White Dwarf Stars}. Physical Review Letters
  104(23):231101, \doi{10.1103/PhysRevLett.104.231101}, \eprint{1005.2441}

\bibitem[{{Howell} et~al.(2014){Howell}, {Sobeck}, {Haas}, {Still}, {Barclay},
  {Mullally}, {Troeltzsch}, {Aigrain}, {Bryson}, {Caldwell}, {Chaplin},
  {Cochran}, {Huber}, {Marcy}, {Miglio}, {Najita}, {Smith}, {Twicken}, and
  {Fortney}}]{2014PASP..126..398H}
{Howell} SB, {Sobeck} C, {Haas} M, {Still} M, {Barclay} T, {Mullally} F,
  {Troeltzsch} J, {Aigrain} S, {Bryson} ST, {Caldwell} D, {Chaplin} WJ,
  {Cochran} WD, {Huber} D, {Marcy} GW, {Miglio} A, {Najita} JR, {Smith} M,
  {Twicken} JD, {Fortney} JJ (2014) {The K2 Mission: Characterization and Early
  Results}. PASP 126:398, \doi{10.1086/676406}, \eprint{1402.5163}

\bibitem[{{Hu} et~al.(2019){Hu}, {Webb}, {Ayres}, {Bainbridge}, {Barrow},
  {Barstow}, {Berengut}, {Carswell}, {Dzuba}, {Flambaum}, {Holberg}, {Lee},
  {Preval}, {Reindl}, and {Tchang-Brillet}}]{2019MNRAS.485.5050H}
{Hu} J, {Webb} JK, {Ayres} TR, {Bainbridge} MB, {Barrow} JD, {Barstow} MA,
  {Berengut} JC, {Carswell} RF, {Dzuba} VA, {Flambaum} VV, {Holberg} JB, {Lee}
  CC, {Preval} SP, {Reindl} N, {Tchang-Brillet} W{\"U}L (2019) {Constraining
  the magnetic field on white dwarf surfaces; Zeeman effects and fine structure
  constant variation}. \mnras 485:5050--5058, \doi{10.1093/mnras/stz739},
  \eprint{1812.11480}

\bibitem[{{Iben}(1984)}]{1984ApJ...277..333I}
{Iben} I Jr (1984) {On the frequency of planetary nebula nuclei powered by
  helium burning and on the frequency of white dwarfs with hydrogen-deficient
  atmospheres}. \apj 277:333--354, \doi{10.1086/161700}

\bibitem[{{Iben} and {MacDonald}(1986)}]{1986ApJ...301..164I}
{Iben} I Jr, {MacDonald} J (1986) {The effects of diffusion due to gravity and
  due to composition gradients on the rate of hydrogen burning in a cooling
  degenerate dwarf. II - Dependence on initial metallicity and on buffer mass}.
  \apj 301:164--176, \doi{10.1086/163884}

\bibitem[{{Iben} and {MacDonald}(1995)}]{1995LNP...443...48I}
{Iben} I Jr, {MacDonald} J (1995) {The Born Again AGB Phenomenon}. In:
  {Koester} D, {Werner} K (eds) White Dwarfs, Lecture Notes in Physics, Berlin
  Springer Verlag, vol 443, p~48

\bibitem[{{Irastorza} and {Redondo}(2018)}]{2018PrPNP.102...89I}
{Irastorza} IG, {Redondo} J (2018) {New experimental approaches in the search
  for axion-like particles}. Progress in Particle and Nuclear Physics
  102:89--159, \doi{10.1016/j.ppnp.2018.05.003}, \eprint{1801.08127}

\bibitem[{{Isern} et~al.(1992){Isern}, {Hernanz}, and
  {Garcia-Berro}}]{1992ApJ...392L..23I}
{Isern} J, {Hernanz} M, {Garcia-Berro} E (1992) {Axion cooling of white
  dwarfs}. ApJl 392:L23--L25, \doi{10.1086/186416}

\bibitem[{{Isern} et~al.(2008){Isern}, {Garc{\'{\i}}a-Berro}, {Torres}, and
  {Catal{\'a}n}}]{2008ApJ...682L.109I}
{Isern} J, {Garc{\'{\i}}a-Berro} E, {Torres} S, {Catal{\'a}n} S (2008) {Axions
  and the Cooling of White Dwarf Stars}. \apjl 682:L109, \doi{10.1086/591042},
  \eprint{0806.2807}

\bibitem[{{Isern} et~al.(2009){Isern}, {Catal{\'a}n}, {Garc{\'{\i}}a-Berro},
  and {Torres}}]{2009JPhCS.172a2005I}
{Isern} J, {Catal{\'a}n} S, {Garc{\'{\i}}a-Berro} E, {Torres} S (2009) {Axions
  and the white dwarf luminosity function}. In: Journal of Physics Conference
  Series, Journal of Physics Conference Series, vol 172, p 012005,
  \doi{10.1088/1742-6596/172/1/012005}, \eprint{0812.3043}

\bibitem[{{Isern} et~al.(2018){Isern}, {Garc{\'{\i}}a-Berro}, {Torres},
  {Cojocaru}, and {Catal{\'a}n}}]{2018MNRAS.478.2569I}
{Isern} J, {Garc{\'{\i}}a-Berro} E, {Torres} S, {Cojocaru} R, {Catal{\'a}n} S
  (2018) {Axions and the luminosity function of white dwarfs: the thin and
  thick discs, and the halo}. \mnras 478:2569--2575,
  \doi{10.1093/mnras/sty1162}, \eprint{1805.00135}

\bibitem[{{Istrate} et~al.(2016{\natexlab{a}}){Istrate}, {Fontaine},
  {Gianninas}, {Grassitelli}, {Marchant}, {Tauris}, and
  {Langer}}]{2016A&A...595L..12I}
{Istrate} AG, {Fontaine} G, {Gianninas} A, {Grassitelli} L, {Marchant} P,
  {Tauris} TM, {Langer} N (2016{\natexlab{a}}) {Asteroseismic test of
  rotational mixing in low-mass white dwarfs}. \aap 595:L12,
  \doi{10.1051/0004-6361/201629876}, \eprint{1610.08513}

\bibitem[{{Istrate} et~al.(2016{\natexlab{b}}){Istrate}, {Marchant}, {Tauris},
  {Langer}, {Stancliffe}, and {Grassitelli}}]{2016A&A...595A..35I}
{Istrate} AG, {Marchant} P, {Tauris} TM, {Langer} N, {Stancliffe} RJ,
  {Grassitelli} L (2016{\natexlab{b}}) {Models of low-mass helium white dwarfs
  including gravitational settling, thermal and chemical diffusion, and
  rotational mixing}. \aap 595:A35, \doi{10.1051/0004-6361/201628874},
  \eprint{1606.04947}

\bibitem[{{Istrate} et~al.(2017){Istrate}, {Fontaine}, and
  {Heuser}}]{2017ApJ...847..130I}
{Istrate} AG, {Fontaine} G, {Heuser} C (2017) {A Model of the Pulsating
  Extremely Low-mass White Dwarf Precursor WASP 0247-25B}. \apj 847:130,
  \doi{10.3847/1538-4357/aa8958}, \eprint{1708.09388}

\bibitem[{{Ivanova} et~al.(2013){Ivanova}, {Justham}, {Chen}, {De Marco},
  {Fryer}, {Gaburov}, {Ge}, {Glebbeek}, {Han}, {Li}, {Lu}, {Marsh},
  {Podsiadlowski}, {Potter}, {Soker}, {Taam}, {Tauris}, {van den Heuvel}, and
  {Webbink}}]{2013A&ARv..21...59I}
{Ivanova} N, {Justham} S, {Chen} X, {De Marco} O, {Fryer} CL, {Gaburov} E, {Ge}
  H, {Glebbeek} E, {Han} Z, {Li} XD, {Lu} G, {Marsh} T, {Podsiadlowski} P,
  {Potter} A, {Soker} N, {Taam} R, {Tauris} TM, {van den Heuvel} EPJ, {Webbink}
  RF (2013) {Common envelope evolution: where we stand and how we can move
  forward}. \aapr 21:59, \doi{10.1007/s00159-013-0059-2}, \eprint{1209.4302}

\bibitem[{{Jeffery} and {Saio}(2013)}]{2013MNRAS.435..885J}
{Jeffery} CS, {Saio} H (2013) {Pulsation in extremely low mass helium stars}.
  \mnras 435:885--892, \doi{10.1093/mnras/stt1360}

\bibitem[{{Kanaan} et~al.(1992){Kanaan}, {Kepler}, {Giovannini}, and
  {Diaz}}]{1992ApJ...390L..89K}
{Kanaan} A, {Kepler} SO, {Giovannini} O, {Diaz} M (1992) {The discovery of a
  new DAV star using IUE temperature determination}. \apjl 390:L89--L91,
  \doi{10.1086/186379}

\bibitem[{{Kanaan} et~al.(2005){Kanaan}, {Nitta}, {Winget}, {Kepler},
  {Montgomery}, {Metcalfe}, {Oliveira}, {Fraga}, {da Costa}, {Costa},
  {Castanheira}, {Giovannini}, {Nather}, {Mukadam}, {Kawaler}, {O'Brien},
  {Reed}, {Kleinman}, {Provencal}, {Watson}, {Kilkenny}, {Sullivan},
  {Sullivan}, {Shobbrook}, {Jiang}, {Ashoka}, {Seetha}, {Leibowitz},
  {Ibbetson}, {Mendelson}, {Mei{\v s}tas}, {Kalytis}, {Ali{\v s}auskas},
  {O'Donoghue}, {Buckley}, {Martinez}, {van Wyk}, {Stobie}, {Marang}, {van
  Zyl}, {Ogloza}, {Krzesinski}, {Zola}, {Moskalik}, {Breger}, {Stankov},
  {Silvotti}, {Piccioni}, {Vauclair}, {Dolez}, {Chevreton}, {Deetjen},
  {Dreizler}, {Schuh}, {Gonzalez Perez}, {{\O}stensen}, {Ulla}, {Manteiga},
  {Suarez}, {Burleigh}, and {Barstow}}]{2005A&A...432..219K}
{Kanaan} A, {Nitta} A, {Winget} DE, {Kepler} SO, {Montgomery} MH, {Metcalfe}
  TS, {Oliveira} H, {Fraga} L, {da Costa} AFM, {Costa} JES, {Castanheira} BG,
  {Giovannini} O, {Nather} RE, {Mukadam} A, {Kawaler} SD, {O'Brien} MS, {Reed}
  MD, {Kleinman} SJ, {Provencal} JL, {Watson} TK, {Kilkenny} D, {Sullivan} DJ,
  {Sullivan} T, {Shobbrook} B, {Jiang} XJ, {Ashoka} BN, {Seetha} S, {Leibowitz}
  E, {Ibbetson} P, {Mendelson} H, {Mei{\v s}tas} EG, {Kalytis} R, {Ali{\v
  s}auskas} D, {O'Donoghue} D, {Buckley} D, {Martinez} P, {van Wyk} F, {Stobie}
  R, {Marang} F, {van Zyl} L, {Ogloza} W, {Krzesinski} J, {Zola} S, {Moskalik}
  P, {Breger} M, {Stankov} A, {Silvotti} R, {Piccioni} A, {Vauclair} G, {Dolez}
  N, {Chevreton} M, {Deetjen} J, {Dreizler} S, {Schuh} S, {Gonzalez Perez} JM,
  {{\O}stensen} R, {Ulla} A, {Manteiga} M, {Suarez} O, {Burleigh} MR, {Barstow}
  MA (2005) {Whole Earth Telescope observations of BPM 37093: A seismological
  test of crystallization theory in white dwarfs}. \aap 432:219--224,
  \doi{10.1051/0004-6361:20041125}, \eprint{astro-ph/0411199}

\bibitem[{{Karakas} and {Lattanzio}(2014)}]{2014PASA...31...30K}
{Karakas} AI, {Lattanzio} JC (2014) {The Dawes Review 2: Nucleosynthesis and
  Stellar Yields of Low- and Intermediate-Mass Single Stars}. \pasa 31:e030,
  \doi{10.1017/pasa.2014.21}, \eprint{1405.0062}

\bibitem[{{Kawaler}(1988)}]{1988ApJ...334..220K}
{Kawaler} SD (1988) {The hydrogen shell game - Pulsational instabilities in
  hydrogen shell-burning planetary nebula nuclei}. \apj 334:220--228,
  \doi{10.1086/166832}

\bibitem[{{Kawaler}(2015)}]{2015ASPC..493...65K}
{Kawaler} SD (2015) {Rotation of White Dwarf Stars}. In: {Dufour} P, {Bergeron}
  P, {Fontaine} G (eds) 19th European Workshop on White Dwarfs, Astronomical
  Society of the Pacific Conference Series, vol 493, p~65, \eprint{1410.6934}

\bibitem[{{Kawaler} et~al.(1986){Kawaler}, {Winget}, {Hansen}, and
  {Iben}}]{1986ApJ...306L..41K}
{Kawaler} SD, {Winget} DE, {Hansen} CJ, {Iben} I Jr (1986) {The helium shell
  game - Nonradial g-mode instabilities in hydrogen-deficient planetary nebula
  nuclei}. \apjl 306:L41--L44, \doi{10.1086/184701}

\bibitem[{{Kawaler} et~al.(1999){Kawaler}, {Sekii}, and
  {Gough}}]{1999ApJ...516..349K}
{Kawaler} SD, {Sekii} T, {Gough} D (1999) {Prospects for Measuring Differential
  Rotation in White Dwarfs through Asteroseismology}. \apj 516:349--365,
  \doi{10.1086/307087}, \eprint{astro-ph/9811286}

\bibitem[{{Kepler}(2012)}]{2012ASPC..462..322K}
{Kepler} SO (2012) {White Dwarf Stars: Pulsations and Magnetism}. In:
  {Shibahashi} H, {Takata} M, {Lynas-Gray} AE (eds) Progress in Solar/Stellar
  Physics with Helio- and Asteroseismology, Astronomical Society of the Pacific
  Conference Series, vol 462, p 322

\bibitem[{{Kepler} and {Bradley}(1995)}]{1995BaltA...4..166K}
{Kepler} SO, {Bradley} PA (1995) {Structure and Evolution of White Dwarfs}.
  Baltic Astronomy 4:166--220, \doi{10.1515/astro-1995-0213}

\bibitem[{{Kepler} and {Romero}(2017)}]{2017EPJWC.15201011K}
{Kepler} SO, {Romero} AD (2017) {Pulsating white dwarfs}. In: European Physical
  Journal Web of Conferences, European Physical Journal Web of Conferences, vol
  152, p 01011, \doi{10.1051/epjconf/201715201011}, \eprint{1706.07020}

\bibitem[{{Kepler} et~al.(2003){Kepler}, {Nather}, {Winget}, {Nitta},
  {Kleinman}, {Metcalfe}, {Sekiguchi}, {Jiang}, {Sullivan}, {Sullivan},
  {Janulis}, {Meistas}, {Kalytis}, {Krzesinski}, {Ogoza}, {Zola}, {O'Donoghue},
  {Romero-Colmenero}, {Martinez}, {Dreizler}, {Deetjen}, {Nagel}, {Schuh},
  {Vauclair}, {Ning}, {Chevreton}, {Solheim}, {Gonzalez Perez}, {Johannessen},
  {Kanaan}, {Costa}, {Murillo Costa}, {Wood}, {Silvestri}, {Ahrens}, {Jones},
  {Collins}, {Boyer}, {Shaw}, {Mukadam}, {Klumpe}, {Larrison}, {Kawaler},
  {Riddle}, {Ulla}, and {Bradley}}]{2003A&A...401..639K}
{Kepler} SO, {Nather} RE, {Winget} DE, {Nitta} A, {Kleinman} SJ, {Metcalfe} T,
  {Sekiguchi} K, {Jiang} X, {Sullivan} D, {Sullivan} T, {Janulis} R, {Meistas}
  E, {Kalytis} R, {Krzesinski} J, {Ogoza} W, {Zola} S, {O'Donoghue} D,
  {Romero-Colmenero} E, {Martinez} P, {Dreizler} S, {Deetjen} J, {Nagel} T,
  {Schuh} SL, {Vauclair} G, {Ning} FJ, {Chevreton} M, {Solheim} JE, {Gonzalez
  Perez} JM, {Johannessen} F, {Kanaan} A, {Costa} JE, {Murillo Costa} AF,
  {Wood} MA, {Silvestri} N, {Ahrens} TJ, {Jones} AK, {Collins} AE, {Boyer} M,
  {Shaw} JS, {Mukadam} A, {Klumpe} EW, {Larrison} J, {Kawaler} S, {Riddle} R,
  {Ulla} A, {Bradley} P (2003) {The everchanging pulsating white dwarf GD358}.
  \aap 401:639--654, \doi{10.1051/0004-6361:20030105},
  \eprint{astro-ph/0301477}

\bibitem[{{Kepler} et~al.(2005){Kepler}, {Costa}, {Castanheira}, {Winget},
  {Mullally}, {Nather}, {Kilic}, {von Hippel}, {Mukadam}, and
  {Sullivan}}]{2005ApJ...634.1311K}
{Kepler} SO, {Costa} JES, {Castanheira} BG, {Winget} DE, {Mullally} F, {Nather}
  RE, {Kilic} M, {von Hippel} T, {Mukadam} AS, {Sullivan} DJ (2005) {Measuring
  the Evolution of the Most Stable Optical Clock G 117-B15A}. \apj
  634:1311--1318, \doi{10.1086/497002}, \eprint{astro-ph/0507487}

\bibitem[{{Kepler} et~al.(2014){Kepler}, {Fraga}, {Winget}, {Bell},
  {C{\'o}rsico}, and {Werner}}]{2014MNRAS.442.2278K}
{Kepler} SO, {Fraga} L, {Winget} DE, {Bell} K, {C{\'o}rsico} AH, {Werner} K
  (2014) {Discovery of a new PG 1159 (GW Vir) pulsator}. \mnras 442:2278--2281,
  \doi{10.1093/mnras/stu1019}, \eprint{1405.5075}

\bibitem[{{Kepler} et~al.(2015){Kepler}, {Pelisoli}, {Koester}, {Ourique},
  {Kleinman}, {Romero}, {Nitta}, {Eisenstein}, {Costa}, {K{\"u}lebi}, {Jordan},
  {Dufour}, {Giommi}, and {Rebassa-Mansergas}}]{2015MNRAS.446.4078K}
{Kepler} SO, {Pelisoli} I, {Koester} D, {Ourique} G, {Kleinman} SJ, {Romero}
  AD, {Nitta} A, {Eisenstein} DJ, {Costa} JES, {K{\"u}lebi} B, {Jordan} S,
  {Dufour} P, {Giommi} P, {Rebassa-Mansergas} A (2015) {New white dwarf stars
  in the Sloan Digital Sky Survey Data Release 10}. \mnras 446:4078--4087,
  \doi{10.1093/mnras/stu2388}, \eprint{1411.4149}

\bibitem[{{Kepler} et~al.(2016{\natexlab{a}}){Kepler}, {Koester}, and
  {Ourique}}]{2016Sci...352...67K}
{Kepler} SO, {Koester} D, {Ourique} G (2016{\natexlab{a}}) {A white dwarf with
  an oxygen atmosphere}. Science 352:67--69, \doi{10.1126/science.aad6705}

\bibitem[{{Kepler} et~al.(2016{\natexlab{b}}){Kepler}, {Pelisoli}, {Koester},
  {Ourique}, {Romero}, {Reindl}, {Kleinman}, {Eisenstein}, {Valois}, and
  {Amaral}}]{2016MNRAS.455.3413K}
{Kepler} SO, {Pelisoli} I, {Koester} D, {Ourique} G, {Romero} AD, {Reindl} N,
  {Kleinman} SJ, {Eisenstein} DJ, {Valois} ADM, {Amaral} LA
  (2016{\natexlab{b}}) {New white dwarf and subdwarf stars in the Sloan Digital
  Sky Survey Data Release 12}. \mnras 455:3413--3423,
  \doi{10.1093/mnras/stv2526}, \eprint{1510.08409}

\bibitem[{{Kepler} et~al.(2019){Kepler}, {Pelisoli}, {Koester}, {Reindl},
  {Geier}, {Romero}, {Ourique}, {Oliveira}, and {Amaral}}]{2019MNRAS.486.2169K}
{Kepler} SO, {Pelisoli} I, {Koester} D, {Reindl} N, {Geier} S, {Romero} AD,
  {Ourique} G, {Oliveira} CdP, {Amaral} LA (2019) {White dwarf and subdwarf
  stars in the Sloan Digital Sky Survey Data Release 14}. \mnras
  486:2169--2183, \doi{10.1093/mnras/stz960}, \eprint{1904.01626}

\bibitem[{{Kilic} et~al.(2011){Kilic}, {Brown}, {Allende Prieto},
  {Ag{\"u}eros}, {Heinke}, and {Kenyon}}]{2011ApJ...727....3K}
{Kilic} M, {Brown} WR, {Allende Prieto} C, {Ag{\"u}eros} MA, {Heinke} C,
  {Kenyon} SJ (2011) {The ELM Survey. II. Twelve Binary White Dwarf Merger
  Systems}. \apj 727:3, \doi{10.1088/0004-637X/727/1/3}, \eprint{1011.4073}

\bibitem[{{Kilic} et~al.(2012){Kilic}, {Brown}, {Allende Prieto}, {Kenyon},
  {Heinke}, {Ag{\"u}eros}, and {Kleinman}}]{2012ApJ...751..141K}
{Kilic} M, {Brown} WR, {Allende Prieto} C, {Kenyon} SJ, {Heinke} CO,
  {Ag{\"u}eros} MA, {Kleinman} SJ (2012) {The ELM Survey. IV. 24 White Dwarf
  Merger Systems}. \apj 751:141, \doi{10.1088/0004-637X/751/2/141},
  \eprint{1204.0028}

\bibitem[{{Kilic} et~al.(2015){Kilic}, {Hermes}, {Gianninas}, and
  {Brown}}]{2015MNRAS.446L..26K}
{Kilic} M, {Hermes} JJ, {Gianninas} A, {Brown} WR (2015) {PSR J1738+0333: the
  first millisecond pulsar + pulsating white dwarf binary*}. \mnras
  446:L26--L30, \doi{10.1093/mnrasl/slu152}, \eprint{1410.4898}

\bibitem[{{Kilic} et~al.(2017){Kilic}, {Munn}, {Harris}, {von Hippel},
  {Liebert}, {Williams}, {Jeffery}, and {DeGennaro}}]{2017ApJ...837..162K}
{Kilic} M, {Munn} JA, {Harris} HC, {von Hippel} T, {Liebert} JW, {Williams} KA,
  {Jeffery} E, {DeGennaro} S (2017) {The Ages of the Thin Disk, Thick Disk, and
  the Halo from Nearby White Dwarfs}. apj 837:162,
  \doi{10.3847/1538-4357/aa62a5}, \eprint{1702.06984}

\bibitem[{{Kilic} et~al.(2018){Kilic}, {Hermes}, {C{\'o}rsico}, {Kosakowski},
  {Brown}, {Antoniadis}, {Calcaferro}, {Gianninas}, {Althaus}, and
  {Green}}]{2018MNRAS.479.1267K}
{Kilic} M, {Hermes} JJ, {C{\'o}rsico} AH, {Kosakowski} A, {Brown} WR,
  {Antoniadis} J, {Calcaferro} LM, {Gianninas} A, {Althaus} LG, {Green} MJ
  (2018) {A refined search for pulsations in white dwarf companions to
  millisecond pulsars}. \mnras 479:1267--1272, \doi{10.1093/mnras/sty1546},
  \eprint{1806.03650}

\bibitem[{{Kim} and {Carosi}(2010)}]{2010RvMP...82..557K}
{Kim} JE, {Carosi} G (2010) {Axions and the strong CP problem}. Reviews of
  Modern Physics 82:557--601, \doi{10.1103/RevModPhys.82.557},
  \eprint{0807.3125}

\bibitem[{{Kippenhahn} et~al.(2012){Kippenhahn}, {Weigert}, and
  {Weiss}}]{2012sse..book.....K}
{Kippenhahn} R, {Weigert} A, {Weiss} A (2012) {Stellar Structure and
  Evolution}. \doi{10.1007/978-3-642-30304-3}

\bibitem[{{Kleinman} et~al.(2013){Kleinman}, {Kepler}, {Koester}, {Pelisoli},
  {Pe{\c c}anha}, {Nitta}, {Costa}, {Krzesinski}, {Dufour}, {Lachapelle},
  {Bergeron}, {Yip}, {Harris}, {Eisenstein}, {Althaus}, and
  {C{\'o}rsico}}]{2013ApJS..204....5K}
{Kleinman} SJ, {Kepler} SO, {Koester} D, {Pelisoli} I, {Pe{\c c}anha} V,
  {Nitta} A, {Costa} JES, {Krzesinski} J, {Dufour} P, {Lachapelle} FR,
  {Bergeron} P, {Yip} CW, {Harris} HC, {Eisenstein} DJ, {Althaus} L,
  {C{\'o}rsico} A (2013) {SDSS DR7 White Dwarf Catalog}. ApJs 204:5,
  \doi{10.1088/0067-0049/204/1/5}, \eprint{1212.1222}

\bibitem[{{Koester}(2002)}]{2002A&ARv..11...33K}
{Koester} D (2002) {White dwarfs: Recent developments}. \aapr 11:33--66,
  \doi{10.1007/s001590100015}

\bibitem[{{Koester} and {Chanmugam}(1990)}]{1990RPPh...53..837K}
{Koester} D, {Chanmugam} G (1990) {REVIEW: Physics of white dwarf stars}.
  Reports on Progress in Physics 53:837--915, \doi{10.1088/0034-4885/53/7/001}

\bibitem[{{Koester} and {Kepler}(2015)}]{2015A&A...583A..86K}
{Koester} D, {Kepler} SO (2015) {DB white dwarfs in the Sloan Digital Sky
  Survey data release 10 and 12}. \aap 583:A86,
  \doi{10.1051/0004-6361/201527169}, \eprint{1509.08244}

\bibitem[{{Koester} et~al.(2009){Koester}, {Voss}, {Napiwotzki}, {Christlieb},
  {Homeier}, {Lisker}, {Reimers}, and {Heber}}]{2009A&A...505..441K}
{Koester} D, {Voss} B, {Napiwotzki} R, {Christlieb} N, {Homeier} D, {Lisker} T,
  {Reimers} D, {Heber} U (2009) {High-resolution UVES/VLT spectra of white
  dwarfs observed for the ESO SN Ia Progenitor Survey. III. DA white dwarfs}.
  \aap 505:441--462, \doi{10.1051/0004-6361/200912531}, \eprint{0908.2322}

\bibitem[{{Koester} et~al.(2014){Koester}, {G{\"a}nsicke}, and
  {Farihi}}]{2014A&A...566A..34K}
{Koester} D, {G{\"a}nsicke} BT, {Farihi} J (2014) {The frequency of planetary
  debris around young white dwarfs}. \aap 566:A34,
  \doi{10.1051/0004-6361/201423691}, \eprint{1404.2617}

\bibitem[{{Kunz} et~al.(2002){Kunz}, {Fey}, {Jaeger}, {Mayer}, {Hammer},
  {Staudt}, {Harissopulos}, and {Paradellis}}]{2002ApJ...567..643K}
{Kunz} R, {Fey} M, {Jaeger} M, {Mayer} A, {Hammer} JW, {Staudt} G,
  {Harissopulos} S, {Paradellis} T (2002) {Astrophysical Reaction Rate of
  $^{12}$C({$\alpha$}, {$\gamma$})$^{16}$O}. \apj 567:643--650,
  \doi{10.1086/338384}

\bibitem[{{Kurtz} et~al.(2008){Kurtz}, {Shibahashi}, {Dhillon}, {Marsh}, and
  {Littlefair}}]{2008MNRAS.389.1771K}
{Kurtz} DW, {Shibahashi} H, {Dhillon} VS, {Marsh} TR, {Littlefair} SP (2008) {A
  search for a new class of pulsating DA white dwarf stars in the DB gap}.
  \mnras 389:1771--1779, \doi{10.1111/j.1365-2966.2008.13664.x}

\bibitem[{{Kurtz} et~al.(2013){Kurtz}, {Shibahashi}, {Dhillon}, {Marsh},
  {Littlefair}, {Copperwheat}, {G{\"a}nsicke}, and
  {Parsons}}]{2013MNRAS.432.1632K}
{Kurtz} DW, {Shibahashi} H, {Dhillon} VS, {Marsh} TR, {Littlefair} SP,
  {Copperwheat} CM, {G{\"a}nsicke} BT, {Parsons} SG (2013) {Hot DAVs: a
  probable new class of pulsating white dwarf stars}. \mnras 432:1632--1639,
  \doi{10.1093/mnras/stt585}

\bibitem[{{Landolt}(1968)}]{1968ApJ...153..151L}
{Landolt} AU (1968) {A New Short-Period Blue Variable}. \apj 153:151,
  \doi{10.1086/149645}

\bibitem[{{Laplace}(1810)}]{Laplace1810}
{Laplace} PS (1810) {M\'emoire sur les approximations des formules qui sont
  fonctions de tr\`es-grands nombres, et sur leur application aux
  probabilit\'es (suite)}. "M\'emoires l'Institut" 12:353--415, 559--565.
  Oeuvres 12 p.301--345, p.349--353

\bibitem[{{Lauffer} et~al.(2018){Lauffer}, {Romero}, and
  {Kepler}}]{2018MNRAS.480.1547L}
{Lauffer} GR, {Romero} AD, {Kepler} SO (2018) {New full evolutionary sequences
  of H- and He-atmosphere massive white dwarf stars using MESA}. \mnras
  480:1547--1562, \doi{10.1093/mnras/sty1925}, \eprint{1807.04774}

\bibitem[{{Lecoanet} et~al.(2016){Lecoanet}, {Schwab}, {Quataert}, {Bildsten},
  {Timmes}, {Burns}, {Vasil}, {Oishi}, and {Brown}}]{2016ApJ...832...71L}
{Lecoanet} D, {Schwab} J, {Quataert} E, {Bildsten} L, {Timmes} FX, {Burns} KJ,
  {Vasil} GM, {Oishi} JS, {Brown} BP (2016) {Turbulent Chemical Diffusion in
  Convectively Bounded Carbon Flames}. \apj 832:71,
  \doi{10.3847/0004-637X/832/1/71}, \eprint{1603.08921}

\bibitem[{{Ledoux} and {Walraven}(1958)}]{1958HDP....51..353L}
{Ledoux} P, {Walraven} T (1958) {Variable Stars.} Handbuch der Physik
  51:353--604

\bibitem[{{Legendre}(1806)}]{Legendre1805}
{Legendre} AM (1806) {Nouvelles Methodes pour la determination des Orbites des
  Cometes}

\bibitem[{{Luan} and {Goldreich}(2018)}]{2018ApJ...863...82L}
{Luan} J, {Goldreich} P (2018) {DAVs: Red Edge and Outbursts}. \apj 863:82,
  \doi{10.3847/1538-4357/aad0f4}, \eprint{1711.06367}

\bibitem[{Luders(1954)}]{Luders54}
Luders G (1954) {On the equivalence of invariance under time reversal and under
  particle-antiparticle conjugation for relativistic field theories}. Dan Mat
  Fys Medd 28(5):1--17, \urlprefix\url{http://cds.cern.ch/record/1071765}

\bibitem[{{Lund} et~al.(2017){Lund}, {Silva Aguirre}, {Davies}, {Chaplin},
  {Christensen-Dalsgaard}, {Houdek}, {White}, {Bedding}, {Ball}, {Huber},
  {Antia}, {Lebreton}, {Latham}, {Handberg}, {Verma}, {Basu}, {Casagrande},
  {Justesen}, {Kjeldsen}, and {Mosumgaard}}]{2017ApJ...835..172L}
{Lund} MN, {Silva Aguirre} V, {Davies} GR, {Chaplin} WJ,
  {Christensen-Dalsgaard} J, {Houdek} G, {White} TR, {Bedding} TR, {Ball} WH,
  {Huber} D, {Antia} HM, {Lebreton} Y, {Latham} DW, {Handberg} R, {Verma} K,
  {Basu} S, {Casagrande} L, {Justesen} AB, {Kjeldsen} H, {Mosumgaard} JR (2017)
  {Standing on the Shoulders of Dwarfs: the Kepler Asteroseismic LEGACY Sample.
  I. Oscillation Mode Parameters}. \apj 835:172,
  \doi{10.3847/1538-4357/835/2/172}, \eprint{1612.00436}

\bibitem[{{Maeda} and {Shibahashi}(2014)}]{2014PASJ...66...76M}
{Maeda} K, {Shibahashi} H (2014) {Pulsations of pre-white dwarfs with
  hydrogen-dominated atmospheres}. \pasj 66:76, \doi{10.1093/pasj/psu051},
  \eprint{1405.4568}

\bibitem[{{Malec} and {Biesiada}(2013)}]{2013ASPC..469...21M}
{Malec} B, {Biesiada} M (2013) {White Dwarf Constraints on Exotic Physical
  Theories}. In: 18th European White Dwarf Workshop., Astronomical Society of
  the Pacific Conference Series, vol 469, p~21

\bibitem[{{Maoz} et~al.(2014){Maoz}, {Mannucci}, and
  {Nelemans}}]{2014ARA&A..52..107M}
{Maoz} D, {Mannucci} F, {Nelemans} G (2014) {Observational Clues to the
  Progenitors of Type Ia Supernovae}. Annula Review Astronomy and Astrophysics
  52:107--170, \doi{10.1146/annurev-astro-082812-141031}, \eprint{1312.0628}

\bibitem[{{Marino} et~al.(2017){Marino}, {Milone}, {Yong}, {Da Costa},
  {Asplund}, {Bedin}, {Jerjen}, {Nardiello}, {Piotto}, {Renzini}, and
  {Shetrone}}]{2017ApJ...843...66M}
{Marino} AF, {Milone} AP, {Yong} D, {Da Costa} G, {Asplund} M, {Bedin} LR,
  {Jerjen} H, {Nardiello} D, {Piotto} G, {Renzini} A, {Shetrone} M (2017)
  {Spectroscopy and Photometry of Multiple Populations along the Asymptotic
  Giant Branch of NGC 2808 and NGC 6121 (M4)}. \apj 843:66,
  \doi{10.3847/1538-4357/aa7852}

\bibitem[{{Maxted} et~al.(2011){Maxted}, {Anderson}, {Burleigh}, {Collier
  Cameron}, {Heber}, {G{\"a}nsicke}, {Geier}, {Kupfer}, {Marsh}, {Nelemans},
  {O'Toole}, {{\O}stensen}, {Smalley}, and {West}}]{2011MNRAS.418.1156M}
{Maxted} PFL, {Anderson} DR, {Burleigh} MR, {Collier Cameron} A, {Heber} U,
  {G{\"a}nsicke} BT, {Geier} S, {Kupfer} T, {Marsh} TR, {Nelemans} G, {O'Toole}
  SJ, {{\O}stensen} RH, {Smalley} B, {West} RG (2011) {Discovery of a stripped
  red giant core in a bright eclipsing binary system}. \mnras 418:1156--1164,
  \doi{10.1111/j.1365-2966.2011.19567.x}, \eprint{1107.4986}

\bibitem[{{Maxted} et~al.(2013){Maxted}, {Serenelli}, {Miglio}, {Marsh},
  {Heber}, {Dhillon}, {Littlefair}, {Copperwheat}, {Smalley}, {Breedt}, and
  {Schaffenroth}}]{2013Natur.498..463M}
{Maxted} PFL, {Serenelli} AM, {Miglio} A, {Marsh} TR, {Heber} U, {Dhillon} VS,
  {Littlefair} S, {Copperwheat} C, {Smalley} B, {Breedt} E, {Schaffenroth} V
  (2013) {Multi-periodic pulsations of a stripped red-giant star in an
  eclipsing binary system}. \nat 498:463--465, \doi{10.1038/nature12192},
  \eprint{1307.1654}

\bibitem[{{Maxted} et~al.(2014{\natexlab{a}}){Maxted}, {Bloemen}, {Heber},
  {Geier}, {Wheatley}, {Marsh}, {Breedt}, {Sebastian}, {Faillace}, {Owen},
  {Pulley}, {Smith}, {Kolb}, {Haswell}, {Southworth}, {Anderson}, {Smalley},
  {Collier Cameron}, {Hebb}, {Simpson}, {West}, {Bochinski}, {Busuttil}, and
  {Hadigal}}]{2014MNRAS.437.1681M}
{Maxted} PFL, {Bloemen} S, {Heber} U, {Geier} S, {Wheatley} PJ, {Marsh} TR,
  {Breedt} E, {Sebastian} D, {Faillace} G, {Owen} C, {Pulley} D, {Smith} D,
  {Kolb} U, {Haswell} CA, {Southworth} J, {Anderson} DR, {Smalley} B, {Collier
  Cameron} A, {Hebb} L, {Simpson} EK, {West} RG, {Bochinski} J, {Busuttil} R,
  {Hadigal} S (2014{\natexlab{a}}) {EL CVn-type binaries - discovery of 17
  helium white dwarf precursors in bright eclipsing binary star systems}.
  \mnras 437:1681--1697, \doi{10.1093/mnras/stt2007}, \eprint{1310.4863}

\bibitem[{{Maxted} et~al.(2014{\natexlab{b}}){Maxted}, {Serenelli}, {Marsh},
  {Catal{\'a}n}, {Mahtani}, and {Dhillon}}]{2014MNRAS.444..208M}
{Maxted} PFL, {Serenelli} AM, {Marsh} TR, {Catal{\'a}n} S, {Mahtani} DP,
  {Dhillon} VS (2014{\natexlab{b}}) {WASP 1628+10 - an EL CVn-type binary with
  a very low mass stripped red giant star and multiperiodic pulsations}. \mnras
  444:208--216, \doi{10.1093/mnras/stu1465}, \eprint{1407.5415}

\bibitem[{{McGraw} et~al.(1979){McGraw}, {Starrfield}, {Liebert}, and
  {Green}}]{1979wdvd.coll..377M}
{McGraw} JT, {Starrfield} SG, {Liebert} J, {Green} R (1979) {PG1159-035: A new,
  hot, non-DA pulsating degenerate}. In: {van Horn} HM, {Weidemann} V,
  {Savedoff} MP (eds) IAU Colloq. 53: White Dwarfs and Variable Degenerate
  Stars, pp 377--381

\bibitem[{{Medin} and {Cumming}(2010)}]{2010PhRvE..81c6107M}
{Medin} Z, {Cumming} A (2010) {Crystallization of classical multicomponent
  plasmas}. \pre 81(3):036107, \doi{10.1103/PhysRevE.81.036107},
  \eprint{1002.3327}

\bibitem[{{Mestel}(1952)}]{1952MNRAS.112..583M}
{Mestel} L (1952) {On the theory of white dwarf stars. I. The energy sources of
  white dwarfs}. \mnras 112:583, \doi{10.1093/mnras/112.6.583}

\bibitem[{{Metcalfe}(2003)}]{2003ApJ...587L..43M}
{Metcalfe} TS (2003) {White Dwarf Asteroseismology and the
  $^{12}$C({$\alpha$},{$\gamma$})$^{16}$O Rate}. \apjl 587:L43--L46,
  \doi{10.1086/375044}, \eprint{astro-ph/0303039}

\bibitem[{{Metcalfe} et~al.(2004){Metcalfe}, {Montgomery}, and
  {Kanaan}}]{2004ApJ...605L.133M}
{Metcalfe} TS, {Montgomery} MH, {Kanaan} A (2004) {Testing White Dwarf
  Crystallization Theory with Asteroseismology of the Massive Pulsating DA Star
  BPM 37093}. \apjl 605:L133--L136, \doi{10.1086/420884},
  \eprint{astro-ph/0402046}

\bibitem[{{Miller Bertolami}(2014)}]{2014A&A...562A.123M}
{Miller Bertolami} MM (2014) {Limits on the neutrino magnetic dipole moment
  from the luminosity function of hot white dwarfs}. \aap 562:A123,
  \doi{10.1051/0004-6361/201322641}, \eprint{1407.1404}

\bibitem[{{Miller Bertolami}(2016)}]{2016A&A...588A..25M}
{Miller Bertolami} MM (2016) {New models for the evolution of post-asymptotic
  giant branch stars and central stars of planetary nebulae}. \aap 588:A25,
  \doi{10.1051/0004-6361/201526577}, \eprint{1512.04129}

\bibitem[{{Miller Bertolami} and {Althaus}(2006)}]{2006A&A...454..845M}
{Miller Bertolami} MM, {Althaus} LG (2006) {Full evolutionary models for PG
  1159 stars. Implications for the helium-rich O(He) stars}. \aap 454:845--854,
  \doi{10.1051/0004-6361:20054723}, \eprint{astro-ph/0603846}

\bibitem[{{Miller Bertolami} and {Althaus}(2007)}]{2007MNRAS.380..763M}
{Miller Bertolami} MM, {Althaus} LG (2007) {The born-again (very late thermal
  pulse) scenario revisited: the mass of the remnants and implications for
  V4334 Sgr}. \mnras 380:763--770, \doi{10.1111/j.1365-2966.2007.12115.x},
  \eprint{0706.0714}

\bibitem[{{Miller Bertolami} et~al.(2011){Miller Bertolami}, {Althaus},
  {Olano}, and {Jim{\'e}nez}}]{2011MNRAS.415.1396M}
{Miller Bertolami} MM, {Althaus} LG, {Olano} C, {Jim{\'e}nez} N (2011) {The
  diffusion-induced nova scenario: CK Vul and PB8 as possible observational
  counterparts}. \mnras 415:1396--1408, \doi{10.1111/j.1365-2966.2011.18790.x},
  \eprint{1103.5455}

\bibitem[{{Miller Bertolami} et~al.(2013){Miller Bertolami}, {Althaus}, and
  {Garc{\'\i}a-Berro}}]{2013ApJ...775L..22M}
{Miller Bertolami} MM, {Althaus} LG, {Garc{\'\i}a-Berro} E (2013) {Quiescent
  Nuclear Burning in Low-metallicity White Dwarfs}. \apj 775(1):L22,
  \doi{10.1088/2041-8205/775/1/L22}, \eprint{1308.2062}

\bibitem[{{Miller Bertolami} et~al.(2014){Miller Bertolami}, {Melendez},
  {Althaus}, and {Isern}}]{2014JCAP...10..069M}
{Miller Bertolami} MM, {Melendez} BE, {Althaus} LG, {Isern} J (2014)
  {Revisiting the axion bounds from the Galactic white dwarf luminosity
  function}. \jcap 10:069, \doi{10.1088/1475-7516/2014/10/069},
  \eprint{1406.7712}

\bibitem[{{Miller Bertolami} et~al.(2017){Miller Bertolami}, {Althaus}, and
  {C{\'o}rsico}}]{2017ASPC..509..435M}
{Miller Bertolami} MM, {Althaus} LG, {C{\'o}rsico} AH (2017) {On the Formation
  of DA White Dwarfs with low Hydrogen Contents: Preliminary Results}. In:
  {Tremblay} PE, {Gaensicke} B, {Marsh} T (eds) 20th European White Dwarf
  Workshop, Astronomical Society of the Pacific Conference Series, vol 509, p
  435, \eprint{1609.08683}

\bibitem[{{Montgomery} and {Winget}(1999)}]{1999ApJ...526..976M}
{Montgomery} MH, {Winget} DE (1999) {The Effect of Crystallization on the
  Pulsations of White Dwarf Stars}. \apj 526:976--990, \doi{10.1086/308044},
  \eprint{astro-ph/9907040}

\bibitem[{{Montgomery} et~al.(2008){Montgomery}, {Williams}, {Winget},
  {Dufour}, {DeGennaro}, and {Liebert}}]{2008ApJ...678L..51M}
{Montgomery} MH, {Williams} KA, {Winget} DE, {Dufour} P, {DeGennaro} S,
  {Liebert} J (2008) {SDSS J142625.71+575218.3: A Prototype for a New Class of
  Variable White Dwarf}. \apjl 678:L51, \doi{10.1086/588286},
  \eprint{0803.2646}

\bibitem[{{Mosser} et~al.(2014){Mosser}, {Benomar}, {Belkacem}, {Goupil},
  {Lagarde}, {Michel}, {Lebreton}, {Stello}, {Vrard}, {Barban}, {Bedding},
  {Deheuvels}, {Chaplin}, {De Ridder}, {Elsworth}, {Montalban}, {Noels},
  {Ouazzani}, {Samadi}, {White}, and {Kjeldsen}}]{2014A&A...572L...5M}
{Mosser} B, {Benomar} O, {Belkacem} K, {Goupil} MJ, {Lagarde} N, {Michel} E,
  {Lebreton} Y, {Stello} D, {Vrard} M, {Barban} C, {Bedding} TR, {Deheuvels} S,
  {Chaplin} WJ, {De Ridder} J, {Elsworth} Y, {Montalban} J, {Noels} A,
  {Ouazzani} RM, {Samadi} R, {White} TR, {Kjeldsen} H (2014) {Mixed modes in
  red giants: a window on stellar evolution}. \aap 572:L5,
  \doi{10.1051/0004-6361/201425039}, \eprint{1411.1082}

\bibitem[{{Mould} and {Uddin}(2014)}]{2014PASA...31...15M}
{Mould} J, {Uddin} SA (2014) {Constraining a Possible Variation of G with Type
  Ia Supernovae}. pasa 31:e015, \doi{10.1017/pasa.2014.9}, \eprint{1402.1534}

\bibitem[{{Mukadam} et~al.(2004){Mukadam}, {Mullally}, {Nather}, {Winget}, {von
  Hippel}, {Kleinman}, {Nitta}, {Krzesi{\'n}ski}, {Kepler}, {Kanaan},
  {Koester}, {Sullivan}, {Homeier}, {Thompson}, {Reaves}, {Cotter},
  {Slaughter}, and {Brinkmann}}]{2004ApJ...607..982M}
{Mukadam} AS, {Mullally} F, {Nather} RE, {Winget} DE, {von Hippel} T,
  {Kleinman} SJ, {Nitta} A, {Krzesi{\'n}ski} J, {Kepler} SO, {Kanaan} A,
  {Koester} D, {Sullivan} DJ, {Homeier} D, {Thompson} SE, {Reaves} D, {Cotter}
  C, {Slaughter} D, {Brinkmann} J (2004) {Thirty-Five New Pulsating DA White
  Dwarf Stars}. \apj 607:982--998, \doi{10.1086/383083}

\bibitem[{{Mukadam} et~al.(2013){Mukadam}, {Bischoff-Kim}, {Fraser},
  {C{\'o}rsico}, {Montgomery}, {Kepler}, {Romero}, {Winget}, {Hermes},
  {Riecken}, {Kronberg}, {Winget}, {Falcon}, {Chandler}, {Kuehne}, {Sullivan},
  {Reaves}, {von Hippel}, {Mullally}, {Shipman}, {Thompson}, {Silvestri}, and
  {Hynes}}]{2013ApJ...771...17M}
{Mukadam} AS, {Bischoff-Kim} A, {Fraser} O, {C{\'o}rsico} AH, {Montgomery} MH,
  {Kepler} SO, {Romero} AD, {Winget} DE, {Hermes} JJ, {Riecken} TS, {Kronberg}
  ME, {Winget} KI, {Falcon} RE, {Chandler} DW, {Kuehne} JW, {Sullivan} DJ,
  {Reaves} D, {von Hippel} T, {Mullally} F, {Shipman} H, {Thompson} SE,
  {Silvestri} NM, {Hynes} RI (2013) {Measuring the Evolutionary Rate of Cooling
  of ZZ Ceti}. \apj 771:17, \doi{10.1088/0004-637X/771/1/17}

\bibitem[{{Mullally} et~al.(2005){Mullally}, {Thompson}, {Castanheira},
  {Winget}, {Kepler}, {Eisenstein}, {Kleinman}, and
  {Nitta}}]{2005ApJ...625..966M}
{Mullally} F, {Thompson} SE, {Castanheira} BG, {Winget} DE, {Kepler} SO,
  {Eisenstein} DJ, {Kleinman} SJ, {Nitta} A (2005) {Eleven New DA White Dwarf
  Variable Stars from the Sloan Digital Sky Survey}. \apj 625:966--972,
  \doi{10.1086/429885}

\bibitem[{{Nandez} and {Ivanova}(2016)}]{2016MNRAS.460.3992N}
{Nandez} JLA, {Ivanova} N (2016) {Common envelope events with low-mass giants:
  understanding the energy budget}. \mnras 460:3992--4002,
  \doi{10.1093/mnras/stw1266}, \eprint{1606.04922}

\bibitem[{{Nather} et~al.(1990){Nather}, {Winget}, {Clemens}, {Hansen}, and
  {Hine}}]{1990ApJ...361..309N}
{Nather} RE, {Winget} DE, {Clemens} JC, {Hansen} CJ, {Hine} BP (1990) {The
  whole earth telescope - A new astronomical instrument}. \apj 361:309--317,
  \doi{10.1086/169196}

\bibitem[{{Nitta} et~al.(2009){Nitta}, {Kleinman}, {Krzesinski}, {Kepler},
  {Metcalfe}, {Mukadam}, {Mullally}, {Nather}, {Sullivan}, {Thompson}, and
  {Winget}}]{2009ApJ...690..560N}
{Nitta} A, {Kleinman} SJ, {Krzesinski} J, {Kepler} SO, {Metcalfe} TS, {Mukadam}
  AS, {Mullally} F, {Nather} RE, {Sullivan} DJ, {Thompson} SE, {Winget} DE
  (2009) {New Pulsating DB White Dwarf Stars from the Sloan Digital Sky
  Survey}. \apj 690:560--565, \doi{10.1088/0004-637X/690/1/560},
  \eprint{0809.0921}

\bibitem[{{Nitta} et~al.(2016){Nitta}, {Kepler}, {Chen{\'e}}, {Koester},
  {Provencal}, {Kleinmani}, {Sullivan}, {Chote}, {Sefako}, {Kanaan}, {Romero},
  {Corti}, {Kilic}, {Montgomery}, and {Winget}}]{2016IAUFM..29B.493N}
{Nitta} A, {Kepler} SO, {Chen{\'e}} AN, {Koester} D, {Provencal} JL,
  {Kleinmani} SJ, {Sullivan} DJ, {Chote} P, {Sefako} R, {Kanaan} A, {Romero} A,
  {Corti} M, {Kilic} M, {Montgomery} MH, {Winget} DE (2016) {Constraining the
  physics of carbon crystallization through pulsations of a massive DAV
  BPM37093}. IAU Focus Meeting 29:493--496, \doi{10.1017/S1743921316005962}

\bibitem[{{Norris}(2004)}]{2004ApJ...612L..25N}
{Norris} JE (2004) {The Helium Abundances of {$\omega$} Centauri}. \apjl
  612:L25--L28, \doi{10.1086/423986}

\bibitem[{{O'Brien} and {Kawaler}(2000)}]{2000ApJ...539..372O}
{O'Brien} MS, {Kawaler} SD (2000) {The Predicted Signature of Neutrino Emission
  in Observations of Pulsating Pre-White Dwarf Stars}. \apj 539:372--378,
  \doi{10.1086/309216}, \eprint{astro-ph/0003261}

\bibitem[{{{\O}stensen} et~al.(2010){{\O}stensen}, {Silvotti}, {Charpinet},
  {Oreiro}, {Handler}, {Green}, {Bloemen}, {Heber}, {G{\"a}nsicke}, {Marsh},
  {Kurtz}, {Telting}, {Reed}, {Kawaler}, {Aerts}, {Rodr{\'{\i}}guez-L{\'o}pez},
  {Vu{\v c}kovi{\'c}}, {Ottosen}, {Liimets}, {Quint}, {Van Grootel}, {Randall},
  {Gilliland}, {Kjeldsen}, {Christensen-Dalsgaard}, {Borucki}, {Koch}, and
  {Quintana}}]{2010MNRAS.409.1470O}
{{\O}stensen} RH, {Silvotti} R, {Charpinet} S, {Oreiro} R, {Handler} G, {Green}
  EM, {Bloemen} S, {Heber} U, {G{\"a}nsicke} BT, {Marsh} TR, {Kurtz} DW,
  {Telting} JH, {Reed} MD, {Kawaler} SD, {Aerts} C,
  {Rodr{\'{\i}}guez-L{\'o}pez} C, {Vu{\v c}kovi{\'c}} M, {Ottosen} TA,
  {Liimets} T, {Quint} AC, {Van Grootel} V, {Randall} SK, {Gilliland} RL,
  {Kjeldsen} H, {Christensen-Dalsgaard} J, {Borucki} WJ, {Koch} D, {Quintana}
  EV (2010) {First Kepler results on compact pulsators - I. Survey target
  selection and the first pulsators}. \mnras 409:1470--1486,
  \doi{10.1111/j.1365-2966.2010.17366.x}, \eprint{1007.3170}

\bibitem[{{{\O}stensen} et~al.(2011{\natexlab{a}}){{\O}stensen}, {Bloemen},
  {Vu{\v c}kovi{\'c}}, {Aerts}, {Oreiro}, {Kinemuchi}, {Still}, and
  {Koester}}]{2011ApJ...736L..39O}
{{\O}stensen} RH, {Bloemen} S, {Vu{\v c}kovi{\'c}} M, {Aerts} C, {Oreiro} R,
  {Kinemuchi} K, {Still} M, {Koester} D (2011{\natexlab{a}}) {At Last A V777
  Her Pulsator in the Kepler Field}. \apjl 736:L39,
  \doi{10.1088/2041-8205/736/2/L39}

\bibitem[{{{\O}stensen} et~al.(2011{\natexlab{b}}){{\O}stensen}, {Silvotti},
  {Charpinet}, {Oreiro}, {Bloemen}, {Baran}, {Reed}, {Kawaler}, {Telting},
  {Green}, {O'Toole}, {Aerts}, {G{\"a}nsicke}, {Marsh}, {Breedt}, {Heber},
  {Koester}, {Quint}, {Kurtz}, {Rodr{\'{\i}}guez-L{\'o}pez}, {Vu{\v
  c}kovi{\'c}}, {Ottosen}, {Frimann}, {Somero}, {Wilson}, {Thygesen},
  {Lindberg}, {Kjeldsen}, {Christensen-Dalsgaard}, {Allen}, {McCauliff}, and
  {Middour}}]{2011MNRAS.414.2860O}
{{\O}stensen} RH, {Silvotti} R, {Charpinet} S, {Oreiro} R, {Bloemen} S, {Baran}
  AS, {Reed} MD, {Kawaler} SD, {Telting} JH, {Green} EM, {O'Toole} SJ, {Aerts}
  C, {G{\"a}nsicke} BT, {Marsh} TR, {Breedt} E, {Heber} U, {Koester} D, {Quint}
  AC, {Kurtz} DW, {Rodr{\'{\i}}guez-L{\'o}pez} C, {Vu{\v c}kovi{\'c}} M,
  {Ottosen} TA, {Frimann} S, {Somero} A, {Wilson} PA, {Thygesen} AO, {Lindberg}
  JE, {Kjeldsen} H, {Christensen-Dalsgaard} J, {Allen} C, {McCauliff} S,
  {Middour} CK (2011{\natexlab{b}}) {First Kepler results on compact pulsators
  - VI. Targets in the final half of the survey phase}. \mnras 414:2860--2870,
  \doi{10.1111/j.1365-2966.2011.18405.x}, \eprint{1101.4150}

\bibitem[{{Paczy{\'n}ski}(1971)}]{1971AcA....21..417P}
{Paczy{\'n}ski} B (1971) {Evolution of Single Stars. VI. Model Nuclei of
  Planetary Nebulae}. \actaa 21:417

\bibitem[{{Pauli} et~al.(1955){Pauli}, {Rosenfeld}, and
  {Weisskopf}}]{1955nbdp.book.....P}
{Pauli} W, {Rosenfeld} L, {Weisskopf} V (1955) {Niels Bohr and the Development
  of Physics}

\bibitem[{{Paxton} et~al.(2011){Paxton}, {Bildsten}, {Dotter}, {Herwig},
  {Lesaffre}, and {Timmes}}]{2011ApJS..192....3P}
{Paxton} B, {Bildsten} L, {Dotter} A, {Herwig} F, {Lesaffre} P, {Timmes} F
  (2011) {Modules for Experiments in Stellar Astrophysics (MESA)}. \apjs 192:3,
  \doi{10.1088/0067-0049/192/1/3}, \eprint{1009.1622}

\bibitem[{{Paxton} et~al.(2013){Paxton}, {Cantiello}, {Arras}, {Bildsten},
  {Brown}, {Dotter}, {Mankovich}, {Montgomery}, {Stello}, {Timmes}, and
  {Townsend}}]{2013ApJS..208....4P}
{Paxton} B, {Cantiello} M, {Arras} P, {Bildsten} L, {Brown} EF, {Dotter} A,
  {Mankovich} C, {Montgomery} MH, {Stello} D, {Timmes} FX, {Townsend} R (2013)
  {Modules for Experiments in Stellar Astrophysics (MESA): Planets,
  Oscillations, Rotation, and Massive Stars}. \apjs 208:4,
  \doi{10.1088/0067-0049/208/1/4}, \eprint{1301.0319}

\bibitem[{{Paxton} et~al.(2015){Paxton}, {Marchant}, {Schwab}, {Bauer},
  {Bildsten}, {Cantiello}, {Dessart}, {Farmer}, {Hu}, {Langer}, {Townsend},
  {Townsley}, and {Timmes}}]{2015ApJS..220...15P}
{Paxton} B, {Marchant} P, {Schwab} J, {Bauer} EB, {Bildsten} L, {Cantiello} M,
  {Dessart} L, {Farmer} R, {Hu} H, {Langer} N, {Townsend} RHD, {Townsley} DM,
  {Timmes} FX (2015) {Modules for Experiments in Stellar Astrophysics (MESA):
  Binaries, Pulsations, and Explosions}. \apjs 220:15,
  \doi{10.1088/0067-0049/220/1/15}, \eprint{1506.03146}

\bibitem[{{Paxton} et~al.(2018){Paxton}, {Schwab}, {Bauer}, {Bildsten},
  {Blinnikov}, {Duffell}, {Farmer}, {Goldberg}, {Marchant}, {Sorokina},
  {Thoul}, {Townsend}, and {Timmes}}]{2018ApJS..234...34P}
{Paxton} B, {Schwab} J, {Bauer} EB, {Bildsten} L, {Blinnikov} S, {Duffell} P,
  {Farmer} R, {Goldberg} JA, {Marchant} P, {Sorokina} E, {Thoul} A, {Townsend}
  RHD, {Timmes} FX (2018) {Modules for Experiments in Stellar Astrophysics
  (MESA): Convective Boundaries, Element Diffusion, and Massive Star
  Explosions}. \apjs 234:34, \doi{10.3847/1538-4365/aaa5a8},
  \eprint{1710.08424}

\bibitem[{{Paxton} et~al.(2019){Paxton}, {Smolec}, {Gautschy}, {Bildsten},
  {Cantiello}, {Dotter}, {Farmer}, {Goldberg}, {Jermyn}, {Kanbur}, {Marchant},
  {Schwab}, {Thoul}, {Townsend}, {Wolf}, {Zhang}, and
  {Timmes}}]{2019arXiv190301426P}
{Paxton} B, {Smolec} R, {Gautschy} A, {Bildsten} L, {Cantiello} M, {Dotter} A,
  {Farmer} R, {Goldberg} JA, {Jermyn} AS, {Kanbur} SM, {Marchant} P, {Schwab}
  J, {Thoul} A, {Townsend} RHD, {Wolf} WM, {Zhang} M, {Timmes} FX (2019)
  {Modules for Experiments in Stellar Astrophysics (MESA): Pulsating Variable
  Stars, Rotation, Convective Boundaries, and Energy Conservation}. arXiv
  e-prints \eprint{1903.01426}

\bibitem[{{Peccei} and {Quinn}(1977)}]{1977PhRvL..38.1440P}
{Peccei} RD, {Quinn} HR (1977) {CP conservation in the presence of
  pseudoparticles}. Physical Review Letters 38:1440--1443,
  \doi{10.1103/PhysRevLett.38.1440}

\bibitem[{{Pelisoli} et~al.(2018{\natexlab{a}}){Pelisoli}, {Kepler}, and
  {Koester}}]{2018MNRAS.475.2480P}
{Pelisoli} I, {Kepler} SO, {Koester} D (2018{\natexlab{a}}) {The sdA problem -
  I. Physical properties}. \mnras 475:2480--2495, \doi{10.1093/mnras/sty011},
  \eprint{1801.00495}

\bibitem[{{Pelisoli} et~al.(2018{\natexlab{b}}){Pelisoli}, {Kepler}, {Koester},
  {Castanheira}, {Romero}, and {Fraga}}]{2018MNRAS.478..867P}
{Pelisoli} I, {Kepler} SO, {Koester} D, {Castanheira} BG, {Romero} AD, {Fraga}
  L (2018{\natexlab{b}}) {The sdA problem - II. Photometric and spectroscopic
  follow-up}. \mnras 478:867--884, \doi{10.1093/mnras/sty1101},
  \eprint{1804.09059}

\bibitem[{{Pelisoli} et~al.(2019){Pelisoli}, {Bell}, {Kepler}, and
  {Koester}}]{2019MNRAS.482.3831P}
{Pelisoli} I, {Bell} KJ, {Kepler} SO, {Koester} D (2019) {The sdA problem -
  III. New extremely low-mass white dwarfs and their precursors from Gaia
  astrometry}. \mnras 482:3831--3842, \doi{10.1093/mnras/sty2979},
  \eprint{1805.04070}

\bibitem[{{Pietrinferni} et~al.(2004){Pietrinferni}, {Cassisi}, {Salaris}, and
  {Castelli}}]{2004ApJ...612..168P}
{Pietrinferni} A, {Cassisi} S, {Salaris} M, {Castelli} F (2004) {A Large
  Stellar Evolution Database for Population Synthesis Studies. I. Scaled Solar
  Models and Isochrones}. \apj 612:168--190, \doi{10.1086/422498},
  \eprint{astro-ph/0405193}

\bibitem[{{Pietrukowicz}(2018)}]{2018pas6.conf..258P}
{Pietrukowicz} P (2018) {On the Properties of Blue Large-Amplitude Pulsators.
  No BLAPs in the Magellanic Clouds}. In: {Smolec} R, {Kinemuchi} K, {Anderson}
  RI (eds) The RR Lyrae 2017 Conference. Revival of the Classical Pulsators:
  from Galactic Structure to Stellar Interior Diagnostics, vol~6, pp 258--262,
  \eprint{1802.04405}

\bibitem[{{Pietrukowicz} et~al.(2017){Pietrukowicz}, {Dziembowski}, {Latour},
  {Angeloni}, {Poleski}, {di Mille}, {Soszy{\'n}ski}, {Udalski},
  {Szyma{\'n}ski}, {Wyrzykowski}, {Koz{\l}owski}, {Skowron}, {Skowron},
  {Mr{\'o}z}, {Pawlak}, and {Ulaczyk}}]{2017NatAs...1E.166P}
{Pietrukowicz} P, {Dziembowski} WA, {Latour} M, {Angeloni} R, {Poleski} R, {di
  Mille} F, {Soszy{\'n}ski} I, {Udalski} A, {Szyma{\'n}ski} MK, {Wyrzykowski}
  {\L}, {Koz{\l}owski} S, {Skowron} J, {Skowron} D, {Mr{\'o}z} P, {Pawlak} M,
  {Ulaczyk} K (2017) {Blue large-amplitude pulsators as a new class of variable
  stars}. Nature Astronomy 1:0166, \doi{10.1038/s41550-017-0166},
  \eprint{1706.07802}

\bibitem[{{Pringle}(1975)}]{1975MNRAS.170..633P}
{Pringle} JE (1975) {Period changes in eruptive binaries}. \mnras 170:633--642,
  \doi{10.1093/mnras/170.3.633}

\bibitem[{{Provencal} et~al.(2009){Provencal}, {Montgomery}, {Kanaan},
  {Shipman}, {Childers}, {Baran}, {Kepler}, {Reed}, {Zhou}, {Eggen}, {Watson},
  {Winget}, {Thompson}, {Riaz}, {Nitta}, {Kleinman}, {Crowe}, {Slivkoff},
  {Sherard}, {Purves}, {Binder}, {Knight}, {Kim}, {Chen}, {Yang}, {Lin}, {Lin},
  {Chen}, {Jiang}, {Sergeev}, {Mkrtichian}, {Andreev}, {Janulis}, {Siwak},
  {Zola}, {Koziel}, {Stachowski}, {Paparo}, {Bognar}, {Handler}, {Lorenz},
  {Steininger}, {Beck}, {Nagel}, {Kusterer}, {Hoffman}, {Reiff}, {Kowalski},
  {Vauclair}, {Charpinet}, {Chevreton}, {Solheim}, {Pakstiene}, {Fraga}, and
  {Dalessio}}]{2009ApJ...693..564P}
{Provencal} JL, {Montgomery} MH, {Kanaan} A, {Shipman} HL, {Childers} D,
  {Baran} A, {Kepler} SO, {Reed} M, {Zhou} A, {Eggen} J, {Watson} TK, {Winget}
  DE, {Thompson} SE, {Riaz} B, {Nitta} A, {Kleinman} SJ, {Crowe} R, {Slivkoff}
  J, {Sherard} P, {Purves} N, {Binder} P, {Knight} R, {Kim} S, {Chen} WP,
  {Yang} M, {Lin} HC, {Lin} CC, {Chen} CW, {Jiang} XJ, {Sergeev} AV,
  {Mkrtichian} D, {Andreev} M, {Janulis} R, {Siwak} M, {Zola} S, {Koziel} D,
  {Stachowski} G, {Paparo} M, {Bognar} Z, {Handler} G, {Lorenz} D, {Steininger}
  B, {Beck} P, {Nagel} T, {Kusterer} D, {Hoffman} A, {Reiff} E, {Kowalski} R,
  {Vauclair} G, {Charpinet} S, {Chevreton} M, {Solheim} JE, {Pakstiene} E,
  {Fraga} L, {Dalessio} J (2009) {2006 Whole Earth Telescope Observations of
  GD358: A New Look at the Prototype DBV}. \apj 693:564--585,
  \doi{10.1088/0004-637X/693/1/564}, \eprint{0811.0768}

\bibitem[{{Pyrzas} et~al.(2015){Pyrzas}, {G{\"a}nsicke}, {Hermes},
  {Copperwheat}, {Rebassa-Mansergas}, {Dhillon}, {Littlefair}, {Marsh},
  {Parsons}, {Savoury}, {Schreiber}, {Barros}, {Bento}, {Breedt}, and
  {Kerry}}]{2015MNRAS.447..691P}
{Pyrzas} S, {G{\"a}nsicke} BT, {Hermes} JJ, {Copperwheat} CM,
  {Rebassa-Mansergas} A, {Dhillon} VS, {Littlefair} SP, {Marsh} TR, {Parsons}
  SG, {Savoury} CDJ, {Schreiber} MR, {Barros} SCC, {Bento} J, {Breedt} E,
  {Kerry} P (2015) {Discovery of ZZ Cetis in detached white dwarf plus
  main-sequence binaries}. \mnras 447:691--697, \doi{10.1093/mnras/stu2412},
  \eprint{1411.5045}

\bibitem[{{Quirion} et~al.(2008){Quirion}, {Dupret}, {Fontaine}, {Brassard},
  and {Grigahc{\`e}ne}}]{2008ASPC..391..183Q}
{Quirion} PO, {Dupret} MA, {Fontaine} G, {Brassard} P, {Grigahc{\`e}ne} A
  (2008) {Hydrogen-Deficient Compact Pulsators: The GW Virginis Stars and the
  Variable DB White Dwarfs}. In: {Werner} A, {Rauch} T (eds) Hydrogen-Deficient
  Stars, Astronomical Society of the Pacific Conference Series, vol 391, p 183

\bibitem[{{Quirion} et~al.(2012){Quirion}, {Fontaine}, and
  {Brassard}}]{2012ApJ...755..128Q}
{Quirion} PO, {Fontaine} G, {Brassard} P (2012) {Wind Competing Against
  Settling: A Coherent Model of the GW Virginis Instability Domain}. \apj
  755:128, \doi{10.1088/0004-637X/755/2/128}

\bibitem[{{Raffelt}(2012)}]{2012arXiv1201.1637R}
{Raffelt} G (2012) {Neutrinos and the stars}. ArXiv e-prints \eprint{1201.1637}

\bibitem[{{Raffelt}(1986)}]{1986PhLB..166..402R}
{Raffelt} GG (1986) {Axion constraints from white dwarf cooling times}. Physics
  Letters B 166:402--406, \doi{10.1016/0370-2693(86)91588-1}

\bibitem[{{Raffelt}(1990)}]{1990PhR...198....1R}
{Raffelt} GG (1990) {Astrophysical methods to constrain axions and other novel
  particle phenomena}. \physrep 198:1--113, \doi{10.1016/0370-1573(90)90054-6}

\bibitem[{{Raffelt}(1996)}]{1996slfp.book.....R}
{Raffelt} GG (1996) {Stars as laboratories for fundamental physics : the
  astrophysics of neutrinos, axions, and other weakly interacting particles}

\bibitem[{{Raffelt}(2007)}]{2007JPhA...40.6607R}
{Raffelt} GG (2007) {Axions---motivation, limits and searches}. Journal of
  Physics A Mathematical General 40:6607--6620,
  \doi{10.1088/1751-8113/40/25/S05}, \eprint{hep-ph/0611118}

\bibitem[{{Ramsay}(2018)}]{2018A&A...620L...9R}
{Ramsay} G (2018) {Identifying blue large-amplitude pulsators in the Galactic
  plane using Gaia DR2: a case study}. \aap 620:L9,
  \doi{10.1051/0004-6361/201834604}, \eprint{1811.09522}

\bibitem[{{Redaelli} et~al.(2011){Redaelli}, {Kepler}, {Costa}, {Winget},
  {Handler}, {Castanheira}, {Kanaan}, {Fraga}, {Henrique}, {Giovannini},
  {Provencal}, {Shipman}, {Dalessio}, {Thompson}, {Mullally}, {Brewer},
  {Childers}, {Oksala}, {Rosen}, {Wood}, {Reed}, {Walter}, {Strickland},
  {Chandler}, {Watson}, {Nather}, {Montgomery}, {Bischoff-Kim}, {Hansen},
  {Nitta}, {Kleinman}, {Claver}, {Brown}, {Sullivan}, {Kim}, {Chen}, {Yang},
  {Shih}, {Zhang}, {Jiang}, {Fu}, {Seetha}, {Ashoka}, {Marar}, {Baliyan},
  {Vats}, {Chernyshev}, {Ibbetson}, {Leibowitz}, {Hemar}, {Sergeev}, {Andreev},
  {Janulis}, {Mei{\v s}tas}, {Moskalik}, {Pajdosz}, {Baran}, {Winiarski},
  {Zola}, {Ogloza}, {Siwak}, {Bogn{\'a}r}, {Solheim}, {Sefako}, {Buckley},
  {O'Donoghue}, {Nagel}, {Silvotti}, {Bruni}, {Fremy}, {Vauclair}, {Chevreton},
  {Dolez}, {Pfeiffer}, {Barstow}, {Creevey}, {Kawaler}, and
  {Clemens}}]{2011MNRAS.415.1220R}
{Redaelli} M, {Kepler} SO, {Costa} JES, {Winget} DE, {Handler} G, {Castanheira}
  BG, {Kanaan} A, {Fraga} L, {Henrique} P, {Giovannini} O, {Provencal} JL,
  {Shipman} HL, {Dalessio} J, {Thompson} SE, {Mullally} F, {Brewer} MM,
  {Childers} D, {Oksala} ME, {Rosen} R, {Wood} MA, {Reed} MD, {Walter} B,
  {Strickland} W, {Chandler} D, {Watson} TK, {Nather} RE, {Montgomery} MH,
  {Bischoff-Kim} A, {Hansen} CJ, {Nitta} A, {Kleinman} SJ, {Claver} CF, {Brown}
  TM, {Sullivan} DJ, {Kim} SL, {Chen} WP, {Yang} M, {Shih} CY, {Zhang} X,
  {Jiang} X, {Fu} JN, {Seetha} S, {Ashoka} BN, {Marar} TMK, {Baliyan} KS,
  {Vats} HO, {Chernyshev} AV, {Ibbetson} P, {Leibowitz} E, {Hemar} S, {Sergeev}
  AV, {Andreev} MV, {Janulis} R, {Mei{\v s}tas} EG, {Moskalik} P, {Pajdosz} G,
  {Baran} A, {Winiarski} M, {Zola} S, {Ogloza} W, {Siwak} M, {Bogn{\'a}r} Z,
  {Solheim} JE, {Sefako} R, {Buckley} D, {O'Donoghue} D, {Nagel} T, {Silvotti}
  R, {Bruni} I, {Fremy} JR, {Vauclair} G, {Chevreton} M, {Dolez} N, {Pfeiffer}
  B, {Barstow} MA, {Creevey} OL, {Kawaler} SD, {Clemens} JC (2011) {The
  pulsations of PG 1351+489}. \mnras 415:1220--1227,
  \doi{10.1111/j.1365-2966.2011.18743.x}

\bibitem[{{Redondo}(2016)}]{2016arXiv160100578R}
{Redondo} J (2016) {Axions at the International Axion Observatory}. ArXiv
  e-prints \eprint{1601.00578}

\bibitem[{{Ricker} et~al.(2015){Ricker}, {Winn}, {Vanderspek}, {Latham},
  {Bakos}, {Bean}, {Berta-Thompson}, {Brown}, {Buchhave}, {Butler}, {Butler},
  {Chaplin}, {Charbonneau}, {Christensen-Dalsgaard}, {Clampin}, {Deming},
  {Doty}, {De Lee}, {Dressing}, {Dunham}, {Endl}, {Fressin}, {Ge}, {Henning},
  {Holman}, {Howard}, {Ida}, {Jenkins}, {Jernigan}, {Johnson}, {Kaltenegger},
  {Kawai}, {Kjeldsen}, {Laughlin}, {Levine}, {Lin}, {Lissauer}, {MacQueen},
  {Marcy}, {McCullough}, {Morton}, {Narita}, {Paegert}, {Palle}, {Pepe},
  {Pepper}, {Quirrenbach}, {Rinehart}, {Sasselov}, {Sato}, {Seager},
  {Sozzetti}, {Stassun}, {Sullivan}, {Szentgyorgyi}, {Torres}, {Udry}, and
  {Villasenor}}]{2015JATIS...1a4003R}
{Ricker} GR, {Winn} JN, {Vanderspek} R, {Latham} DW, {Bakos} G{\'A}, {Bean} JL,
  {Berta-Thompson} ZK, {Brown} TM, {Buchhave} L, {Butler} NR, {Butler} RP,
  {Chaplin} WJ, {Charbonneau} D, {Christensen-Dalsgaard} J, {Clampin} M,
  {Deming} D, {Doty} J, {De Lee} N, {Dressing} C, {Dunham} EW, {Endl} M,
  {Fressin} F, {Ge} J, {Henning} T, {Holman} MJ, {Howard} AW, {Ida} S,
  {Jenkins} JM, {Jernigan} G, {Johnson} JA, {Kaltenegger} L, {Kawai} N,
  {Kjeldsen} H, {Laughlin} G, {Levine} AM, {Lin} D, {Lissauer} JJ, {MacQueen}
  P, {Marcy} G, {McCullough} PR, {Morton} TD, {Narita} N, {Paegert} M, {Palle}
  E, {Pepe} F, {Pepper} J, {Quirrenbach} A, {Rinehart} SA, {Sasselov} D, {Sato}
  B, {Seager} S, {Sozzetti} A, {Stassun} KG, {Sullivan} P, {Szentgyorgyi} A,
  {Torres} G, {Udry} S, {Villasenor} J (2015) {Transiting Exoplanet Survey
  Satellite (TESS)}. Journal of Astronomical Telescopes, Instruments, and
  Systems 1(1):014003, \doi{10.1117/1.JATIS.1.1.014003}

\bibitem[{{Ritossa} et~al.(1996){Ritossa}, {Garcia-Berro}, and
  {Iben}}]{1996ApJ...460..489R}
{Ritossa} C, {Garcia-Berro} E, {Iben} I Jr (1996) {On the Evolution of Stars
  That Form Electron-degenerate Cores Processed by Carbon Burning. II. Isotope
  Abundances and Thermal Pulses in a 10 M$_{sun}$ Model with an ONe Core and
  Applications to Long-Period Variables, Classical Novae, and Accretion-induced
  Collapse}. \apj 460:489, \doi{10.1086/176987}

\bibitem[{{Romero} et~al.(2012){Romero}, {C{\'o}rsico}, {Althaus}, {Kepler},
  {Castanheira}, and {Miller Bertolami}}]{2012MNRAS.420.1462R}
{Romero} AD, {C{\'o}rsico} AH, {Althaus} LG, {Kepler} SO, {Castanheira} BG,
  {Miller Bertolami} MM (2012) {Toward ensemble asteroseismology of ZZ Ceti
  stars with fully evolutionary models}. \mnras 420:1462--1480,
  \doi{10.1111/j.1365-2966.2011.20134.x}

\bibitem[{{Romero} et~al.(2013){Romero}, {Kepler}, {C{\'o}rsico}, {Althaus},
  and {Fraga}}]{2013ApJ...779...58R}
{Romero} AD, {Kepler} SO, {C{\'o}rsico} AH, {Althaus} LG, {Fraga} L (2013)
  {Asteroseismological Study of Massive ZZ Ceti Stars with Fully Evolutionary
  Models}. \apj 779:58, \doi{10.1088/0004-637X/779/1/58}, \eprint{1310.4137}

\bibitem[{{Romero} et~al.(2017){Romero}, {C{\'o}rsico}, {Castanheira}, {De
  Ger{\'o}nimo}, {Kepler}, {Koester}, {Kawka}, {Althaus}, {Hermes}, {Bonato},
  and {Gianninas}}]{2017ApJ...851...60R}
{Romero} AD, {C{\'o}rsico} AH, {Castanheira} BG, {De Ger{\'o}nimo} FC, {Kepler}
  SO, {Koester} D, {Kawka} A, {Althaus} LG, {Hermes} JJ, {Bonato} C,
  {Gianninas} A (2017) {Probing the Structure of Kepler ZZ Ceti Stars with Full
  Evolutionary Models-based Asteroseismology}. \apj 851:60,
  \doi{10.3847/1538-4357/aa9899}, \eprint{1711.01338}

\bibitem[{{Romero} et~al.(2018){Romero}, {C{\'o}rsico}, {Althaus}, {Pelisoli},
  and {Kepler}}]{2018MNRAS.477L..30R}
{Romero} AD, {C{\'o}rsico} AH, {Althaus} LG, {Pelisoli} I, {Kepler} SO (2018)
  {On the evolutionary status and pulsations of the recently discovered blue
  large-amplitude pulsators (BLAPs)}. \mnras 477:L30--L34,
  \doi{10.1093/mnrasl/sly051}, \eprint{1803.09600}

\bibitem[{{Rowan} et~al.(2019){Rowan}, {Tucker}, {Shappee}, and
  {Hermes}}]{2019MNRAS.486.4574R}
{Rowan} DM, {Tucker} MA, {Shappee} BJ, {Hermes} JJ (2019) {Detections and
  constraints on white dwarf variability from time-series GALEX observations}.
  \mnras 486:4574--4589, \doi{10.1093/mnras/stz1116}, \eprint{1812.05614}

\bibitem[{{Saio}(2019)}]{2019MNRAS.tmp.1346S}
{Saio} H (2019) {R-mode oscillations in accreting white dwarfs in cataclysmic
  variables}. \mnras \doi{10.1093/mnras/stz1407}, \eprint{1905.08390}

\bibitem[{{Salaris} and {Cassisi}(2018)}]{2018PhyS...93d4002S}
{Salaris} M, {Cassisi} S (2018) {White dwarf stars: cosmic chronometers and
  dark matter probes}. Physica Scripta 93(4):044002,
  \doi{10.1088/1402-4896/aaaef4}

\bibitem[{{Salaris} et~al.(2009){Salaris}, {Serenelli}, {Weiss}, and {Miller
  Bertolami}}]{2009ApJ...692.1013S}
{Salaris} M, {Serenelli} A, {Weiss} A, {Miller Bertolami} M (2009)
  {Semi-empirical White Dwarf Initial-Final Mass Relationships: A Thorough
  Analysis of Systematic Uncertainties Due to Stellar Evolution Models}. \apj
  692:1013--1032, \doi{10.1088/0004-637X/692/2/1013}, \eprint{0807.3567}

\bibitem[{{Schoenberner}(1987)}]{1987ASSL..132..337S}
{Schoenberner} D (1987) {Mass loss and the transformation of AGB stars into
  central stars of planetary nebulae}. In: {Kwok} S, {Pottasch} SR (eds) Late
  Stages of Stellar Evolution, Astrophysics and Space Science Library, vol 132

\bibitem[{{Schwarzschild}(1958)}]{1958ses..book.....S}
{Schwarzschild} M (1958) {Structure and evolution of the stars.}

\bibitem[{{Scuflaire}(1974)}]{1974A&A....36..107S}
{Scuflaire} R (1974) {The Non Radial Oscillations of Condensed Polytropes}.
  \aap 36:107

\bibitem[{{Segretain} and {Chabrier}(1993)}]{1993A&A...271L..13S}
{Segretain} L, {Chabrier} G (1993) {Crystallization of binary ionic mixtures in
  dense stellar plasmas}. \aap 271:L13

\bibitem[{{Serenelli} et~al.(2002){Serenelli}, {Althaus}, {Rohrmann}, and
  {Benvenuto}}]{2002MNRAS.337.1091S}
{Serenelli} AM, {Althaus} LG, {Rohrmann} RD, {Benvenuto} OG (2002) {Evolution
  and colours of helium-core white dwarf stars: the case of low-metallicity
  progenitors}. \mnras 337:1091--1104, \doi{10.1046/j.1365-8711.2002.05994.x},
  \eprint{astro-ph/0208408}

\bibitem[{{Shibahashi}(2005)}]{2005EAS....17..143S}
{Shibahashi} H (2005) {The DB gap and pulsations of white dwarfs}. In:
  {Alecian} G, {Richard} O, {Vauclair} S (eds) EAS Publications Series, EAS
  Publications Series, vol~17, pp 143--148, \doi{10.1051/eas:2005108}

\bibitem[{{Shibahashi}(2007)}]{2007AIPC..948...35S}
{Shibahashi} H (2007) {The DB Gap and Pulsations of White Dwarfs}. In:
  {Stancliffe} RJ, {Houdek} G, {Martin} RG, {Tout} CA (eds) Unsolved Problems
  in Stellar Physics: A Conference in Honor of Douglas Gough, American
  Institute of Physics Conference Series, vol 948, pp 35--42,
  \doi{10.1063/1.2818994}

\bibitem[{{Shibahashi}(2013)}]{2013EAS....63..185S}
{Shibahashi} H (2013) {A new kind of pulsator in the DB valley of white dwarf
  stars}. In: {Alecian} G, {Lebreton} Y, {Richard} O, {Vauclair} G (eds) EAS
  Publications Series, EAS Publications Series, vol~63, pp 185--190,
  \doi{10.1051/eas/1363021}

\bibitem[{{Siess}(2010)}]{2010A&A...512A..10S}
{Siess} L (2010) {Evolution of massive AGB stars. III. the thermally pulsing
  super-AGB phase}. \aap 512:A10, \doi{10.1051/0004-6361/200913556}

\bibitem[{{Silvotti} et~al.(2012){Silvotti}, {{\O}stensen}, {Bloemen},
  {Telting}, {Heber}, {Oreiro}, {Reed}, {Farris}, {O'Toole}, {Lanteri},
  {Degroote}, {Hu}, {Baran}, {Hermes}, {Althaus}, {Marsh}, {Charpinet}, {Li},
  {Morris}, and {Sanderfer}}]{2012MNRAS.424.1752S}
{Silvotti} R, {{\O}stensen} RH, {Bloemen} S, {Telting} JH, {Heber} U, {Oreiro}
  R, {Reed} MD, {Farris} LE, {O'Toole} SJ, {Lanteri} L, {Degroote} P, {Hu} H,
  {Baran} AS, {Hermes} JJ, {Althaus} LG, {Marsh} TR, {Charpinet} S, {Li} J,
  {Morris} RL, {Sanderfer} DT (2012) {Orbital properties of an unusually
  low-mass sdB star in a close binary system with a white dwarf}. \mnras
  424:1752--1761, \doi{10.1111/j.1365-2966.2012.21232.x}, \eprint{1205.2457}

\bibitem[{{Sion}(2011)}]{2011wdac.book....1S}
{Sion} EM (2011) {Hot White Dwarfs}, pp 1--24

\bibitem[{{Smolec} et~al.(2013){Smolec}, {Pietrzy{\'n}ski}, {Graczyk},
  {Pilecki}, {Gieren}, {Thompson}, {St{\c e}pie{\'n}}, {Karczmarek},
  {Konorski}, {G{\'o}rski}, {Suchomska}, {Bono}, {Prada}, and
  {Nardetto}}]{2013MNRAS.428.3034S}
{Smolec} R, {Pietrzy{\'n}ski} G, {Graczyk} D, {Pilecki} B, {Gieren} W,
  {Thompson} I, {St{\c e}pie{\'n}} K, {Karczmarek} P, {Konorski} P,
  {G{\'o}rski} M, {Suchomska} K, {Bono} G, {Prada} PGM, {Nardetto} N (2013)
  {Pulsation models for the 0.26 M$_{\odot}$ star mimicking RR Lyrae pulsator.
  Model survey for the new class of variable stars}. \mnras 428:3034--3047,
  \doi{10.1093/mnras/sts258}, \eprint{1210.6030}

\bibitem[{{Sowicka} et~al.(2018){Sowicka}, {Handler}, and
  {Jones}}]{2018MNRAS.479.2476S}
{Sowicka} P, {Handler} G, {Jones} D (2018) {On {$\epsilon$}-mechanism-driven
  pulsations in VV 47}. \mnras 479:2476--2480, \doi{10.1093/mnras/sty1660},
  \eprint{1806.07935}

\bibitem[{{Steinfadt} et~al.(2010){Steinfadt}, {Bildsten}, and
  {Arras}}]{2010ApJ...718..441S}
{Steinfadt} JDR, {Bildsten} L, {Arras} P (2010) {Pulsations in Hydrogen Burning
  Low-mass Helium White Dwarfs}. \apj 718:441--445,
  \doi{10.1088/0004-637X/718/1/441}, \eprint{1005.5423}

\bibitem[{{Steinfadt} et~al.(2012){Steinfadt}, {Bildsten}, {Kaplan}, {Fulton},
  {Howell}, {Marsh}, {Ofek}, and {Shporer}}]{2012PASP..124....1S}
{Steinfadt} JDR, {Bildsten} L, {Kaplan} DL, {Fulton} BJ, {Howell} SB, {Marsh}
  TR, {Ofek} EO, {Shporer} A (2012) {A Search for Pulsations in Helium White
  Dwarfs}. PASP 124:1--13, \doi{10.1086/663865}, \eprint{1105.0472}

\bibitem[{{Strickler} et~al.(2009){Strickler}, {Cool}, {Anderson}, {Cohn},
  {Lugger}, and {Serenelli}}]{2009ApJ...699...40S}
{Strickler} RR, {Cool} AM, {Anderson} J, {Cohn} HN, {Lugger} PM, {Serenelli} AM
  (2009) {Helium-core White Dwarfs in the Globular Cluster NGC 6397}. \apj
  699:40--55, \doi{10.1088/0004-637X/699/1/40}, \eprint{0904.3496}

\bibitem[{{Sullivan}(2017)}]{2017ASPC..509..315S}
{Sullivan} DJ (2017) {Time-Series Spectroscopy and Photometry of the Helium
  Atmosphere Pulsating White Dwarf EC 20058-5234}. In: {Tremblay} PE,
  {Gaensicke} B, {Marsh} T (eds) 20th European White Dwarf Workshop,
  Astronomical Society of the Pacific Conference Series, vol 509, p 315

\bibitem[{{Sullivan} and {Chote}(2015)}]{2015ASPC..493..199S}
{Sullivan} DJ, {Chote} P (2015) {The Frequency Stability of the Pulsating White
  Dwarf L19-2}. In: {Dufour} P, {Bergeron} P, {Fontaine} G (eds) 19th European
  Workshop on White Dwarfs, Astronomical Society of the Pacific Conference
  Series, vol 493, p 199

\bibitem[{{Szkody} et~al.(2010){Szkody}, {Mukadam}, {G{\"a}nsicke}, {Henden},
  {Templeton}, {Holtzman}, {Montgomery}, {Howell}, {Nitta}, {Sion}, {Schwartz},
  and {Dillon}}]{2010ApJ...710...64S}
{Szkody} P, {Mukadam} A, {G{\"a}nsicke} BT, {Henden} A, {Templeton} M,
  {Holtzman} J, {Montgomery} MH, {Howell} SB, {Nitta} A, {Sion} EM, {Schwartz}
  RD, {Dillon} W (2010) {Finding the Instability Strip for Accreting Pulsating
  White Dwarfs From Hubble Space Telescope and Optical Observations}. \apj
  710:64--77, \doi{10.1088/0004-637X/710/1/64}, \eprint{1001.0192}

\bibitem[{{Szkody} et~al.(2013){Szkody}, {Mukadam}, {G{\"a}nsicke}, {Sion},
  {Townsley}, and {Henden}}]{2013ASPC..469...31S}
{Szkody} P, {Mukadam} AS, {G{\"a}nsicke} BT, {Sion} EM, {Townsley} D, {Henden}
  A (2013) {Enigmas of Accreting Pulsating White Dwarfs}. In: 18th European
  White Dwarf Workshop., Astronomical Society of the Pacific Conference Series,
  vol 469, p~31

\bibitem[{{Szkody} et~al.(2015){Szkody}, {Mukadam}, {G{\"a}nsicke}, {Hermes},
  and {Toloza}}]{2015ASPC..493..205S}
{Szkody} P, {Mukadam} AS, {G{\"a}nsicke} BT, {Hermes} JJ, {Toloza} O (2015) {An
  Update on the Quirks of Pulsating, Accreting White Dwarfs}. In: {Dufour} P,
  {Bergeron} P, {Fontaine} G (eds) 19th European Workshop on White Dwarfs,
  Astronomical Society of the Pacific Conference Series, vol 493, p 205

\bibitem[{{Tassoul} et~al.(1990){Tassoul}, {Fontaine}, and
  {Winget}}]{1990ApJS...72..335T}
{Tassoul} M, {Fontaine} G, {Winget} DE (1990) {Evolutionary models for
  pulsation studies of white dwarfs}. ApJs 72:335--386, \doi{10.1086/191420}

\bibitem[{{Tayar} and {Pinsonneault}(2013)}]{2013ApJ...775L...1T}
{Tayar} J, {Pinsonneault} MH (2013) {Implications of Rapid Core Rotation in Red
  Giants for Internal Angular Momentum Transport in Stars}. \apjl 775:L1,
  \doi{10.1088/2041-8205/775/1/L1}, \eprint{1306.3986}

\bibitem[{{Thompson} et~al.(2009){Thompson}, {Proven{\c c}al}, {Kanaan},
  {Montgomery}, {Bishoff-Kim}, {Shipman}, and {Wet Team}}]{2009JPhCS.172a2067T}
{Thompson} SE, {Proven{\c c}al} JL, {Kanaan} A, {Montgomery} MH, {Bishoff-Kim}
  A, {Shipman} H, {Wet Team} (2009) {Whole Earth Telescope observations of the
  DAVs R808 and G38-29}. In: Journal of Physics Conference Series, Journal of
  Physics Conference Series, vol 172, p 012067,
  \doi{10.1088/1742-6596/172/1/012067}

\bibitem[{{Toloza} et~al.(2016){Toloza}, {G{\"a}nsicke}, {Hermes}, {Townsley},
  {Schreiber}, {Szkody}, {Pala}, {Beuermann}, {Bildsten}, {Breedt}, {Cook},
  {Godon}, {Henden}, {Hubeny}, {Knigge}, {Long}, {Marsh}, {de Martino},
  {Mukadam}, {Myers}, {Nelson}, {Oksanen}, {Patterson}, {Sion}, and
  {Zorotovic}}]{2016MNRAS.459.3929T}
{Toloza} O, {G{\"a}nsicke} BT, {Hermes} JJ, {Townsley} DM, {Schreiber} MR,
  {Szkody} P, {Pala} A, {Beuermann} K, {Bildsten} L, {Breedt} E, {Cook} M,
  {Godon} P, {Henden} AA, {Hubeny} I, {Knigge} C, {Long} KS, {Marsh} TR, {de
  Martino} D, {Mukadam} AS, {Myers} G, {Nelson} P, {Oksanen} A, {Patterson} J,
  {Sion} EM, {Zorotovic} M (2016) {GW Librae: a unique laboratory for
  pulsations in an accreting white dwarf}. \mnras 459:3929--3938,
  \doi{10.1093/mnras/stw838}, \eprint{1604.02162}

\bibitem[{{Tremblay} and {Bergeron}(2008)}]{2008ApJ...672.1144T}
{Tremblay} PE, {Bergeron} P (2008) {The Ratio of Helium- to Hydrogen-Atmosphere
  White Dwarfs: Direct Evidence for Convective Mixing}. \apj 672:1144--1152,
  \doi{10.1086/524134}, \eprint{0710.1073}

\bibitem[{{Tremblay} et~al.(2013{\natexlab{a}}){Tremblay}, {Ludwig}, {Steffen},
  and {Freytag}}]{2013A&A...552A..13T}
{Tremblay} PE, {Ludwig} HG, {Steffen} M, {Freytag} B (2013{\natexlab{a}})
  {Pure-hydrogen 3D model atmospheres of cool white dwarfs}. \aap 552:A13,
  \doi{10.1051/0004-6361/201220813}, \eprint{1302.2013}

\bibitem[{{Tremblay} et~al.(2013{\natexlab{b}}){Tremblay}, {Ludwig}, {Steffen},
  and {Freytag}}]{2013A&A...559A.104T}
{Tremblay} PE, {Ludwig} HG, {Steffen} M, {Freytag} B (2013{\natexlab{b}})
  {Spectroscopic analysis of DA white dwarfs with 3D model atmospheres}. \aap
  559:A104, \doi{10.1051/0004-6361/201322318}, \eprint{1309.0886}

\bibitem[{{Tremblay} et~al.(2015){Tremblay}, {Gianninas}, {Kilic}, {Ludwig},
  {Steffen}, {Freytag}, and {Hermes}}]{2015ApJ...809..148T}
{Tremblay} PE, {Gianninas} A, {Kilic} M, {Ludwig} HG, {Steffen} M, {Freytag} B,
  {Hermes} JJ (2015) {3D Model Atmospheres for Extremely Low-mass White
  Dwarfs}. \apj 809:148, \doi{10.1088/0004-637X/809/2/148}, \eprint{1507.01927}

\bibitem[{{Tremblay} et~al.(2019){Tremblay}, {Fontaine}, {Fusillo}, {Dunlap},
  {G{\"a}nsicke}, {Hollands}, {Hermes}, {Marsh}, {Cukanovaite}, and
  {Cunningham}}]{2019Natur.565..202T}
{Tremblay} PE, {Fontaine} G, {Fusillo} NPG, {Dunlap} BH, {G{\"a}nsicke} BT,
  {Hollands} MA, {Hermes} JJ, {Marsh} TR, {Cukanovaite} E, {Cunningham} T
  (2019) {Core crystallization and pile-up in the cooling sequence of evolving
  white dwarfs}. \nat 565:202--205, \doi{10.1038/s41586-018-0791-x}

\bibitem[{{Udalski} et~al.(2015){Udalski}, {Szyma{\'n}ski}, and
  {Szyma{\'n}ski}}]{2015AcA....65....1U}
{Udalski} A, {Szyma{\'n}ski} MK, {Szyma{\'n}ski} G (2015) {OGLE-IV: Fourth
  Phase of the Optical Gravitational Lensing Experiment}. AcA 65:1--38,
  \eprint{1504.05966}

\bibitem[{{Unno} et~al.(1989){Unno}, {Osaki}, {Ando}, {Saio}, and
  {Shibahashi}}]{1989nos..book.....U}
{Unno} W, {Osaki} Y, {Ando} H, {Saio} H, {Shibahashi} H (1989) {Nonradial
  oscillations of stars}

\bibitem[{{Uzan}(2003)}]{2003RvMP...75..403U}
{Uzan} JP (2003) {The fundamental constants and their variation: observational
  and theoretical status}. Reviews of Modern Physics 75:403--455,
  \doi{10.1103/RevModPhys.75.403}, \eprint{hep-ph/0205340}

\bibitem[{{Van Grootel} et~al.(2012){Van Grootel}, {Dupret}, {Fontaine},
  {Brassard}, {Grigahc{\`e}ne}, and {Quirion}}]{2012A&A...539A..87V}
{Van Grootel} V, {Dupret} MA, {Fontaine} G, {Brassard} P, {Grigahc{\`e}ne} A,
  {Quirion} PO (2012) {The instability strip of ZZ Ceti white dwarfs. I.
  Introduction of time-dependent convection}. \aap 539:A87,
  \doi{10.1051/0004-6361/201118371}

\bibitem[{{Van Grootel} et~al.(2013){Van Grootel}, {Fontaine}, {Brassard}, and
  {Dupret}}]{2013ApJ...762...57V}
{Van Grootel} V, {Fontaine} G, {Brassard} P, {Dupret} MA (2013) {The Newly
  Discovered Pulsating Low-mass White Dwarfs: An Extension of the ZZ Ceti
  Instability Strip}. \apj 762:57, \doi{10.1088/0004-637X/762/1/57}

\bibitem[{{Van Grootel} et~al.(2015){Van Grootel}, {Fontaine}, {Brassard}, and
  {Dupret}}]{2015A&A...575A.125V}
{Van Grootel} V, {Fontaine} G, {Brassard} P, {Dupret} MA (2015) {A connection
  between the instability strips of ZZ Ceti and V777 Herculis white dwarfs.
  Pulsating accreting GW Lib white dwarfs}. \aap 575:A125,
  \doi{10.1051/0004-6361/201425386}

\bibitem[{{Van Grootel} et~al.(2017){Van Grootel}, {Fontaine}, {Brassard}, and
  {Dupret}}]{2017ASPC..509..321V}
{Van Grootel} V, {Fontaine} G, {Brassard} P, {Dupret} MA (2017) {The
  Theoretical Instability Strip of V777 Her White Dwarfs}. In: {Tremblay} PE,
  {Gaensicke} B, {Marsh} T (eds) 20th European White Dwarf Workshop,
  Astronomical Society of the Pacific Conference Series, vol 509, p 321

\bibitem[{{van Horn}(1968)}]{1968ApJ...151..227V}
{van Horn} HM (1968) {Crystallization of White Dwarfs}. \apj 151:227,
  \doi{10.1086/149432}

\bibitem[{{Vanderbosch} et~al.(2018){Vanderbosch}, {Winget}, and
  {Winget}}]{zach18}
{Vanderbosch} ZP, {Winget} KI, {Winget} DE (2018) {New DBVs}. In: 21th European
  White Dwarf Workshop

\bibitem[{{Vauclair}(2013)}]{2013EAS....63..175V}
{Vauclair} G (2013) {Constraints on white dwarfs structure and evolution from
  asteroseismology}. In: {Alecian} G, {Lebreton} Y, {Richard} O, {Vauclair} G
  (eds) EAS Publications Series, EAS Publications Series, vol~63, pp 175--183,
  \doi{10.1051/eas/1363020}

\bibitem[{{Vauclair} et~al.(2011){Vauclair}, {Fu}, {Solheim}, {Kim}, {Dolez},
  {Chevreton}, {Chen}, {Wood}, {Silver}, {Bogn{\'a}r}, {Papar{\'o}}, and
  {C{\'o}rsico}}]{2011A&A...528A...5V}
{Vauclair} G, {Fu} JN, {Solheim} JE, {Kim} SL, {Dolez} N, {Chevreton} M, {Chen}
  L, {Wood} MA, {Silver} IM, {Bogn{\'a}r} Z, {Papar{\'o}} M, {C{\'o}rsico} AH
  (2011) {The period and amplitude changes in the coolest GW Virginis variable
  star (PG 1159-type) PG 0122+200}. \aap 528:A5,
  \doi{10.1051/0004-6361/201014457}

\bibitem[{{Vennes} et~al.(2017){Vennes}, {Nemeth}, {Kawka}, {Thorstensen},
  {Khalack}, {Ferrario}, and {Alper}}]{2017Sci...357..680V}
{Vennes} S, {Nemeth} P, {Kawka} A, {Thorstensen} JR, {Khalack} V, {Ferrario} L,
  {Alper} EH (2017) {An unusual white dwarf star may be a surviving remnant of
  a subluminous Type Ia supernova}. Science 357:680--683,
  \doi{10.1126/science.aam8378}, \eprint{1708.05568}

\bibitem[{{Ventura} et~al.(2001){Ventura}, {D'Antona}, {Mazzitelli}, and
  {Gratton}}]{2001ApJ...550L..65V}
{Ventura} P, {D'Antona} F, {Mazzitelli} I, {Gratton} R (2001) {Predictions for
  Self-Pollution in Globular Cluster Stars}. \apjl 550:L65--L69,
  \doi{10.1086/319496}, \eprint{astro-ph/0103337}

\bibitem[{{Viaux} et~al.(2013{\natexlab{a}}){Viaux}, {Catelan}, {Stetson},
  {Raffelt}, {Redondo}, {Valcarce}, and {Weiss}}]{2013PhRvL.111w1301V}
{Viaux} N, {Catelan} M, {Stetson} PB, {Raffelt} GG, {Redondo} J, {Valcarce}
  AAR, {Weiss} A (2013{\natexlab{a}}) {Neutrino and Axion Bounds from the
  Globular Cluster M5 (NGC 5904)}. Physical Review Letters 111(23):231301,
  \doi{10.1103/PhysRevLett.111.231301}, \eprint{1311.1669}

\bibitem[{{Viaux} et~al.(2013{\natexlab{b}}){Viaux}, {Catelan}, {Stetson},
  {Raffelt}, {Redondo}, {Valcarce}, and {Weiss}}]{2013A&A...558A..12V}
{Viaux} N, {Catelan} M, {Stetson} PB, {Raffelt} GG, {Redondo} J, {Valcarce}
  AAR, {Weiss} A (2013{\natexlab{b}}) {Particle-physics constraints from the
  globular cluster M5: neutrino dipole moments}. \aap 558:A12,
  \doi{10.1051/0004-6361/201322004}, \eprint{1308.4627}

\bibitem[{{Voss} et~al.(2007){Voss}, {Koester}, {{\O}stensen}, {Napiwotzki},
  {Homeier}, and {Reimers}}]{2007ASPC..372..583V}
{Voss} B, {Koester} D, {{\O}stensen} R, {Napiwotzki} R, {Homeier} D, {Reimers}
  D (2007) {Six New ZZ Ceti Stars from the SPY and the HQS Surveys}. In:
  {Napiwotzki} R, {Burleigh} MR (eds) 15th European Workshop on White Dwarfs,
  Astronomical Society of the Pacific Conference Series, vol 372, p 583,
  \eprint{0704.2710}

\bibitem[{Vysotsky et~al.(1978)Vysotsky, Zeldovich, Khlopov, and
  Chechetkin}]{Vysotsky:1978dc}
Vysotsky MI, Zeldovich {\relax Ya}B, Khlopov M{\relax Yu}, Chechetkin VM (1978)
  {Some Astrophysical Limitations on Axion Mass}. Pisma Zh Eksp Teor Fiz
  27:533--536, [JETP Lett.27,502(1978)]

\bibitem[{{Wachlin} et~al.(2017){Wachlin}, {Vauclair}, {Vauclair}, and
  {Althaus}}]{2017A&A...601A..13W}
{Wachlin} FC, {Vauclair} G, {Vauclair} S, {Althaus} LG (2017) {Importance of
  fingering convection for accreting white dwarfs in the framework of full
  evolutionary calculations: the case of the hydrogen-rich white dwarfs GD 133
  and G 29-38}. \aap 601:A13, \doi{10.1051/0004-6361/201630094},
  \eprint{1612.09320}

\bibitem[{{Warner} and {van Zyl}(1998)}]{1998IAUS..185..321W}
{Warner} B, {van Zyl} L (1998) {Discovery of non-radial pulsations in the white
  dwarf primary of a cataclysmic variable star}. In: {Deubner} FL,
  {Christensen-Dalsgaard} J, {Kurtz} D (eds) New Eyes to See Inside the Sun and
  Stars, IAU Symposium, vol 185, p 321, \eprint{cond-mat/9701105}

\bibitem[{{Weinberg}(1978)}]{1978PhRvL..40..223W}
{Weinberg} S (1978) {A new light boson?} Physical Review Letters 40:223--226,
  \doi{10.1103/PhysRevLett.40.223}

\bibitem[{{Werner} and {Herwig}(2006)}]{2006PASP..118..183W}
{Werner} K, {Herwig} F (2006) {The Elemental Abundances in Bare Planetary
  Nebula Central Stars and the Shell Burning in AGB Stars}. PASP 118:183--204,
  \doi{10.1086/500443}, \eprint{astro-ph/0512320}

\bibitem[{{Wilczek}(1978)}]{1978PhRvL..40..279W}
{Wilczek} F (1978) {Problem of strong P and T invariance in the presence of
  instantons}. Physical Review Letters 40:279--282,
  \doi{10.1103/PhysRevLett.40.279}

\bibitem[{{Williams} et~al.(2013){Williams}, {Winget}, {Montgomery}, {Dufour},
  {Kepler}, {Hermes}, {Falcon}, {Winget}, {Bolte}, {Rubin}, and
  {Liebert}}]{2013ApJ...769..123W}
{Williams} KA, {Winget} DE, {Montgomery} MH, {Dufour} P, {Kepler} SO, {Hermes}
  JJ, {Falcon} RE, {Winget} KI, {Bolte} M, {Rubin} KHR, {Liebert} J (2013)
  {Photometric Variability in a Warm, Strongly Magnetic DQ White Dwarf, SDSS
  J103655.39+652252.2}. \apj 769:123, \doi{10.1088/0004-637X/769/2/123},
  \eprint{1304.3165}

\bibitem[{{Williams} et~al.(2016){Williams}, {Montgomery}, {Winget}, {Falcon},
  and {Bierwagen}}]{2016ApJ...817...27W}
{Williams} KA, {Montgomery} MH, {Winget} DE, {Falcon} RE, {Bierwagen} M (2016)
  {Variability in Hot Carbon-dominated Atmosphere (Hot DQ) White Dwarfs: Rapid
  Rotation?} \apj 817:27, \doi{10.3847/0004-637X/817/1/27}, \eprint{1511.08834}

\bibitem[{{Winget}(1982)}]{1982PhDT........27W}
{Winget} DE (1982) {Gravity Mode Instabilities in DA White Dwarfs.} PhD thesis,
  THE UNIVERSITY OF ROCHESTER.

\bibitem[{{Winget}(1988)}]{1988IAUS..123..305W}
{Winget} DE (1988) {Seismological Investigations of Compact Stars}. In:
  {Christensen-Dalsgaard} J, {Frandsen} S (eds) Advances in Helio- and
  Asteroseismology, IAU Symposium, vol 123, p 305

\bibitem[{{Winget} and {Kepler}(2008)}]{2008ARA&A..46..157W}
{Winget} DE, {Kepler} SO (2008) {Pulsating White Dwarf Stars and Precision
  Asteroseismology}. \araa 46:157--199,
  \doi{10.1146/annurev.astro.46.060407.145250}, \eprint{0806.2573}

\bibitem[{{Winget} et~al.(1982{\natexlab{a}}){Winget}, {Robinson}, {Nather},
  and {Fontaine}}]{1982ApJ...262L..11W}
{Winget} DE, {Robinson} EL, {Nather} RD, {Fontaine} G (1982{\natexlab{a}})
  {Photometric observations of GD 358 - DB white dwarfs do pulsate}. \apjl
  262:L11--L15, \doi{10.1086/183902}

\bibitem[{{Winget} et~al.(1982{\natexlab{b}}){Winget}, {van Horn}, {Tassoul},
  {Fontaine}, {Hansen}, and {Carroll}}]{1982ApJ...252L..65W}
{Winget} DE, {van Horn} HM, {Tassoul} M, {Fontaine} G, {Hansen} CJ, {Carroll}
  BW (1982{\natexlab{b}}) {Hydrogen-driving and the blue edge of
  compositionally stratified ZZ Ceti star models}. \apjl 252:L65--L68,
  \doi{10.1086/183721}

\bibitem[{{Winget} et~al.(1983){Winget}, {Hansen}, and {van
  Horn}}]{1983Natur.303..781W}
{Winget} DE, {Hansen} CJ, {van Horn} HM (1983) {Do pulsating PG1159-035 stars
  put constraints on stellar evolution?} \nat 303:781--782,
  \doi{10.1038/303781a0}

\bibitem[{{Winget} et~al.(1991){Winget}, {Nather}, {Clemens}, {Provencal},
  {Kleinman}, {Bradley}, {Wood}, {Claver}, {Frueh}, {Grauer}, {Hine}, {Hansen},
  {Fontaine}, {Achilleos}, {Wickramasinghe}, {Marar}, {Seetha}, {Ashoka},
  {O'Donoghue}, {Warner}, {Kurtz}, {Buckley}, {Brickhill}, {Vauclair}, {Dolez},
  {Chevreton}, {Barstow}, {Solheim}, {Kanaan}, {Kepler}, {Henry}, and
  {Kawaler}}]{1991ApJ...378..326W}
{Winget} DE, {Nather} RE, {Clemens} JC, {Provencal} J, {Kleinman} SJ, {Bradley}
  PA, {Wood} MA, {Claver} CF, {Frueh} ML, {Grauer} AD, {Hine} BP, {Hansen} CJ,
  {Fontaine} G, {Achilleos} N, {Wickramasinghe} DT, {Marar} TMK, {Seetha} S,
  {Ashoka} BN, {O'Donoghue} D, {Warner} B, {Kurtz} DW, {Buckley} DA,
  {Brickhill} J, {Vauclair} G, {Dolez} N, {Chevreton} M, {Barstow} MA,
  {Solheim} JE, {Kanaan} A, {Kepler} SO, {Henry} GW, {Kawaler} SD (1991)
  {Asteroseismology of the DOV star PG 1159 - 035 with the Whole Earth
  Telescope}. \apj 378:326--346, \doi{10.1086/170434}

\bibitem[{{Winget} et~al.(1994){Winget}, {Nather}, {Clemens}, {Provencal},
  {Kleinman}, {Bradley}, {Claver}, {Dixson}, {Montgomery}, {Hansen}, {Hine},
  {Birch}, {Candy}, {Marar}, {Seetha}, {Ashoka}, {Leibowitz}, {O'Donoghue},
  {Warner}, {Buckley}, {Tripe}, {Vauclair}, {Dolez}, {Chevreton}, {Serre},
  {Garrido}, {Kepler}, {Kanaan}, {Augusteijn}, {Wood}, {Bergeron}, and
  {Grauer}}]{1994ApJ...430..839W}
{Winget} DE, {Nather} RE, {Clemens} JC, {Provencal} JL, {Kleinman} SJ,
  {Bradley} PA, {Claver} CF, {Dixson} JS, {Montgomery} MH, {Hansen} CJ, {Hine}
  BP, {Birch} P, {Candy} M, {Marar} TMK, {Seetha} S, {Ashoka} BN, {Leibowitz}
  EM, {O'Donoghue} D, {Warner} B, {Buckley} DAH, {Tripe} P, {Vauclair} G,
  {Dolez} N, {Chevreton} M, {Serre} T, {Garrido} R, {Kepler} SO, {Kanaan} A,
  {Augusteijn} T, {Wood} MA, {Bergeron} P, {Grauer} AD (1994) {Whole earth
  telescope observations of the DBV white dwarf GD 358}. \apj 430:839--849,
  \doi{10.1086/174455}

\bibitem[{{Winget} et~al.(2004){Winget}, {Sullivan}, {Metcalfe}, {Kawaler}, and
  {Montgomery}}]{2004ApJ...602L.109W}
{Winget} DE, {Sullivan} DJ, {Metcalfe} TS, {Kawaler} SD, {Montgomery} MH (2004)
  {A Strong Test of Electroweak Theory Using Pulsating DB White Dwarf Stars as
  Plasmon Neutrino Detectors}. \apjl 602:L109--L112, \doi{10.1086/382591},
  \eprint{astro-ph/0312303}

\bibitem[{{Winget} et~al.(2009){Winget}, {Kepler}, {Campos}, {Montgomery},
  {Girardi}, {Bergeron}, and {Williams}}]{2009ApJ...693L...6W}
{Winget} DE, {Kepler} SO, {Campos} F, {Montgomery} MH, {Girardi} L, {Bergeron}
  P, {Williams} K (2009) {The Physics of Crystallization From Globular Cluster
  White Dwarf Stars in NGC 6397}. \apjl 693:L6--L10,
  \doi{10.1088/0004-637X/693/1/L6}, \eprint{0901.2950}

\bibitem[{{Woosley} and {Heger}(2015)}]{2015ApJ...810...34W}
{Woosley} SE, {Heger} A (2015) {The Remarkable Deaths of 9-11 Solar Mass
  Stars}. \apj 810:34, \doi{10.1088/0004-637X/810/1/34}, \eprint{1505.06712}

\bibitem[{{Woudt} et~al.(2012){Woudt}, {Warner}, and
  {Zietsman}}]{2012MNRAS.426.2137W}
{Woudt} PA, {Warner} B, {Zietsman} E (2012) {SDSS J0349-0059 is a GW Virginis
  star}. \mnras 426:2137--2141, \doi{10.1111/j.1365-2966.2012.21899.x},
  \eprint{1208.1844}

\bibitem[{{Wu} and {Goldreich}(2001)}]{2001ApJ...546..469W}
{Wu} Y, {Goldreich} P (2001) {Gravity Modes in ZZ Ceti Stars. IV. Amplitude
  Saturation by Parametric Instability}. \apj 546:469--483,
  \doi{10.1086/318234}, \eprint{astro-ph/0003163}

\bibitem[{{York} et~al.(2000){York}, {Adelman}, {Anderson}, {Anderson},
  {Annis}, {Bahcall}, {Bakken}, {Barkhouser}, {Bastian}, {Berman}, {Boroski},
  {Bracker}, {Briegel}, {Briggs}, {Brinkmann}, {Brunner}, {Burles}, {Carey},
  {Carr}, {Castander}, {Chen}, {Colestock}, {Connolly}, {Crocker}, {Csabai},
  {Czarapata}, {Davis}, {Doi}, {Dombeck}, {Eisenstein}, {Ellman}, {Elms},
  {Evans}, {Fan}, {Federwitz}, {Fiscelli}, {Friedman}, {Frieman}, {Fukugita},
  {Gillespie}, {Gunn}, {Gurbani}, {de Haas}, {Haldeman}, {Harris}, {Hayes},
  {Heckman}, {Hennessy}, {Hindsley}, {Holm}, {Holmgren}, {Huang}, {Hull},
  {Husby}, {Ichikawa}, {Ichikawa}, {Ivezi{\'c}}, {Kent}, {Kim}, {Kinney},
  {Klaene}, {Kleinman}, {Kleinman}, {Knapp}, {Korienek}, {Kron}, {Kunszt},
  {Lamb}, {Lee}, {Leger}, {Limmongkol}, {Lindenmeyer}, {Long}, {Loomis},
  {Loveday}, {Lucinio}, {Lupton}, {MacKinnon}, {Mannery}, {Mantsch}, {Margon},
  {McGehee}, {McKay}, {Meiksin}, {Merelli}, {Monet}, {Munn}, {Narayanan},
  {Nash}, {Neilsen}, {Neswold}, {Newberg}, {Nichol}, {Nicinski}, {Nonino},
  {Okada}, {Okamura}, {Ostriker}, {Owen}, {Pauls}, {Peoples}, {Peterson},
  {Petravick}, {Pier}, {Pope}, {Pordes}, {Prosapio}, {Rechenmacher}, {Quinn},
  {Richards}, {Richmond}, {Rivetta}, {Rockosi}, {Ruthmansdorfer}, {Sandford},
  {Schlegel}, {Schneider}, {Sekiguchi}, {Sergey}, {Shimasaku}, {Siegmund},
  {Smee}, {Smith}, {Snedden}, {Stone}, {Stoughton}, {Strauss}, {Stubbs},
  {SubbaRao}, {Szalay}, {Szapudi}, {Szokoly}, {Thakar}, {Tremonti}, {Tucker},
  {Uomoto}, {Vanden Berk}, {Vogeley}, {Waddell}, {Wang}, {Watanabe},
  {Weinberg}, {Yanny}, {Yasuda}, and {SDSS
  Collaboration}}]{2000AJ....120.1579Y}
{York} DG, {Adelman} J, {Anderson} JE Jr, {Anderson} SF, {Annis} J, {Bahcall}
  NA, {Bakken} JA, {Barkhouser} R, {Bastian} S, {Berman} E, {Boroski} WN,
  {Bracker} S, {Briegel} C, {Briggs} JW, {Brinkmann} J, {Brunner} R, {Burles}
  S, {Carey} L, {Carr} MA, {Castander} FJ, {Chen} B, {Colestock} PL, {Connolly}
  AJ, {Crocker} JH, {Csabai} I, {Czarapata} PC, {Davis} JE, {Doi} M, {Dombeck}
  T, {Eisenstein} D, {Ellman} N, {Elms} BR, {Evans} ML, {Fan} X, {Federwitz}
  GR, {Fiscelli} L, {Friedman} S, {Frieman} JA, {Fukugita} M, {Gillespie} B,
  {Gunn} JE, {Gurbani} VK, {de Haas} E, {Haldeman} M, {Harris} FH, {Hayes} J,
  {Heckman} TM, {Hennessy} GS, {Hindsley} RB, {Holm} S, {Holmgren} DJ, {Huang}
  Ch, {Hull} C, {Husby} D, {Ichikawa} SI, {Ichikawa} T, {Ivezi{\'c}} {\v Z},
  {Kent} S, {Kim} RSJ, {Kinney} E, {Klaene} M, {Kleinman} AN, {Kleinman} S,
  {Knapp} GR, {Korienek} J, {Kron} RG, {Kunszt} PZ, {Lamb} DQ, {Lee} B, {Leger}
  RF, {Limmongkol} S, {Lindenmeyer} C, {Long} DC, {Loomis} C, {Loveday} J,
  {Lucinio} R, {Lupton} RH, {MacKinnon} B, {Mannery} EJ, {Mantsch} PM, {Margon}
  B, {McGehee} P, {McKay} TA, {Meiksin} A, {Merelli} A, {Monet} DG, {Munn} JA,
  {Narayanan} VK, {Nash} T, {Neilsen} E, {Neswold} R, {Newberg} HJ, {Nichol}
  RC, {Nicinski} T, {Nonino} M, {Okada} N, {Okamura} S, {Ostriker} JP, {Owen}
  R, {Pauls} AG, {Peoples} J, {Peterson} RL, {Petravick} D, {Pier} JR, {Pope}
  A, {Pordes} R, {Prosapio} A, {Rechenmacher} R, {Quinn} TR, {Richards} GT,
  {Richmond} MW, {Rivetta} CH, {Rockosi} CM, {Ruthmansdorfer} K, {Sandford} D,
  {Schlegel} DJ, {Schneider} DP, {Sekiguchi} M, {Sergey} G, {Shimasaku} K,
  {Siegmund} WA, {Smee} S, {Smith} JA, {Snedden} S, {Stone} R, {Stoughton} C,
  {Strauss} MA, {Stubbs} C, {SubbaRao} M, {Szalay} AS, {Szapudi} I, {Szokoly}
  GP, {Thakar} AR, {Tremonti} C, {Tucker} DL, {Uomoto} A, {Vanden Berk} D,
  {Vogeley} MS, {Waddell} P, {Wang} Si, {Watanabe} M, {Weinberg} DH, {Yanny} B,
  {Yasuda} N, {SDSS Collaboration} (2000) {The Sloan Digital Sky Survey:
  Technical Summary}. \aj 120:1579--1587, \doi{10.1086/301513},
  \eprint{astro-ph/0006396}

\bibitem[{{Zhang} et~al.(2016){Zhang}, {Fu}, {Li}, {Ren}, and
  {Luo}}]{2016ApJ...821L..32Z}
{Zhang} XB, {Fu} JN, {Li} Y, {Ren} AB, {Luo} CQ (2016) {Multi-period g-mode
  Pulsations of a Pre-He-WD Star in the Eclipsing Binary KIC 9164561}. ApJl
  821:L32, \doi{10.3847/2041-8205/821/2/L32}

\bibitem[{Zhitnitsky(1980)}]{Zhitnitsky:1980tq}
Zhitnitsky AR (1980) {On Possible Suppression of the Axion Hadron Interactions.
  (In Russian)}. Sov J Nucl Phys 31:260, [Yad. Fiz.31,497(1980)]

\bibitem[{{Zong} et~al.(2016{\natexlab{a}}){Zong}, {Charpinet}, and
  {Vauclair}}]{2016A&A...594A..46Z}
{Zong} W, {Charpinet} S, {Vauclair} G (2016{\natexlab{a}}) {Signatures of
  nonlinear mode interactions in the pulsating hot B subdwarf star KIC
  10139564}. \aap 594:A46, \doi{10.1051/0004-6361/201629132},
  \eprint{1607.06621}

\bibitem[{{Zong} et~al.(2016{\natexlab{b}}){Zong}, {Charpinet}, {Vauclair},
  {Giammichele}, and {Van Grootel}}]{2016A&A...585A..22Z}
{Zong} W, {Charpinet} S, {Vauclair} G, {Giammichele} N, {Van Grootel} V
  (2016{\natexlab{b}}) {Amplitude and frequency variations of oscillation modes
  in the pulsating DB white dwarf star KIC 08626021. The likely signature of
  nonlinear resonant mode coupling}. \aap 585:A22,
  \doi{10.1051/0004-6361/201526300}, \eprint{1510.06884}

\end{thebibliography}

%\begin{thebibliography}{}
% etc
%\end{thebibliography}
%\bibitem[\protect\citeauthoryear{Kepler \& Romero}{2017}]{KeplerRomero17} Kepler S.~O., Romero A.~D., 2017, EPJWC, 152, 01011 

\end{document}